%% file: alusiani-tau21-procs.tex
\pdfoutput=1
\documentclass[submission, Proceedings, x11names, svgnames]{SciPost}

\newcommand{\nobibtexflag}{}

\hypersetup{colorlinks,
linkcolor={red!50!black},
citecolor={blue!50!black},
urlcolor={blue!80!black}}

\usepackage[bitstream-charter]{mathdesign}
\usepackage[T1]{fontenc}
\usepackage[utf8]{inputenc}

\usepackage{calc}
\usepackage{overpic}
\usepackage{relsize}
\usepackage{aluautobfmath}
\usepackage{ifthen}
\usepackage{booktabs}
\usepackage{tikz}
\usepackage{xspace}

\newcommand{\etal}{\emph{et al.}\xspace}

\newenvironment{ensuredisplaymath}%
{\(\displaystyle}
{\)}

\newcommand{\gev}{\ensuremath{\,\text{Ge\kern -0.1em V}}\xspace}
\newcommand{\gevc}{\ensuremath{\,\text{Ge\kern -0.1em V}\!/c}\xspace}
\newcommand{\gevcc}{\ensuremath{\,\text{Ge\kern -0.1em V}\!/c^2}\xspace}
\newcommand{\mev}{\ensuremath{\,\text{Me\kern -0.1em V}}\xspace}
\newcommand{\mevc}{\ensuremath{\,\text{Me\kern -0.1em V}\!/c}\xspace}
\newcommand{\mevcc}{\ensuremath{\,\text{Me\kern -0.1em V}\!/c^2}\xspace}

\newcommand{\babar}{\mbox{%
    \slshape B\kern-0.1em{\smaller A}\kern-0.1em
    B\kern-0.1em{\smaller A\kern-0.2em R}}\xspace}

\newcommand{\nut}       {\ensuremath{\nu_\tau}\xspace}

\newcommand{\Bbar}   {\kern 0.18em\overline{\kern -0.18em B}{}\xspace}

\newcommand{\Bu}     {\ensuremath{B^+}\xspace}
\newcommand{\Bub}    {\ensuremath{B^-}\xspace}

\newcommand{\BpBm}   {\ensuremath{\Bu {\kern -0.16em \Bub}}\xspace}
\newcommand{\Bz}     {\ensuremath{B^0}\xspace}
\newcommand{\Bzb}    {\ensuremath{\Bbar^0}\xspace}
\newcommand{\BzBzb}  {\ensuremath{\Bz {\kern -0.16em \Bzb}}\xspace}

\newcommand{\BR}{{\cal B}\xspace}

\def\Y#1S{\ensuremath{\Upsilon{(#1S)}}\xspace}

\newcommand{\Vud}{\ensuremath{\left|V_{ud}\right|}\xspace}
\newcommand{\Vus}{\ensuremath{\left|V_{us}\right|}\xspace}
\newcommand{\Vub}{\ensuremath{\left|V_{ub}\right|}\xspace}

\makeatletter
\def\ht@base{htbase@def@}
\newcommand{\htset}[1]{%
  \def\ht@base{htbase@#1@}%
}

\newcommand{\htdef}[2]{%
  \@namedef{\ht@base#1}{#2}%
}

\newcommand{\htuse}[1]{%
  \ifcsname \ht@base#1\endcsname
  \@nameuse{\ht@base#1}%
  \else
  \@latex@error{Undefined name \ht@base#1}\@eha
  \fi
}

\newcommand{\htuseb}[2]{%
  \ifcsname htbase@#1@#2\endcsname
  \@nameuse{htbase@#1@#2}%
  \else
  \@latex@error{Undefined name htbase@#1@#2}\@eha
  \fi
}

\newcommand{\htquantdef}[6]{%
  \ifx&#2&\else
  \@namedef{\ht@base#1.gn}{\ensuremath{#2}}%
  \fi
  \ifx&#3&\else
  \@namedef{\ht@base#1.td}{\ensuremath{#3}}%
  \fi
  \ifx&#6&%
    \@namedef{\ht@base#1}{\ensuremath{#5}}%
  \else
    \ifthenelse{\equal{#6}{0}}{%
      \@namedef{\ht@base#1}{\ensuremath{#5}}%
    }{%
      \@namedef{\ht@base#1}{\ensuremath{#4}}%
      \@namedef{\ht@base#1.v}{\ensuremath{#5}}%
      \@namedef{\ht@base#1.e}{\ensuremath{#6}}%
    }%
  \fi
}

\newcommand{\htmeasdef}[8]{%
  \@namedef{\ht@base#1,quant}{\ensuremath{#2}}%
  \@namedef{\ht@base#1,exp}{#3}%
  \@namedef{\ht@base#1,ref}{\cite{#4}}%
  \@namedef{\ht@base#1}{\ensuremath{#5}}%
  \@namedef{\ht@base#1,val}{\ensuremath{#6}}%
  \@namedef{\ht@base#1,stat}{\ensuremath{#7}}%
  \@namedef{\ht@base#1,syst}{\ensuremath{#8}}%
}

\newcommand{\htconstrdef}[4]{%
  \@namedef{\ht@base#1.left}{\ensuremath{#2}}%
  \@namedef{\ht@base#1.right}{\ensuremath{#3}}%
  \@namedef{\ht@base#1.right.split}{\ensuremath{#4}}%
  \@namedef{\ht@base#1.constr.eq}{\htuse{#1.left} ={}& \htuse{#1.right}}%
}

\newcommand{\htQuantLine}[3]{\ensuremath{\htuse{#1.td}}&\ensuremath{#2}\\}
\makeatother

\newif\ifhevea\heveafalse

\colorlet{xbarcolor}{RoyalBlue}
\newlength\xbarBaseWidth
\newlength\xbarHeight
\colorlet{magentaEm}{DarkMagenta}

\input{tau-defs}

\htset{tau21}%
\input{hflav-2018-upd2021.tex}
\input{hflav-2018-upd2021-elab-vadirect.tex}

\begin{document}

\begin{center}\Large\bfseries
  Updated determinations of \Vus with tau decays using new recent estimates of radiative corrections for light-meson leptonic decay rates
\end{center}

\begin{center}
  A.\ Lusiani\textsuperscript{1*},
\end{center}

\begin{center}
  {\bf 1} Scuola Normale Superiore and INFN sezione di Pisa, Italy
  \\
  * alberto.lusiani@pi.infn.it
\end{center}

\begin{center}
  December 17, 2024
\end{center}


\definecolor{palegray}{gray}{0.95}
\begin{center}
  \colorbox{palegray}{\begin{minipage}{0.95\textwidth}
      \begin{center}
        {\it 16th International Workshop on Tau Lepton Physics (TAU2021),}\\
        {\it September 27 – October 1, 2021} \\
        \doi{10.21468/SciPostPhysProc.?}\\
      \end{center}
    \end{minipage}}
\end{center}

\section*{Abstract}
{\bfseries
We update the \Vus determinations using the HFLAV 2018 report tau branching fraction results with recent new estimates of the $\pi\ell2$ and $K\ell2$ radiative corrections. There are minor changes of the central values and uncertainties.
}

\vspace{10pt}
\noindent\rule{\textwidth}{1pt}
\tableofcontents\thispagestyle{fancy}
\noindent\rule{\textwidth}{1pt}
\vspace{10pt}

\section{Introduction}
\label{sec:intro}

Recent measurements of \Vud, \Vus and \Vub are not consistent with the unitarity condition on the first row of the CKM matrix~\cite{Hardy:2020qwl, Seng:2021nar}. Tau decay measurements are used to determine \Vus~\cite{Amhis:2019ckw, Lusiani:2021xuv}, supplementing the more precise determinations that are obtained using kaon decays. All these estimates also rely on lattice QCD estimates of form factors and decay constants\cite{Aoki:2019cca, FLAG-2019-web}, with the exception of the \Vus determinations based on the total branching fraction of the tau lepton into strange final states. Two of the \Vus determinations using tau measurements rely on estimates of the radiative corrections for the branching fractions $\BR(\tau \to \pi / K \nu)$, which are computed using also the radiative corrections for $\BR(\pi \to \ell \nu)$ and $\BR(K \to \ell \nu)$~\cite{Cirigliano:2011tm}. New estimates of these radiative corrections have been computed with a novel approach using lattice QCD+QED~\cite{DiCarlo:2019thl}. We evaluate in the following the impact of these new estimates on the \Vus determinations using tau decay measurements.

\section{\texorpdfstring{%
  \Vus from $\BR(\tau \to K\nu) / \BR(\tau \to \pi\nu)$ and $\BR(\tau \to K\nu)$}{%
  Vus from B(tau -> K nu) / B(tau -> pi nu and Vus from B(tau -> K nu)}}
\label{sec:tau:vus:taukpi}

We use the tau branching fractions of the HFLAV 2018 report fit~\cite{Amhis:2019ckw} and we compute \Vus using the updated external inputs provided by the Review of Particle Physics~\cite{zyla:2020zbs}, by the FLAG review of lattice QCD calculations~\cite{Aoki:2019cca, FLAG-2019-web}, by CODATA 2018~\cite{Tiesinga:2021myr}. In updating the CODATA constants from the values used in the HFLAV 2018 report, a numerical trascription error that slightly affected this \Vus determination in the HFLAV 2018 report has been corrected. Finally, we obtain \Vus using both the original~\cite{Cirigliano:2011tm} and the recently published~\cite{DiCarlo:2019thl} radiative corrections.

\sloppypar
We compute \Vus as in the HFLAV 2018 report from the ratio of branching fractions $\BR(\tau^- \to \htuse{Gamma10.td}) / \BR(\tau^- \to \htuse{Gamma9.td})$ and from the branching fraction $\BFtautoknu$ using the equations
\begin{align*}
  \frac{\BR(\tau^- \to \htuse{Gamma10.td})}{\BR(\tau^- \to \htuse{Gamma9.td})} & =
  \frac{f_{K\pm}^2 \Vus^2}{f_{\pi\pm}^2 \Vud^2} \frac{\left( m_\tau^2 - m_K^2 \right)^2}{\left( m_\tau^2 -  m_\pi^2 \right)^2}
  \frac{1+\delta R_{\tau/K}}{1+\delta R_{\tau/\pi}}(1+\delta R_{K\mu2/\pi\mu2})~,
  \\
  \BR(\tau^- \to K^-\nu_\tau)                                                  & =
  \frac{1}{16\pi}
  \left(\frac{G_F}{(\hslash c)^3}\right)^2
  f^2_{K\pm} \Vus^2
  \frac{\tau_{\tau}}{\hslash}
  (m_{\tau} c^2)^3 \left(1 - \frac{m_K^2}{m_\tau^2}\right)^2
  (1+\delta R_{\tau/K}) (1+\delta R_{K\mu2})~,
  %
  %
\end{align*}
respectively.
$\Vud = \htuse{Vud}$ is taken from a 2020 updated determination~\cite{Hardy:2020qwl}.

We use Refs.~\cite{Marciano:1993sh, Decker:1994dd, Decker:1994ea,
Decker:1994kw} to get $\delta R_{\tau/K} = \htuse{dRrad_tauK_by_Kmu}$,
$\delta R_{\tau/\pi} = \htuse{dRrad_taupi_by_pimu}$. The radiation
correction terms $\delta R_{K\mu2/\pi\mu2} =
\htuse{dRrad_kmunu_by_pimunu}$ and $\delta
R_{K\mu2}=\htuse{dRrad_K_munu}$ are provided without the
isospin-breaking corrections by
Refs.~\cite{Pich:2013lsa, Cirigliano:2011tm, Marciano:2004uf,
Tanabashi:2018oca, Rosner:2015wva}. The same sources report also $\delta
R_{\pi\mu2}=\htuse{dRrad_pi_munu}$. The three estimates are consistent
with a correlation of \htuse{corr_dRrad_pi_munu_K_munu} between $\delta
R_{\pi\mu2}$ and $\delta R_{K\mu2}$, which is used when computing \Vus.
These radiative correction terms are used with the lattice QCD decay
constants that include isospin-breaking corrections from the FLAG 2019
lattice QCD averages with $N_f=2+1+1$: $f_{K\pm}/f_{\pi\pm} =
\htuse{f_Kpm_by_f_pipm}$~\cite{Aoki:2019cca, Miller:2020xhy,
Dowdall:2013rya, Carrasco:2014poa, Bazavov:2017lyh} and $f_{K\pm} =
\htuse{f_Kpm}\,\mev$~\cite{Aoki:2019cca, Dowdall:2013rya,
Bazavov:2014wgs, Carrasco:2014poa}.

New estimates of the radiation correction terms inclusive of
isospin-breaking corrections $\delta R^\prime_{K\mu2/\pi\mu2} =
\htuse{dRrad_kmunu_by_pimunu_dicarlo2019}$ and $\delta
R^\prime_{K\mu2}=\htuse{dRrad_K_munu_dicarlo2019}$ are provided by
Ref.~\cite{DiCarlo:2019thl}. The same source reports also $\delta
R^\prime_{\pi\mu2}=\htuse{dRrad_pi_munu_dicarlo2019}$. The three
estimates are consistent with a correlation of
\htuse{corr_dRrad_pi_munu_K_munu_dicarlo2019} between $\delta
R^\prime_{\pi\mu2}$ and $\delta R^\prime_{K\mu2}$, which originates from
a correlation of 0.794 between $\delta R^\prime_{\pi\mu2}$ and the
non-isospin-breaking component of $\delta
R^\prime_{K\mu2}$~\cite{Simula:email-jul2020}. These radiative
correction terms are used with the isospin-symmetric lattice QCD decay
constants with $N_f=2+1+1$ provided by Ref.~\cite{DiCarlo:2019thl}:
$f_{K}/f_{\pi} = \htuse{f_K_by_f_pi_dicarlo2019}$ and $f_{K} =
\htuse{f_K_dicarlo2019}\,\mev$.

The above radiative correction factors
\begin{align*}
  &\frac{1+\delta R_{\tau/K}}{1+\delta R_{\tau/\pi}}
  (1+\delta R_{K\mu2/\pi\mu2})~,
  &
  (1+\delta R_{\tau/K}) (1+\delta R_{K\mu2})
\end{align*}
have recently been evaluated~\cite{Arroyo-Urena:2021nil} as
\begin{alignat*}{6}
  &(1 + \delta_{\tau K/\tau\pi})~,\quad &&\text{ with }\delta_{\tau
  K/\tau\pi} = (0.10 \pm 0.80)\%~, \\
  &(1 + \delta_{\tau K})~,\quad &&\text{ with }\delta_{\tau K} = (-0.15
  \pm 0.57)\%~,
\end{alignat*}
respectively, with larger and more conservative uncertainties. We do not
use these last values in the following.

\begin{figure}[tb]
  \begin{center}
    \fbox{\begin{overpic}[trim=0 10 0 0, width=0.7\linewidth-2\fboxsep-2\fboxrule,clip]{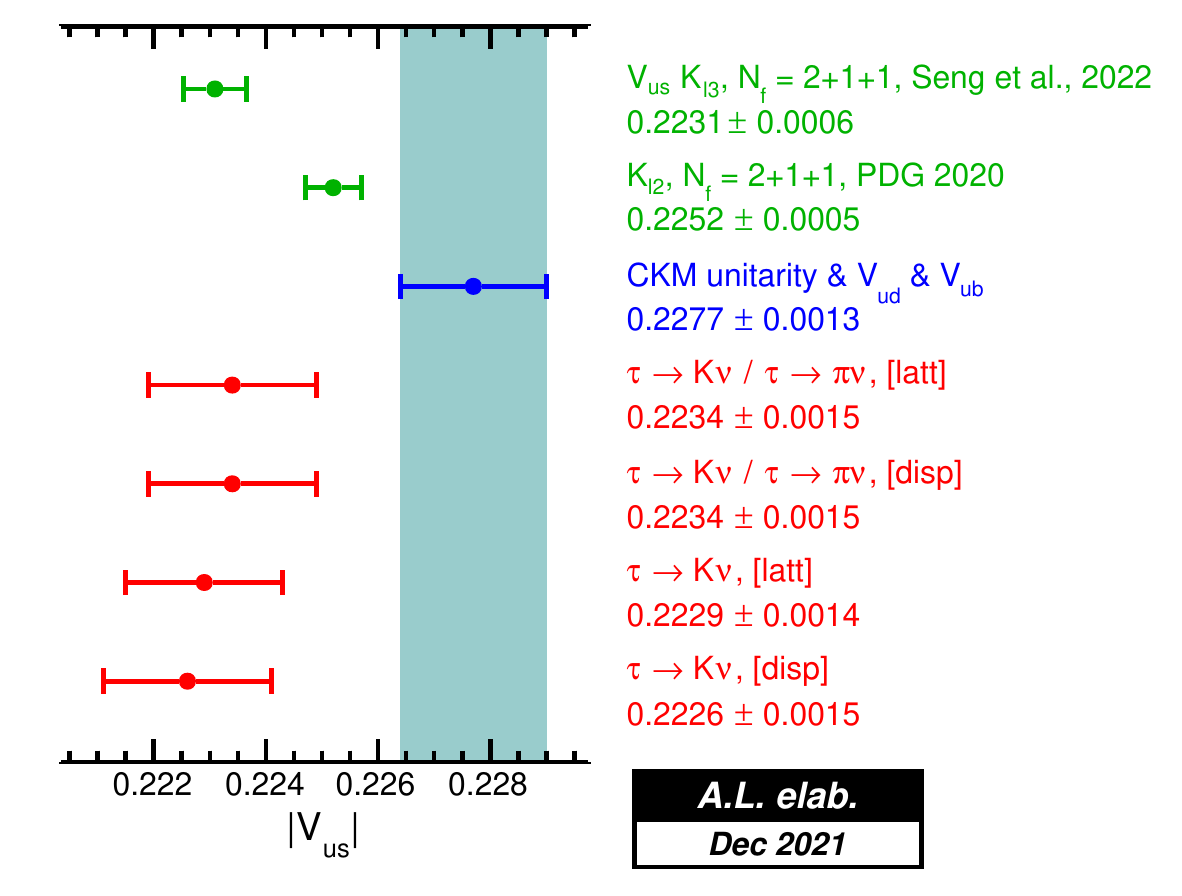}
      \end{overpic}}
    \caption{%
    \Vus determinations using tau decay measurements with the old
    $\pi\ell2$ and $K\ell2$ radiative
    corrections ([disp]) and with the new ones ([latt]), compared with \Vus from the
    CKM matrix unitarity and \Vus from kaon
    measurements.\label{fig:vus-summary}%
    }
  \end{center}
\end{figure}

Table~\ref{tab:Vus} reports the values of \VusTauKpi from $\BR(\tau \to
K\nu) / \BR(\tau \to \pi\nu)$ and \VusTauKnu from $\BR(\tau \to K\nu)$
using both the original and the new $\pi\ell2$ and $K\ell2$ radiative
corrections. The improvements in the precision of the radiative
corrections results in minor improvements on the \Vus determinations
because of other larger contributing uncertainties. The changes of the
central values are also minor compared with the total uncertainties.
Figure~\ref{fig:vus-summary} reports the \Vus determinations in this
document compared with \Vus from kaon decay measurements
$K\ell3$~\cite{Seng:2021nar} and $K\ell3$~\cite{zyla:2020zbs}, and \Vus
implied by the CKM matrix unitarity given the measured values of
\Vud~\cite{Hardy:2020qwl} and \Vub~\cite{zyla:2020zbs}.

\begin{table}[tb]
  \caption{%
  \Vus determinations. \VusUni denotes the value of \Vus assuming that
  the CKM matrix is unitary.
  The determinations of
  \VusTauKpi and \VusTauKnu using the exclusive tau branching fractions
  are reported using both the old and the new estimates of the
  $\pi\ell2$ and $K\ell2$ radiative corrections.
  The signed
  difference with respect to \VusUni in standard deviations is reported
  for all \Vus determinations.}
  \label{tab:Vus}
  \begin{center}
    \begin{tabular}{lrrrr}
      \toprule
                                     & \multicolumn{4}{c}{\Vus}                                                                                        \\
      \midrule
      \VusUni                        & \multicolumn{4}{c}{\htuse{Vus_uni}\quad$\hphantom{-}0.0\,\sigma$}                                               \\
      \midrule
                                     & \multicolumn{2}{l}{Cirigliano \& Neufeld 2011}                        & \multicolumn{2}{l}{Di Carlo~\etal 2019} \\
      \midrule
      \VusTauKpi                     &
      \htuse{Vus_tauKpi}             & $\htuse{Vus_tauKpi_mism_sigma}\,\sigma$                               &
      \htuse{Vus_tauKpi_dicarlo2019} & $\htuse{Vus_tauKpi_mism_sigma_dicarlo2019}\,\sigma$                                                             \\
      \VusTauKnu                     &
      \htuse{Vus_tauKnu}             & $\htuse{Vus_tauKnu_mism_sigma}\,\sigma$                               &
      \htuse{Vus_tauKnu_dicarlo2019} & $\htuse{Vus_tauKnu_mism_sigma_dicarlo2019}\,\sigma$                                                             \\
      \bottomrule
    \end{tabular}
  \end{center}
\end{table}

\section{Conclusions}

Recent new slightly more precise estimates of the $\pi\ell2$ and $K\ell2$ radiative corrections have been used to update the \Vus determination using the tau branching fractions of the HFLAV 2018 report, resulting in minor changes of the central values and uncertainties.




\ifdefined\nobibtexflag

\input{alusiani-tau21-procs.bbl}
\else
  \iffalse
    \bibliography{%
      alusiani-tau21-procs%
      ,bibtex/pub-2021%
      ,bibtex/pub-2020%
      ,bibtex/pub-2018%
      ,bibtex/pub-2017%
      ,bibtex/pub-2016%
      ,bibtex/oth-tau-lepton%
      ,lattice%
      ,tau-refs%
      ,tau-refs-pdg%
      ,bibtex/pub-norefereed.bib%
    }
  \else
    \bibliography{%
      alusiani-tau21-procs-bib%
    }
  \fi
\fi

\end{document}

%% file: tau-defs.tex
\newcommand{\BRF}[2]{#2}
\renewcommand{\BR}{\ensuremath{B}\xspace}
\ifhevea
\renewcommand{\babar}{BaBar\xspace}
\else
\providecommand{\babar}{\mbox{\slshape B\kern-0.1em{\smaller A}\kern-0.1em
    B\kern-0.1em{\smaller A\kern-0.2em R}}\xspace}
\fi

\newcommand{\tauknu}{\ensuremath{\tau^{-} \to K^{-} \nut}\xspace}

\newcommand{\BFtautoknu}{\ensuremath{\BR(\tauknu)}\xspace}

\newcommand{\VusUni}{\ensuremath{\Vus_{\text{uni}}}\xspace}

\newcommand{\VusTauKpi}{\ensuremath{\Vus_{\tau K/\pi}}\xspace}
\newcommand{\VusTauKnu}{\ensuremath{\Vus_{\tau K}}\xspace}

\providecommand{\slashLikeH}{%
  \raisebox{.9ex}{%
    \scalebox{.7}{%
      \rotatebox[origin=c]{18}{$-$}%
    }%
  }%
}
\providecommand{\hslash}{%
  {%
   \vphantom{h}%
   \ooalign{\kern.05em\smash{\slashLikeH}\hidewidth\cr$h$\cr}%
   \kern.05em
  }%
}

%% file: hflav-2018-upd2021.tex
\htdef{UnitarityResid}{(0.03 \pm 0.10)\%}%
\htdef{MeasNum}{176}%
\htdef{QuantNum}{136}%
\htdef{QuantNumNonRatio}{119}%
\htdef{QuantNumRatio}{17}%
\htdef{QuantNumWithMeas}{86}%
\htdef{QuantNumNonRatioWithMeas}{73}%
\htdef{QuantNumRatioWithMeas}{13}%
\htdef{QuantNumPdg}{130}%
\htdef{QuantNumNonRatioPdg}{113}%
\htdef{QuantNumRatioPdg}{17}%
\htdef{QuantNumWithMeasPdg}{84}%
\htdef{QuantNumNonRatioWithMeasPdg}{71}%
\htdef{QuantNumRatioWithMeasPdg}{13}%
\htdef{IndepQuantNum}{47}%
\htdef{BaseQuantNum}{47}%
\htdef{UnitarityQuantNum}{48}%
\htdef{ConstrNum}{89}%
\htdef{ConstrNumPdg}{83}%
\htdef{Chisq}{142}%
\htdef{Dof}{129}%
\htdef{ChisqProb}{20.13\%}%
\htdef{ChisqProbRound}{20\%}%
\htmeasdef{ALEPH.Gamma10.pub.BARATE.99K}{Gamma10}{ALEPH}{Barate:1999hi}{0.00696 \pm 0.00025 \pm 0.00014}{0.00696}{\pm 0.00025}{0.00014}%
\htmeasdef{ALEPH.Gamma103.pub.SCHAEL.05C}{Gamma103}{ALEPH}{Schael:2005am}{0.00072 \pm 0.00009 \pm 0.00012}{0.00072}{\pm 0.00009}{0.00012}%
\htmeasdef{ALEPH.Gamma104.pub.SCHAEL.05C}{Gamma104}{ALEPH}{Schael:2005am}{( 0.021 \pm 0.007 \pm 0.009 ) \cdot 10^{ -2 }}{0.021e-2}{\pm 0.007e-2}{0.009e-2}%
\htmeasdef{ALEPH.Gamma126.pub.BUSKULIC.97C}{Gamma126}{ALEPH}{Buskulic:1996qs}{0.0018 \pm 0.0004 \pm 0.0002}{0.0018}{\pm 0.0004}{0.0002}%
\htmeasdef{ALEPH.Gamma128.pub.BUSKULIC.97C}{Gamma128}{ALEPH}{Buskulic:1996qs}{( 2.9 {}^{+1.3\cdot 10^{-4}}_{-1.2} \pm 0.7 ) \cdot 10^{ -4 }}{2.9e-4}{{}^{+1.3e-4}_{-1.2e-4}}{0.7e-4}%
\htmeasdef{ALEPH.Gamma13.pub.SCHAEL.05C}{Gamma13}{ALEPH}{Schael:2005am}{( 25.924 \pm 0.097 \pm 0.085 ) \cdot 10^{ -2 }}{25.924e-2}{\pm 0.097e-2}{0.085e-2}%
\htmeasdef{ALEPH.Gamma150.pub.BUSKULIC.97C}{Gamma150}{ALEPH}{Buskulic:1996qs}{0.0191 \pm 0.0007 \pm 0.0006}{0.0191}{\pm 0.0007}{0.0006}%
\htmeasdef{ALEPH.Gamma150by66.pub.BUSKULIC.96}{Gamma150by66}{ALEPH}{Buskulic:1995ty}{0.431 \pm 0.033}{0.431}{\pm 0.033}{0}%
\htmeasdef{ALEPH.Gamma152.pub.BUSKULIC.97C}{Gamma152}{ALEPH}{Buskulic:1996qs}{0.0043 \pm 0.0006 \pm 0.0005}{0.0043}{\pm 0.0006}{0.0005}%
\htmeasdef{ALEPH.Gamma16.pub.BARATE.99K}{Gamma16}{ALEPH}{Barate:1999hi}{0.00444 \pm 0.00026 \pm 0.00024}{0.00444}{\pm 0.00026}{0.00024}%
\htmeasdef{ALEPH.Gamma19.pub.SCHAEL.05C}{Gamma19}{ALEPH}{Schael:2005am}{( 9.295 \pm 0.084 \pm 0.088 ) \cdot 10^{ -2 }}{9.295e-2}{\pm 0.084e-2}{0.088e-2}%
\htmeasdef{ALEPH.Gamma23.pub.BARATE.99K}{Gamma23}{ALEPH}{Barate:1999hi}{0.00056 \pm 0.0002 \pm 0.00015}{0.00056}{\pm 0.0002}{0.00015}%
\htmeasdef{ALEPH.Gamma26.pub.SCHAEL.05C}{Gamma26}{ALEPH}{Schael:2005am}{( 1.08200 \pm 0.0709295 \pm 0.0594643 ) \cdot 10^{ -2 }}{1.08200e-2}{\pm 0.0709295e-2}{0.0594643e-2}%
\htmeasdef{ALEPH.Gamma28.pub.BARATE.99K}{Gamma28}{ALEPH}{Barate:1999hi}{0.00037 \pm 0.00021 \pm 0.00011}{0.00037}{\pm 0.00021}{0.00011}%
\htmeasdef{ALEPH.Gamma3.pub.SCHAEL.05C}{Gamma3}{ALEPH}{Schael:2005am}{0.17319 \pm 0.0007 \pm 0.00032}{0.17319}{\pm 0.0007}{0.00032}%
\htmeasdef{ALEPH.Gamma30.pub.SCHAEL.05C}{Gamma30}{ALEPH}{Schael:2005am}{0.00112 \pm 0.00037 \pm 0.00035}{0.00112}{\pm 0.00037}{0.00035}%
\htmeasdef{ALEPH.Gamma33.pub.BARATE.98E}{Gamma33}{ALEPH}{Barate:1997tt}{0.0097 \pm 0.00058 \pm 0.00062}{0.0097}{\pm 0.00058}{0.00062}%
\htmeasdef{ALEPH.Gamma35.pub.BARATE.99K}{Gamma35}{ALEPH}{Barate:1999hi}{0.00928 \pm 0.00045 \pm 0.00034}{0.00928}{\pm 0.00045}{0.00034}%
\htmeasdef{ALEPH.Gamma37.pub.BARATE.98E}{Gamma37}{ALEPH}{Barate:1997tt}{0.00158 \pm 0.00042 \pm 0.00017}{0.00158}{\pm 0.00042}{0.00017}%
\htmeasdef{ALEPH.Gamma37.pub.BARATE.99K}{Gamma37}{ALEPH}{Barate:1999hi}{0.00162 \pm 0.00021 \pm 0.00011}{0.00162}{\pm 0.00021}{0.00011}%
\htmeasdef{ALEPH.Gamma40.pub.BARATE.98E}{Gamma40}{ALEPH}{Barate:1997tt}{0.00294 \pm 0.00073 \pm 0.00037}{0.00294}{\pm 0.00073}{0.00037}%
\htmeasdef{ALEPH.Gamma40.pub.BARATE.99K}{Gamma40}{ALEPH}{Barate:1999hi}{0.00347 \pm 0.00053 \pm 0.00037}{0.00347}{\pm 0.00053}{0.00037}%
\htmeasdef{ALEPH.Gamma42.pub.BARATE.98E}{Gamma42}{ALEPH}{Barate:1997tt}{0.00152 \pm 0.00076 \pm 0.00021}{0.00152}{\pm 0.00076}{0.00021}%
\htmeasdef{ALEPH.Gamma42.pub.BARATE.99K}{Gamma42}{ALEPH}{Barate:1999hi}{0.00143 \pm 0.00025 \pm 0.00015}{0.00143}{\pm 0.00025}{0.00015}%
\htmeasdef{ALEPH.Gamma44.pub.BARATE.99R}{Gamma44}{ALEPH}{Barate:1999hj}{0.00026 \pm 0.00024}{0.00026}{\pm 0.00024}{0}%
\htmeasdef{ALEPH.Gamma47.pub.BARATE.98E}{Gamma47}{ALEPH}{Barate:1997tt}{0.00026 \pm 0.0001 \pm 5\cdot 10^{-5}}{0.00026}{\pm 0.0001}{5e-05}%
\htmeasdef{ALEPH.Gamma48.pub.BARATE.98E}{Gamma48}{ALEPH}{Barate:1997tt}{0.00101 \pm 0.00023 \pm 0.00013}{0.00101}{\pm 0.00023}{0.00013}%
\htmeasdef{ALEPH.Gamma5.pub.SCHAEL.05C}{Gamma5}{ALEPH}{Schael:2005am}{0.17837 \pm 0.00072 \pm 0.00036}{0.17837}{\pm 0.00072}{0.00036}%
\htmeasdef{ALEPH.Gamma51.pub.BARATE.98E}{Gamma51}{ALEPH}{Barate:1997tt}{( 3.1 \pm 1.1 \pm 0.5 ) \cdot 10^{ -4 }}{3.1e-4}{\pm 1.1e-4}{0.5e-4}%
\htmeasdef{ALEPH.Gamma53.pub.BARATE.98E}{Gamma53}{ALEPH}{Barate:1997tt}{0.00023 \pm 0.00019 \pm 0.00007}{0.00023}{\pm 0.00019}{0.00007}%
\htmeasdef{ALEPH.Gamma58.pub.SCHAEL.05C}{Gamma58}{ALEPH}{Schael:2005am}{0.09469 \pm 0.00062 \pm 0.00073}{0.09469}{\pm 0.00062}{0.00073}%
\htmeasdef{ALEPH.Gamma66.pub.SCHAEL.05C}{Gamma66}{ALEPH}{Schael:2005am}{0.04734 \pm 0.00059 \pm 0.00049}{0.04734}{\pm 0.00059}{0.00049}%
\htmeasdef{ALEPH.Gamma76.pub.SCHAEL.05C}{Gamma76}{ALEPH}{Schael:2005am}{0.00435 \pm 0.0003 \pm 0.00035}{0.00435}{\pm 0.0003}{0.00035}%
\htmeasdef{ALEPH.Gamma8.pub.SCHAEL.05C}{Gamma8}{ALEPH}{Schael:2005am}{( 11.524 \pm 0.070 \pm 0.078 ) \cdot 10^{ -2 }}{11.524e-2}{\pm 0.070e-2}{0.078e-2}%
\htmeasdef{ALEPH.Gamma805.pub.SCHAEL.05C}{Gamma805}{ALEPH}{Schael:2005am}{( 4 \pm 2 ) \cdot 10^{ -4 }}{4e-04}{\pm 2e-04}{0}%
\htmeasdef{ALEPH.Gamma85.pub.BARATE.98}{Gamma85}{ALEPH}{Barate:1997ma}{0.00214 \pm 0.00037 \pm 0.00029}{0.00214}{\pm 0.00037}{0.00029}%
\htmeasdef{ALEPH.Gamma88.pub.BARATE.98}{Gamma88}{ALEPH}{Barate:1997ma}{0.00061 \pm 0.00039 \pm 0.00018}{0.00061}{\pm 0.00039}{0.00018}%
\htmeasdef{ALEPH.Gamma93.pub.BARATE.98}{Gamma93}{ALEPH}{Barate:1997ma}{0.00163 \pm 0.00021 \pm 0.00017}{0.00163}{\pm 0.00021}{0.00017}%
\htmeasdef{ALEPH.Gamma94.pub.BARATE.98}{Gamma94}{ALEPH}{Barate:1997ma}{0.00075 \pm 0.00029 \pm 0.00015}{0.00075}{\pm 0.00029}{0.00015}%
\htmeasdef{ARGUS.Gamma103.pub.ALBRECHT.88B}{Gamma103}{ARGUS}{Albrecht:1987zf}{0.00064 \pm 0.00023 \pm 0.0001}{0.00064}{\pm 0.00023}{0.0001}%
\htmeasdef{ARGUS.Gamma3by5.pub.ALBRECHT.92D}{Gamma3by5}{ARGUS}{Albrecht:1991rh}{0.997 \pm 0.035 \pm 0.04}{0.997}{\pm 0.035}{0.04}%
\htmeasdef{BaBar.Gamma10.prelim.ICHEP2018}{Gamma10}{\babar}{Lusiani:2019mis}{( 7.17 \pm 0.031 \pm 0.21 ) \cdot 10^{ -3 }}{7.17e-03}{\pm 0.031e-03}{0.21e-03}%
\htmeasdef{BaBar.Gamma10by5.pub.AUBERT.10F}{Gamma10by5}{\babar}{Aubert:2009qj}{0.03882 \pm 0.00032 \pm 0.00057}{0.03882}{\pm 0.00032}{0.00057}%
\htmeasdef{BaBar.Gamma128.pub.DEL-AMO-SANCHEZ.11E}{Gamma128}{\babar}{delAmoSanchez:2010pc}{0.000142 \pm 1.1\cdot 10^{-5} \pm 7\cdot 10^{-6}}{0.000142}{\pm 1.1e-05}{7e-06}%
\htmeasdef{BaBar.Gamma16.prelim.ICHEP2018}{Gamma16}{\babar}{Lusiani:2019mis}{( 5.05 \pm 0.02 \pm 0.15 ) \cdot 10^{ -3 }}{5.05e-03}{\pm 0.02e-03}{0.15e-03}%
\htmeasdef{BaBar.Gamma23.prelim.ICHEP2018}{Gamma23}{\babar}{Lusiani:2019mis}{( 6.15 \pm 0.12 \pm 0.34 ) \cdot 10^{ -4 }}{6.15e-04}{\pm 0.12e-04}{0.34e-04}%
\htmeasdef{BaBar.Gamma27.prelim.ICHEP2018}{Gamma27}{\babar}{Lusiani:2019mis}{( 1.168 \pm 0.006 \pm 0.038 ) \cdot 10^{ -2 }}{1.168e-02}{\pm 0.006e-02}{0.038e-02}%
\htmeasdef{BaBar.Gamma28.prelim.ICHEP2018}{Gamma28}{\babar}{Lusiani:2019mis}{( 1.25 \pm 0.16 \pm 0.24 ) \cdot 10^{ -4 }}{1.25e-04}{\pm 0.16e-04}{0.24e-04}%
\htmeasdef{BaBar.Gamma37.pub.LEES.18B}{Gamma37}{\babar}{BaBar:2018qry}{( 14.78 \pm 0.22 \pm 0.40 ) \cdot 10^{ -4 }}{14.78e-4}{\pm 0.22e-4}{0.40e-4}%
\htmeasdef{BaBar.Gamma3by5.pub.AUBERT.10F}{Gamma3by5}{\babar}{Aubert:2009qj}{0.9796 \pm 0.0016 \pm 0.0036}{0.9796}{\pm 0.0016}{0.0036}%
\htmeasdef{BaBar.Gamma47.pub.LEES.12Y}{Gamma47}{\babar}{Lees:2012de}{( 2.31 \pm 0.04 \pm 0.08 ) \cdot 10^{ -4 }}{2.31e-4}{\pm 0.04e-4}{0.08e-4}%
\htmeasdef{BaBar.Gamma50.pub.LEES.12Y}{Gamma50}{\babar}{Lees:2012de}{( 1.60 \pm 0.20 \pm 0.22 ) \cdot 10^{ -5 }}{1.60e-5}{\pm 0.20e-5}{0.22e-5}%
\htmeasdef{BaBar.Gamma60.pub.AUBERT.08}{Gamma60}{\babar}{Aubert:2007mh}{0.0883 \pm 0.0001 \pm 0.0013}{0.0883}{\pm 0.0001}{0.0013}%
\htmeasdef{BaBar.Gamma809.prelim.ICHEP2018}{Gamma809}{\babar}{Lusiani:2019mis}{( 9.02 \pm 0.40 \pm 0.65 ) \cdot 10^{ -4 }}{9.02e-04}{\pm 0.40e-04}{0.65e-04}%
\htmeasdef{BaBar.Gamma811.pub.LEES.12X}{Gamma811}{\babar}{Lees:2012ks}{( 7.3 \pm 1.2 \pm 1.2 ) \cdot 10^{ -5 }}{7.3e-5}{\pm 1.2e-5}{1.2e-5}%
\htmeasdef{BaBar.Gamma812.pub.LEES.12X}{Gamma812}{\babar}{Lees:2012ks}{( 0.1 \pm 0.08 \pm 0.30 ) \cdot 10^{ -4 }}{0.1e-4}{\pm 0.08e-4}{0.30e-4}%
\htmeasdef{BaBar.Gamma821.pub.LEES.12X}{Gamma821}{\babar}{Lees:2012ks}{( 7.68 \pm 0.04 \pm 0.40 ) \cdot 10^{ -4 }}{7.68e-4}{\pm 0.04e-4}{0.40e-4}%
\htmeasdef{BaBar.Gamma822.pub.LEES.12X}{Gamma822}{\babar}{Lees:2012ks}{( 0.6 \pm 0.5 \pm 1.1 ) \cdot 10^{ -6 }}{0.6e-06}{\pm 0.5e-06}{1.1e-06}%
\htmeasdef{BaBar.Gamma831.pub.LEES.12X}{Gamma831}{\babar}{Lees:2012ks}{( 8.4 \pm 0.4 \pm 0.6 ) \cdot 10^{ -5 }}{8.4e-5}{\pm 0.4e-5}{0.6e-5}%
\htmeasdef{BaBar.Gamma832.pub.LEES.12X}{Gamma832}{\babar}{Lees:2012ks}{( 0.36 \pm 0.03 \pm 0.09 ) \cdot 10^{ -4 }}{0.36e-4}{\pm 0.03e-4}{0.09e-4}%
\htmeasdef{BaBar.Gamma833.pub.LEES.12X}{Gamma833}{\babar}{Lees:2012ks}{( 1.1 \pm 0.4 \pm 0.4 ) \cdot 10^{ -6 }}{1.1e-6}{\pm 0.4e-6}{0.4e-6}%
\htmeasdef{BaBar.Gamma85.pub.AUBERT.08}{Gamma85}{\babar}{Aubert:2007mh}{0.00273 \pm 2\cdot 10^{-5} \pm 9\cdot 10^{-5}}{0.00273}{\pm 2e-05}{9e-05}%
\htmeasdef{BaBar.Gamma910.pub.LEES.12X}{Gamma910}{\babar}{Lees:2012ks}{( 8.27 \pm 0.88 \pm 0.81 ) \cdot 10^{ -5 }}{8.27e-5}{\pm 0.88e-5}{0.81e-5}%
\htmeasdef{BaBar.Gamma911.pub.LEES.12X}{Gamma911}{\babar}{Lees:2012ks}{( 4.57 \pm 0.77 \pm 0.50 ) \cdot 10^{ -5 }}{4.57e-5}{\pm 0.77e-5}{0.50e-5}%
\htmeasdef{BaBar.Gamma920.pub.LEES.12X}{Gamma920}{\babar}{Lees:2012ks}{( 5.20 \pm 0.31 \pm 0.37 ) \cdot 10^{ -5 }}{5.20e-5}{\pm 0.31e-5}{0.37e-5}%
\htmeasdef{BaBar.Gamma93.pub.AUBERT.08}{Gamma93}{\babar}{Aubert:2007mh}{0.001346 \pm 1\cdot 10^{-5} \pm 3.6\cdot 10^{-5}}{0.001346}{\pm 1e-05}{3.6e-05}%
\htmeasdef{BaBar.Gamma930.pub.LEES.12X}{Gamma930}{\babar}{Lees:2012ks}{( 5.39 \pm 0.27 \pm 0.41 ) \cdot 10^{ -5 }}{5.39e-5}{\pm 0.27e-5}{0.41e-5}%
\htmeasdef{BaBar.Gamma944.pub.LEES.12X}{Gamma944}{\babar}{Lees:2012ks}{( 8.26 \pm 0.35 \pm 0.51 ) \cdot 10^{ -5 }}{8.26e-5}{\pm 0.35e-5}{0.51e-5}%
\htmeasdef{BaBar.Gamma96.pub.AUBERT.08}{Gamma96}{\babar}{Aubert:2007mh}{1.5777\cdot 10^{-5} \pm 1.3\cdot 10^{-6} \pm 1.2308\cdot 10^{-6}}{1.5777e-05}{\pm 1.3e-06}{1.2308e-06}%
\htmeasdef{BaBar.Gamma9by5.pub.AUBERT.10F}{Gamma9by5}{\babar}{Aubert:2009qj}{0.5945 \pm 0.0014 \pm 0.0061}{0.5945}{\pm 0.0014}{0.0061}%
\htmeasdef{Belle.Gamma126.pub.INAMI.09}{Gamma126}{Belle}{Inami:2008ar}{0.00135 \pm 3\cdot 10^{-5} \pm 7\cdot 10^{-5}}{0.00135}{\pm 3e-05}{7e-05}%
\htmeasdef{Belle.Gamma128.pub.INAMI.09}{Gamma128}{Belle}{Inami:2008ar}{0.000158 \pm 5\cdot 10^{-6} \pm 9\cdot 10^{-6}}{0.000158}{\pm 5e-06}{9e-06}%
\htmeasdef{Belle.Gamma13.pub.FUJIKAWA.08}{Gamma13}{Belle}{Fujikawa:2008ma}{0.2567 \pm 1\cdot 10^{-4} \pm 0.0039}{0.2567}{\pm 1e-04}{0.0039}%
\htmeasdef{Belle.Gamma130.pub.INAMI.09}{Gamma130}{Belle}{Inami:2008ar}{4.6\cdot 10^{-5} \pm 1.1\cdot 10^{-5} \pm 4\cdot 10^{-6}}{4.6e-05}{\pm 1.1e-05}{4e-06}%
\htmeasdef{Belle.Gamma132.pub.INAMI.09}{Gamma132}{Belle}{Inami:2008ar}{8.8\cdot 10^{-5} \pm 1.4\cdot 10^{-5} \pm 6\cdot 10^{-6}}{8.8e-05}{\pm 1.4e-05}{6e-06}%
\htmeasdef{Belle.Gamma35.pub.RYU.14vpc}{Gamma35}{Belle}{Ryu:2014vpc}{8.32\cdot 10^{-3} \pm 0.3\% \pm 1.8\%}{8.32e-03}{\pm 0.3\%}{1.8\%}%
\htmeasdef{Belle.Gamma37.pub.RYU.14vpc}{Gamma37}{Belle}{Ryu:2014vpc}{14.8\cdot 10^{-4} \pm 0.9\% \pm 3.7\%}{14.8e-04}{\pm 0.9\%}{3.7\%}%
\htmeasdef{Belle.Gamma40.pub.RYU.14vpc}{Gamma40}{Belle}{Ryu:2014vpc}{3.86\cdot 10^{-3} \pm 0.8\% \pm 3.5\%}{3.86e-03}{\pm 0.8\%}{3.5\%}%
\htmeasdef{Belle.Gamma42.pub.RYU.14vpc}{Gamma42}{Belle}{Ryu:2014vpc}{14.96\cdot 10^{-4} \pm 1.3\% \pm 4.9\%}{14.96e-04}{\pm 1.3\%}{4.9\%}%
\htmeasdef{Belle.Gamma47.pub.RYU.14vpc}{Gamma47}{Belle}{Ryu:2014vpc}{2.33\cdot 10^{-4} \pm 1.4\% \pm 4.0\%}{2.33e-04}{\pm 1.4\%}{4.0\%}%
\htmeasdef{Belle.Gamma50.pub.RYU.14vpc}{Gamma50}{Belle}{Ryu:2014vpc}{2.00\cdot 10^{-5} \pm 10.8\% \pm 10.1\%}{2.00e-05}{\pm 10.8\%}{10.1\%}%
\htmeasdef{Belle.Gamma60.pub.LEE.10}{Gamma60}{Belle}{Lee:2010tc}{0.0842 \pm 0 {}^{+0.0026}_{-0.0025}}{0.0842}{\pm 0}{{}^{+0.0026}_{-0.0025}}%
\htmeasdef{Belle.Gamma85.pub.LEE.10}{Gamma85}{Belle}{Lee:2010tc}{0.0033 \pm 1\cdot 10^{-5} {}^{+0.00016}_{-0.00017}}{0.0033}{\pm 1e-05}{{}^{+0.00016}_{-0.00017}}%
\htmeasdef{Belle.Gamma93.pub.LEE.10}{Gamma93}{Belle}{Lee:2010tc}{0.00155 \pm 1\cdot 10^{-5} {}^{+6\cdot 10^{-5}}_{-5\cdot 10^{-5}}}{0.00155}{\pm 1e-05}{{}^{+6e-05}_{-5e-05}}%
\htmeasdef{Belle.Gamma96.pub.LEE.10}{Gamma96}{Belle}{Lee:2010tc}{3.29\cdot 10^{-5} \pm 1.7\cdot 10^{-6} {}^{+1.9\cdot 10^{-6}}_{-2.0\cdot 10^{-6}}}{3.29e-05}{\pm 1.7e-06}{{}^{+1.9e-06}_{-2.0e-06}}%
\htmeasdef{CELLO.Gamma54.pub.BEHREND.89B}{Gamma54}{CELLO}{Behrend:1989wc}{0.15 \pm 0.004 \pm 0.003}{0.15}{\pm 0.004}{0.003}%
\htmeasdef{CLEO.Gamma10.pub.BATTLE.94}{Gamma10}{CLEO}{Battle:1994by}{0.0066 \pm 0.0007 \pm 0.0009}{0.0066}{\pm 0.0007}{0.0009}%
\htmeasdef{CLEO.Gamma102.pub.GIBAUT.94B}{Gamma102}{CLEO}{Gibaut:1994ik}{0.00097 \pm 5\cdot 10^{-5} \pm 0.00011}{0.00097}{\pm 5e-05}{0.00011}%
\htmeasdef{CLEO.Gamma103.pub.GIBAUT.94B}{Gamma103}{CLEO}{Gibaut:1994ik}{0.00077 \pm 5\cdot 10^{-5} \pm 9\cdot 10^{-5}}{0.00077}{\pm 5e-05}{9e-05}%
\htmeasdef{CLEO.Gamma104.pub.ANASTASSOV.01}{Gamma104}{CLEO}{Anastassov:2000xu}{0.00017 \pm 2\cdot 10^{-5} \pm 2\cdot 10^{-5}}{0.00017}{\pm 2e-05}{2e-05}%
\htmeasdef{CLEO.Gamma126.pub.ARTUSO.92}{Gamma126}{CLEO}{Artuso:1992qu}{0.0017 \pm 0.0002 \pm 0.0002}{0.0017}{\pm 0.0002}{0.0002}%
\htmeasdef{CLEO.Gamma128.pub.BARTELT.96}{Gamma128}{CLEO}{Bartelt:1996iv}{( 2.6 \pm 0.5 \pm 0.5 ) \cdot 10^{ -4 }}{2.6e-4}{\pm 0.5e-4}{0.5e-4}%
\htmeasdef{CLEO.Gamma13.pub.ARTUSO.94}{Gamma13}{CLEO}{Artuso:1994ii}{0.2587 \pm 0.0012 \pm 0.0042}{0.2587}{\pm 0.0012}{0.0042}%
\htmeasdef{CLEO.Gamma130.pub.BISHAI.99}{Gamma130}{CLEO}{Bishai:1998gf}{( 1.77 \pm 0.56 \pm 0.71 ) \cdot 10^{ -4 }}{1.77e-4}{\pm 0.56e-4}{0.71e-4}%
\htmeasdef{CLEO.Gamma132.pub.BISHAI.99}{Gamma132}{CLEO}{Bishai:1998gf}{( 2.2 \pm 0.70 \pm 0.22 ) \cdot 10^{ -4 }}{2.2e-4}{\pm 0.70e-4}{0.22e-4}%
\htmeasdef{CLEO.Gamma150.pub.BARINGER.87}{Gamma150}{CLEO}{Baringer:1987tr}{0.016 \pm 0.0027 \pm 0.0041}{0.016}{\pm 0.0027}{0.0041}%
\htmeasdef{CLEO.Gamma150by66.pub.BALEST.95C}{Gamma150by66}{CLEO}{Balest:1995kq}{0.464 \pm 0.016 \pm 0.017}{0.464}{\pm 0.016}{0.017}%
\htmeasdef{CLEO.Gamma152by76.pub.BORTOLETTO.93}{Gamma152by76}{CLEO}{Bortoletto:1993px}{0.81 \pm 0.06 \pm 0.06}{0.81}{\pm 0.06}{0.06}%
\htmeasdef{CLEO.Gamma16.pub.BATTLE.94}{Gamma16}{CLEO}{Battle:1994by}{0.0051 \pm 0.001 \pm 0.0007}{0.0051}{\pm 0.001}{0.0007}%
\htmeasdef{CLEO.Gamma19by13.pub.PROCARIO.93}{Gamma19by13}{CLEO}{Procario:1992hd}{0.342 \pm 0.006 \pm 0.016}{0.342}{\pm 0.006}{0.016}%
\htmeasdef{CLEO.Gamma23.pub.BATTLE.94}{Gamma23}{CLEO}{Battle:1994by}{0.0009 \pm 0.001 \pm 0.0003}{0.0009}{\pm 0.001}{0.0003}%
\htmeasdef{CLEO.Gamma26by13.pub.PROCARIO.93}{Gamma26by13}{CLEO}{Procario:1992hd}{0.044 \pm 0.003 \pm 0.005}{0.044}{\pm 0.003}{0.005}%
\htmeasdef{CLEO.Gamma29.pub.PROCARIO.93}{Gamma29}{CLEO}{Procario:1992hd}{0.0016 \pm 0.0005 \pm 0.0005}{0.0016}{\pm 0.0005}{0.0005}%
\htmeasdef{CLEO.Gamma31.pub.BATTLE.94}{Gamma31}{CLEO}{Battle:1994by}{0.017 \pm 0.0012 \pm 0.0019}{0.017}{\pm 0.0012}{0.0019}%
\htmeasdef{CLEO.Gamma34.pub.COAN.96}{Gamma34}{CLEO}{Coan:1996iu}{0.00855 \pm 0.00036 \pm 0.00073}{0.00855}{\pm 0.00036}{0.00073}%
\htmeasdef{CLEO.Gamma37.pub.COAN.96}{Gamma37}{CLEO}{Coan:1996iu}{0.00151 \pm 0.00021 \pm 0.00022}{0.00151}{\pm 0.00021}{0.00022}%
\htmeasdef{CLEO.Gamma39.pub.COAN.96}{Gamma39}{CLEO}{Coan:1996iu}{0.00562 \pm 0.0005 \pm 0.00048}{0.00562}{\pm 0.0005}{0.00048}%
\htmeasdef{CLEO.Gamma3by5.pub.ANASTASSOV.97}{Gamma3by5}{CLEO}{Anastassov:1996tc}{0.9777 \pm 0.0063 \pm 0.0087}{0.9777}{\pm 0.0063}{0.0087}%
\htmeasdef{CLEO.Gamma42.pub.COAN.96}{Gamma42}{CLEO}{Coan:1996iu}{0.00145 \pm 0.00036 \pm 0.0002}{0.00145}{\pm 0.00036}{0.0002}%
\htmeasdef{CLEO.Gamma47.pub.COAN.96}{Gamma47}{CLEO}{Coan:1996iu}{0.00023 \pm 5\cdot 10^{-5} \pm 3\cdot 10^{-5}}{0.00023}{\pm 5e-05}{3e-05}%
\htmeasdef{CLEO.Gamma5.pub.ANASTASSOV.97}{Gamma5}{CLEO}{Anastassov:1996tc}{0.1776 \pm 0.0006 \pm 0.0017}{0.1776}{\pm 0.0006}{0.0017}%
\htmeasdef{CLEO.Gamma57.pub.BALEST.95C}{Gamma57}{CLEO}{Balest:1995kq}{0.0951 \pm 0.0007 \pm 0.002}{0.0951}{\pm 0.0007}{0.002}%
\htmeasdef{CLEO.Gamma66.pub.BALEST.95C}{Gamma66}{CLEO}{Balest:1995kq}{0.0423 \pm 0.0006 \pm 0.0022}{0.0423}{\pm 0.0006}{0.0022}%
\htmeasdef{CLEO.Gamma69.pub.EDWARDS.00A}{Gamma69}{CLEO}{Edwards:1999fj}{0.0419 \pm 0.001 \pm 0.0021}{0.0419}{\pm 0.001}{0.0021}%
\htmeasdef{CLEO.Gamma76by54.pub.BORTOLETTO.93}{Gamma76by54}{CLEO}{Bortoletto:1993px}{0.034 \pm 0.002 \pm 0.003}{0.034}{\pm 0.002}{0.003}%
\htmeasdef{CLEO.Gamma78.pub.ANASTASSOV.01}{Gamma78}{CLEO}{Anastassov:2000xu}{0.00022 \pm 3\cdot 10^{-5} \pm 4\cdot 10^{-5}}{0.00022}{\pm 3e-05}{4e-05}%
\htmeasdef{CLEO.Gamma8.pub.ANASTASSOV.97}{Gamma8}{CLEO}{Anastassov:1996tc}{0.1152 \pm 0.0005 \pm 0.0012}{0.1152}{\pm 0.0005}{0.0012}%
\htmeasdef{CLEO.Gamma80by60.pub.RICHICHI.99}{Gamma80by60}{CLEO}{Richichi:1998bc}{0.0544 \pm 0.0021 \pm 0.0053}{0.0544}{\pm 0.0021}{0.0053}%
\htmeasdef{CLEO.Gamma81by69.pub.RICHICHI.99}{Gamma81by69}{CLEO}{Richichi:1998bc}{0.0261 \pm 0.0045 \pm 0.0042}{0.0261}{\pm 0.0045}{0.0042}%
\htmeasdef{CLEO.Gamma93by60.pub.RICHICHI.99}{Gamma93by60}{CLEO}{Richichi:1998bc}{0.016 \pm 0.0015 \pm 0.003}{0.016}{\pm 0.0015}{0.003}%
\htmeasdef{CLEO.Gamma94by69.pub.RICHICHI.99}{Gamma94by69}{CLEO}{Richichi:1998bc}{0.0079 \pm 0.0044 \pm 0.0016}{0.0079}{\pm 0.0044}{0.0016}%
\htmeasdef{CLEO3.Gamma151.pub.ARMS.05}{Gamma151}{CLEO3}{Arms:2005qg}{( 4.1 \pm 0.6 \pm 0.7 ) \cdot 10^{ -4 }}{4.1e-4}{\pm 0.6e-4}{0.7e-4}%
\htmeasdef{CLEO3.Gamma60.pub.BRIERE.03}{Gamma60}{CLEO3}{Briere:2003fr}{0.0913 \pm 0.0005 \pm 0.0046}{0.0913}{\pm 0.0005}{0.0046}%
\htmeasdef{CLEO3.Gamma85.pub.BRIERE.03}{Gamma85}{CLEO3}{Briere:2003fr}{0.00384 \pm 0.00014 \pm 0.00038}{0.00384}{\pm 0.00014}{0.00038}%
\htmeasdef{CLEO3.Gamma88.pub.ARMS.05}{Gamma88}{CLEO3}{Arms:2005qg}{0.00074 \pm 8\cdot 10^{-5} \pm 0.00011}{0.00074}{\pm 8e-05}{0.00011}%
\htmeasdef{CLEO3.Gamma93.pub.BRIERE.03}{Gamma93}{CLEO3}{Briere:2003fr}{0.00155 \pm 6\cdot 10^{-5} \pm 9\cdot 10^{-5}}{0.00155}{\pm 6e-05}{9e-05}%
\htmeasdef{CLEO3.Gamma94.pub.ARMS.05}{Gamma94}{CLEO3}{Arms:2005qg}{( 5.5 \pm 1.4 \pm 1.2 ) \cdot 10^{ -5 }}{5.5e-05}{\pm 1.4e-05}{1.2e-05}%
\htmeasdef{DELPHI.Gamma10.pub.ABREU.94K}{Gamma10}{DELPHI}{Abreu:1994fi}{0.0085 \pm 0.0018}{0.0085}{\pm 0.0018}{0}%
\htmeasdef{DELPHI.Gamma103.pub.ABDALLAH.06A}{Gamma103}{DELPHI}{Abdallah:2003cw}{0.00097 \pm 0.00015 \pm 5\cdot 10^{-5}}{0.00097}{\pm 0.00015}{5e-05}%
\htmeasdef{DELPHI.Gamma104.pub.ABDALLAH.06A}{Gamma104}{DELPHI}{Abdallah:2003cw}{0.00016 \pm 0.00012 \pm 6\cdot 10^{-5}}{0.00016}{\pm 0.00012}{6e-05}%
\htmeasdef{DELPHI.Gamma13.pub.ABDALLAH.06A}{Gamma13}{DELPHI}{Abdallah:2003cw}{0.2574 \pm 0.00201 \pm 0.00138}{0.2574}{\pm 0.00201}{0.00138}%
\htmeasdef{DELPHI.Gamma19.pub.ABDALLAH.06A}{Gamma19}{DELPHI}{Abdallah:2003cw}{0.09498 \pm 0.0032 \pm 0.00275}{0.09498}{\pm 0.0032}{0.00275}%
\htmeasdef{DELPHI.Gamma25.pub.ABDALLAH.06A}{Gamma25}{DELPHI}{Abdallah:2003cw}{0.01403 \pm 0.00214 \pm 0.00224}{0.01403}{\pm 0.00214}{0.00224}%
\htmeasdef{DELPHI.Gamma3.pub.ABREU.99X}{Gamma3}{DELPHI}{Abreu:1999rb}{0.17325 \pm 0.00095 \pm 0.00077}{0.17325}{\pm 0.00095}{0.00077}%
\htmeasdef{DELPHI.Gamma31.pub.ABREU.94K}{Gamma31}{DELPHI}{Abreu:1994fi}{0.0154 \pm 0.0024}{0.0154}{\pm 0.0024}{0}%
\htmeasdef{DELPHI.Gamma5.pub.ABREU.99X}{Gamma5}{DELPHI}{Abreu:1999rb}{0.17877 \pm 0.00109 \pm 0.0011}{0.17877}{\pm 0.00109}{0.0011}%
\htmeasdef{DELPHI.Gamma57.pub.ABDALLAH.06A}{Gamma57}{DELPHI}{Abdallah:2003cw}{0.09317 \pm 0.0009 \pm 0.00082}{0.09317}{\pm 0.0009}{0.00082}%
\htmeasdef{DELPHI.Gamma66.pub.ABDALLAH.06A}{Gamma66}{DELPHI}{Abdallah:2003cw}{0.04545 \pm 0.00106 \pm 0.00103}{0.04545}{\pm 0.00106}{0.00103}%
\htmeasdef{DELPHI.Gamma7.pub.ABREU.92N}{Gamma7}{DELPHI}{Abreu:1992gn}{0.124 \pm 0.007 \pm 0.007}{0.124}{\pm 0.007}{0.007}%
\htmeasdef{DELPHI.Gamma74.pub.ABDALLAH.06A}{Gamma74}{DELPHI}{Abdallah:2003cw}{0.00561 \pm 0.00068 \pm 0.00095}{0.00561}{\pm 0.00068}{0.00095}%
\htmeasdef{DELPHI.Gamma8.pub.ABDALLAH.06A}{Gamma8}{DELPHI}{Abdallah:2003cw}{0.11571 \pm 0.0012 \pm 0.00114}{0.11571}{\pm 0.0012}{0.00114}%
\htmeasdef{HRS.Gamma102.pub.BYLSMA.87}{Gamma102}{HRS}{Bylsma:1986zy}{0.00102 \pm 0.00029}{0.00102}{\pm 0.00029}{0}%
\htmeasdef{HRS.Gamma103.pub.BYLSMA.87}{Gamma103}{HRS}{Bylsma:1986zy}{0.00051 \pm 0.0002}{0.00051}{\pm 0.0002}{0}%
\htmeasdef{L3.Gamma102.pub.ACHARD.01D}{Gamma102}{L3}{Achard:2001pk}{0.0017 \pm 0.00022 \pm 0.00026}{0.0017}{\pm 0.00022}{0.00026}%
\htmeasdef{L3.Gamma13.pub.ACCIARRI.95}{Gamma13}{L3}{Acciarri:1994vr}{0.2505 \pm 0.0035 \pm 0.005}{0.2505}{\pm 0.0035}{0.005}%
\htmeasdef{L3.Gamma19.pub.ACCIARRI.95}{Gamma19}{L3}{Acciarri:1994vr}{0.0888 \pm 0.0037 \pm 0.0042}{0.0888}{\pm 0.0037}{0.0042}%
\htmeasdef{L3.Gamma26.pub.ACCIARRI.95}{Gamma26}{L3}{Acciarri:1994vr}{0.017 \pm 0.0024 \pm 0.0038}{0.017}{\pm 0.0024}{0.0038}%
\htmeasdef{L3.Gamma3.pub.ACCIARRI.01F}{Gamma3}{L3}{Acciarri:2001sg}{0.17342 \pm 0.0011 \pm 0.00067}{0.17342}{\pm 0.0011}{0.00067}%
\htmeasdef{L3.Gamma35.pub.ACCIARRI.95F}{Gamma35}{L3}{Acciarri:1995kx}{0.0095 \pm 0.0015 \pm 0.0006}{0.0095}{\pm 0.0015}{0.0006}%
\htmeasdef{L3.Gamma40.pub.ACCIARRI.95F}{Gamma40}{L3}{Acciarri:1995kx}{0.0041 \pm 0.0012 \pm 0.0003}{0.0041}{\pm 0.0012}{0.0003}%
\htmeasdef{L3.Gamma5.pub.ACCIARRI.01F}{Gamma5}{L3}{Acciarri:2001sg}{0.17806 \pm 0.00104 \pm 0.00076}{0.17806}{\pm 0.00104}{0.00076}%
\htmeasdef{L3.Gamma54.pub.ADEVA.91F}{Gamma54}{L3}{Adeva:1991qq}{0.144 \pm 0.006 \pm 0.003}{0.144}{\pm 0.006}{0.003}%
\htmeasdef{L3.Gamma55.pub.ACHARD.01D}{Gamma55}{L3}{Achard:2001pk}{0.14556 \pm 0.00105 \pm 0.00076}{0.14556}{\pm 0.00105}{0.00076}%
\htmeasdef{L3.Gamma7.pub.ACCIARRI.95}{Gamma7}{L3}{Acciarri:1994vr}{0.1247 \pm 0.0026 \pm 0.0043}{0.1247}{\pm 0.0026}{0.0043}%
\htmeasdef{OPAL.Gamma10.pub.ABBIENDI.01J}{Gamma10}{OPAL}{Abbiendi:2000ee}{0.00658 \pm 0.00027 \pm 0.00029}{0.00658}{\pm 0.00027}{0.00029}%
\htmeasdef{OPAL.Gamma103.pub.ACKERSTAFF.99E}{Gamma103}{OPAL}{Ackerstaff:1998ia}{0.00091 \pm 0.00014 \pm 6\cdot 10^{-5}}{0.00091}{\pm 0.00014}{6e-05}%
\htmeasdef{OPAL.Gamma104.pub.ACKERSTAFF.99E}{Gamma104}{OPAL}{Ackerstaff:1998ia}{0.00027 \pm 0.00018 \pm 9\cdot 10^{-5}}{0.00027}{\pm 0.00018}{9e-05}%
\htmeasdef{OPAL.Gamma13.pub.ACKERSTAFF.98M}{Gamma13}{OPAL}{Ackerstaff:1997tx}{0.2589 \pm 0.0017 \pm 0.0029}{0.2589}{\pm 0.0017}{0.0029}%
\htmeasdef{OPAL.Gamma16.pub.ABBIENDI.04J}{Gamma16}{OPAL}{Abbiendi:2004xa}{0.00471 \pm 0.00059 \pm 0.00023}{0.00471}{\pm 0.00059}{0.00023}%
\htmeasdef{OPAL.Gamma17.pub.ACKERSTAFF.98M}{Gamma17}{OPAL}{Ackerstaff:1997tx}{0.0991 \pm 0.0031 \pm 0.0027}{0.0991}{\pm 0.0031}{0.0027}%
\htmeasdef{OPAL.Gamma3.pub.ABBIENDI.03}{Gamma3}{OPAL}{Abbiendi:2002jw}{0.1734 \pm 0.0009 \pm 0.0006}{0.1734}{\pm 0.0009}{0.0006}%
\htmeasdef{OPAL.Gamma31.pub.ABBIENDI.01J}{Gamma31}{OPAL}{Abbiendi:2000ee}{0.01528 \pm 0.00039 \pm 0.0004}{0.01528}{\pm 0.00039}{0.0004}%
\htmeasdef{OPAL.Gamma33.pub.AKERS.94G}{Gamma33}{OPAL}{Akers:1994td}{0.0097 \pm 0.0009 \pm 0.0006}{0.0097}{\pm 0.0009}{0.0006}%
\htmeasdef{OPAL.Gamma35.pub.ABBIENDI.00C}{Gamma35}{OPAL}{Abbiendi:1999pm}{0.00933 \pm 0.00068 \pm 0.00049}{0.00933}{\pm 0.00068}{0.00049}%
\htmeasdef{OPAL.Gamma38.pub.ABBIENDI.00C}{Gamma38}{OPAL}{Abbiendi:1999pm}{0.0033 \pm 0.00055 \pm 0.00039}{0.0033}{\pm 0.00055}{0.00039}%
\htmeasdef{OPAL.Gamma43.pub.ABBIENDI.00C}{Gamma43}{OPAL}{Abbiendi:1999pm}{0.00324 \pm 0.00074 \pm 0.00066}{0.00324}{\pm 0.00074}{0.00066}%
\htmeasdef{OPAL.Gamma5.pub.ABBIENDI.99H}{Gamma5}{OPAL}{Abbiendi:1998cx}{0.1781 \pm 0.0009 \pm 0.0006}{0.1781}{\pm 0.0009}{0.0006}%
\htmeasdef{OPAL.Gamma55.pub.AKERS.95Y}{Gamma55}{OPAL}{Akers:1995ry}{0.1496 \pm 0.0009 \pm 0.0022}{0.1496}{\pm 0.0009}{0.0022}%
\htmeasdef{OPAL.Gamma57by55.pub.AKERS.95Y}{Gamma57by55}{OPAL}{Akers:1995ry}{0.66 \pm 0.004 \pm 0.014}{0.66}{\pm 0.004}{0.014}%
\htmeasdef{OPAL.Gamma7.pub.ALEXANDER.91D}{Gamma7}{OPAL}{Alexander:1991am}{0.121 \pm 0.007 \pm 0.005}{0.121}{\pm 0.007}{0.005}%
\htmeasdef{OPAL.Gamma8.pub.ACKERSTAFF.98M}{Gamma8}{OPAL}{Ackerstaff:1997tx}{0.1198 \pm 0.0013 \pm 0.0016}{0.1198}{\pm 0.0013}{0.0016}%
\htmeasdef{OPAL.Gamma85.pub.ABBIENDI.04J}{Gamma85}{OPAL}{Abbiendi:2004xa}{0.00415 \pm 0.00053 \pm 0.0004}{0.00415}{\pm 0.00053}{0.0004}%
\htmeasdef{OPAL.Gamma92.pub.ABBIENDI.00D}{Gamma92}{OPAL}{Abbiendi:1999cq}{0.00159 \pm 0.00053 \pm 0.0002}{0.00159}{\pm 0.00053}{0.0002}%
\htmeasdef{TPC.Gamma54.pub.AIHARA.87B}{Gamma54}{TPC}{Aihara:1986mw}{0.151 \pm 0.008 \pm 0.006}{0.151}{\pm 0.008}{0.006}%
\htmeasdef{TPC.Gamma82.pub.BAUER.94}{Gamma82}{TPC}{Bauer:1993wn}{0.0058 {}^{+0.0015}_{-0.0013} \pm 0.0012}{0.0058}{{}^{+0.0015}_{-0.0013}}{0.0012}%
\htmeasdef{TPC.Gamma92.pub.BAUER.94}{Gamma92}{TPC}{Bauer:1993wn}{0.0015 {}^{+0.0009}_{-0.0007} \pm 0.0003}{0.0015}{{}^{+0.0009}_{-0.0007}}{0.0003}%
\htdef{Gamma1.qt}{\ensuremath{0.8521 \pm 0.0011}}%
\htdef{Gamma2.qt}{\ensuremath{0.8455 \pm 0.0010}}%
\htdef{Gamma3.qt}{\ensuremath{0.17392 \pm 0.00039}}%
\htdef{ALEPH.Gamma3.pub.SCHAEL.05C,qt}{\ensuremath{0.17319 \pm 0.00070 \pm 0.00032}}%
\htdef{DELPHI.Gamma3.pub.ABREU.99X,qt}{\ensuremath{0.17325 \pm 0.00095 \pm 0.00077}}%
\htdef{L3.Gamma3.pub.ACCIARRI.01F,qt}{\ensuremath{0.17342 \pm 0.00110 \pm 0.00067}}%
\htdef{OPAL.Gamma3.pub.ABBIENDI.03,qt}{\ensuremath{0.17340 \pm 0.00090 \pm 0.00060}}%
\htdef{Gamma3by5.qt}{\ensuremath{0.9761 \pm 0.0028}}%
\htdef{ARGUS.Gamma3by5.pub.ALBRECHT.92D,qt}{\ensuremath{0.9970 \pm 0.0350 \pm 0.0400}}%
\htdef{BaBar.Gamma3by5.pub.AUBERT.10F,qt}{\ensuremath{0.9796 \pm 0.0016 \pm 0.0036}}%
\htdef{CLEO.Gamma3by5.pub.ANASTASSOV.97,qt}{\ensuremath{0.9777 \pm 0.0063 \pm 0.0087}}%
\htdef{Gamma5.qt}{\ensuremath{0.17817 \pm 0.00041}}%
\htdef{ALEPH.Gamma5.pub.SCHAEL.05C,qt}{\ensuremath{0.17837 \pm 0.00072 \pm 0.00036}}%
\htdef{CLEO.Gamma5.pub.ANASTASSOV.97,qt}{\ensuremath{0.17760 \pm 0.00060 \pm 0.00170}}%
\htdef{DELPHI.Gamma5.pub.ABREU.99X,qt}{\ensuremath{0.17877 \pm 0.00109 \pm 0.00110}}%
\htdef{L3.Gamma5.pub.ACCIARRI.01F,qt}{\ensuremath{0.17806 \pm 0.00104 \pm 0.00076}}%
\htdef{OPAL.Gamma5.pub.ABBIENDI.99H,qt}{\ensuremath{0.17810 \pm 0.00090 \pm 0.00060}}%
\htdef{Gamma7.qt}{\ensuremath{0.12019 \pm 0.00053}}%
\htdef{DELPHI.Gamma7.pub.ABREU.92N,qt}{\ensuremath{0.12400 \pm 0.00700 \pm 0.00700}}%
\htdef{L3.Gamma7.pub.ACCIARRI.95,qt}{\ensuremath{0.12470 \pm 0.00260 \pm 0.00430}}%
\htdef{OPAL.Gamma7.pub.ALEXANDER.91D,qt}{\ensuremath{0.12100 \pm 0.00700 \pm 0.00500}}%
\htdef{Gamma8.qt}{\ensuremath{0.11502 \pm 0.00053}}%
\htdef{ALEPH.Gamma8.pub.SCHAEL.05C,qt}{\ensuremath{0.11524 \pm 0.00070 \pm 0.00078}}%
\htdef{CLEO.Gamma8.pub.ANASTASSOV.97,qt}{\ensuremath{0.11520 \pm 0.00050 \pm 0.00120}}%
\htdef{DELPHI.Gamma8.pub.ABDALLAH.06A,qt}{\ensuremath{0.11571 \pm 0.00120 \pm 0.00114}}%
\htdef{OPAL.Gamma8.pub.ACKERSTAFF.98M,qt}{\ensuremath{0.11980 \pm 0.00130 \pm 0.00160}}%
\htdef{Gamma8by5.qt}{\ensuremath{0.6456 \pm 0.0033}}%
\htdef{Gamma9.qt}{\ensuremath{0.10804 \pm 0.00052}}%
\htdef{Gamma9by5.qt}{\ensuremath{0.6064 \pm 0.0032}}%
\htdef{BaBar.Gamma9by5.pub.AUBERT.10F,qt}{\ensuremath{0.5945 \pm 0.0014 \pm 0.0061}}%
\htdef{Gamma10.qt}{\ensuremath{(0.6986 \pm 0.0085) \cdot 10^{-2}}}%
\htdef{ALEPH.Gamma10.pub.BARATE.99K,qt}{\ensuremath{(0.6960 \pm 0.0250 \pm 0.0140) \cdot 10^{-2} }}%
\htdef{BaBar.Gamma10.prelim.ICHEP2018,qt}{\ensuremath{(0.7170 \pm 0.0031 \pm 0.0210) \cdot 10^{-2} }}%
\htdef{CLEO.Gamma10.pub.BATTLE.94,qt}{\ensuremath{(0.6600 \pm 0.0700 \pm 0.0900) \cdot 10^{-2} }}%
\htdef{DELPHI.Gamma10.pub.ABREU.94K,qt}{\ensuremath{(0.8500 \pm 0.1800 \pm 0.0000) \cdot 10^{-2} }}%
\htdef{OPAL.Gamma10.pub.ABBIENDI.01J,qt}{\ensuremath{(0.6580 \pm 0.0270 \pm 0.0290) \cdot 10^{-2} }}%
\htdef{Gamma10by5.qt}{\ensuremath{(3.921 \pm 0.048) \cdot 10^{-2}}}%
\htdef{BaBar.Gamma10by5.pub.AUBERT.10F,qt}{\ensuremath{(3.882 \pm 0.032 \pm 0.057) \cdot 10^{-2} }}%
\htdef{Gamma10by9.qt}{\ensuremath{(6.467 \pm 0.084) \cdot 10^{-2}}}%
\htdef{Gamma11.qt}{\ensuremath{0.36996 \pm 0.00094}}%
\htdef{Gamma12.qt}{\ensuremath{0.36495 \pm 0.00094}}%
\htdef{Gamma13.qt}{\ensuremath{0.25938 \pm 0.00090}}%
\htdef{ALEPH.Gamma13.pub.SCHAEL.05C,qt}{\ensuremath{0.25924 \pm 0.00097 \pm 0.00085}}%
\htdef{Belle.Gamma13.pub.FUJIKAWA.08,qt}{\ensuremath{0.25670 \pm 0.00010 \pm 0.00390}}%
\htdef{CLEO.Gamma13.pub.ARTUSO.94,qt}{\ensuremath{0.25870 \pm 0.00120 \pm 0.00420}}%
\htdef{DELPHI.Gamma13.pub.ABDALLAH.06A,qt}{\ensuremath{0.25740 \pm 0.00201 \pm 0.00138}}%
\htdef{L3.Gamma13.pub.ACCIARRI.95,qt}{\ensuremath{0.25050 \pm 0.00350 \pm 0.00500}}%
\htdef{OPAL.Gamma13.pub.ACKERSTAFF.98M,qt}{\ensuremath{0.25890 \pm 0.00170 \pm 0.00290}}%
\htdef{Gamma14.qt}{\ensuremath{0.25447 \pm 0.00091}}%
\htdef{Gamma16.qt}{\ensuremath{(0.4904 \pm 0.0092) \cdot 10^{-2}}}%
\htdef{ALEPH.Gamma16.pub.BARATE.99K,qt}{\ensuremath{(0.4440 \pm 0.0260 \pm 0.0240) \cdot 10^{-2} }}%
\htdef{BaBar.Gamma16.prelim.ICHEP2018,qt}{\ensuremath{(0.5050 \pm 0.0020 \pm 0.0150) \cdot 10^{-2} }}%
\htdef{CLEO.Gamma16.pub.BATTLE.94,qt}{\ensuremath{(0.5100 \pm 0.1000 \pm 0.0700) \cdot 10^{-2} }}%
\htdef{OPAL.Gamma16.pub.ABBIENDI.04J,qt}{\ensuremath{(0.4710 \pm 0.0590 \pm 0.0230) \cdot 10^{-2} }}%
\htdef{Gamma17.qt}{\ensuremath{0.10793 \pm 0.00091}}%
\htdef{OPAL.Gamma17.pub.ACKERSTAFF.98M,qt}{\ensuremath{0.09910 \pm 0.00310 \pm 0.00270}}%
\htdef{Gamma18.qt}{\ensuremath{(9.421 \pm 0.092) \cdot 10^{-2}}}%
\htdef{Gamma19.qt}{\ensuremath{(9.270 \pm 0.092) \cdot 10^{-2}}}%
\htdef{ALEPH.Gamma19.pub.SCHAEL.05C,qt}{\ensuremath{(9.295 \pm 0.084 \pm 0.088) \cdot 10^{-2} }}%
\htdef{DELPHI.Gamma19.pub.ABDALLAH.06A,qt}{\ensuremath{(9.498 \pm 0.320 \pm 0.275) \cdot 10^{-2} }}%
\htdef{L3.Gamma19.pub.ACCIARRI.95,qt}{\ensuremath{(8.880 \pm 0.370 \pm 0.420) \cdot 10^{-2} }}%
\htdef{Gamma19by13.qt}{\ensuremath{0.3574 \pm 0.0042}}%
\htdef{CLEO.Gamma19by13.pub.PROCARIO.93,qt}{\ensuremath{0.3420 \pm 0.0060 \pm 0.0160}}%
\htdef{Gamma20.qt}{\ensuremath{(9.211 \pm 0.092) \cdot 10^{-2}}}%
\htdef{Gamma23.qt}{\ensuremath{(0.0585 \pm 0.0027) \cdot 10^{-2}}}%
\htdef{ALEPH.Gamma23.pub.BARATE.99K,qt}{\ensuremath{(0.0560 \pm 0.0200 \pm 0.0150) \cdot 10^{-2} }}%
\htdef{BaBar.Gamma23.prelim.ICHEP2018,qt}{\ensuremath{(0.0615 \pm 0.0012 \pm 0.0034) \cdot 10^{-2} }}%
\htdef{CLEO.Gamma23.pub.BATTLE.94,qt}{\ensuremath{(0.0900 \pm 0.1000 \pm 0.0300) \cdot 10^{-2} }}%
\htdef{Gamma24.qt}{\ensuremath{(1.372 \pm 0.034) \cdot 10^{-2}}}%
\htdef{Gamma25.qt}{\ensuremath{(1.288 \pm 0.034) \cdot 10^{-2}}}%
\htdef{DELPHI.Gamma25.pub.ABDALLAH.06A,qt}{\ensuremath{(1.403 \pm 0.214 \pm 0.224) \cdot 10^{-2} }}%
\htdef{Gamma26.qt}{\ensuremath{(1.236 \pm 0.030) \cdot 10^{-2}}}%
\htdef{ALEPH.Gamma26.pub.SCHAEL.05C,qt}{\ensuremath{(1.082 \pm 0.071 \pm 0.059) \cdot 10^{-2} }}%
\htdef{L3.Gamma26.pub.ACCIARRI.95,qt}{\ensuremath{(1.700 \pm 0.240 \pm 0.380) \cdot 10^{-2} }}%
\htdef{Gamma26by13.qt}{\ensuremath{(4.764 \pm 0.118) \cdot 10^{-2}}}%
\htdef{CLEO.Gamma26by13.pub.PROCARIO.93,qt}{\ensuremath{(4.400 \pm 0.300 \pm 0.500) \cdot 10^{-2} }}%
\htdef{Gamma27.qt}{\ensuremath{(1.1381 \pm 0.0292) \cdot 10^{-2}}}%
\htdef{BaBar.Gamma27.prelim.ICHEP2018,qt}{\ensuremath{(1.1680 \pm 0.0060 \pm 0.0380) \cdot 10^{-2} }}%
\htdef{Gamma28.qt}{\ensuremath{(1.127 \pm 0.263) \cdot 10^{-4}}}%
\htdef{ALEPH.Gamma28.pub.BARATE.99K,qt}{\ensuremath{(3.700 \pm 2.100 \pm 1.100) \cdot 10^{-4} }}%
\htdef{BaBar.Gamma28.prelim.ICHEP2018,qt}{\ensuremath{(1.250 \pm 0.160 \pm 0.240) \cdot 10^{-4} }}%
\htdef{Gamma29.qt}{\ensuremath{(0.1333 \pm 0.0071) \cdot 10^{-2}}}%
\htdef{CLEO.Gamma29.pub.PROCARIO.93,qt}{\ensuremath{(0.1600 \pm 0.0500 \pm 0.0500) \cdot 10^{-2} }}%
\htdef{Gamma30.qt}{\ensuremath{(0.0864 \pm 0.0067) \cdot 10^{-2}}}%
\htdef{ALEPH.Gamma30.pub.SCHAEL.05C,qt}{\ensuremath{(0.1120 \pm 0.0370 \pm 0.0350) \cdot 10^{-2} }}%
\htdef{Gamma31.qt}{\ensuremath{(1.568 \pm 0.018) \cdot 10^{-2}}}%
\htdef{CLEO.Gamma31.pub.BATTLE.94,qt}{\ensuremath{(1.700 \pm 0.120 \pm 0.190) \cdot 10^{-2} }}%
\htdef{DELPHI.Gamma31.pub.ABREU.94K,qt}{\ensuremath{(1.540 \pm 0.240 \pm 0.000) \cdot 10^{-2} }}%
\htdef{OPAL.Gamma31.pub.ABBIENDI.01J,qt}{\ensuremath{(1.528 \pm 0.039 \pm 0.040) \cdot 10^{-2} }}%
\htdef{Gamma32.qt}{\ensuremath{(0.8729 \pm 0.0141) \cdot 10^{-2}}}%
\htdef{Gamma33.qt}{\ensuremath{(0.9366 \pm 0.0292) \cdot 10^{-2}}}%
\htdef{ALEPH.Gamma33.pub.BARATE.98E,qt}{\ensuremath{(0.9700 \pm 0.0580 \pm 0.0620) \cdot 10^{-2} }}%
\htdef{OPAL.Gamma33.pub.AKERS.94G,qt}{\ensuremath{(0.9700 \pm 0.0900 \pm 0.0600) \cdot 10^{-2} }}%
\htdef{Gamma34.qt}{\ensuremath{(0.9860 \pm 0.0138) \cdot 10^{-2}}}%
\htdef{CLEO.Gamma34.pub.COAN.96,qt}{\ensuremath{(0.8550 \pm 0.0360 \pm 0.0730) \cdot 10^{-2} }}%
\htdef{Gamma35.qt}{\ensuremath{(0.8378 \pm 0.0139) \cdot 10^{-2}}}%
\htdef{ALEPH.Gamma35.pub.BARATE.99K,qt}{\ensuremath{(0.9280 \pm 0.0450 \pm 0.0340) \cdot 10^{-2} }}%
\htdef{Belle.Gamma35.pub.RYU.14vpc,qt}{\ensuremath{(0.8320 \pm 0.0025 \pm 0.0150) \cdot 10^{-2} }}%
\htdef{L3.Gamma35.pub.ACCIARRI.95F,qt}{\ensuremath{(0.9500 \pm 0.1500 \pm 0.0600) \cdot 10^{-2} }}%
\htdef{OPAL.Gamma35.pub.ABBIENDI.00C,qt}{\ensuremath{(0.9330 \pm 0.0680 \pm 0.0490) \cdot 10^{-2} }}%
\htdef{Gamma37.qt}{\ensuremath{(0.1483 \pm 0.0034) \cdot 10^{-2}}}%
\htdef{ALEPH.Gamma37.pub.BARATE.98E,qt}{\ensuremath{(0.1580 \pm 0.0420 \pm 0.0170) \cdot 10^{-2} }}%
\htdef{ALEPH.Gamma37.pub.BARATE.99K,qt}{\ensuremath{(0.1620 \pm 0.0210 \pm 0.0110) \cdot 10^{-2} }}%
\htdef{BaBar.Gamma37.pub.LEES.18B,qt}{\ensuremath{(0.1478 \pm 0.0022 \pm 0.0040) \cdot 10^{-2} }}%
\htdef{Belle.Gamma37.pub.RYU.14vpc,qt}{\ensuremath{(0.1480 \pm 0.0013 \pm 0.0055) \cdot 10^{-2} }}%
\htdef{CLEO.Gamma37.pub.COAN.96,qt}{\ensuremath{(0.1510 \pm 0.0210 \pm 0.0220) \cdot 10^{-2} }}%
\htdef{Gamma38.qt}{\ensuremath{(0.2977 \pm 0.0073) \cdot 10^{-2}}}%
\htdef{OPAL.Gamma38.pub.ABBIENDI.00C,qt}{\ensuremath{(0.3300 \pm 0.0550 \pm 0.0390) \cdot 10^{-2} }}%
\htdef{Gamma39.qt}{\ensuremath{(0.5302 \pm 0.0134) \cdot 10^{-2}}}%
\htdef{CLEO.Gamma39.pub.COAN.96,qt}{\ensuremath{(0.5620 \pm 0.0500 \pm 0.0480) \cdot 10^{-2} }}%
\htdef{Gamma40.qt}{\ensuremath{(0.3807 \pm 0.0129) \cdot 10^{-2}}}%
\htdef{ALEPH.Gamma40.pub.BARATE.98E,qt}{\ensuremath{(0.2940 \pm 0.0730 \pm 0.0370) \cdot 10^{-2} }}%
\htdef{ALEPH.Gamma40.pub.BARATE.99K,qt}{\ensuremath{(0.3470 \pm 0.0530 \pm 0.0370) \cdot 10^{-2} }}%
\htdef{Belle.Gamma40.pub.RYU.14vpc,qt}{\ensuremath{(0.3860 \pm 0.0031 \pm 0.0135) \cdot 10^{-2} }}%
\htdef{L3.Gamma40.pub.ACCIARRI.95F,qt}{\ensuremath{(0.4100 \pm 0.1200 \pm 0.0300) \cdot 10^{-2} }}%
\htdef{Gamma42.qt}{\ensuremath{(0.1494 \pm 0.0070) \cdot 10^{-2}}}%
\htdef{ALEPH.Gamma42.pub.BARATE.98E,qt}{\ensuremath{(0.1520 \pm 0.0760 \pm 0.0210) \cdot 10^{-2} }}%
\htdef{ALEPH.Gamma42.pub.BARATE.99K,qt}{\ensuremath{(0.1430 \pm 0.0250 \pm 0.0150) \cdot 10^{-2} }}%
\htdef{Belle.Gamma42.pub.RYU.14vpc,qt}{\ensuremath{(0.1496 \pm 0.0019 \pm 0.0073) \cdot 10^{-2} }}%
\htdef{CLEO.Gamma42.pub.COAN.96,qt}{\ensuremath{(0.1450 \pm 0.0360 \pm 0.0200) \cdot 10^{-2} }}%
\htdef{Gamma43.qt}{\ensuremath{(0.4042 \pm 0.0260) \cdot 10^{-2}}}%
\htdef{OPAL.Gamma43.pub.ABBIENDI.00C,qt}{\ensuremath{(0.3240 \pm 0.0740 \pm 0.0660) \cdot 10^{-2} }}%
\htdef{Gamma44.qt}{\ensuremath{(2.346 \pm 2.306) \cdot 10^{-4}}}%
\htdef{ALEPH.Gamma44.pub.BARATE.99R,qt}{\ensuremath{(2.600 \pm 2.400 \pm 0.000) \cdot 10^{-4} }}%
\htdef{Gamma46.qt}{\ensuremath{(0.1516 \pm 0.0247) \cdot 10^{-2}}}%
\htdef{Gamma47.qt}{\ensuremath{(2.342 \pm 0.065) \cdot 10^{-4}}}%
\htdef{ALEPH.Gamma47.pub.BARATE.98E,qt}{\ensuremath{(2.600 \pm 1.000 \pm 0.500) \cdot 10^{-4} }}%
\htdef{BaBar.Gamma47.pub.LEES.12Y,qt}{\ensuremath{(2.310 \pm 0.040 \pm 0.080) \cdot 10^{-4} }}%
\htdef{Belle.Gamma47.pub.RYU.14vpc,qt}{\ensuremath{(2.330 \pm 0.033 \pm 0.093) \cdot 10^{-4} }}%
\htdef{CLEO.Gamma47.pub.COAN.96,qt}{\ensuremath{(2.300 \pm 0.500 \pm 0.300) \cdot 10^{-4} }}%
\htdef{Gamma48.qt}{\ensuremath{(0.1048 \pm 0.0247) \cdot 10^{-2}}}%
\htdef{ALEPH.Gamma48.pub.BARATE.98E,qt}{\ensuremath{(0.1010 \pm 0.0230 \pm 0.0130) \cdot 10^{-2} }}%
\htdef{Gamma49.qt}{\ensuremath{(3.543 \pm 1.193) \cdot 10^{-4}}}%
\htdef{Gamma50.qt}{\ensuremath{(1.816 \pm 0.207) \cdot 10^{-5}}}%
\htdef{BaBar.Gamma50.pub.LEES.12Y,qt}{\ensuremath{(1.600 \pm 0.200 \pm 0.220) \cdot 10^{-5} }}%
\htdef{Belle.Gamma50.pub.RYU.14vpc,qt}{\ensuremath{(2.000 \pm 0.216 \pm 0.202) \cdot 10^{-5} }}%
\htdef{Gamma51.qt}{\ensuremath{(3.179 \pm 1.192) \cdot 10^{-4}}}%
\htdef{ALEPH.Gamma51.pub.BARATE.98E,qt}{\ensuremath{(3.100 \pm 1.100 \pm 0.500) \cdot 10^{-4} }}%
\htdef{Gamma53.qt}{\ensuremath{(2.220 \pm 2.024) \cdot 10^{-4}}}%
\htdef{ALEPH.Gamma53.pub.BARATE.98E,qt}{\ensuremath{(2.300 \pm 1.900 \pm 0.700) \cdot 10^{-4} }}%
\htdef{Gamma54.qt}{\ensuremath{0.15206 \pm 0.00061}}%
\htdef{CELLO.Gamma54.pub.BEHREND.89B,qt}{\ensuremath{0.15000 \pm 0.00400 \pm 0.00300}}%
\htdef{L3.Gamma54.pub.ADEVA.91F,qt}{\ensuremath{0.14400 \pm 0.00600 \pm 0.00300}}%
\htdef{TPC.Gamma54.pub.AIHARA.87B,qt}{\ensuremath{0.15100 \pm 0.00800 \pm 0.00600}}%
\htdef{Gamma55.qt}{\ensuremath{0.14558 \pm 0.00056}}%
\htdef{L3.Gamma55.pub.ACHARD.01D,qt}{\ensuremath{0.14556 \pm 0.00105 \pm 0.00076}}%
\htdef{OPAL.Gamma55.pub.AKERS.95Y,qt}{\ensuremath{0.14960 \pm 0.00090 \pm 0.00220}}%
\htdef{Gamma56.qt}{\ensuremath{(9.769 \pm 0.053) \cdot 10^{-2}}}%
\htdef{Gamma57.qt}{\ensuremath{(9.428 \pm 0.053) \cdot 10^{-2}}}%
\htdef{CLEO.Gamma57.pub.BALEST.95C,qt}{\ensuremath{(9.510 \pm 0.070 \pm 0.200) \cdot 10^{-2} }}%
\htdef{DELPHI.Gamma57.pub.ABDALLAH.06A,qt}{\ensuremath{(9.317 \pm 0.090 \pm 0.082) \cdot 10^{-2} }}%
\htdef{Gamma57by55.qt}{\ensuremath{0.6476 \pm 0.0029}}%
\htdef{OPAL.Gamma57by55.pub.AKERS.95Y,qt}{\ensuremath{0.6600 \pm 0.0040 \pm 0.0140}}%
\htdef{Gamma58.qt}{\ensuremath{(9.397 \pm 0.053) \cdot 10^{-2}}}%
\htdef{ALEPH.Gamma58.pub.SCHAEL.05C,qt}{\ensuremath{(9.469 \pm 0.062 \pm 0.073) \cdot 10^{-2} }}%
\htdef{Gamma59.qt}{\ensuremath{(9.279 \pm 0.051) \cdot 10^{-2}}}%
\htdef{Gamma60.qt}{\ensuremath{(8.990 \pm 0.051) \cdot 10^{-2}}}%
\htdef{BaBar.Gamma60.pub.AUBERT.08,qt}{\ensuremath{(8.830 \pm 0.010 \pm 0.130) \cdot 10^{-2} }}%
\htdef{Belle.Gamma60.pub.LEE.10,qt}{\ensuremath{(8.420 \pm 0.000 {}^{+0.260}_{-0.250}) \cdot 10^{-2} }}%
\htdef{CLEO3.Gamma60.pub.BRIERE.03,qt}{\ensuremath{(9.130 \pm 0.050 \pm 0.460) \cdot 10^{-2} }}%
\htdef{Gamma62.qt}{\ensuremath{(8.960 \pm 0.051) \cdot 10^{-2}}}%
\htdef{Gamma63.qt}{\ensuremath{(5.327 \pm 0.049) \cdot 10^{-2}}}%
\htdef{Gamma64.qt}{\ensuremath{(5.122 \pm 0.049) \cdot 10^{-2}}}%
\htdef{Gamma65.qt}{\ensuremath{(4.791 \pm 0.052) \cdot 10^{-2}}}%
\htdef{Gamma66.qt}{\ensuremath{(4.607 \pm 0.051) \cdot 10^{-2}}}%
\htdef{ALEPH.Gamma66.pub.SCHAEL.05C,qt}{\ensuremath{(4.734 \pm 0.059 \pm 0.049) \cdot 10^{-2} }}%
\htdef{CLEO.Gamma66.pub.BALEST.95C,qt}{\ensuremath{(4.230 \pm 0.060 \pm 0.220) \cdot 10^{-2} }}%
\htdef{DELPHI.Gamma66.pub.ABDALLAH.06A,qt}{\ensuremath{(4.545 \pm 0.106 \pm 0.103) \cdot 10^{-2} }}%
\htdef{Gamma67.qt}{\ensuremath{(2.819 \pm 0.070) \cdot 10^{-2}}}%
\htdef{Gamma68.qt}{\ensuremath{(4.652 \pm 0.053) \cdot 10^{-2}}}%
\htdef{Gamma69.qt}{\ensuremath{(4.520 \pm 0.052) \cdot 10^{-2}}}%
\htdef{CLEO.Gamma69.pub.EDWARDS.00A,qt}{\ensuremath{(4.190 \pm 0.100 \pm 0.210) \cdot 10^{-2} }}%
\htdef{Gamma70.qt}{\ensuremath{(2.768 \pm 0.071) \cdot 10^{-2}}}%
\htdef{Gamma74.qt}{\ensuremath{(0.5149 \pm 0.0311) \cdot 10^{-2}}}%
\htdef{DELPHI.Gamma74.pub.ABDALLAH.06A,qt}{\ensuremath{(0.5610 \pm 0.0680 \pm 0.0950) \cdot 10^{-2} }}%
\htdef{Gamma75.qt}{\ensuremath{(0.5037 \pm 0.0309) \cdot 10^{-2}}}%
\htdef{Gamma76.qt}{\ensuremath{(0.4937 \pm 0.0309) \cdot 10^{-2}}}%
\htdef{ALEPH.Gamma76.pub.SCHAEL.05C,qt}{\ensuremath{(0.4350 \pm 0.0300 \pm 0.0350) \cdot 10^{-2} }}%
\htdef{Gamma76by54.qt}{\ensuremath{(3.247 \pm 0.202) \cdot 10^{-2}}}%
\htdef{CLEO.Gamma76by54.pub.BORTOLETTO.93,qt}{\ensuremath{(3.400 \pm 0.200 \pm 0.300) \cdot 10^{-2} }}%
\htdef{Gamma77.qt}{\ensuremath{(9.772 \pm 3.558) \cdot 10^{-4}}}%
\htdef{Gamma78.qt}{\ensuremath{(2.115 \pm 0.299) \cdot 10^{-4}}}%
\htdef{CLEO.Gamma78.pub.ANASTASSOV.01,qt}{\ensuremath{(2.200 \pm 0.300 \pm 0.400) \cdot 10^{-4} }}%
\htdef{Gamma79.qt}{\ensuremath{(0.6293 \pm 0.0140) \cdot 10^{-2}}}%
\htdef{Gamma80.qt}{\ensuremath{(0.4361 \pm 0.0072) \cdot 10^{-2}}}%
\htdef{Gamma80by60.qt}{\ensuremath{(4.851 \pm 0.080) \cdot 10^{-2}}}%
\htdef{CLEO.Gamma80by60.pub.RICHICHI.99,qt}{\ensuremath{(5.440 \pm 0.210 \pm 0.530) \cdot 10^{-2} }}%
\htdef{Gamma81.qt}{\ensuremath{(8.727 \pm 1.177) \cdot 10^{-4}}}%
\htdef{Gamma81by69.qt}{\ensuremath{(1.931 \pm 0.266) \cdot 10^{-2}}}%
\htdef{CLEO.Gamma81by69.pub.RICHICHI.99,qt}{\ensuremath{(2.610 \pm 0.450 \pm 0.420) \cdot 10^{-2} }}%
\htdef{Gamma82.qt}{\ensuremath{(0.4779 \pm 0.0137) \cdot 10^{-2}}}%
\htdef{TPC.Gamma82.pub.BAUER.94,qt}{\ensuremath{(0.5800 {}^{+0.1500}_{-0.1300} \pm 0.1200) \cdot 10^{-2} }}%
\htdef{Gamma83.qt}{\ensuremath{(0.3741 \pm 0.0135) \cdot 10^{-2}}}%
\htdef{Gamma84.qt}{\ensuremath{(0.3442 \pm 0.0068) \cdot 10^{-2}}}%
\htdef{Gamma85.qt}{\ensuremath{(0.2929 \pm 0.0067) \cdot 10^{-2}}}%
\htdef{ALEPH.Gamma85.pub.BARATE.98,qt}{\ensuremath{(0.2140 \pm 0.0370 \pm 0.0290) \cdot 10^{-2} }}%
\htdef{BaBar.Gamma85.pub.AUBERT.08,qt}{\ensuremath{(0.2730 \pm 0.0020 \pm 0.0090) \cdot 10^{-2} }}%
\htdef{Belle.Gamma85.pub.LEE.10,qt}{\ensuremath{(0.3300 \pm 0.0010 {}^{+0.0160}_{-0.0170}) \cdot 10^{-2} }}%
\htdef{CLEO3.Gamma85.pub.BRIERE.03,qt}{\ensuremath{(0.3840 \pm 0.0140 \pm 0.0380) \cdot 10^{-2} }}%
\htdef{OPAL.Gamma85.pub.ABBIENDI.04J,qt}{\ensuremath{(0.4150 \pm 0.0530 \pm 0.0400) \cdot 10^{-2} }}%
\htdef{Gamma85by60.qt}{\ensuremath{(3.258 \pm 0.074) \cdot 10^{-2}}}%
\htdef{Gamma87.qt}{\ensuremath{(0.1329 \pm 0.0119) \cdot 10^{-2}}}%
\htdef{Gamma88.qt}{\ensuremath{(8.116 \pm 1.168) \cdot 10^{-4}}}%
\htdef{ALEPH.Gamma88.pub.BARATE.98,qt}{\ensuremath{(6.100 \pm 3.900 \pm 1.800) \cdot 10^{-4} }}%
\htdef{CLEO3.Gamma88.pub.ARMS.05,qt}{\ensuremath{(7.400 \pm 0.800 \pm 1.100) \cdot 10^{-4} }}%
\htdef{Gamma89.qt}{\ensuremath{(7.762 \pm 1.168) \cdot 10^{-4}}}%
\htdef{Gamma92.qt}{\ensuremath{(0.1493 \pm 0.0033) \cdot 10^{-2}}}%
\htdef{OPAL.Gamma92.pub.ABBIENDI.00D,qt}{\ensuremath{(0.1590 \pm 0.0530 \pm 0.0200) \cdot 10^{-2} }}%
\htdef{TPC.Gamma92.pub.BAUER.94,qt}{\ensuremath{(0.1500 {}^{+0.0900}_{-0.0700} \pm 0.0300) \cdot 10^{-2} }}%
\htdef{Gamma93.qt}{\ensuremath{(0.1432 \pm 0.0027) \cdot 10^{-2}}}%
\htdef{ALEPH.Gamma93.pub.BARATE.98,qt}{\ensuremath{(0.1630 \pm 0.0210 \pm 0.0170) \cdot 10^{-2} }}%
\htdef{BaBar.Gamma93.pub.AUBERT.08,qt}{\ensuremath{(0.1346 \pm 0.0010 \pm 0.0036) \cdot 10^{-2} }}%
\htdef{Belle.Gamma93.pub.LEE.10,qt}{\ensuremath{(0.1550 \pm 0.0010 {}^{+0.0060}_{-0.0050}) \cdot 10^{-2} }}%
\htdef{CLEO3.Gamma93.pub.BRIERE.03,qt}{\ensuremath{(0.1550 \pm 0.0060 \pm 0.0090) \cdot 10^{-2} }}%
\htdef{Gamma93by60.qt}{\ensuremath{(1.592 \pm 0.030) \cdot 10^{-2}}}%
\htdef{CLEO.Gamma93by60.pub.RICHICHI.99,qt}{\ensuremath{(1.600 \pm 0.150 \pm 0.300) \cdot 10^{-2} }}%
\htdef{Gamma94.qt}{\ensuremath{(0.611 \pm 0.183) \cdot 10^{-4}}}%
\htdef{ALEPH.Gamma94.pub.BARATE.98,qt}{\ensuremath{(7.500 \pm 2.900 \pm 1.500) \cdot 10^{-4} }}%
\htdef{CLEO3.Gamma94.pub.ARMS.05,qt}{\ensuremath{(0.550 \pm 0.140 \pm 0.120) \cdot 10^{-4} }}%
\htdef{Gamma94by69.qt}{\ensuremath{(0.1353 \pm 0.0405) \cdot 10^{-2}}}%
\htdef{CLEO.Gamma94by69.pub.RICHICHI.99,qt}{\ensuremath{(0.7900 \pm 0.4400 \pm 0.1600) \cdot 10^{-2} }}%
\htdef{Gamma96.qt}{\ensuremath{(2.169 \pm 0.800) \cdot 10^{-5}}}%
\htdef{BaBar.Gamma96.pub.AUBERT.08,qt}{\ensuremath{(1.578 \pm 0.130 \pm 0.123) \cdot 10^{-5} }}%
\htdef{Belle.Gamma96.pub.LEE.10,qt}{\ensuremath{(3.290 \pm 0.170 {}^{+0.190}_{-0.200}) \cdot 10^{-5} }}%
\htdef{Gamma102.qt}{\ensuremath{(0.0990 \pm 0.0037) \cdot 10^{-2}}}%
\htdef{CLEO.Gamma102.pub.GIBAUT.94B,qt}{\ensuremath{(0.0970 \pm 0.0050 \pm 0.0110) \cdot 10^{-2} }}%
\htdef{HRS.Gamma102.pub.BYLSMA.87,qt}{\ensuremath{(0.1020 \pm 0.0290 \pm 0.0000) \cdot 10^{-2} }}%
\htdef{L3.Gamma102.pub.ACHARD.01D,qt}{\ensuremath{(0.1700 \pm 0.0220 \pm 0.0260) \cdot 10^{-2} }}%
\htdef{Gamma103.qt}{\ensuremath{(8.260 \pm 0.314) \cdot 10^{-4}}}%
\htdef{ALEPH.Gamma103.pub.SCHAEL.05C,qt}{\ensuremath{(7.200 \pm 0.900 \pm 1.200) \cdot 10^{-4} }}%
\htdef{ARGUS.Gamma103.pub.ALBRECHT.88B,qt}{\ensuremath{(6.400 \pm 2.300 \pm 1.000) \cdot 10^{-4} }}%
\htdef{CLEO.Gamma103.pub.GIBAUT.94B,qt}{\ensuremath{(7.700 \pm 0.500 \pm 0.900) \cdot 10^{-4} }}%
\htdef{DELPHI.Gamma103.pub.ABDALLAH.06A,qt}{\ensuremath{(9.700 \pm 1.500 \pm 0.500) \cdot 10^{-4} }}%
\htdef{HRS.Gamma103.pub.BYLSMA.87,qt}{\ensuremath{(5.100 \pm 2.000 \pm 0.000) \cdot 10^{-4} }}%
\htdef{OPAL.Gamma103.pub.ACKERSTAFF.99E,qt}{\ensuremath{(9.100 \pm 1.400 \pm 0.600) \cdot 10^{-4} }}%
\htdef{Gamma104.qt}{\ensuremath{(1.642 \pm 0.114) \cdot 10^{-4}}}%
\htdef{ALEPH.Gamma104.pub.SCHAEL.05C,qt}{\ensuremath{(2.100 \pm 0.700 \pm 0.900) \cdot 10^{-4} }}%
\htdef{CLEO.Gamma104.pub.ANASTASSOV.01,qt}{\ensuremath{(1.700 \pm 0.200 \pm 0.200) \cdot 10^{-4} }}%
\htdef{DELPHI.Gamma104.pub.ABDALLAH.06A,qt}{\ensuremath{(1.600 \pm 1.200 \pm 0.600) \cdot 10^{-4} }}%
\htdef{OPAL.Gamma104.pub.ACKERSTAFF.99E,qt}{\ensuremath{(2.700 \pm 1.800 \pm 0.900) \cdot 10^{-4} }}%
\htdef{Gamma106.qt}{\ensuremath{(0.7532 \pm 0.0356) \cdot 10^{-2}}}%
\htdef{Gamma110.qt}{\ensuremath{(2.931 \pm 0.041) \cdot 10^{-2}}}%
\htdef{Gamma126.qt}{\ensuremath{(0.1386 \pm 0.0072) \cdot 10^{-2}}}%
\htdef{ALEPH.Gamma126.pub.BUSKULIC.97C,qt}{\ensuremath{(0.1800 \pm 0.0400 \pm 0.0200) \cdot 10^{-2} }}%
\htdef{Belle.Gamma126.pub.INAMI.09,qt}{\ensuremath{(0.1350 \pm 0.0030 \pm 0.0070) \cdot 10^{-2} }}%
\htdef{CLEO.Gamma126.pub.ARTUSO.92,qt}{\ensuremath{(0.1700 \pm 0.0200 \pm 0.0200) \cdot 10^{-2} }}%
\htdef{Gamma128.qt}{\ensuremath{(1.543 \pm 0.080) \cdot 10^{-4}}}%
\htdef{ALEPH.Gamma128.pub.BUSKULIC.97C,qt}{\ensuremath{(2.900 {}^{+1.300}_{-1.200} \pm 0.700) \cdot 10^{-4} }}%
\htdef{BaBar.Gamma128.pub.DEL-AMO-SANCHEZ.11E,qt}{\ensuremath{(1.420 \pm 0.110 \pm 0.070) \cdot 10^{-4} }}%
\htdef{Belle.Gamma128.pub.INAMI.09,qt}{\ensuremath{(1.580 \pm 0.050 \pm 0.090) \cdot 10^{-4} }}%
\htdef{CLEO.Gamma128.pub.BARTELT.96,qt}{\ensuremath{(2.600 \pm 0.500 \pm 0.500) \cdot 10^{-4} }}%
\htdef{Gamma130.qt}{\ensuremath{(0.483 \pm 0.116) \cdot 10^{-4}}}%
\htdef{Belle.Gamma130.pub.INAMI.09,qt}{\ensuremath{(0.460 \pm 0.110 \pm 0.040) \cdot 10^{-4} }}%
\htdef{CLEO.Gamma130.pub.BISHAI.99,qt}{\ensuremath{(1.770 \pm 0.560 \pm 0.710) \cdot 10^{-4} }}%
\htdef{Gamma132.qt}{\ensuremath{(0.936 \pm 0.149) \cdot 10^{-4}}}%
\htdef{Belle.Gamma132.pub.INAMI.09,qt}{\ensuremath{(0.880 \pm 0.140 \pm 0.060) \cdot 10^{-4} }}%
\htdef{CLEO.Gamma132.pub.BISHAI.99,qt}{\ensuremath{(2.200 \pm 0.700 \pm 0.220) \cdot 10^{-4} }}%
\htdef{Gamma136.qt}{\ensuremath{(2.196 \pm 0.129) \cdot 10^{-4}}}%
\htdef{Gamma149.qt}{\ensuremath{(2.402 \pm 0.075) \cdot 10^{-2}}}%
\htdef{Gamma150.qt}{\ensuremath{(1.996 \pm 0.064) \cdot 10^{-2}}}%
\htdef{ALEPH.Gamma150.pub.BUSKULIC.97C,qt}{\ensuremath{(1.910 \pm 0.070 \pm 0.060) \cdot 10^{-2} }}%
\htdef{CLEO.Gamma150.pub.BARINGER.87,qt}{\ensuremath{(1.600 \pm 0.270 \pm 0.410) \cdot 10^{-2} }}%
\htdef{Gamma150by66.qt}{\ensuremath{0.4331 \pm 0.0139}}%
\htdef{ALEPH.Gamma150by66.pub.BUSKULIC.96,qt}{\ensuremath{0.4310 \pm 0.0330 \pm 0.0000}}%
\htdef{CLEO.Gamma150by66.pub.BALEST.95C,qt}{\ensuremath{0.4640 \pm 0.0160 \pm 0.0170}}%
\htdef{Gamma151.qt}{\ensuremath{(4.100 \pm 0.922) \cdot 10^{-4}}}%
\htdef{CLEO3.Gamma151.pub.ARMS.05,qt}{\ensuremath{(4.100 \pm 0.600 \pm 0.700) \cdot 10^{-4} }}%
\htdef{Gamma152.qt}{\ensuremath{(0.4066 \pm 0.0419) \cdot 10^{-2}}}%
\htdef{ALEPH.Gamma152.pub.BUSKULIC.97C,qt}{\ensuremath{(0.4300 \pm 0.0600 \pm 0.0500) \cdot 10^{-2} }}%
\htdef{Gamma152by54.qt}{\ensuremath{(2.674 \pm 0.275) \cdot 10^{-2}}}%
\htdef{Gamma152by76.qt}{\ensuremath{0.8236 \pm 0.0757}}%
\htdef{CLEO.Gamma152by76.pub.BORTOLETTO.93,qt}{\ensuremath{0.8100 \pm 0.0600 \pm 0.0600}}%
\htdef{Gamma167.qt}{\ensuremath{(4.409 \pm 1.626) \cdot 10^{-5}}}%
\htdef{Gamma168.qt}{\ensuremath{(2.169 \pm 0.800) \cdot 10^{-5}}}%
\htdef{Gamma169.qt}{\ensuremath{(1.499 \pm 0.553) \cdot 10^{-5}}}%
\htdef{Gamma800.qt}{\ensuremath{(1.955 \pm 0.065) \cdot 10^{-2}}}%
\htdef{Gamma802.qt}{\ensuremath{(0.2923 \pm 0.0067) \cdot 10^{-2}}}%
\htdef{Gamma803.qt}{\ensuremath{(4.101 \pm 1.429) \cdot 10^{-4}}}%
\htdef{Gamma804.qt}{\ensuremath{(2.342 \pm 0.065) \cdot 10^{-4}}}%
\htdef{Gamma805.qt}{\ensuremath{(4.000 \pm 2.000) \cdot 10^{-4}}}%
\htdef{ALEPH.Gamma805.pub.SCHAEL.05C,qt}{\ensuremath{(4.000 \pm 2.000 \pm 0.000) \cdot 10^{-4} }}%
\htdef{Gamma806.qt}{\ensuremath{(1.816 \pm 0.207) \cdot 10^{-5}}}%
\htdef{Gamma809.qt}{\ensuremath{(8.640 \pm 0.670) \cdot 10^{-4}}}%
\htdef{BaBar.Gamma809.prelim.ICHEP2018,qt}{\ensuremath{(9.020 \pm 0.400 \pm 0.650) \cdot 10^{-4} }}%
\htdef{Gamma810.qt}{\ensuremath{(1.932 \pm 0.298) \cdot 10^{-4}}}%
\htdef{Gamma811.qt}{\ensuremath{(7.139 \pm 1.586) \cdot 10^{-5}}}%
\htdef{BaBar.Gamma811.pub.LEES.12X,qt}{\ensuremath{(7.300 \pm 1.200 \pm 1.200) \cdot 10^{-5} }}%
\htdef{Gamma812.qt}{\ensuremath{(1.323 \pm 2.683) \cdot 10^{-5}}}%
\htdef{BaBar.Gamma812.pub.LEES.12X,qt}{\ensuremath{(1.000 \pm 0.800 \pm 3.000) \cdot 10^{-5} }}%
\htdef{Gamma820.qt}{\ensuremath{(8.241 \pm 0.313) \cdot 10^{-4}}}%
\htdef{Gamma821.qt}{\ensuremath{(7.719 \pm 0.295) \cdot 10^{-4}}}%
\htdef{BaBar.Gamma821.pub.LEES.12X,qt}{\ensuremath{(7.680 \pm 0.040 \pm 0.400) \cdot 10^{-4} }}%
\htdef{Gamma822.qt}{\ensuremath{(0.594 \pm 1.208) \cdot 10^{-6}}}%
\htdef{BaBar.Gamma822.pub.LEES.12X,qt}{\ensuremath{(0.600 \pm 0.500 \pm 1.100) \cdot 10^{-6} }}%
\htdef{Gamma830.qt}{\ensuremath{(1.631 \pm 0.113) \cdot 10^{-4}}}%
\htdef{Gamma831.qt}{\ensuremath{(8.399 \pm 0.624) \cdot 10^{-5}}}%
\htdef{BaBar.Gamma831.pub.LEES.12X,qt}{\ensuremath{(8.400 \pm 0.400 \pm 0.600) \cdot 10^{-5} }}%
\htdef{Gamma832.qt}{\ensuremath{(3.774 \pm 0.874) \cdot 10^{-5}}}%
\htdef{BaBar.Gamma832.pub.LEES.12X,qt}{\ensuremath{(3.600 \pm 0.300 \pm 0.900) \cdot 10^{-5} }}%
\htdef{Gamma833.qt}{\ensuremath{(1.107 \pm 0.566) \cdot 10^{-6}}}%
\htdef{BaBar.Gamma833.pub.LEES.12X,qt}{\ensuremath{(1.100 \pm 0.400 \pm 0.400) \cdot 10^{-6} }}%
\htdef{Gamma910.qt}{\ensuremath{(7.176 \pm 0.422) \cdot 10^{-5}}}%
\htdef{BaBar.Gamma910.pub.LEES.12X,qt}{\ensuremath{(8.270 \pm 0.880 \pm 0.810) \cdot 10^{-5} }}%
\htdef{Gamma911.qt}{\ensuremath{(4.443 \pm 0.867) \cdot 10^{-5}}}%
\htdef{BaBar.Gamma911.pub.LEES.12X,qt}{\ensuremath{(4.570 \pm 0.770 \pm 0.500) \cdot 10^{-5} }}%
\htdef{Gamma920.qt}{\ensuremath{(5.225 \pm 0.444) \cdot 10^{-5}}}%
\htdef{BaBar.Gamma920.pub.LEES.12X,qt}{\ensuremath{(5.200 \pm 0.310 \pm 0.370) \cdot 10^{-5} }}%
\htdef{Gamma930.qt}{\ensuremath{(5.033 \pm 0.296) \cdot 10^{-5}}}%
\htdef{BaBar.Gamma930.pub.LEES.12X,qt}{\ensuremath{(5.390 \pm 0.270 \pm 0.410) \cdot 10^{-5} }}%
\htdef{Gamma944.qt}{\ensuremath{(8.653 \pm 0.509) \cdot 10^{-5}}}%
\htdef{BaBar.Gamma944.pub.LEES.12X,qt}{\ensuremath{(8.260 \pm 0.350 \pm 0.510) \cdot 10^{-5} }}%
\htdef{Gamma945.qt}{\ensuremath{(1.939 \pm 0.378) \cdot 10^{-4}}}%
\htdef{Gamma998.qt}{\ensuremath{(0.0298 \pm 0.1026) \cdot 10^{-2}}}%
\htdef{Gamma1.qm}{%
\begin{ensuredisplaymath}
\htuse{Gamma1.gn} = \htuse{Gamma1.td}
\end{ensuredisplaymath} & & \\
\htuse{Gamma1.qt} & average &}%
\htdef{Gamma2.qm}{%
\begin{ensuredisplaymath}
\htuse{Gamma2.gn} = \htuse{Gamma2.td}
\end{ensuredisplaymath} & & \\
\htuse{Gamma2.qt} & average &}%
\htdef{Gamma3.qm}{%
\begin{ensuredisplaymath}
\htuse{Gamma3.gn} = \htuse{Gamma3.td}
\end{ensuredisplaymath} & & \\
\htuse{Gamma3.qt} & average & \\
\htuse{ALEPH.Gamma3.pub.SCHAEL.05C,qt} & \htuse{ALEPH.Gamma3.pub.SCHAEL.05C,exp} & \htuse{ALEPH.Gamma3.pub.SCHAEL.05C,ref} \\
\htuse{DELPHI.Gamma3.pub.ABREU.99X,qt} & \htuse{DELPHI.Gamma3.pub.ABREU.99X,exp} & \htuse{DELPHI.Gamma3.pub.ABREU.99X,ref} \\
\htuse{L3.Gamma3.pub.ACCIARRI.01F,qt} & \htuse{L3.Gamma3.pub.ACCIARRI.01F,exp} & \htuse{L3.Gamma3.pub.ACCIARRI.01F,ref} \\
\htuse{OPAL.Gamma3.pub.ABBIENDI.03,qt} & \htuse{OPAL.Gamma3.pub.ABBIENDI.03,exp} & \htuse{OPAL.Gamma3.pub.ABBIENDI.03,ref}
}%
\htdef{Gamma3by5.qm}{%
\begin{ensuredisplaymath}
\htuse{Gamma3by5.gn} = \htuse{Gamma3by5.td}
\end{ensuredisplaymath} & & \\
\htuse{Gamma3by5.qt} & average & \\
\htuse{ARGUS.Gamma3by5.pub.ALBRECHT.92D,qt} & \htuse{ARGUS.Gamma3by5.pub.ALBRECHT.92D,exp} & \htuse{ARGUS.Gamma3by5.pub.ALBRECHT.92D,ref} \\
\htuse{BaBar.Gamma3by5.pub.AUBERT.10F,qt} & \htuse{BaBar.Gamma3by5.pub.AUBERT.10F,exp} & \htuse{BaBar.Gamma3by5.pub.AUBERT.10F,ref} \\
\htuse{CLEO.Gamma3by5.pub.ANASTASSOV.97,qt} & \htuse{CLEO.Gamma3by5.pub.ANASTASSOV.97,exp} & \htuse{CLEO.Gamma3by5.pub.ANASTASSOV.97,ref}
}%
\htdef{Gamma5.qm}{%
\begin{ensuredisplaymath}
\htuse{Gamma5.gn} = \htuse{Gamma5.td}
\end{ensuredisplaymath} & & \\
\htuse{Gamma5.qt} & average & \\
\htuse{ALEPH.Gamma5.pub.SCHAEL.05C,qt} & \htuse{ALEPH.Gamma5.pub.SCHAEL.05C,exp} & \htuse{ALEPH.Gamma5.pub.SCHAEL.05C,ref} \\
\htuse{CLEO.Gamma5.pub.ANASTASSOV.97,qt} & \htuse{CLEO.Gamma5.pub.ANASTASSOV.97,exp} & \htuse{CLEO.Gamma5.pub.ANASTASSOV.97,ref} \\
\htuse{DELPHI.Gamma5.pub.ABREU.99X,qt} & \htuse{DELPHI.Gamma5.pub.ABREU.99X,exp} & \htuse{DELPHI.Gamma5.pub.ABREU.99X,ref} \\
\htuse{L3.Gamma5.pub.ACCIARRI.01F,qt} & \htuse{L3.Gamma5.pub.ACCIARRI.01F,exp} & \htuse{L3.Gamma5.pub.ACCIARRI.01F,ref} \\
\htuse{OPAL.Gamma5.pub.ABBIENDI.99H,qt} & \htuse{OPAL.Gamma5.pub.ABBIENDI.99H,exp} & \htuse{OPAL.Gamma5.pub.ABBIENDI.99H,ref}
}%
\htdef{Gamma7.qm}{%
\begin{ensuredisplaymath}
\htuse{Gamma7.gn} = \htuse{Gamma7.td}
\end{ensuredisplaymath} & & \\
\htuse{Gamma7.qt} & average & \\
\htuse{DELPHI.Gamma7.pub.ABREU.92N,qt} & \htuse{DELPHI.Gamma7.pub.ABREU.92N,exp} & \htuse{DELPHI.Gamma7.pub.ABREU.92N,ref} \\
\htuse{L3.Gamma7.pub.ACCIARRI.95,qt} & \htuse{L3.Gamma7.pub.ACCIARRI.95,exp} & \htuse{L3.Gamma7.pub.ACCIARRI.95,ref} \\
\htuse{OPAL.Gamma7.pub.ALEXANDER.91D,qt} & \htuse{OPAL.Gamma7.pub.ALEXANDER.91D,exp} & \htuse{OPAL.Gamma7.pub.ALEXANDER.91D,ref}
}%
\htdef{Gamma8.qm}{%
\begin{ensuredisplaymath}
\htuse{Gamma8.gn} = \htuse{Gamma8.td}
\end{ensuredisplaymath} & & \\
\htuse{Gamma8.qt} & average & \\
\htuse{ALEPH.Gamma8.pub.SCHAEL.05C,qt} & \htuse{ALEPH.Gamma8.pub.SCHAEL.05C,exp} & \htuse{ALEPH.Gamma8.pub.SCHAEL.05C,ref} \\
\htuse{CLEO.Gamma8.pub.ANASTASSOV.97,qt} & \htuse{CLEO.Gamma8.pub.ANASTASSOV.97,exp} & \htuse{CLEO.Gamma8.pub.ANASTASSOV.97,ref} \\
\htuse{DELPHI.Gamma8.pub.ABDALLAH.06A,qt} & \htuse{DELPHI.Gamma8.pub.ABDALLAH.06A,exp} & \htuse{DELPHI.Gamma8.pub.ABDALLAH.06A,ref} \\
\htuse{OPAL.Gamma8.pub.ACKERSTAFF.98M,qt} & \htuse{OPAL.Gamma8.pub.ACKERSTAFF.98M,exp} & \htuse{OPAL.Gamma8.pub.ACKERSTAFF.98M,ref}
}%
\htdef{Gamma8by5.qm}{%
\begin{ensuredisplaymath}
\htuse{Gamma8by5.gn} = \htuse{Gamma8by5.td}
\end{ensuredisplaymath} & & \\
\htuse{Gamma8by5.qt} & average &}%
\htdef{Gamma9.qm}{%
\begin{ensuredisplaymath}
\htuse{Gamma9.gn} = \htuse{Gamma9.td}
\end{ensuredisplaymath} & & \\
\htuse{Gamma9.qt} & average &}%
\htdef{Gamma9by5.qm}{%
\begin{ensuredisplaymath}
\htuse{Gamma9by5.gn} = \htuse{Gamma9by5.td}
\end{ensuredisplaymath} & & \\
\htuse{Gamma9by5.qt} & average & \\
\htuse{BaBar.Gamma9by5.pub.AUBERT.10F,qt} & \htuse{BaBar.Gamma9by5.pub.AUBERT.10F,exp} & \htuse{BaBar.Gamma9by5.pub.AUBERT.10F,ref}
}%
\htdef{Gamma10.qm}{%
\begin{ensuredisplaymath}
\htuse{Gamma10.gn} = \htuse{Gamma10.td}
\end{ensuredisplaymath} & & \\
\htuse{Gamma10.qt} & average & \\
\htuse{ALEPH.Gamma10.pub.BARATE.99K,qt} & \htuse{ALEPH.Gamma10.pub.BARATE.99K,exp} & \htuse{ALEPH.Gamma10.pub.BARATE.99K,ref} \\
\htuse{BaBar.Gamma10.prelim.ICHEP2018,qt} & \htuse{BaBar.Gamma10.prelim.ICHEP2018,exp} & \htuse{BaBar.Gamma10.prelim.ICHEP2018,ref} \\
\htuse{CLEO.Gamma10.pub.BATTLE.94,qt} & \htuse{CLEO.Gamma10.pub.BATTLE.94,exp} & \htuse{CLEO.Gamma10.pub.BATTLE.94,ref} \\
\htuse{DELPHI.Gamma10.pub.ABREU.94K,qt} & \htuse{DELPHI.Gamma10.pub.ABREU.94K,exp} & \htuse{DELPHI.Gamma10.pub.ABREU.94K,ref} \\
\htuse{OPAL.Gamma10.pub.ABBIENDI.01J,qt} & \htuse{OPAL.Gamma10.pub.ABBIENDI.01J,exp} & \htuse{OPAL.Gamma10.pub.ABBIENDI.01J,ref}
}%
\htdef{Gamma10by5.qm}{%
\begin{ensuredisplaymath}
\htuse{Gamma10by5.gn} = \htuse{Gamma10by5.td}
\end{ensuredisplaymath} & & \\
\htuse{Gamma10by5.qt} & average & \\
\htuse{BaBar.Gamma10by5.pub.AUBERT.10F,qt} & \htuse{BaBar.Gamma10by5.pub.AUBERT.10F,exp} & \htuse{BaBar.Gamma10by5.pub.AUBERT.10F,ref}
}%
\htdef{Gamma10by9.qm}{%
\begin{ensuredisplaymath}
\htuse{Gamma10by9.gn} = \htuse{Gamma10by9.td}
\end{ensuredisplaymath} & & \\
\htuse{Gamma10by9.qt} & average &}%
\htdef{Gamma11.qm}{%
\begin{ensuredisplaymath}
\htuse{Gamma11.gn} = \htuse{Gamma11.td}
\end{ensuredisplaymath} & & \\
\htuse{Gamma11.qt} & average &}%
\htdef{Gamma12.qm}{%
\begin{ensuredisplaymath}
\htuse{Gamma12.gn} = \htuse{Gamma12.td}
\end{ensuredisplaymath} & & \\
\htuse{Gamma12.qt} & average &}%
\htdef{Gamma13.qm}{%
\begin{ensuredisplaymath}
\htuse{Gamma13.gn} = \htuse{Gamma13.td}
\end{ensuredisplaymath} & & \\
\htuse{Gamma13.qt} & average & \\
\htuse{ALEPH.Gamma13.pub.SCHAEL.05C,qt} & \htuse{ALEPH.Gamma13.pub.SCHAEL.05C,exp} & \htuse{ALEPH.Gamma13.pub.SCHAEL.05C,ref} \\
\htuse{Belle.Gamma13.pub.FUJIKAWA.08,qt} & \htuse{Belle.Gamma13.pub.FUJIKAWA.08,exp} & \htuse{Belle.Gamma13.pub.FUJIKAWA.08,ref} \\
\htuse{CLEO.Gamma13.pub.ARTUSO.94,qt} & \htuse{CLEO.Gamma13.pub.ARTUSO.94,exp} & \htuse{CLEO.Gamma13.pub.ARTUSO.94,ref} \\
\htuse{DELPHI.Gamma13.pub.ABDALLAH.06A,qt} & \htuse{DELPHI.Gamma13.pub.ABDALLAH.06A,exp} & \htuse{DELPHI.Gamma13.pub.ABDALLAH.06A,ref} \\
\htuse{L3.Gamma13.pub.ACCIARRI.95,qt} & \htuse{L3.Gamma13.pub.ACCIARRI.95,exp} & \htuse{L3.Gamma13.pub.ACCIARRI.95,ref} \\
\htuse{OPAL.Gamma13.pub.ACKERSTAFF.98M,qt} & \htuse{OPAL.Gamma13.pub.ACKERSTAFF.98M,exp} & \htuse{OPAL.Gamma13.pub.ACKERSTAFF.98M,ref}
}%
\htdef{Gamma14.qm}{%
\begin{ensuredisplaymath}
\htuse{Gamma14.gn} = \htuse{Gamma14.td}
\end{ensuredisplaymath} & & \\
\htuse{Gamma14.qt} & average &}%
\htdef{Gamma16.qm}{%
\begin{ensuredisplaymath}
\htuse{Gamma16.gn} = \htuse{Gamma16.td}
\end{ensuredisplaymath} & & \\
\htuse{Gamma16.qt} & average & \\
\htuse{ALEPH.Gamma16.pub.BARATE.99K,qt} & \htuse{ALEPH.Gamma16.pub.BARATE.99K,exp} & \htuse{ALEPH.Gamma16.pub.BARATE.99K,ref} \\
\htuse{BaBar.Gamma16.prelim.ICHEP2018,qt} & \htuse{BaBar.Gamma16.prelim.ICHEP2018,exp} & \htuse{BaBar.Gamma16.prelim.ICHEP2018,ref} \\
\htuse{CLEO.Gamma16.pub.BATTLE.94,qt} & \htuse{CLEO.Gamma16.pub.BATTLE.94,exp} & \htuse{CLEO.Gamma16.pub.BATTLE.94,ref} \\
\htuse{OPAL.Gamma16.pub.ABBIENDI.04J,qt} & \htuse{OPAL.Gamma16.pub.ABBIENDI.04J,exp} & \htuse{OPAL.Gamma16.pub.ABBIENDI.04J,ref}
}%
\htdef{Gamma17.qm}{%
\begin{ensuredisplaymath}
\htuse{Gamma17.gn} = \htuse{Gamma17.td}
\end{ensuredisplaymath} & & \\
\htuse{Gamma17.qt} & average & \\
\htuse{OPAL.Gamma17.pub.ACKERSTAFF.98M,qt} & \htuse{OPAL.Gamma17.pub.ACKERSTAFF.98M,exp} & \htuse{OPAL.Gamma17.pub.ACKERSTAFF.98M,ref}
}%
\htdef{Gamma18.qm}{%
\begin{ensuredisplaymath}
\htuse{Gamma18.gn} = \htuse{Gamma18.td}
\end{ensuredisplaymath} & & \\
\htuse{Gamma18.qt} & average &}%
\htdef{Gamma19.qm}{%
\begin{ensuredisplaymath}
\htuse{Gamma19.gn} = \htuse{Gamma19.td}
\end{ensuredisplaymath} & & \\
\htuse{Gamma19.qt} & average & \\
\htuse{ALEPH.Gamma19.pub.SCHAEL.05C,qt} & \htuse{ALEPH.Gamma19.pub.SCHAEL.05C,exp} & \htuse{ALEPH.Gamma19.pub.SCHAEL.05C,ref} \\
\htuse{DELPHI.Gamma19.pub.ABDALLAH.06A,qt} & \htuse{DELPHI.Gamma19.pub.ABDALLAH.06A,exp} & \htuse{DELPHI.Gamma19.pub.ABDALLAH.06A,ref} \\
\htuse{L3.Gamma19.pub.ACCIARRI.95,qt} & \htuse{L3.Gamma19.pub.ACCIARRI.95,exp} & \htuse{L3.Gamma19.pub.ACCIARRI.95,ref}
}%
\htdef{Gamma19by13.qm}{%
\begin{ensuredisplaymath}
\htuse{Gamma19by13.gn} = \htuse{Gamma19by13.td}
\end{ensuredisplaymath} & & \\
\htuse{Gamma19by13.qt} & average & \\
\htuse{CLEO.Gamma19by13.pub.PROCARIO.93,qt} & \htuse{CLEO.Gamma19by13.pub.PROCARIO.93,exp} & \htuse{CLEO.Gamma19by13.pub.PROCARIO.93,ref}
}%
\htdef{Gamma20.qm}{%
\begin{ensuredisplaymath}
\htuse{Gamma20.gn} = \htuse{Gamma20.td}
\end{ensuredisplaymath} & & \\
\htuse{Gamma20.qt} & average &}%
\htdef{Gamma23.qm}{%
\begin{ensuredisplaymath}
\htuse{Gamma23.gn} = \htuse{Gamma23.td}
\end{ensuredisplaymath} & & \\
\htuse{Gamma23.qt} & average & \\
\htuse{ALEPH.Gamma23.pub.BARATE.99K,qt} & \htuse{ALEPH.Gamma23.pub.BARATE.99K,exp} & \htuse{ALEPH.Gamma23.pub.BARATE.99K,ref} \\
\htuse{BaBar.Gamma23.prelim.ICHEP2018,qt} & \htuse{BaBar.Gamma23.prelim.ICHEP2018,exp} & \htuse{BaBar.Gamma23.prelim.ICHEP2018,ref} \\
\htuse{CLEO.Gamma23.pub.BATTLE.94,qt} & \htuse{CLEO.Gamma23.pub.BATTLE.94,exp} & \htuse{CLEO.Gamma23.pub.BATTLE.94,ref}
}%
\htdef{Gamma24.qm}{%
\begin{ensuredisplaymath}
\htuse{Gamma24.gn} = \htuse{Gamma24.td}
\end{ensuredisplaymath} & & \\
\htuse{Gamma24.qt} & average &}%
\htdef{Gamma25.qm}{%
\begin{ensuredisplaymath}
\htuse{Gamma25.gn} = \htuse{Gamma25.td}
\end{ensuredisplaymath} & & \\
\htuse{Gamma25.qt} & average & \\
\htuse{DELPHI.Gamma25.pub.ABDALLAH.06A,qt} & \htuse{DELPHI.Gamma25.pub.ABDALLAH.06A,exp} & \htuse{DELPHI.Gamma25.pub.ABDALLAH.06A,ref}
}%
\htdef{Gamma26.qm}{%
\begin{ensuredisplaymath}
\htuse{Gamma26.gn} = \htuse{Gamma26.td}
\end{ensuredisplaymath} & & \\
\htuse{Gamma26.qt} & average & \\
\htuse{ALEPH.Gamma26.pub.SCHAEL.05C,qt} & \htuse{ALEPH.Gamma26.pub.SCHAEL.05C,exp} & \htuse{ALEPH.Gamma26.pub.SCHAEL.05C,ref} \\
\htuse{L3.Gamma26.pub.ACCIARRI.95,qt} & \htuse{L3.Gamma26.pub.ACCIARRI.95,exp} & \htuse{L3.Gamma26.pub.ACCIARRI.95,ref}
}%
\htdef{Gamma26by13.qm}{%
\begin{ensuredisplaymath}
\htuse{Gamma26by13.gn} = \htuse{Gamma26by13.td}
\end{ensuredisplaymath} & & \\
\htuse{Gamma26by13.qt} & average & \\
\htuse{CLEO.Gamma26by13.pub.PROCARIO.93,qt} & \htuse{CLEO.Gamma26by13.pub.PROCARIO.93,exp} & \htuse{CLEO.Gamma26by13.pub.PROCARIO.93,ref}
}%
\htdef{Gamma27.qm}{%
\begin{ensuredisplaymath}
\htuse{Gamma27.gn} = \htuse{Gamma27.td}
\end{ensuredisplaymath} & & \\
\htuse{Gamma27.qt} & average & \\
\htuse{BaBar.Gamma27.prelim.ICHEP2018,qt} & \htuse{BaBar.Gamma27.prelim.ICHEP2018,exp} & \htuse{BaBar.Gamma27.prelim.ICHEP2018,ref}
}%
\htdef{Gamma28.qm}{%
\begin{ensuredisplaymath}
\htuse{Gamma28.gn} = \htuse{Gamma28.td}
\end{ensuredisplaymath} & & \\
\htuse{Gamma28.qt} & average & \\
\htuse{ALEPH.Gamma28.pub.BARATE.99K,qt} & \htuse{ALEPH.Gamma28.pub.BARATE.99K,exp} & \htuse{ALEPH.Gamma28.pub.BARATE.99K,ref} \\
\htuse{BaBar.Gamma28.prelim.ICHEP2018,qt} & \htuse{BaBar.Gamma28.prelim.ICHEP2018,exp} & \htuse{BaBar.Gamma28.prelim.ICHEP2018,ref}
}%
\htdef{Gamma29.qm}{%
\begin{ensuredisplaymath}
\htuse{Gamma29.gn} = \htuse{Gamma29.td}
\end{ensuredisplaymath} & & \\
\htuse{Gamma29.qt} & average & \\
\htuse{CLEO.Gamma29.pub.PROCARIO.93,qt} & \htuse{CLEO.Gamma29.pub.PROCARIO.93,exp} & \htuse{CLEO.Gamma29.pub.PROCARIO.93,ref}
}%
\htdef{Gamma30.qm}{%
\begin{ensuredisplaymath}
\htuse{Gamma30.gn} = \htuse{Gamma30.td}
\end{ensuredisplaymath} & & \\
\htuse{Gamma30.qt} & average & \\
\htuse{ALEPH.Gamma30.pub.SCHAEL.05C,qt} & \htuse{ALEPH.Gamma30.pub.SCHAEL.05C,exp} & \htuse{ALEPH.Gamma30.pub.SCHAEL.05C,ref}
}%
\htdef{Gamma31.qm}{%
\begin{ensuredisplaymath}
\htuse{Gamma31.gn} = \htuse{Gamma31.td}
\end{ensuredisplaymath} & & \\
\htuse{Gamma31.qt} & average & \\
\htuse{CLEO.Gamma31.pub.BATTLE.94,qt} & \htuse{CLEO.Gamma31.pub.BATTLE.94,exp} & \htuse{CLEO.Gamma31.pub.BATTLE.94,ref} \\
\htuse{DELPHI.Gamma31.pub.ABREU.94K,qt} & \htuse{DELPHI.Gamma31.pub.ABREU.94K,exp} & \htuse{DELPHI.Gamma31.pub.ABREU.94K,ref} \\
\htuse{OPAL.Gamma31.pub.ABBIENDI.01J,qt} & \htuse{OPAL.Gamma31.pub.ABBIENDI.01J,exp} & \htuse{OPAL.Gamma31.pub.ABBIENDI.01J,ref}
}%
\htdef{Gamma32.qm}{%
\begin{ensuredisplaymath}
\htuse{Gamma32.gn} = \htuse{Gamma32.td}
\end{ensuredisplaymath} & & \\
\htuse{Gamma32.qt} & average &}%
\htdef{Gamma33.qm}{%
\begin{ensuredisplaymath}
\htuse{Gamma33.gn} = \htuse{Gamma33.td}
\end{ensuredisplaymath} & & \\
\htuse{Gamma33.qt} & average & \\
\htuse{ALEPH.Gamma33.pub.BARATE.98E,qt} & \htuse{ALEPH.Gamma33.pub.BARATE.98E,exp} & \htuse{ALEPH.Gamma33.pub.BARATE.98E,ref} \\
\htuse{OPAL.Gamma33.pub.AKERS.94G,qt} & \htuse{OPAL.Gamma33.pub.AKERS.94G,exp} & \htuse{OPAL.Gamma33.pub.AKERS.94G,ref}
}%
\htdef{Gamma34.qm}{%
\begin{ensuredisplaymath}
\htuse{Gamma34.gn} = \htuse{Gamma34.td}
\end{ensuredisplaymath} & & \\
\htuse{Gamma34.qt} & average & \\
\htuse{CLEO.Gamma34.pub.COAN.96,qt} & \htuse{CLEO.Gamma34.pub.COAN.96,exp} & \htuse{CLEO.Gamma34.pub.COAN.96,ref}
}%
\htdef{Gamma35.qm}{%
\begin{ensuredisplaymath}
\htuse{Gamma35.gn} = \htuse{Gamma35.td}
\end{ensuredisplaymath} & & \\
\htuse{Gamma35.qt} & average & \\
\htuse{ALEPH.Gamma35.pub.BARATE.99K,qt} & \htuse{ALEPH.Gamma35.pub.BARATE.99K,exp} & \htuse{ALEPH.Gamma35.pub.BARATE.99K,ref} \\
\htuse{Belle.Gamma35.pub.RYU.14vpc,qt} & \htuse{Belle.Gamma35.pub.RYU.14vpc,exp} & \htuse{Belle.Gamma35.pub.RYU.14vpc,ref} \\
\htuse{L3.Gamma35.pub.ACCIARRI.95F,qt} & \htuse{L3.Gamma35.pub.ACCIARRI.95F,exp} & \htuse{L3.Gamma35.pub.ACCIARRI.95F,ref} \\
\htuse{OPAL.Gamma35.pub.ABBIENDI.00C,qt} & \htuse{OPAL.Gamma35.pub.ABBIENDI.00C,exp} & \htuse{OPAL.Gamma35.pub.ABBIENDI.00C,ref}
}%
\htdef{Gamma37.qm}{%
\begin{ensuredisplaymath}
\htuse{Gamma37.gn} = \htuse{Gamma37.td}
\end{ensuredisplaymath} & & \\
\htuse{Gamma37.qt} & average & \\
\htuse{ALEPH.Gamma37.pub.BARATE.98E,qt} & \htuse{ALEPH.Gamma37.pub.BARATE.98E,exp} & \htuse{ALEPH.Gamma37.pub.BARATE.98E,ref} \\
\htuse{ALEPH.Gamma37.pub.BARATE.99K,qt} & \htuse{ALEPH.Gamma37.pub.BARATE.99K,exp} & \htuse{ALEPH.Gamma37.pub.BARATE.99K,ref} \\
\htuse{BaBar.Gamma37.pub.LEES.18B,qt} & \htuse{BaBar.Gamma37.pub.LEES.18B,exp} & \htuse{BaBar.Gamma37.pub.LEES.18B,ref} \\
\htuse{Belle.Gamma37.pub.RYU.14vpc,qt} & \htuse{Belle.Gamma37.pub.RYU.14vpc,exp} & \htuse{Belle.Gamma37.pub.RYU.14vpc,ref} \\
\htuse{CLEO.Gamma37.pub.COAN.96,qt} & \htuse{CLEO.Gamma37.pub.COAN.96,exp} & \htuse{CLEO.Gamma37.pub.COAN.96,ref}
}%
\htdef{Gamma38.qm}{%
\begin{ensuredisplaymath}
\htuse{Gamma38.gn} = \htuse{Gamma38.td}
\end{ensuredisplaymath} & & \\
\htuse{Gamma38.qt} & average & \\
\htuse{OPAL.Gamma38.pub.ABBIENDI.00C,qt} & \htuse{OPAL.Gamma38.pub.ABBIENDI.00C,exp} & \htuse{OPAL.Gamma38.pub.ABBIENDI.00C,ref}
}%
\htdef{Gamma39.qm}{%
\begin{ensuredisplaymath}
\htuse{Gamma39.gn} = \htuse{Gamma39.td}
\end{ensuredisplaymath} & & \\
\htuse{Gamma39.qt} & average & \\
\htuse{CLEO.Gamma39.pub.COAN.96,qt} & \htuse{CLEO.Gamma39.pub.COAN.96,exp} & \htuse{CLEO.Gamma39.pub.COAN.96,ref}
}%
\htdef{Gamma40.qm}{%
\begin{ensuredisplaymath}
\htuse{Gamma40.gn} = \htuse{Gamma40.td}
\end{ensuredisplaymath} & & \\
\htuse{Gamma40.qt} & average & \\
\htuse{ALEPH.Gamma40.pub.BARATE.98E,qt} & \htuse{ALEPH.Gamma40.pub.BARATE.98E,exp} & \htuse{ALEPH.Gamma40.pub.BARATE.98E,ref} \\
\htuse{ALEPH.Gamma40.pub.BARATE.99K,qt} & \htuse{ALEPH.Gamma40.pub.BARATE.99K,exp} & \htuse{ALEPH.Gamma40.pub.BARATE.99K,ref} \\
\htuse{Belle.Gamma40.pub.RYU.14vpc,qt} & \htuse{Belle.Gamma40.pub.RYU.14vpc,exp} & \htuse{Belle.Gamma40.pub.RYU.14vpc,ref} \\
\htuse{L3.Gamma40.pub.ACCIARRI.95F,qt} & \htuse{L3.Gamma40.pub.ACCIARRI.95F,exp} & \htuse{L3.Gamma40.pub.ACCIARRI.95F,ref}
}%
\htdef{Gamma42.qm}{%
\begin{ensuredisplaymath}
\htuse{Gamma42.gn} = \htuse{Gamma42.td}
\end{ensuredisplaymath} & & \\
\htuse{Gamma42.qt} & average & \\
\htuse{ALEPH.Gamma42.pub.BARATE.98E,qt} & \htuse{ALEPH.Gamma42.pub.BARATE.98E,exp} & \htuse{ALEPH.Gamma42.pub.BARATE.98E,ref} \\
\htuse{ALEPH.Gamma42.pub.BARATE.99K,qt} & \htuse{ALEPH.Gamma42.pub.BARATE.99K,exp} & \htuse{ALEPH.Gamma42.pub.BARATE.99K,ref} \\
\htuse{Belle.Gamma42.pub.RYU.14vpc,qt} & \htuse{Belle.Gamma42.pub.RYU.14vpc,exp} & \htuse{Belle.Gamma42.pub.RYU.14vpc,ref} \\
\htuse{CLEO.Gamma42.pub.COAN.96,qt} & \htuse{CLEO.Gamma42.pub.COAN.96,exp} & \htuse{CLEO.Gamma42.pub.COAN.96,ref}
}%
\htdef{Gamma43.qm}{%
\begin{ensuredisplaymath}
\htuse{Gamma43.gn} = \htuse{Gamma43.td}
\end{ensuredisplaymath} & & \\
\htuse{Gamma43.qt} & average & \\
\htuse{OPAL.Gamma43.pub.ABBIENDI.00C,qt} & \htuse{OPAL.Gamma43.pub.ABBIENDI.00C,exp} & \htuse{OPAL.Gamma43.pub.ABBIENDI.00C,ref}
}%
\htdef{Gamma44.qm}{%
\begin{ensuredisplaymath}
\htuse{Gamma44.gn} = \htuse{Gamma44.td}
\end{ensuredisplaymath} & & \\
\htuse{Gamma44.qt} & average & \\
\htuse{ALEPH.Gamma44.pub.BARATE.99R,qt} & \htuse{ALEPH.Gamma44.pub.BARATE.99R,exp} & \htuse{ALEPH.Gamma44.pub.BARATE.99R,ref}
}%
\htdef{Gamma46.qm}{%
\begin{ensuredisplaymath}
\htuse{Gamma46.gn} = \htuse{Gamma46.td}
\end{ensuredisplaymath} & & \\
\htuse{Gamma46.qt} & average &}%
\htdef{Gamma47.qm}{%
\begin{ensuredisplaymath}
\htuse{Gamma47.gn} = \htuse{Gamma47.td}
\end{ensuredisplaymath} & & \\
\htuse{Gamma47.qt} & average & \\
\htuse{ALEPH.Gamma47.pub.BARATE.98E,qt} & \htuse{ALEPH.Gamma47.pub.BARATE.98E,exp} & \htuse{ALEPH.Gamma47.pub.BARATE.98E,ref} \\
\htuse{BaBar.Gamma47.pub.LEES.12Y,qt} & \htuse{BaBar.Gamma47.pub.LEES.12Y,exp} & \htuse{BaBar.Gamma47.pub.LEES.12Y,ref} \\
\htuse{Belle.Gamma47.pub.RYU.14vpc,qt} & \htuse{Belle.Gamma47.pub.RYU.14vpc,exp} & \htuse{Belle.Gamma47.pub.RYU.14vpc,ref} \\
\htuse{CLEO.Gamma47.pub.COAN.96,qt} & \htuse{CLEO.Gamma47.pub.COAN.96,exp} & \htuse{CLEO.Gamma47.pub.COAN.96,ref}
}%
\htdef{Gamma48.qm}{%
\begin{ensuredisplaymath}
\htuse{Gamma48.gn} = \htuse{Gamma48.td}
\end{ensuredisplaymath} & & \\
\htuse{Gamma48.qt} & average & \\
\htuse{ALEPH.Gamma48.pub.BARATE.98E,qt} & \htuse{ALEPH.Gamma48.pub.BARATE.98E,exp} & \htuse{ALEPH.Gamma48.pub.BARATE.98E,ref}
}%
\htdef{Gamma49.qm}{%
\begin{ensuredisplaymath}
\htuse{Gamma49.gn} = \htuse{Gamma49.td}
\end{ensuredisplaymath} & & \\
\htuse{Gamma49.qt} & average &}%
\htdef{Gamma50.qm}{%
\begin{ensuredisplaymath}
\htuse{Gamma50.gn} = \htuse{Gamma50.td}
\end{ensuredisplaymath} & & \\
\htuse{Gamma50.qt} & average & \\
\htuse{BaBar.Gamma50.pub.LEES.12Y,qt} & \htuse{BaBar.Gamma50.pub.LEES.12Y,exp} & \htuse{BaBar.Gamma50.pub.LEES.12Y,ref} \\
\htuse{Belle.Gamma50.pub.RYU.14vpc,qt} & \htuse{Belle.Gamma50.pub.RYU.14vpc,exp} & \htuse{Belle.Gamma50.pub.RYU.14vpc,ref}
}%
\htdef{Gamma51.qm}{%
\begin{ensuredisplaymath}
\htuse{Gamma51.gn} = \htuse{Gamma51.td}
\end{ensuredisplaymath} & & \\
\htuse{Gamma51.qt} & average & \\
\htuse{ALEPH.Gamma51.pub.BARATE.98E,qt} & \htuse{ALEPH.Gamma51.pub.BARATE.98E,exp} & \htuse{ALEPH.Gamma51.pub.BARATE.98E,ref}
}%
\htdef{Gamma53.qm}{%
\begin{ensuredisplaymath}
\htuse{Gamma53.gn} = \htuse{Gamma53.td}
\end{ensuredisplaymath} & & \\
\htuse{Gamma53.qt} & average & \\
\htuse{ALEPH.Gamma53.pub.BARATE.98E,qt} & \htuse{ALEPH.Gamma53.pub.BARATE.98E,exp} & \htuse{ALEPH.Gamma53.pub.BARATE.98E,ref}
}%
\htdef{Gamma54.qm}{%
\begin{ensuredisplaymath}
\htuse{Gamma54.gn} = \htuse{Gamma54.td}
\end{ensuredisplaymath} & & \\
\htuse{Gamma54.qt} & average & \\
\htuse{CELLO.Gamma54.pub.BEHREND.89B,qt} & \htuse{CELLO.Gamma54.pub.BEHREND.89B,exp} & \htuse{CELLO.Gamma54.pub.BEHREND.89B,ref} \\
\htuse{L3.Gamma54.pub.ADEVA.91F,qt} & \htuse{L3.Gamma54.pub.ADEVA.91F,exp} & \htuse{L3.Gamma54.pub.ADEVA.91F,ref} \\
\htuse{TPC.Gamma54.pub.AIHARA.87B,qt} & \htuse{TPC.Gamma54.pub.AIHARA.87B,exp} & \htuse{TPC.Gamma54.pub.AIHARA.87B,ref}
}%
\htdef{Gamma55.qm}{%
\begin{ensuredisplaymath}
\htuse{Gamma55.gn} = \htuse{Gamma55.td}
\end{ensuredisplaymath} & & \\
\htuse{Gamma55.qt} & average & \\
\htuse{L3.Gamma55.pub.ACHARD.01D,qt} & \htuse{L3.Gamma55.pub.ACHARD.01D,exp} & \htuse{L3.Gamma55.pub.ACHARD.01D,ref} \\
\htuse{OPAL.Gamma55.pub.AKERS.95Y,qt} & \htuse{OPAL.Gamma55.pub.AKERS.95Y,exp} & \htuse{OPAL.Gamma55.pub.AKERS.95Y,ref}
}%
\htdef{Gamma56.qm}{%
\begin{ensuredisplaymath}
\htuse{Gamma56.gn} = \htuse{Gamma56.td}
\end{ensuredisplaymath} & & \\
\htuse{Gamma56.qt} & average &}%
\htdef{Gamma57.qm}{%
\begin{ensuredisplaymath}
\htuse{Gamma57.gn} = \htuse{Gamma57.td}
\end{ensuredisplaymath} & & \\
\htuse{Gamma57.qt} & average & \\
\htuse{CLEO.Gamma57.pub.BALEST.95C,qt} & \htuse{CLEO.Gamma57.pub.BALEST.95C,exp} & \htuse{CLEO.Gamma57.pub.BALEST.95C,ref} \\
\htuse{DELPHI.Gamma57.pub.ABDALLAH.06A,qt} & \htuse{DELPHI.Gamma57.pub.ABDALLAH.06A,exp} & \htuse{DELPHI.Gamma57.pub.ABDALLAH.06A,ref}
}%
\htdef{Gamma57by55.qm}{%
\begin{ensuredisplaymath}
\htuse{Gamma57by55.gn} = \htuse{Gamma57by55.td}
\end{ensuredisplaymath} & & \\
\htuse{Gamma57by55.qt} & average & \\
\htuse{OPAL.Gamma57by55.pub.AKERS.95Y,qt} & \htuse{OPAL.Gamma57by55.pub.AKERS.95Y,exp} & \htuse{OPAL.Gamma57by55.pub.AKERS.95Y,ref}
}%
\htdef{Gamma58.qm}{%
\begin{ensuredisplaymath}
\htuse{Gamma58.gn} = \htuse{Gamma58.td}
\end{ensuredisplaymath} & & \\
\htuse{Gamma58.qt} & average & \\
\htuse{ALEPH.Gamma58.pub.SCHAEL.05C,qt} & \htuse{ALEPH.Gamma58.pub.SCHAEL.05C,exp} & \htuse{ALEPH.Gamma58.pub.SCHAEL.05C,ref}
}%
\htdef{Gamma59.qm}{%
\begin{ensuredisplaymath}
\htuse{Gamma59.gn} = \htuse{Gamma59.td}
\end{ensuredisplaymath} & & \\
\htuse{Gamma59.qt} & average &}%
\htdef{Gamma60.qm}{%
\begin{ensuredisplaymath}
\htuse{Gamma60.gn} = \htuse{Gamma60.td}
\end{ensuredisplaymath} & & \\
\htuse{Gamma60.qt} & average & \\
\htuse{BaBar.Gamma60.pub.AUBERT.08,qt} & \htuse{BaBar.Gamma60.pub.AUBERT.08,exp} & \htuse{BaBar.Gamma60.pub.AUBERT.08,ref} \\
\htuse{Belle.Gamma60.pub.LEE.10,qt} & \htuse{Belle.Gamma60.pub.LEE.10,exp} & \htuse{Belle.Gamma60.pub.LEE.10,ref} \\
\htuse{CLEO3.Gamma60.pub.BRIERE.03,qt} & \htuse{CLEO3.Gamma60.pub.BRIERE.03,exp} & \htuse{CLEO3.Gamma60.pub.BRIERE.03,ref}
}%
\htdef{Gamma62.qm}{%
\begin{ensuredisplaymath}
\htuse{Gamma62.gn} = \htuse{Gamma62.td}
\end{ensuredisplaymath} & & \\
\htuse{Gamma62.qt} & average &}%
\htdef{Gamma63.qm}{%
\begin{ensuredisplaymath}
\htuse{Gamma63.gn} = \htuse{Gamma63.td}
\end{ensuredisplaymath} & & \\
\htuse{Gamma63.qt} & average &}%
\htdef{Gamma64.qm}{%
\begin{ensuredisplaymath}
\htuse{Gamma64.gn} = \htuse{Gamma64.td}
\end{ensuredisplaymath} & & \\
\htuse{Gamma64.qt} & average &}%
\htdef{Gamma65.qm}{%
\begin{ensuredisplaymath}
\htuse{Gamma65.gn} = \htuse{Gamma65.td}
\end{ensuredisplaymath} & & \\
\htuse{Gamma65.qt} & average &}%
\htdef{Gamma66.qm}{%
\begin{ensuredisplaymath}
\htuse{Gamma66.gn} = \htuse{Gamma66.td}
\end{ensuredisplaymath} & & \\
\htuse{Gamma66.qt} & average & \\
\htuse{ALEPH.Gamma66.pub.SCHAEL.05C,qt} & \htuse{ALEPH.Gamma66.pub.SCHAEL.05C,exp} & \htuse{ALEPH.Gamma66.pub.SCHAEL.05C,ref} \\
\htuse{CLEO.Gamma66.pub.BALEST.95C,qt} & \htuse{CLEO.Gamma66.pub.BALEST.95C,exp} & \htuse{CLEO.Gamma66.pub.BALEST.95C,ref} \\
\htuse{DELPHI.Gamma66.pub.ABDALLAH.06A,qt} & \htuse{DELPHI.Gamma66.pub.ABDALLAH.06A,exp} & \htuse{DELPHI.Gamma66.pub.ABDALLAH.06A,ref}
}%
\htdef{Gamma67.qm}{%
\begin{ensuredisplaymath}
\htuse{Gamma67.gn} = \htuse{Gamma67.td}
\end{ensuredisplaymath} & & \\
\htuse{Gamma67.qt} & average &}%
\htdef{Gamma68.qm}{%
\begin{ensuredisplaymath}
\htuse{Gamma68.gn} = \htuse{Gamma68.td}
\end{ensuredisplaymath} & & \\
\htuse{Gamma68.qt} & average &}%
\htdef{Gamma69.qm}{%
\begin{ensuredisplaymath}
\htuse{Gamma69.gn} = \htuse{Gamma69.td}
\end{ensuredisplaymath} & & \\
\htuse{Gamma69.qt} & average & \\
\htuse{CLEO.Gamma69.pub.EDWARDS.00A,qt} & \htuse{CLEO.Gamma69.pub.EDWARDS.00A,exp} & \htuse{CLEO.Gamma69.pub.EDWARDS.00A,ref}
}%
\htdef{Gamma70.qm}{%
\begin{ensuredisplaymath}
\htuse{Gamma70.gn} = \htuse{Gamma70.td}
\end{ensuredisplaymath} & & \\
\htuse{Gamma70.qt} & average &}%
\htdef{Gamma74.qm}{%
\begin{ensuredisplaymath}
\htuse{Gamma74.gn} = \htuse{Gamma74.td}
\end{ensuredisplaymath} & & \\
\htuse{Gamma74.qt} & average & \\
\htuse{DELPHI.Gamma74.pub.ABDALLAH.06A,qt} & \htuse{DELPHI.Gamma74.pub.ABDALLAH.06A,exp} & \htuse{DELPHI.Gamma74.pub.ABDALLAH.06A,ref}
}%
\htdef{Gamma75.qm}{%
\begin{ensuredisplaymath}
\htuse{Gamma75.gn} = \htuse{Gamma75.td}
\end{ensuredisplaymath} & & \\
\htuse{Gamma75.qt} & average &}%
\htdef{Gamma76.qm}{%
\begin{ensuredisplaymath}
\htuse{Gamma76.gn} = \htuse{Gamma76.td}
\end{ensuredisplaymath} & & \\
\htuse{Gamma76.qt} & average & \\
\htuse{ALEPH.Gamma76.pub.SCHAEL.05C,qt} & \htuse{ALEPH.Gamma76.pub.SCHAEL.05C,exp} & \htuse{ALEPH.Gamma76.pub.SCHAEL.05C,ref}
}%
\htdef{Gamma76by54.qm}{%
\begin{ensuredisplaymath}
\htuse{Gamma76by54.gn} = \htuse{Gamma76by54.td}
\end{ensuredisplaymath} & & \\
\htuse{Gamma76by54.qt} & average & \\
\htuse{CLEO.Gamma76by54.pub.BORTOLETTO.93,qt} & \htuse{CLEO.Gamma76by54.pub.BORTOLETTO.93,exp} & \htuse{CLEO.Gamma76by54.pub.BORTOLETTO.93,ref}
}%
\htdef{Gamma77.qm}{%
\begin{ensuredisplaymath}
\htuse{Gamma77.gn} = \htuse{Gamma77.td}
\end{ensuredisplaymath} & & \\
\htuse{Gamma77.qt} & average &}%
\htdef{Gamma78.qm}{%
\begin{ensuredisplaymath}
\htuse{Gamma78.gn} = \htuse{Gamma78.td}
\end{ensuredisplaymath} & & \\
\htuse{Gamma78.qt} & average & \\
\htuse{CLEO.Gamma78.pub.ANASTASSOV.01,qt} & \htuse{CLEO.Gamma78.pub.ANASTASSOV.01,exp} & \htuse{CLEO.Gamma78.pub.ANASTASSOV.01,ref}
}%
\htdef{Gamma79.qm}{%
\begin{ensuredisplaymath}
\htuse{Gamma79.gn} = \htuse{Gamma79.td}
\end{ensuredisplaymath} & & \\
\htuse{Gamma79.qt} & average &}%
\htdef{Gamma80.qm}{%
\begin{ensuredisplaymath}
\htuse{Gamma80.gn} = \htuse{Gamma80.td}
\end{ensuredisplaymath} & & \\
\htuse{Gamma80.qt} & average &}%
\htdef{Gamma80by60.qm}{%
\begin{ensuredisplaymath}
\htuse{Gamma80by60.gn} = \htuse{Gamma80by60.td}
\end{ensuredisplaymath} & & \\
\htuse{Gamma80by60.qt} & average & \\
\htuse{CLEO.Gamma80by60.pub.RICHICHI.99,qt} & \htuse{CLEO.Gamma80by60.pub.RICHICHI.99,exp} & \htuse{CLEO.Gamma80by60.pub.RICHICHI.99,ref}
}%
\htdef{Gamma81.qm}{%
\begin{ensuredisplaymath}
\htuse{Gamma81.gn} = \htuse{Gamma81.td}
\end{ensuredisplaymath} & & \\
\htuse{Gamma81.qt} & average &}%
\htdef{Gamma81by69.qm}{%
\begin{ensuredisplaymath}
\htuse{Gamma81by69.gn} = \htuse{Gamma81by69.td}
\end{ensuredisplaymath} & & \\
\htuse{Gamma81by69.qt} & average & \\
\htuse{CLEO.Gamma81by69.pub.RICHICHI.99,qt} & \htuse{CLEO.Gamma81by69.pub.RICHICHI.99,exp} & \htuse{CLEO.Gamma81by69.pub.RICHICHI.99,ref}
}%
\htdef{Gamma82.qm}{%
\begin{ensuredisplaymath}
\htuse{Gamma82.gn} = \htuse{Gamma82.td}
\end{ensuredisplaymath} & & \\
\htuse{Gamma82.qt} & average & \\
\htuse{TPC.Gamma82.pub.BAUER.94,qt} & \htuse{TPC.Gamma82.pub.BAUER.94,exp} & \htuse{TPC.Gamma82.pub.BAUER.94,ref}
}%
\htdef{Gamma83.qm}{%
\begin{ensuredisplaymath}
\htuse{Gamma83.gn} = \htuse{Gamma83.td}
\end{ensuredisplaymath} & & \\
\htuse{Gamma83.qt} & average &}%
\htdef{Gamma84.qm}{%
\begin{ensuredisplaymath}
\htuse{Gamma84.gn} = \htuse{Gamma84.td}
\end{ensuredisplaymath} & & \\
\htuse{Gamma84.qt} & average &}%
\htdef{Gamma85.qm}{%
\begin{ensuredisplaymath}
\htuse{Gamma85.gn} = \htuse{Gamma85.td}
\end{ensuredisplaymath} & & \\
\htuse{Gamma85.qt} & average & \\
\htuse{ALEPH.Gamma85.pub.BARATE.98,qt} & \htuse{ALEPH.Gamma85.pub.BARATE.98,exp} & \htuse{ALEPH.Gamma85.pub.BARATE.98,ref} \\
\htuse{BaBar.Gamma85.pub.AUBERT.08,qt} & \htuse{BaBar.Gamma85.pub.AUBERT.08,exp} & \htuse{BaBar.Gamma85.pub.AUBERT.08,ref} \\
\htuse{Belle.Gamma85.pub.LEE.10,qt} & \htuse{Belle.Gamma85.pub.LEE.10,exp} & \htuse{Belle.Gamma85.pub.LEE.10,ref} \\
\htuse{CLEO3.Gamma85.pub.BRIERE.03,qt} & \htuse{CLEO3.Gamma85.pub.BRIERE.03,exp} & \htuse{CLEO3.Gamma85.pub.BRIERE.03,ref} \\
\htuse{OPAL.Gamma85.pub.ABBIENDI.04J,qt} & \htuse{OPAL.Gamma85.pub.ABBIENDI.04J,exp} & \htuse{OPAL.Gamma85.pub.ABBIENDI.04J,ref}
}%
\htdef{Gamma85by60.qm}{%
\begin{ensuredisplaymath}
\htuse{Gamma85by60.gn} = \htuse{Gamma85by60.td}
\end{ensuredisplaymath} & & \\
\htuse{Gamma85by60.qt} & average &}%
\htdef{Gamma87.qm}{%
\begin{ensuredisplaymath}
\htuse{Gamma87.gn} = \htuse{Gamma87.td}
\end{ensuredisplaymath} & & \\
\htuse{Gamma87.qt} & average &}%
\htdef{Gamma88.qm}{%
\begin{ensuredisplaymath}
\htuse{Gamma88.gn} = \htuse{Gamma88.td}
\end{ensuredisplaymath} & & \\
\htuse{Gamma88.qt} & average & \\
\htuse{ALEPH.Gamma88.pub.BARATE.98,qt} & \htuse{ALEPH.Gamma88.pub.BARATE.98,exp} & \htuse{ALEPH.Gamma88.pub.BARATE.98,ref} \\
\htuse{CLEO3.Gamma88.pub.ARMS.05,qt} & \htuse{CLEO3.Gamma88.pub.ARMS.05,exp} & \htuse{CLEO3.Gamma88.pub.ARMS.05,ref}
}%
\htdef{Gamma89.qm}{%
\begin{ensuredisplaymath}
\htuse{Gamma89.gn} = \htuse{Gamma89.td}
\end{ensuredisplaymath} & & \\
\htuse{Gamma89.qt} & average &}%
\htdef{Gamma92.qm}{%
\begin{ensuredisplaymath}
\htuse{Gamma92.gn} = \htuse{Gamma92.td}
\end{ensuredisplaymath} & & \\
\htuse{Gamma92.qt} & average & \\
\htuse{OPAL.Gamma92.pub.ABBIENDI.00D,qt} & \htuse{OPAL.Gamma92.pub.ABBIENDI.00D,exp} & \htuse{OPAL.Gamma92.pub.ABBIENDI.00D,ref} \\
\htuse{TPC.Gamma92.pub.BAUER.94,qt} & \htuse{TPC.Gamma92.pub.BAUER.94,exp} & \htuse{TPC.Gamma92.pub.BAUER.94,ref}
}%
\htdef{Gamma93.qm}{%
\begin{ensuredisplaymath}
\htuse{Gamma93.gn} = \htuse{Gamma93.td}
\end{ensuredisplaymath} & & \\
\htuse{Gamma93.qt} & average & \\
\htuse{ALEPH.Gamma93.pub.BARATE.98,qt} & \htuse{ALEPH.Gamma93.pub.BARATE.98,exp} & \htuse{ALEPH.Gamma93.pub.BARATE.98,ref} \\
\htuse{BaBar.Gamma93.pub.AUBERT.08,qt} & \htuse{BaBar.Gamma93.pub.AUBERT.08,exp} & \htuse{BaBar.Gamma93.pub.AUBERT.08,ref} \\
\htuse{Belle.Gamma93.pub.LEE.10,qt} & \htuse{Belle.Gamma93.pub.LEE.10,exp} & \htuse{Belle.Gamma93.pub.LEE.10,ref} \\
\htuse{CLEO3.Gamma93.pub.BRIERE.03,qt} & \htuse{CLEO3.Gamma93.pub.BRIERE.03,exp} & \htuse{CLEO3.Gamma93.pub.BRIERE.03,ref}
}%
\htdef{Gamma93by60.qm}{%
\begin{ensuredisplaymath}
\htuse{Gamma93by60.gn} = \htuse{Gamma93by60.td}
\end{ensuredisplaymath} & & \\
\htuse{Gamma93by60.qt} & average & \\
\htuse{CLEO.Gamma93by60.pub.RICHICHI.99,qt} & \htuse{CLEO.Gamma93by60.pub.RICHICHI.99,exp} & \htuse{CLEO.Gamma93by60.pub.RICHICHI.99,ref}
}%
\htdef{Gamma94.qm}{%
\begin{ensuredisplaymath}
\htuse{Gamma94.gn} = \htuse{Gamma94.td}
\end{ensuredisplaymath} & & \\
\htuse{Gamma94.qt} & average & \\
\htuse{ALEPH.Gamma94.pub.BARATE.98,qt} & \htuse{ALEPH.Gamma94.pub.BARATE.98,exp} & \htuse{ALEPH.Gamma94.pub.BARATE.98,ref} \\
\htuse{CLEO3.Gamma94.pub.ARMS.05,qt} & \htuse{CLEO3.Gamma94.pub.ARMS.05,exp} & \htuse{CLEO3.Gamma94.pub.ARMS.05,ref}
}%
\htdef{Gamma94by69.qm}{%
\begin{ensuredisplaymath}
\htuse{Gamma94by69.gn} = \htuse{Gamma94by69.td}
\end{ensuredisplaymath} & & \\
\htuse{Gamma94by69.qt} & average & \\
\htuse{CLEO.Gamma94by69.pub.RICHICHI.99,qt} & \htuse{CLEO.Gamma94by69.pub.RICHICHI.99,exp} & \htuse{CLEO.Gamma94by69.pub.RICHICHI.99,ref}
}%
\htdef{Gamma96.qm}{%
\begin{ensuredisplaymath}
\htuse{Gamma96.gn} = \htuse{Gamma96.td}
\end{ensuredisplaymath} & & \\
\htuse{Gamma96.qt} & average & \\
\htuse{BaBar.Gamma96.pub.AUBERT.08,qt} & \htuse{BaBar.Gamma96.pub.AUBERT.08,exp} & \htuse{BaBar.Gamma96.pub.AUBERT.08,ref} \\
\htuse{Belle.Gamma96.pub.LEE.10,qt} & \htuse{Belle.Gamma96.pub.LEE.10,exp} & \htuse{Belle.Gamma96.pub.LEE.10,ref}
}%
\htdef{Gamma102.qm}{%
\begin{ensuredisplaymath}
\htuse{Gamma102.gn} = \htuse{Gamma102.td}
\end{ensuredisplaymath} & & \\
\htuse{Gamma102.qt} & average & \\
\htuse{CLEO.Gamma102.pub.GIBAUT.94B,qt} & \htuse{CLEO.Gamma102.pub.GIBAUT.94B,exp} & \htuse{CLEO.Gamma102.pub.GIBAUT.94B,ref} \\
\htuse{HRS.Gamma102.pub.BYLSMA.87,qt} & \htuse{HRS.Gamma102.pub.BYLSMA.87,exp} & \htuse{HRS.Gamma102.pub.BYLSMA.87,ref} \\
\htuse{L3.Gamma102.pub.ACHARD.01D,qt} & \htuse{L3.Gamma102.pub.ACHARD.01D,exp} & \htuse{L3.Gamma102.pub.ACHARD.01D,ref}
}%
\htdef{Gamma103.qm}{%
\begin{ensuredisplaymath}
\htuse{Gamma103.gn} = \htuse{Gamma103.td}
\end{ensuredisplaymath} & & \\
\htuse{Gamma103.qt} & average & \\
\htuse{ALEPH.Gamma103.pub.SCHAEL.05C,qt} & \htuse{ALEPH.Gamma103.pub.SCHAEL.05C,exp} & \htuse{ALEPH.Gamma103.pub.SCHAEL.05C,ref} \\
\htuse{ARGUS.Gamma103.pub.ALBRECHT.88B,qt} & \htuse{ARGUS.Gamma103.pub.ALBRECHT.88B,exp} & \htuse{ARGUS.Gamma103.pub.ALBRECHT.88B,ref} \\
\htuse{CLEO.Gamma103.pub.GIBAUT.94B,qt} & \htuse{CLEO.Gamma103.pub.GIBAUT.94B,exp} & \htuse{CLEO.Gamma103.pub.GIBAUT.94B,ref} \\
\htuse{DELPHI.Gamma103.pub.ABDALLAH.06A,qt} & \htuse{DELPHI.Gamma103.pub.ABDALLAH.06A,exp} & \htuse{DELPHI.Gamma103.pub.ABDALLAH.06A,ref} \\
\htuse{HRS.Gamma103.pub.BYLSMA.87,qt} & \htuse{HRS.Gamma103.pub.BYLSMA.87,exp} & \htuse{HRS.Gamma103.pub.BYLSMA.87,ref} \\
\htuse{OPAL.Gamma103.pub.ACKERSTAFF.99E,qt} & \htuse{OPAL.Gamma103.pub.ACKERSTAFF.99E,exp} & \htuse{OPAL.Gamma103.pub.ACKERSTAFF.99E,ref}
}%
\htdef{Gamma104.qm}{%
\begin{ensuredisplaymath}
\htuse{Gamma104.gn} = \htuse{Gamma104.td}
\end{ensuredisplaymath} & & \\
\htuse{Gamma104.qt} & average & \\
\htuse{ALEPH.Gamma104.pub.SCHAEL.05C,qt} & \htuse{ALEPH.Gamma104.pub.SCHAEL.05C,exp} & \htuse{ALEPH.Gamma104.pub.SCHAEL.05C,ref} \\
\htuse{CLEO.Gamma104.pub.ANASTASSOV.01,qt} & \htuse{CLEO.Gamma104.pub.ANASTASSOV.01,exp} & \htuse{CLEO.Gamma104.pub.ANASTASSOV.01,ref} \\
\htuse{DELPHI.Gamma104.pub.ABDALLAH.06A,qt} & \htuse{DELPHI.Gamma104.pub.ABDALLAH.06A,exp} & \htuse{DELPHI.Gamma104.pub.ABDALLAH.06A,ref} \\
\htuse{OPAL.Gamma104.pub.ACKERSTAFF.99E,qt} & \htuse{OPAL.Gamma104.pub.ACKERSTAFF.99E,exp} & \htuse{OPAL.Gamma104.pub.ACKERSTAFF.99E,ref}
}%
\htdef{Gamma106.qm}{%
\begin{ensuredisplaymath}
\htuse{Gamma106.gn} = \htuse{Gamma106.td}
\end{ensuredisplaymath} & & \\
\htuse{Gamma106.qt} & average &}%
\htdef{Gamma110.qm}{%
\begin{ensuredisplaymath}
\htuse{Gamma110.gn} = \htuse{Gamma110.td}
\end{ensuredisplaymath} & & \\
\htuse{Gamma110.qt} & average &}%
\htdef{Gamma126.qm}{%
\begin{ensuredisplaymath}
\htuse{Gamma126.gn} = \htuse{Gamma126.td}
\end{ensuredisplaymath} & & \\
\htuse{Gamma126.qt} & average & \\
\htuse{ALEPH.Gamma126.pub.BUSKULIC.97C,qt} & \htuse{ALEPH.Gamma126.pub.BUSKULIC.97C,exp} & \htuse{ALEPH.Gamma126.pub.BUSKULIC.97C,ref} \\
\htuse{Belle.Gamma126.pub.INAMI.09,qt} & \htuse{Belle.Gamma126.pub.INAMI.09,exp} & \htuse{Belle.Gamma126.pub.INAMI.09,ref} \\
\htuse{CLEO.Gamma126.pub.ARTUSO.92,qt} & \htuse{CLEO.Gamma126.pub.ARTUSO.92,exp} & \htuse{CLEO.Gamma126.pub.ARTUSO.92,ref}
}%
\htdef{Gamma128.qm}{%
\begin{ensuredisplaymath}
\htuse{Gamma128.gn} = \htuse{Gamma128.td}
\end{ensuredisplaymath} & & \\
\htuse{Gamma128.qt} & average & \\
\htuse{ALEPH.Gamma128.pub.BUSKULIC.97C,qt} & \htuse{ALEPH.Gamma128.pub.BUSKULIC.97C,exp} & \htuse{ALEPH.Gamma128.pub.BUSKULIC.97C,ref} \\
\htuse{BaBar.Gamma128.pub.DEL-AMO-SANCHEZ.11E,qt} & \htuse{BaBar.Gamma128.pub.DEL-AMO-SANCHEZ.11E,exp} & \htuse{BaBar.Gamma128.pub.DEL-AMO-SANCHEZ.11E,ref} \\
\htuse{Belle.Gamma128.pub.INAMI.09,qt} & \htuse{Belle.Gamma128.pub.INAMI.09,exp} & \htuse{Belle.Gamma128.pub.INAMI.09,ref} \\
\htuse{CLEO.Gamma128.pub.BARTELT.96,qt} & \htuse{CLEO.Gamma128.pub.BARTELT.96,exp} & \htuse{CLEO.Gamma128.pub.BARTELT.96,ref}
}%
\htdef{Gamma130.qm}{%
\begin{ensuredisplaymath}
\htuse{Gamma130.gn} = \htuse{Gamma130.td}
\end{ensuredisplaymath} & & \\
\htuse{Gamma130.qt} & average & \\
\htuse{Belle.Gamma130.pub.INAMI.09,qt} & \htuse{Belle.Gamma130.pub.INAMI.09,exp} & \htuse{Belle.Gamma130.pub.INAMI.09,ref} \\
\htuse{CLEO.Gamma130.pub.BISHAI.99,qt} & \htuse{CLEO.Gamma130.pub.BISHAI.99,exp} & \htuse{CLEO.Gamma130.pub.BISHAI.99,ref}
}%
\htdef{Gamma132.qm}{%
\begin{ensuredisplaymath}
\htuse{Gamma132.gn} = \htuse{Gamma132.td}
\end{ensuredisplaymath} & & \\
\htuse{Gamma132.qt} & average & \\
\htuse{Belle.Gamma132.pub.INAMI.09,qt} & \htuse{Belle.Gamma132.pub.INAMI.09,exp} & \htuse{Belle.Gamma132.pub.INAMI.09,ref} \\
\htuse{CLEO.Gamma132.pub.BISHAI.99,qt} & \htuse{CLEO.Gamma132.pub.BISHAI.99,exp} & \htuse{CLEO.Gamma132.pub.BISHAI.99,ref}
}%
\htdef{Gamma136.qm}{%
\begin{ensuredisplaymath}
\htuse{Gamma136.gn} = \htuse{Gamma136.td}
\end{ensuredisplaymath} & & \\
\htuse{Gamma136.qt} & average &}%
\htdef{Gamma149.qm}{%
\begin{ensuredisplaymath}
\htuse{Gamma149.gn} = \htuse{Gamma149.td}
\end{ensuredisplaymath} & & \\
\htuse{Gamma149.qt} & average &}%
\htdef{Gamma150.qm}{%
\begin{ensuredisplaymath}
\htuse{Gamma150.gn} = \htuse{Gamma150.td}
\end{ensuredisplaymath} & & \\
\htuse{Gamma150.qt} & average & \\
\htuse{ALEPH.Gamma150.pub.BUSKULIC.97C,qt} & \htuse{ALEPH.Gamma150.pub.BUSKULIC.97C,exp} & \htuse{ALEPH.Gamma150.pub.BUSKULIC.97C,ref} \\
\htuse{CLEO.Gamma150.pub.BARINGER.87,qt} & \htuse{CLEO.Gamma150.pub.BARINGER.87,exp} & \htuse{CLEO.Gamma150.pub.BARINGER.87,ref}
}%
\htdef{Gamma150by66.qm}{%
\begin{ensuredisplaymath}
\htuse{Gamma150by66.gn} = \htuse{Gamma150by66.td}
\end{ensuredisplaymath} & & \\
\htuse{Gamma150by66.qt} & average & \\
\htuse{ALEPH.Gamma150by66.pub.BUSKULIC.96,qt} & \htuse{ALEPH.Gamma150by66.pub.BUSKULIC.96,exp} & \htuse{ALEPH.Gamma150by66.pub.BUSKULIC.96,ref} \\
\htuse{CLEO.Gamma150by66.pub.BALEST.95C,qt} & \htuse{CLEO.Gamma150by66.pub.BALEST.95C,exp} & \htuse{CLEO.Gamma150by66.pub.BALEST.95C,ref}
}%
\htdef{Gamma151.qm}{%
\begin{ensuredisplaymath}
\htuse{Gamma151.gn} = \htuse{Gamma151.td}
\end{ensuredisplaymath} & & \\
\htuse{Gamma151.qt} & average & \\
\htuse{CLEO3.Gamma151.pub.ARMS.05,qt} & \htuse{CLEO3.Gamma151.pub.ARMS.05,exp} & \htuse{CLEO3.Gamma151.pub.ARMS.05,ref}
}%
\htdef{Gamma152.qm}{%
\begin{ensuredisplaymath}
\htuse{Gamma152.gn} = \htuse{Gamma152.td}
\end{ensuredisplaymath} & & \\
\htuse{Gamma152.qt} & average & \\
\htuse{ALEPH.Gamma152.pub.BUSKULIC.97C,qt} & \htuse{ALEPH.Gamma152.pub.BUSKULIC.97C,exp} & \htuse{ALEPH.Gamma152.pub.BUSKULIC.97C,ref}
}%
\htdef{Gamma152by54.qm}{%
\begin{ensuredisplaymath}
\htuse{Gamma152by54.gn} = \htuse{Gamma152by54.td}
\end{ensuredisplaymath} & & \\
\htuse{Gamma152by54.qt} & average &}%
\htdef{Gamma152by76.qm}{%
\begin{ensuredisplaymath}
\htuse{Gamma152by76.gn} = \htuse{Gamma152by76.td}
\end{ensuredisplaymath} & & \\
\htuse{Gamma152by76.qt} & average & \\
\htuse{CLEO.Gamma152by76.pub.BORTOLETTO.93,qt} & \htuse{CLEO.Gamma152by76.pub.BORTOLETTO.93,exp} & \htuse{CLEO.Gamma152by76.pub.BORTOLETTO.93,ref}
}%
\htdef{Gamma167.qm}{%
\begin{ensuredisplaymath}
\htuse{Gamma167.gn} = \htuse{Gamma167.td}
\end{ensuredisplaymath} & & \\
\htuse{Gamma167.qt} & average &}%
\htdef{Gamma168.qm}{%
\begin{ensuredisplaymath}
\htuse{Gamma168.gn} = \htuse{Gamma168.td}
\end{ensuredisplaymath} & & \\
\htuse{Gamma168.qt} & average &}%
\htdef{Gamma169.qm}{%
\begin{ensuredisplaymath}
\htuse{Gamma169.gn} = \htuse{Gamma169.td}
\end{ensuredisplaymath} & & \\
\htuse{Gamma169.qt} & average &}%
\htdef{Gamma800.qm}{%
\begin{ensuredisplaymath}
\htuse{Gamma800.gn} = \htuse{Gamma800.td}
\end{ensuredisplaymath} & & \\
\htuse{Gamma800.qt} & average &}%
\htdef{Gamma802.qm}{%
\begin{ensuredisplaymath}
\htuse{Gamma802.gn} = \htuse{Gamma802.td}
\end{ensuredisplaymath} & & \\
\htuse{Gamma802.qt} & average &}%
\htdef{Gamma803.qm}{%
\begin{ensuredisplaymath}
\htuse{Gamma803.gn} = \htuse{Gamma803.td}
\end{ensuredisplaymath} & & \\
\htuse{Gamma803.qt} & average &}%
\htdef{Gamma804.qm}{%
\begin{ensuredisplaymath}
\htuse{Gamma804.gn} = \htuse{Gamma804.td}
\end{ensuredisplaymath} & & \\
\htuse{Gamma804.qt} & average &}%
\htdef{Gamma805.qm}{%
\begin{ensuredisplaymath}
\htuse{Gamma805.gn} = \htuse{Gamma805.td}
\end{ensuredisplaymath} & & \\
\htuse{Gamma805.qt} & average & \\
\htuse{ALEPH.Gamma805.pub.SCHAEL.05C,qt} & \htuse{ALEPH.Gamma805.pub.SCHAEL.05C,exp} & \htuse{ALEPH.Gamma805.pub.SCHAEL.05C,ref}
}%
\htdef{Gamma806.qm}{%
\begin{ensuredisplaymath}
\htuse{Gamma806.gn} = \htuse{Gamma806.td}
\end{ensuredisplaymath} & & \\
\htuse{Gamma806.qt} & average &}%
\htdef{Gamma809.qm}{%
\begin{ensuredisplaymath}
\htuse{Gamma809.gn} = \htuse{Gamma809.td}
\end{ensuredisplaymath} & & \\
\htuse{Gamma809.qt} & average & \\
\htuse{BaBar.Gamma809.prelim.ICHEP2018,qt} & \htuse{BaBar.Gamma809.prelim.ICHEP2018,exp} & \htuse{BaBar.Gamma809.prelim.ICHEP2018,ref}
}%
\htdef{Gamma810.qm}{%
\begin{ensuredisplaymath}
\htuse{Gamma810.gn} = \htuse{Gamma810.td}
\end{ensuredisplaymath} & & \\
\htuse{Gamma810.qt} & average &}%
\htdef{Gamma811.qm}{%
\begin{ensuredisplaymath}
\htuse{Gamma811.gn} = \htuse{Gamma811.td}
\end{ensuredisplaymath} & & \\
\htuse{Gamma811.qt} & average & \\
\htuse{BaBar.Gamma811.pub.LEES.12X,qt} & \htuse{BaBar.Gamma811.pub.LEES.12X,exp} & \htuse{BaBar.Gamma811.pub.LEES.12X,ref}
}%
\htdef{Gamma812.qm}{%
\begin{ensuredisplaymath}
\htuse{Gamma812.gn} = \htuse{Gamma812.td}
\end{ensuredisplaymath} & & \\
\htuse{Gamma812.qt} & average & \\
\htuse{BaBar.Gamma812.pub.LEES.12X,qt} & \htuse{BaBar.Gamma812.pub.LEES.12X,exp} & \htuse{BaBar.Gamma812.pub.LEES.12X,ref}
}%
\htdef{Gamma820.qm}{%
\begin{ensuredisplaymath}
\htuse{Gamma820.gn} = \htuse{Gamma820.td}
\end{ensuredisplaymath} & & \\
\htuse{Gamma820.qt} & average &}%
\htdef{Gamma821.qm}{%
\begin{ensuredisplaymath}
\htuse{Gamma821.gn} = \htuse{Gamma821.td}
\end{ensuredisplaymath} & & \\
\htuse{Gamma821.qt} & average & \\
\htuse{BaBar.Gamma821.pub.LEES.12X,qt} & \htuse{BaBar.Gamma821.pub.LEES.12X,exp} & \htuse{BaBar.Gamma821.pub.LEES.12X,ref}
}%
\htdef{Gamma822.qm}{%
\begin{ensuredisplaymath}
\htuse{Gamma822.gn} = \htuse{Gamma822.td}
\end{ensuredisplaymath} & & \\
\htuse{Gamma822.qt} & average & \\
\htuse{BaBar.Gamma822.pub.LEES.12X,qt} & \htuse{BaBar.Gamma822.pub.LEES.12X,exp} & \htuse{BaBar.Gamma822.pub.LEES.12X,ref}
}%
\htdef{Gamma830.qm}{%
\begin{ensuredisplaymath}
\htuse{Gamma830.gn} = \htuse{Gamma830.td}
\end{ensuredisplaymath} & & \\
\htuse{Gamma830.qt} & average &}%
\htdef{Gamma831.qm}{%
\begin{ensuredisplaymath}
\htuse{Gamma831.gn} = \htuse{Gamma831.td}
\end{ensuredisplaymath} & & \\
\htuse{Gamma831.qt} & average & \\
\htuse{BaBar.Gamma831.pub.LEES.12X,qt} & \htuse{BaBar.Gamma831.pub.LEES.12X,exp} & \htuse{BaBar.Gamma831.pub.LEES.12X,ref}
}%
\htdef{Gamma832.qm}{%
\begin{ensuredisplaymath}
\htuse{Gamma832.gn} = \htuse{Gamma832.td}
\end{ensuredisplaymath} & & \\
\htuse{Gamma832.qt} & average & \\
\htuse{BaBar.Gamma832.pub.LEES.12X,qt} & \htuse{BaBar.Gamma832.pub.LEES.12X,exp} & \htuse{BaBar.Gamma832.pub.LEES.12X,ref}
}%
\htdef{Gamma833.qm}{%
\begin{ensuredisplaymath}
\htuse{Gamma833.gn} = \htuse{Gamma833.td}
\end{ensuredisplaymath} & & \\
\htuse{Gamma833.qt} & average & \\
\htuse{BaBar.Gamma833.pub.LEES.12X,qt} & \htuse{BaBar.Gamma833.pub.LEES.12X,exp} & \htuse{BaBar.Gamma833.pub.LEES.12X,ref}
}%
\htdef{Gamma910.qm}{%
\begin{ensuredisplaymath}
\htuse{Gamma910.gn} = \htuse{Gamma910.td}
\end{ensuredisplaymath} & & \\
\htuse{Gamma910.qt} & average & \\
\htuse{BaBar.Gamma910.pub.LEES.12X,qt} & \htuse{BaBar.Gamma910.pub.LEES.12X,exp} & \htuse{BaBar.Gamma910.pub.LEES.12X,ref}
}%
\htdef{Gamma911.qm}{%
\begin{ensuredisplaymath}
\htuse{Gamma911.gn} = \htuse{Gamma911.td}
\end{ensuredisplaymath} & & \\
\htuse{Gamma911.qt} & average & \\
\htuse{BaBar.Gamma911.pub.LEES.12X,qt} & \htuse{BaBar.Gamma911.pub.LEES.12X,exp} & \htuse{BaBar.Gamma911.pub.LEES.12X,ref}
}%
\htdef{Gamma920.qm}{%
\begin{ensuredisplaymath}
\htuse{Gamma920.gn} = \htuse{Gamma920.td}
\end{ensuredisplaymath} & & \\
\htuse{Gamma920.qt} & average & \\
\htuse{BaBar.Gamma920.pub.LEES.12X,qt} & \htuse{BaBar.Gamma920.pub.LEES.12X,exp} & \htuse{BaBar.Gamma920.pub.LEES.12X,ref}
}%
\htdef{Gamma930.qm}{%
\begin{ensuredisplaymath}
\htuse{Gamma930.gn} = \htuse{Gamma930.td}
\end{ensuredisplaymath} & & \\
\htuse{Gamma930.qt} & average & \\
\htuse{BaBar.Gamma930.pub.LEES.12X,qt} & \htuse{BaBar.Gamma930.pub.LEES.12X,exp} & \htuse{BaBar.Gamma930.pub.LEES.12X,ref}
}%
\htdef{Gamma944.qm}{%
\begin{ensuredisplaymath}
\htuse{Gamma944.gn} = \htuse{Gamma944.td}
\end{ensuredisplaymath} & & \\
\htuse{Gamma944.qt} & average & \\
\htuse{BaBar.Gamma944.pub.LEES.12X,qt} & \htuse{BaBar.Gamma944.pub.LEES.12X,exp} & \htuse{BaBar.Gamma944.pub.LEES.12X,ref}
}%
\htdef{Gamma945.qm}{%
\begin{ensuredisplaymath}
\htuse{Gamma945.gn} = \htuse{Gamma945.td}
\end{ensuredisplaymath} & & \\
\htuse{Gamma945.qt} & average &}%
\htdef{Gamma998.qm}{%
\begin{ensuredisplaymath}
\htuse{Gamma998.gn} = \htuse{Gamma998.td}
\end{ensuredisplaymath} & & \\
\htuse{Gamma998.qt} & average &}%
\htdef{BrVal}{%
\htuse{Gamma1.qm} \\
\midrule
\htuse{Gamma2.qm} \\
\midrule
\htuse{Gamma3.qm} \\
\midrule
\htuse{Gamma3by5.qm} \\
\midrule
\htuse{Gamma5.qm} \\
\midrule
\htuse{Gamma7.qm} \\
\midrule
\htuse{Gamma8.qm} \\
\midrule
\htuse{Gamma8by5.qm} \\
\midrule
\htuse{Gamma9.qm} \\
\midrule
\htuse{Gamma9by5.qm} \\
\midrule
\htuse{Gamma10.qm} \\
\midrule
\htuse{Gamma10by5.qm} \\
\midrule
\htuse{Gamma10by9.qm} \\
\midrule
\htuse{Gamma11.qm} \\
\midrule
\htuse{Gamma12.qm} \\
\midrule
\htuse{Gamma13.qm} \\
\midrule
\htuse{Gamma14.qm} \\
\midrule
\htuse{Gamma16.qm} \\
\midrule
\htuse{Gamma17.qm} \\
\midrule
\htuse{Gamma18.qm} \\
\midrule
\htuse{Gamma19.qm} \\
\midrule
\htuse{Gamma19by13.qm} \\
\midrule
\htuse{Gamma20.qm} \\
\midrule
\htuse{Gamma23.qm} \\
\midrule
\htuse{Gamma24.qm} \\
\midrule
\htuse{Gamma25.qm} \\
\midrule
\htuse{Gamma26.qm} \\
\midrule
\htuse{Gamma26by13.qm} \\
\midrule
\htuse{Gamma27.qm} \\
\midrule
\htuse{Gamma28.qm} \\
\midrule
\htuse{Gamma29.qm} \\
\midrule
\htuse{Gamma30.qm} \\
\midrule
\htuse{Gamma31.qm} \\
\midrule
\htuse{Gamma32.qm} \\
\midrule
\htuse{Gamma33.qm} \\
\midrule
\htuse{Gamma34.qm} \\
\midrule
\htuse{Gamma35.qm} \\
\midrule
\htuse{Gamma37.qm} \\
\midrule
\htuse{Gamma38.qm} \\
\midrule
\htuse{Gamma39.qm} \\
\midrule
\htuse{Gamma40.qm} \\
\midrule
\htuse{Gamma42.qm} \\
\midrule
\htuse{Gamma43.qm} \\
\midrule
\htuse{Gamma44.qm} \\
\midrule
\htuse{Gamma46.qm} \\
\midrule
\htuse{Gamma47.qm} \\
\midrule
\htuse{Gamma48.qm} \\
\midrule
\htuse{Gamma49.qm} \\
\midrule
\htuse{Gamma50.qm} \\
\midrule
\htuse{Gamma51.qm} \\
\midrule
\htuse{Gamma53.qm} \\
\midrule
\htuse{Gamma54.qm} \\
\midrule
\htuse{Gamma55.qm} \\
\midrule
\htuse{Gamma56.qm} \\
\midrule
\htuse{Gamma57.qm} \\
\midrule
\htuse{Gamma57by55.qm} \\
\midrule
\htuse{Gamma58.qm} \\
\midrule
\htuse{Gamma59.qm} \\
\midrule
\htuse{Gamma60.qm} \\
\midrule
\htuse{Gamma62.qm} \\
\midrule
\htuse{Gamma63.qm} \\
\midrule
\htuse{Gamma64.qm} \\
\midrule
\htuse{Gamma65.qm} \\
\midrule
\htuse{Gamma66.qm} \\
\midrule
\htuse{Gamma67.qm} \\
\midrule
\htuse{Gamma68.qm} \\
\midrule
\htuse{Gamma69.qm} \\
\midrule
\htuse{Gamma70.qm} \\
\midrule
\htuse{Gamma74.qm} \\
\midrule
\htuse{Gamma75.qm} \\
\midrule
\htuse{Gamma76.qm} \\
\midrule
\htuse{Gamma76by54.qm} \\
\midrule
\htuse{Gamma77.qm} \\
\midrule
\htuse{Gamma78.qm} \\
\midrule
\htuse{Gamma79.qm} \\
\midrule
\htuse{Gamma80.qm} \\
\midrule
\htuse{Gamma80by60.qm} \\
\midrule
\htuse{Gamma81.qm} \\
\midrule
\htuse{Gamma81by69.qm} \\
\midrule
\htuse{Gamma82.qm} \\
\midrule
\htuse{Gamma83.qm} \\
\midrule
\htuse{Gamma84.qm} \\
\midrule
\htuse{Gamma85.qm} \\
\midrule
\htuse{Gamma85by60.qm} \\
\midrule
\htuse{Gamma87.qm} \\
\midrule
\htuse{Gamma88.qm} \\
\midrule
\htuse{Gamma89.qm} \\
\midrule
\htuse{Gamma92.qm} \\
\midrule
\htuse{Gamma93.qm} \\
\midrule
\htuse{Gamma93by60.qm} \\
\midrule
\htuse{Gamma94.qm} \\
\midrule
\htuse{Gamma94by69.qm} \\
\midrule
\htuse{Gamma96.qm} \\
\midrule
\htuse{Gamma102.qm} \\
\midrule
\htuse{Gamma103.qm} \\
\midrule
\htuse{Gamma104.qm} \\
\midrule
\htuse{Gamma106.qm} \\
\midrule
\htuse{Gamma110.qm} \\
\midrule
\htuse{Gamma126.qm} \\
\midrule
\htuse{Gamma128.qm} \\
\midrule
\htuse{Gamma130.qm} \\
\midrule
\htuse{Gamma132.qm} \\
\midrule
\htuse{Gamma136.qm} \\
\midrule
\htuse{Gamma149.qm} \\
\midrule
\htuse{Gamma150.qm} \\
\midrule
\htuse{Gamma150by66.qm} \\
\midrule
\htuse{Gamma151.qm} \\
\midrule
\htuse{Gamma152.qm} \\
\midrule
\htuse{Gamma152by54.qm} \\
\midrule
\htuse{Gamma152by76.qm} \\
\midrule
\htuse{Gamma167.qm} \\
\midrule
\htuse{Gamma168.qm} \\
\midrule
\htuse{Gamma169.qm} \\
\midrule
\htuse{Gamma800.qm} \\
\midrule
\htuse{Gamma802.qm} \\
\midrule
\htuse{Gamma803.qm} \\
\midrule
\htuse{Gamma804.qm} \\
\midrule
\htuse{Gamma805.qm} \\
\midrule
\htuse{Gamma806.qm} \\
\midrule
\htuse{Gamma809.qm} \\
\midrule
\htuse{Gamma810.qm} \\
\midrule
\htuse{Gamma811.qm} \\
\midrule
\htuse{Gamma812.qm} \\
\midrule
\htuse{Gamma820.qm} \\
\midrule
\htuse{Gamma821.qm} \\
\midrule
\htuse{Gamma822.qm} \\
\midrule
\htuse{Gamma830.qm} \\
\midrule
\htuse{Gamma831.qm} \\
\midrule
\htuse{Gamma832.qm} \\
\midrule
\htuse{Gamma833.qm} \\
\midrule
\htuse{Gamma910.qm} \\
\midrule
\htuse{Gamma911.qm} \\
\midrule
\htuse{Gamma920.qm} \\
\midrule
\htuse{Gamma930.qm} \\
\midrule
\htuse{Gamma944.qm} \\
\midrule
\htuse{Gamma945.qm} \\
\midrule
\htuse{Gamma998.qm}}%
\htdef{BARATE 98.cite}{\cite{Barate:1997ma}}%
\htdef{BARATE 98.collab}{ALEPH}%
\htdef{BARATE 98.ref}{BARATE 98 (ALEPH) \cite{Barate:1997ma}}%
\htdef{BARATE 98.meas}{%
\begin{ensuredisplaymath}
\htuse{Gamma85.gn} = \htuse{Gamma85.td}
\end{ensuredisplaymath} & \htuse{ALEPH.Gamma85.pub.BARATE.98}
\\
\begin{ensuredisplaymath}
\htuse{Gamma88.gn} = \htuse{Gamma88.td}
\end{ensuredisplaymath} & \htuse{ALEPH.Gamma88.pub.BARATE.98}
\\
\begin{ensuredisplaymath}
\htuse{Gamma93.gn} = \htuse{Gamma93.td}
\end{ensuredisplaymath} & \htuse{ALEPH.Gamma93.pub.BARATE.98}
\\
\begin{ensuredisplaymath}
\htuse{Gamma94.gn} = \htuse{Gamma94.td}
\end{ensuredisplaymath} & \htuse{ALEPH.Gamma94.pub.BARATE.98}}%
\htdef{BARATE 98E.cite}{\cite{Barate:1997tt}}%
\htdef{BARATE 98E.collab}{ALEPH}%
\htdef{BARATE 98E.ref}{BARATE 98E (ALEPH) \cite{Barate:1997tt}}%
\htdef{BARATE 98E.meas}{%
\begin{ensuredisplaymath}
\htuse{Gamma33.gn} = \htuse{Gamma33.td}
\end{ensuredisplaymath} & \htuse{ALEPH.Gamma33.pub.BARATE.98E}
\\
\begin{ensuredisplaymath}
\htuse{Gamma37.gn} = \htuse{Gamma37.td}
\end{ensuredisplaymath} & \htuse{ALEPH.Gamma37.pub.BARATE.98E}
\\
\begin{ensuredisplaymath}
\htuse{Gamma40.gn} = \htuse{Gamma40.td}
\end{ensuredisplaymath} & \htuse{ALEPH.Gamma40.pub.BARATE.98E}
\\
\begin{ensuredisplaymath}
\htuse{Gamma42.gn} = \htuse{Gamma42.td}
\end{ensuredisplaymath} & \htuse{ALEPH.Gamma42.pub.BARATE.98E}
\\
\begin{ensuredisplaymath}
\htuse{Gamma47.gn} = \htuse{Gamma47.td}
\end{ensuredisplaymath} & \htuse{ALEPH.Gamma47.pub.BARATE.98E}
\\
\begin{ensuredisplaymath}
\htuse{Gamma48.gn} = \htuse{Gamma48.td}
\end{ensuredisplaymath} & \htuse{ALEPH.Gamma48.pub.BARATE.98E}
\\
\begin{ensuredisplaymath}
\htuse{Gamma51.gn} = \htuse{Gamma51.td}
\end{ensuredisplaymath} & \htuse{ALEPH.Gamma51.pub.BARATE.98E}
\\
\begin{ensuredisplaymath}
\htuse{Gamma53.gn} = \htuse{Gamma53.td}
\end{ensuredisplaymath} & \htuse{ALEPH.Gamma53.pub.BARATE.98E}}%
\htdef{BARATE 99K.cite}{\cite{Barate:1999hi}}%
\htdef{BARATE 99K.collab}{ALEPH}%
\htdef{BARATE 99K.ref}{BARATE 99K (ALEPH) \cite{Barate:1999hi}}%
\htdef{BARATE 99K.meas}{%
\begin{ensuredisplaymath}
\htuse{Gamma10.gn} = \htuse{Gamma10.td}
\end{ensuredisplaymath} & \htuse{ALEPH.Gamma10.pub.BARATE.99K}
\\
\begin{ensuredisplaymath}
\htuse{Gamma16.gn} = \htuse{Gamma16.td}
\end{ensuredisplaymath} & \htuse{ALEPH.Gamma16.pub.BARATE.99K}
\\
\begin{ensuredisplaymath}
\htuse{Gamma23.gn} = \htuse{Gamma23.td}
\end{ensuredisplaymath} & \htuse{ALEPH.Gamma23.pub.BARATE.99K}
\\
\begin{ensuredisplaymath}
\htuse{Gamma28.gn} = \htuse{Gamma28.td}
\end{ensuredisplaymath} & \htuse{ALEPH.Gamma28.pub.BARATE.99K}
\\
\begin{ensuredisplaymath}
\htuse{Gamma35.gn} = \htuse{Gamma35.td}
\end{ensuredisplaymath} & \htuse{ALEPH.Gamma35.pub.BARATE.99K}
\\
\begin{ensuredisplaymath}
\htuse{Gamma37.gn} = \htuse{Gamma37.td}
\end{ensuredisplaymath} & \htuse{ALEPH.Gamma37.pub.BARATE.99K}
\\
\begin{ensuredisplaymath}
\htuse{Gamma40.gn} = \htuse{Gamma40.td}
\end{ensuredisplaymath} & \htuse{ALEPH.Gamma40.pub.BARATE.99K}
\\
\begin{ensuredisplaymath}
\htuse{Gamma42.gn} = \htuse{Gamma42.td}
\end{ensuredisplaymath} & \htuse{ALEPH.Gamma42.pub.BARATE.99K}}%
\htdef{BARATE 99R.cite}{\cite{Barate:1999hj}}%
\htdef{BARATE 99R.collab}{ALEPH}%
\htdef{BARATE 99R.ref}{BARATE 99R (ALEPH) \cite{Barate:1999hj}}%
\htdef{BARATE 99R.meas}{%
\begin{ensuredisplaymath}
\htuse{Gamma44.gn} = \htuse{Gamma44.td}
\end{ensuredisplaymath} & \htuse{ALEPH.Gamma44.pub.BARATE.99R}}%
\htdef{BUSKULIC 96.cite}{\cite{Buskulic:1995ty}}%
\htdef{BUSKULIC 96.collab}{ALEPH}%
\htdef{BUSKULIC 96.ref}{BUSKULIC 96 (ALEPH) \cite{Buskulic:1995ty}}%
\htdef{BUSKULIC 96.meas}{%
\begin{ensuredisplaymath}
\htuse{Gamma150by66.gn} = \htuse{Gamma150by66.td}
\end{ensuredisplaymath} & \htuse{ALEPH.Gamma150by66.pub.BUSKULIC.96}}%
\htdef{BUSKULIC 97C.cite}{\cite{Buskulic:1996qs}}%
\htdef{BUSKULIC 97C.collab}{ALEPH}%
\htdef{BUSKULIC 97C.ref}{BUSKULIC 97C (ALEPH) \cite{Buskulic:1996qs}}%
\htdef{BUSKULIC 97C.meas}{%
\begin{ensuredisplaymath}
\htuse{Gamma126.gn} = \htuse{Gamma126.td}
\end{ensuredisplaymath} & \htuse{ALEPH.Gamma126.pub.BUSKULIC.97C}
\\
\begin{ensuredisplaymath}
\htuse{Gamma128.gn} = \htuse{Gamma128.td}
\end{ensuredisplaymath} & \htuse{ALEPH.Gamma128.pub.BUSKULIC.97C}
\\
\begin{ensuredisplaymath}
\htuse{Gamma150.gn} = \htuse{Gamma150.td}
\end{ensuredisplaymath} & \htuse{ALEPH.Gamma150.pub.BUSKULIC.97C}
\\
\begin{ensuredisplaymath}
\htuse{Gamma152.gn} = \htuse{Gamma152.td}
\end{ensuredisplaymath} & \htuse{ALEPH.Gamma152.pub.BUSKULIC.97C}}%
\htdef{SCHAEL 05C.cite}{\cite{Schael:2005am}}%
\htdef{SCHAEL 05C.collab}{ALEPH}%
\htdef{SCHAEL 05C.ref}{SCHAEL 05C (ALEPH) \cite{Schael:2005am}}%
\htdef{SCHAEL 05C.meas}{%
\begin{ensuredisplaymath}
\htuse{Gamma3.gn} = \htuse{Gamma3.td}
\end{ensuredisplaymath} & \htuse{ALEPH.Gamma3.pub.SCHAEL.05C}
\\
\begin{ensuredisplaymath}
\htuse{Gamma5.gn} = \htuse{Gamma5.td}
\end{ensuredisplaymath} & \htuse{ALEPH.Gamma5.pub.SCHAEL.05C}
\\
\begin{ensuredisplaymath}
\htuse{Gamma8.gn} = \htuse{Gamma8.td}
\end{ensuredisplaymath} & \htuse{ALEPH.Gamma8.pub.SCHAEL.05C}
\\
\begin{ensuredisplaymath}
\htuse{Gamma13.gn} = \htuse{Gamma13.td}
\end{ensuredisplaymath} & \htuse{ALEPH.Gamma13.pub.SCHAEL.05C}
\\
\begin{ensuredisplaymath}
\htuse{Gamma19.gn} = \htuse{Gamma19.td}
\end{ensuredisplaymath} & \htuse{ALEPH.Gamma19.pub.SCHAEL.05C}
\\
\begin{ensuredisplaymath}
\htuse{Gamma26.gn} = \htuse{Gamma26.td}
\end{ensuredisplaymath} & \htuse{ALEPH.Gamma26.pub.SCHAEL.05C}
\\
\begin{ensuredisplaymath}
\htuse{Gamma30.gn} = \htuse{Gamma30.td}
\end{ensuredisplaymath} & \htuse{ALEPH.Gamma30.pub.SCHAEL.05C}
\\
\begin{ensuredisplaymath}
\htuse{Gamma58.gn} = \htuse{Gamma58.td}
\end{ensuredisplaymath} & \htuse{ALEPH.Gamma58.pub.SCHAEL.05C}
\\
\begin{ensuredisplaymath}
\htuse{Gamma66.gn} = \htuse{Gamma66.td}
\end{ensuredisplaymath} & \htuse{ALEPH.Gamma66.pub.SCHAEL.05C}
\\
\begin{ensuredisplaymath}
\htuse{Gamma76.gn} = \htuse{Gamma76.td}
\end{ensuredisplaymath} & \htuse{ALEPH.Gamma76.pub.SCHAEL.05C}
\\
\begin{ensuredisplaymath}
\htuse{Gamma103.gn} = \htuse{Gamma103.td}
\end{ensuredisplaymath} & \htuse{ALEPH.Gamma103.pub.SCHAEL.05C}
\\
\begin{ensuredisplaymath}
\htuse{Gamma104.gn} = \htuse{Gamma104.td}
\end{ensuredisplaymath} & \htuse{ALEPH.Gamma104.pub.SCHAEL.05C}
\\
\begin{ensuredisplaymath}
\htuse{Gamma805.gn} = \htuse{Gamma805.td}
\end{ensuredisplaymath} & \htuse{ALEPH.Gamma805.pub.SCHAEL.05C}}%
\htdef{ALBRECHT 88B.cite}{\cite{Albrecht:1987zf}}%
\htdef{ALBRECHT 88B.collab}{ARGUS}%
\htdef{ALBRECHT 88B.ref}{ALBRECHT 88B (ARGUS) \cite{Albrecht:1987zf}}%
\htdef{ALBRECHT 88B.meas}{%
\begin{ensuredisplaymath}
\htuse{Gamma103.gn} = \htuse{Gamma103.td}
\end{ensuredisplaymath} & \htuse{ARGUS.Gamma103.pub.ALBRECHT.88B}}%
\htdef{ALBRECHT 92D.cite}{\cite{Albrecht:1991rh}}%
\htdef{ALBRECHT 92D.collab}{ARGUS}%
\htdef{ALBRECHT 92D.ref}{ALBRECHT 92D (ARGUS) \cite{Albrecht:1991rh}}%
\htdef{ALBRECHT 92D.meas}{%
\begin{ensuredisplaymath}
\htuse{Gamma3by5.gn} = \htuse{Gamma3by5.td}
\end{ensuredisplaymath} & \htuse{ARGUS.Gamma3by5.pub.ALBRECHT.92D}}%
\htdef{AUBERT 08.cite}{\cite{Aubert:2007mh}}%
\htdef{AUBERT 08.collab}{\babar}%
\htdef{AUBERT 08.ref}{AUBERT 08 (\babar) \cite{Aubert:2007mh}}%
\htdef{AUBERT 08.meas}{%
\begin{ensuredisplaymath}
\htuse{Gamma60.gn} = \htuse{Gamma60.td}
\end{ensuredisplaymath} & \htuse{BaBar.Gamma60.pub.AUBERT.08}
\\
\begin{ensuredisplaymath}
\htuse{Gamma85.gn} = \htuse{Gamma85.td}
\end{ensuredisplaymath} & \htuse{BaBar.Gamma85.pub.AUBERT.08}
\\
\begin{ensuredisplaymath}
\htuse{Gamma93.gn} = \htuse{Gamma93.td}
\end{ensuredisplaymath} & \htuse{BaBar.Gamma93.pub.AUBERT.08}
\\
\begin{ensuredisplaymath}
\htuse{Gamma96.gn} = \htuse{Gamma96.td}
\end{ensuredisplaymath} & \htuse{BaBar.Gamma96.pub.AUBERT.08}}%
\htdef{AUBERT 10F.cite}{\cite{Aubert:2009qj}}%
\htdef{AUBERT 10F.collab}{\babar}%
\htdef{AUBERT 10F.ref}{AUBERT 10F (\babar) \cite{Aubert:2009qj}}%
\htdef{AUBERT 10F.meas}{%
\begin{ensuredisplaymath}
\htuse{Gamma3by5.gn} = \htuse{Gamma3by5.td}
\end{ensuredisplaymath} & \htuse{BaBar.Gamma3by5.pub.AUBERT.10F}
\\
\begin{ensuredisplaymath}
\htuse{Gamma9by5.gn} = \htuse{Gamma9by5.td}
\end{ensuredisplaymath} & \htuse{BaBar.Gamma9by5.pub.AUBERT.10F}
\\
\begin{ensuredisplaymath}
\htuse{Gamma10by5.gn} = \htuse{Gamma10by5.td}
\end{ensuredisplaymath} & \htuse{BaBar.Gamma10by5.pub.AUBERT.10F}}%
\htdef{DEL-AMO-SANCHEZ 11E.cite}{\cite{delAmoSanchez:2010pc}}%
\htdef{DEL-AMO-SANCHEZ 11E.collab}{\babar}%
\htdef{DEL-AMO-SANCHEZ 11E.ref}{DEL-AMO-SANCHEZ 11E (\babar) \cite{delAmoSanchez:2010pc}}%
\htdef{DEL-AMO-SANCHEZ 11E.meas}{%
\begin{ensuredisplaymath}
\htuse{Gamma128.gn} = \htuse{Gamma128.td}
\end{ensuredisplaymath} & \htuse{BaBar.Gamma128.pub.DEL-AMO-SANCHEZ.11E}}%
\htdef{LEES 12X.cite}{\cite{Lees:2012ks}}%
\htdef{LEES 12X.collab}{\babar}%
\htdef{LEES 12X.ref}{LEES 12X (\babar) \cite{Lees:2012ks}}%
\htdef{LEES 12X.meas}{%
\begin{ensuredisplaymath}
\htuse{Gamma811.gn} = \htuse{Gamma811.td}
\end{ensuredisplaymath} & \htuse{BaBar.Gamma811.pub.LEES.12X}
\\
\begin{ensuredisplaymath}
\htuse{Gamma812.gn} = \htuse{Gamma812.td}
\end{ensuredisplaymath} & \htuse{BaBar.Gamma812.pub.LEES.12X}
\\
\begin{ensuredisplaymath}
\htuse{Gamma821.gn} = \htuse{Gamma821.td}
\end{ensuredisplaymath} & \htuse{BaBar.Gamma821.pub.LEES.12X}
\\
\begin{ensuredisplaymath}
\htuse{Gamma822.gn} = \htuse{Gamma822.td}
\end{ensuredisplaymath} & \htuse{BaBar.Gamma822.pub.LEES.12X}
\\
\begin{ensuredisplaymath}
\htuse{Gamma831.gn} = \htuse{Gamma831.td}
\end{ensuredisplaymath} & \htuse{BaBar.Gamma831.pub.LEES.12X}
\\
\begin{ensuredisplaymath}
\htuse{Gamma832.gn} = \htuse{Gamma832.td}
\end{ensuredisplaymath} & \htuse{BaBar.Gamma832.pub.LEES.12X}
\\
\begin{ensuredisplaymath}
\htuse{Gamma833.gn} = \htuse{Gamma833.td}
\end{ensuredisplaymath} & \htuse{BaBar.Gamma833.pub.LEES.12X}
\\
\begin{ensuredisplaymath}
\htuse{Gamma910.gn} = \htuse{Gamma910.td}
\end{ensuredisplaymath} & \htuse{BaBar.Gamma910.pub.LEES.12X}
\\
\begin{ensuredisplaymath}
\htuse{Gamma911.gn} = \htuse{Gamma911.td}
\end{ensuredisplaymath} & \htuse{BaBar.Gamma911.pub.LEES.12X}
\\
\begin{ensuredisplaymath}
\htuse{Gamma920.gn} = \htuse{Gamma920.td}
\end{ensuredisplaymath} & \htuse{BaBar.Gamma920.pub.LEES.12X}
\\
\begin{ensuredisplaymath}
\htuse{Gamma930.gn} = \htuse{Gamma930.td}
\end{ensuredisplaymath} & \htuse{BaBar.Gamma930.pub.LEES.12X}
\\
\begin{ensuredisplaymath}
\htuse{Gamma944.gn} = \htuse{Gamma944.td}
\end{ensuredisplaymath} & \htuse{BaBar.Gamma944.pub.LEES.12X}}%
\htdef{LEES 12Y.cite}{\cite{Lees:2012de}}%
\htdef{LEES 12Y.collab}{\babar}%
\htdef{LEES 12Y.ref}{LEES 12Y (\babar) \cite{Lees:2012de}}%
\htdef{LEES 12Y.meas}{%
\begin{ensuredisplaymath}
\htuse{Gamma47.gn} = \htuse{Gamma47.td}
\end{ensuredisplaymath} & \htuse{BaBar.Gamma47.pub.LEES.12Y}
\\
\begin{ensuredisplaymath}
\htuse{Gamma50.gn} = \htuse{Gamma50.td}
\end{ensuredisplaymath} & \htuse{BaBar.Gamma50.pub.LEES.12Y}}%
\htdef{LEES 18B.cite}{\cite{BaBar:2018qry}}%
\htdef{LEES 18B.collab}{\babar}%
\htdef{LEES 18B.ref}{LEES 18B (\babar) \cite{BaBar:2018qry}}%
\htdef{LEES 18B.meas}{%
\begin{ensuredisplaymath}
\htuse{Gamma37.gn} = \htuse{Gamma37.td}
\end{ensuredisplaymath} & \htuse{BaBar.Gamma37.pub.LEES.18B}}%
\htdef{BaBar prelim. ICHEP2018.cite}{\cite{Lusiani:2019mis}}%
\htdef{BaBar prelim. ICHEP2018.collab}{BaBar}%
\htdef{BaBar prelim. ICHEP2018.ref}{\babar prelim. ICHEP2018 \cite{Lusiani:2019mis}}%
\htdef{BaBar prelim. ICHEP2018.meas}{%
\begin{ensuredisplaymath}
\htuse{Gamma10.gn} = \htuse{Gamma10.td}
\end{ensuredisplaymath} & \htuse{BaBar.Gamma10.prelim.ICHEP2018}
\\
\begin{ensuredisplaymath}
\htuse{Gamma16.gn} = \htuse{Gamma16.td}
\end{ensuredisplaymath} & \htuse{BaBar.Gamma16.prelim.ICHEP2018}
\\
\begin{ensuredisplaymath}
\htuse{Gamma23.gn} = \htuse{Gamma23.td}
\end{ensuredisplaymath} & \htuse{BaBar.Gamma23.prelim.ICHEP2018}
\\
\begin{ensuredisplaymath}
\htuse{Gamma27.gn} = \htuse{Gamma27.td}
\end{ensuredisplaymath} & \htuse{BaBar.Gamma27.prelim.ICHEP2018}
\\
\begin{ensuredisplaymath}
\htuse{Gamma28.gn} = \htuse{Gamma28.td}
\end{ensuredisplaymath} & \htuse{BaBar.Gamma28.prelim.ICHEP2018}
\\
\begin{ensuredisplaymath}
\htuse{Gamma809.gn} = \htuse{Gamma809.td}
\end{ensuredisplaymath} & \htuse{BaBar.Gamma809.prelim.ICHEP2018}}%
\htdef{FUJIKAWA 08.cite}{\cite{Fujikawa:2008ma}}%
\htdef{FUJIKAWA 08.collab}{Belle}%
\htdef{FUJIKAWA 08.ref}{FUJIKAWA 08 (Belle) \cite{Fujikawa:2008ma}}%
\htdef{FUJIKAWA 08.meas}{%
\begin{ensuredisplaymath}
\htuse{Gamma13.gn} = \htuse{Gamma13.td}
\end{ensuredisplaymath} & \htuse{Belle.Gamma13.pub.FUJIKAWA.08}}%
\htdef{INAMI 09.cite}{\cite{Inami:2008ar}}%
\htdef{INAMI 09.collab}{Belle}%
\htdef{INAMI 09.ref}{INAMI 09 (Belle) \cite{Inami:2008ar}}%
\htdef{INAMI 09.meas}{%
\begin{ensuredisplaymath}
\htuse{Gamma126.gn} = \htuse{Gamma126.td}
\end{ensuredisplaymath} & \htuse{Belle.Gamma126.pub.INAMI.09}
\\
\begin{ensuredisplaymath}
\htuse{Gamma128.gn} = \htuse{Gamma128.td}
\end{ensuredisplaymath} & \htuse{Belle.Gamma128.pub.INAMI.09}
\\
\begin{ensuredisplaymath}
\htuse{Gamma130.gn} = \htuse{Gamma130.td}
\end{ensuredisplaymath} & \htuse{Belle.Gamma130.pub.INAMI.09}
\\
\begin{ensuredisplaymath}
\htuse{Gamma132.gn} = \htuse{Gamma132.td}
\end{ensuredisplaymath} & \htuse{Belle.Gamma132.pub.INAMI.09}}%
\htdef{LEE 10.cite}{\cite{Lee:2010tc}}%
\htdef{LEE 10.collab}{Belle}%
\htdef{LEE 10.ref}{LEE 10 (Belle) \cite{Lee:2010tc}}%
\htdef{LEE 10.meas}{%
\begin{ensuredisplaymath}
\htuse{Gamma60.gn} = \htuse{Gamma60.td}
\end{ensuredisplaymath} & \htuse{Belle.Gamma60.pub.LEE.10}
\\
\begin{ensuredisplaymath}
\htuse{Gamma85.gn} = \htuse{Gamma85.td}
\end{ensuredisplaymath} & \htuse{Belle.Gamma85.pub.LEE.10}
\\
\begin{ensuredisplaymath}
\htuse{Gamma93.gn} = \htuse{Gamma93.td}
\end{ensuredisplaymath} & \htuse{Belle.Gamma93.pub.LEE.10}
\\
\begin{ensuredisplaymath}
\htuse{Gamma96.gn} = \htuse{Gamma96.td}
\end{ensuredisplaymath} & \htuse{Belle.Gamma96.pub.LEE.10}}%
\htdef{RYU 14vpc.cite}{\cite{Ryu:2014vpc}}%
\htdef{RYU 14vpc.collab}{Belle}%
\htdef{RYU 14vpc.ref}{RYU 14vpc (Belle) \cite{Ryu:2014vpc}}%
\htdef{RYU 14vpc.meas}{%
\begin{ensuredisplaymath}
\htuse{Gamma35.gn} = \htuse{Gamma35.td}
\end{ensuredisplaymath} & \htuse{Belle.Gamma35.pub.RYU.14vpc}
\\
\begin{ensuredisplaymath}
\htuse{Gamma37.gn} = \htuse{Gamma37.td}
\end{ensuredisplaymath} & \htuse{Belle.Gamma37.pub.RYU.14vpc}
\\
\begin{ensuredisplaymath}
\htuse{Gamma40.gn} = \htuse{Gamma40.td}
\end{ensuredisplaymath} & \htuse{Belle.Gamma40.pub.RYU.14vpc}
\\
\begin{ensuredisplaymath}
\htuse{Gamma42.gn} = \htuse{Gamma42.td}
\end{ensuredisplaymath} & \htuse{Belle.Gamma42.pub.RYU.14vpc}
\\
\begin{ensuredisplaymath}
\htuse{Gamma47.gn} = \htuse{Gamma47.td}
\end{ensuredisplaymath} & \htuse{Belle.Gamma47.pub.RYU.14vpc}
\\
\begin{ensuredisplaymath}
\htuse{Gamma50.gn} = \htuse{Gamma50.td}
\end{ensuredisplaymath} & \htuse{Belle.Gamma50.pub.RYU.14vpc}}%
\htdef{BEHREND 89B.cite}{\cite{Behrend:1989wc}}%
\htdef{BEHREND 89B.collab}{CELLO}%
\htdef{BEHREND 89B.ref}{BEHREND 89B (CELLO) \cite{Behrend:1989wc}}%
\htdef{BEHREND 89B.meas}{%
\begin{ensuredisplaymath}
\htuse{Gamma54.gn} = \htuse{Gamma54.td}
\end{ensuredisplaymath} & \htuse{CELLO.Gamma54.pub.BEHREND.89B}}%
\htdef{ANASTASSOV 01.cite}{\cite{Anastassov:2000xu}}%
\htdef{ANASTASSOV 01.collab}{CLEO}%
\htdef{ANASTASSOV 01.ref}{ANASTASSOV 01 (CLEO) \cite{Anastassov:2000xu}}%
\htdef{ANASTASSOV 01.meas}{%
\begin{ensuredisplaymath}
\htuse{Gamma78.gn} = \htuse{Gamma78.td}
\end{ensuredisplaymath} & \htuse{CLEO.Gamma78.pub.ANASTASSOV.01}
\\
\begin{ensuredisplaymath}
\htuse{Gamma104.gn} = \htuse{Gamma104.td}
\end{ensuredisplaymath} & \htuse{CLEO.Gamma104.pub.ANASTASSOV.01}}%
\htdef{ANASTASSOV 97.cite}{\cite{Anastassov:1996tc}}%
\htdef{ANASTASSOV 97.collab}{CLEO}%
\htdef{ANASTASSOV 97.ref}{ANASTASSOV 97 (CLEO) \cite{Anastassov:1996tc}}%
\htdef{ANASTASSOV 97.meas}{%
\begin{ensuredisplaymath}
\htuse{Gamma3by5.gn} = \htuse{Gamma3by5.td}
\end{ensuredisplaymath} & \htuse{CLEO.Gamma3by5.pub.ANASTASSOV.97}
\\
\begin{ensuredisplaymath}
\htuse{Gamma5.gn} = \htuse{Gamma5.td}
\end{ensuredisplaymath} & \htuse{CLEO.Gamma5.pub.ANASTASSOV.97}
\\
\begin{ensuredisplaymath}
\htuse{Gamma8.gn} = \htuse{Gamma8.td}
\end{ensuredisplaymath} & \htuse{CLEO.Gamma8.pub.ANASTASSOV.97}}%
\htdef{ARTUSO 92.cite}{\cite{Artuso:1992qu}}%
\htdef{ARTUSO 92.collab}{CLEO}%
\htdef{ARTUSO 92.ref}{ARTUSO 92 (CLEO) \cite{Artuso:1992qu}}%
\htdef{ARTUSO 92.meas}{%
\begin{ensuredisplaymath}
\htuse{Gamma126.gn} = \htuse{Gamma126.td}
\end{ensuredisplaymath} & \htuse{CLEO.Gamma126.pub.ARTUSO.92}}%
\htdef{ARTUSO 94.cite}{\cite{Artuso:1994ii}}%
\htdef{ARTUSO 94.collab}{CLEO}%
\htdef{ARTUSO 94.ref}{ARTUSO 94 (CLEO) \cite{Artuso:1994ii}}%
\htdef{ARTUSO 94.meas}{%
\begin{ensuredisplaymath}
\htuse{Gamma13.gn} = \htuse{Gamma13.td}
\end{ensuredisplaymath} & \htuse{CLEO.Gamma13.pub.ARTUSO.94}}%
\htdef{BALEST 95C.cite}{\cite{Balest:1995kq}}%
\htdef{BALEST 95C.collab}{CLEO}%
\htdef{BALEST 95C.ref}{BALEST 95C (CLEO) \cite{Balest:1995kq}}%
\htdef{BALEST 95C.meas}{%
\begin{ensuredisplaymath}
\htuse{Gamma57.gn} = \htuse{Gamma57.td}
\end{ensuredisplaymath} & \htuse{CLEO.Gamma57.pub.BALEST.95C}
\\
\begin{ensuredisplaymath}
\htuse{Gamma66.gn} = \htuse{Gamma66.td}
\end{ensuredisplaymath} & \htuse{CLEO.Gamma66.pub.BALEST.95C}
\\
\begin{ensuredisplaymath}
\htuse{Gamma150by66.gn} = \htuse{Gamma150by66.td}
\end{ensuredisplaymath} & \htuse{CLEO.Gamma150by66.pub.BALEST.95C}}%
\htdef{BARINGER 87.cite}{\cite{Baringer:1987tr}}%
\htdef{BARINGER 87.collab}{CLEO}%
\htdef{BARINGER 87.ref}{BARINGER 87 (CLEO) \cite{Baringer:1987tr}}%
\htdef{BARINGER 87.meas}{%
\begin{ensuredisplaymath}
\htuse{Gamma150.gn} = \htuse{Gamma150.td}
\end{ensuredisplaymath} & \htuse{CLEO.Gamma150.pub.BARINGER.87}}%
\htdef{BARTELT 96.cite}{\cite{Bartelt:1996iv}}%
\htdef{BARTELT 96.collab}{CLEO}%
\htdef{BARTELT 96.ref}{BARTELT 96 (CLEO) \cite{Bartelt:1996iv}}%
\htdef{BARTELT 96.meas}{%
\begin{ensuredisplaymath}
\htuse{Gamma128.gn} = \htuse{Gamma128.td}
\end{ensuredisplaymath} & \htuse{CLEO.Gamma128.pub.BARTELT.96}}%
\htdef{BATTLE 94.cite}{\cite{Battle:1994by}}%
\htdef{BATTLE 94.collab}{CLEO}%
\htdef{BATTLE 94.ref}{BATTLE 94 (CLEO) \cite{Battle:1994by}}%
\htdef{BATTLE 94.meas}{%
\begin{ensuredisplaymath}
\htuse{Gamma10.gn} = \htuse{Gamma10.td}
\end{ensuredisplaymath} & \htuse{CLEO.Gamma10.pub.BATTLE.94}
\\
\begin{ensuredisplaymath}
\htuse{Gamma16.gn} = \htuse{Gamma16.td}
\end{ensuredisplaymath} & \htuse{CLEO.Gamma16.pub.BATTLE.94}
\\
\begin{ensuredisplaymath}
\htuse{Gamma23.gn} = \htuse{Gamma23.td}
\end{ensuredisplaymath} & \htuse{CLEO.Gamma23.pub.BATTLE.94}
\\
\begin{ensuredisplaymath}
\htuse{Gamma31.gn} = \htuse{Gamma31.td}
\end{ensuredisplaymath} & \htuse{CLEO.Gamma31.pub.BATTLE.94}}%
\htdef{BISHAI 99.cite}{\cite{Bishai:1998gf}}%
\htdef{BISHAI 99.collab}{CLEO}%
\htdef{BISHAI 99.ref}{BISHAI 99 (CLEO) \cite{Bishai:1998gf}}%
\htdef{BISHAI 99.meas}{%
\begin{ensuredisplaymath}
\htuse{Gamma130.gn} = \htuse{Gamma130.td}
\end{ensuredisplaymath} & \htuse{CLEO.Gamma130.pub.BISHAI.99}
\\
\begin{ensuredisplaymath}
\htuse{Gamma132.gn} = \htuse{Gamma132.td}
\end{ensuredisplaymath} & \htuse{CLEO.Gamma132.pub.BISHAI.99}}%
\htdef{BORTOLETTO 93.cite}{\cite{Bortoletto:1993px}}%
\htdef{BORTOLETTO 93.collab}{CLEO}%
\htdef{BORTOLETTO 93.ref}{BORTOLETTO 93 (CLEO) \cite{Bortoletto:1993px}}%
\htdef{BORTOLETTO 93.meas}{%
\begin{ensuredisplaymath}
\htuse{Gamma76by54.gn} = \htuse{Gamma76by54.td}
\end{ensuredisplaymath} & \htuse{CLEO.Gamma76by54.pub.BORTOLETTO.93}
\\
\begin{ensuredisplaymath}
\htuse{Gamma152by76.gn} = \htuse{Gamma152by76.td}
\end{ensuredisplaymath} & \htuse{CLEO.Gamma152by76.pub.BORTOLETTO.93}}%
\htdef{COAN 96.cite}{\cite{Coan:1996iu}}%
\htdef{COAN 96.collab}{CLEO}%
\htdef{COAN 96.ref}{COAN 96 (CLEO) \cite{Coan:1996iu}}%
\htdef{COAN 96.meas}{%
\begin{ensuredisplaymath}
\htuse{Gamma34.gn} = \htuse{Gamma34.td}
\end{ensuredisplaymath} & \htuse{CLEO.Gamma34.pub.COAN.96}
\\
\begin{ensuredisplaymath}
\htuse{Gamma37.gn} = \htuse{Gamma37.td}
\end{ensuredisplaymath} & \htuse{CLEO.Gamma37.pub.COAN.96}
\\
\begin{ensuredisplaymath}
\htuse{Gamma39.gn} = \htuse{Gamma39.td}
\end{ensuredisplaymath} & \htuse{CLEO.Gamma39.pub.COAN.96}
\\
\begin{ensuredisplaymath}
\htuse{Gamma42.gn} = \htuse{Gamma42.td}
\end{ensuredisplaymath} & \htuse{CLEO.Gamma42.pub.COAN.96}
\\
\begin{ensuredisplaymath}
\htuse{Gamma47.gn} = \htuse{Gamma47.td}
\end{ensuredisplaymath} & \htuse{CLEO.Gamma47.pub.COAN.96}}%
\htdef{EDWARDS 00A.cite}{\cite{Edwards:1999fj}}%
\htdef{EDWARDS 00A.collab}{CLEO}%
\htdef{EDWARDS 00A.ref}{EDWARDS 00A (CLEO) \cite{Edwards:1999fj}}%
\htdef{EDWARDS 00A.meas}{%
\begin{ensuredisplaymath}
\htuse{Gamma69.gn} = \htuse{Gamma69.td}
\end{ensuredisplaymath} & \htuse{CLEO.Gamma69.pub.EDWARDS.00A}}%
\htdef{GIBAUT 94B.cite}{\cite{Gibaut:1994ik}}%
\htdef{GIBAUT 94B.collab}{CLEO}%
\htdef{GIBAUT 94B.ref}{GIBAUT 94B (CLEO) \cite{Gibaut:1994ik}}%
\htdef{GIBAUT 94B.meas}{%
\begin{ensuredisplaymath}
\htuse{Gamma102.gn} = \htuse{Gamma102.td}
\end{ensuredisplaymath} & \htuse{CLEO.Gamma102.pub.GIBAUT.94B}
\\
\begin{ensuredisplaymath}
\htuse{Gamma103.gn} = \htuse{Gamma103.td}
\end{ensuredisplaymath} & \htuse{CLEO.Gamma103.pub.GIBAUT.94B}}%
\htdef{PROCARIO 93.cite}{\cite{Procario:1992hd}}%
\htdef{PROCARIO 93.collab}{CLEO}%
\htdef{PROCARIO 93.ref}{PROCARIO 93 (CLEO) \cite{Procario:1992hd}}%
\htdef{PROCARIO 93.meas}{%
\begin{ensuredisplaymath}
\htuse{Gamma19by13.gn} = \htuse{Gamma19by13.td}
\end{ensuredisplaymath} & \htuse{CLEO.Gamma19by13.pub.PROCARIO.93}
\\
\begin{ensuredisplaymath}
\htuse{Gamma26by13.gn} = \htuse{Gamma26by13.td}
\end{ensuredisplaymath} & \htuse{CLEO.Gamma26by13.pub.PROCARIO.93}
\\
\begin{ensuredisplaymath}
\htuse{Gamma29.gn} = \htuse{Gamma29.td}
\end{ensuredisplaymath} & \htuse{CLEO.Gamma29.pub.PROCARIO.93}}%
\htdef{RICHICHI 99.cite}{\cite{Richichi:1998bc}}%
\htdef{RICHICHI 99.collab}{CLEO}%
\htdef{RICHICHI 99.ref}{RICHICHI 99 (CLEO) \cite{Richichi:1998bc}}%
\htdef{RICHICHI 99.meas}{%
\begin{ensuredisplaymath}
\htuse{Gamma80by60.gn} = \htuse{Gamma80by60.td}
\end{ensuredisplaymath} & \htuse{CLEO.Gamma80by60.pub.RICHICHI.99}
\\
\begin{ensuredisplaymath}
\htuse{Gamma81by69.gn} = \htuse{Gamma81by69.td}
\end{ensuredisplaymath} & \htuse{CLEO.Gamma81by69.pub.RICHICHI.99}
\\
\begin{ensuredisplaymath}
\htuse{Gamma93by60.gn} = \htuse{Gamma93by60.td}
\end{ensuredisplaymath} & \htuse{CLEO.Gamma93by60.pub.RICHICHI.99}
\\
\begin{ensuredisplaymath}
\htuse{Gamma94by69.gn} = \htuse{Gamma94by69.td}
\end{ensuredisplaymath} & \htuse{CLEO.Gamma94by69.pub.RICHICHI.99}}%
\htdef{ARMS 05.cite}{\cite{Arms:2005qg}}%
\htdef{ARMS 05.collab}{CLEO3}%
\htdef{ARMS 05.ref}{ARMS 05 (CLEO3) \cite{Arms:2005qg}}%
\htdef{ARMS 05.meas}{%
\begin{ensuredisplaymath}
\htuse{Gamma88.gn} = \htuse{Gamma88.td}
\end{ensuredisplaymath} & \htuse{CLEO3.Gamma88.pub.ARMS.05}
\\
\begin{ensuredisplaymath}
\htuse{Gamma94.gn} = \htuse{Gamma94.td}
\end{ensuredisplaymath} & \htuse{CLEO3.Gamma94.pub.ARMS.05}
\\
\begin{ensuredisplaymath}
\htuse{Gamma151.gn} = \htuse{Gamma151.td}
\end{ensuredisplaymath} & \htuse{CLEO3.Gamma151.pub.ARMS.05}}%
\htdef{BRIERE 03.cite}{\cite{Briere:2003fr}}%
\htdef{BRIERE 03.collab}{CLEO3}%
\htdef{BRIERE 03.ref}{BRIERE 03 (CLEO3) \cite{Briere:2003fr}}%
\htdef{BRIERE 03.meas}{%
\begin{ensuredisplaymath}
\htuse{Gamma60.gn} = \htuse{Gamma60.td}
\end{ensuredisplaymath} & \htuse{CLEO3.Gamma60.pub.BRIERE.03}
\\
\begin{ensuredisplaymath}
\htuse{Gamma85.gn} = \htuse{Gamma85.td}
\end{ensuredisplaymath} & \htuse{CLEO3.Gamma85.pub.BRIERE.03}
\\
\begin{ensuredisplaymath}
\htuse{Gamma93.gn} = \htuse{Gamma93.td}
\end{ensuredisplaymath} & \htuse{CLEO3.Gamma93.pub.BRIERE.03}}%
\htdef{ABDALLAH 06A.cite}{\cite{Abdallah:2003cw}}%
\htdef{ABDALLAH 06A.collab}{DELPHI}%
\htdef{ABDALLAH 06A.ref}{ABDALLAH 06A (DELPHI) \cite{Abdallah:2003cw}}%
\htdef{ABDALLAH 06A.meas}{%
\begin{ensuredisplaymath}
\htuse{Gamma8.gn} = \htuse{Gamma8.td}
\end{ensuredisplaymath} & \htuse{DELPHI.Gamma8.pub.ABDALLAH.06A}
\\
\begin{ensuredisplaymath}
\htuse{Gamma13.gn} = \htuse{Gamma13.td}
\end{ensuredisplaymath} & \htuse{DELPHI.Gamma13.pub.ABDALLAH.06A}
\\
\begin{ensuredisplaymath}
\htuse{Gamma19.gn} = \htuse{Gamma19.td}
\end{ensuredisplaymath} & \htuse{DELPHI.Gamma19.pub.ABDALLAH.06A}
\\
\begin{ensuredisplaymath}
\htuse{Gamma25.gn} = \htuse{Gamma25.td}
\end{ensuredisplaymath} & \htuse{DELPHI.Gamma25.pub.ABDALLAH.06A}
\\
\begin{ensuredisplaymath}
\htuse{Gamma57.gn} = \htuse{Gamma57.td}
\end{ensuredisplaymath} & \htuse{DELPHI.Gamma57.pub.ABDALLAH.06A}
\\
\begin{ensuredisplaymath}
\htuse{Gamma66.gn} = \htuse{Gamma66.td}
\end{ensuredisplaymath} & \htuse{DELPHI.Gamma66.pub.ABDALLAH.06A}
\\
\begin{ensuredisplaymath}
\htuse{Gamma74.gn} = \htuse{Gamma74.td}
\end{ensuredisplaymath} & \htuse{DELPHI.Gamma74.pub.ABDALLAH.06A}
\\
\begin{ensuredisplaymath}
\htuse{Gamma103.gn} = \htuse{Gamma103.td}
\end{ensuredisplaymath} & \htuse{DELPHI.Gamma103.pub.ABDALLAH.06A}
\\
\begin{ensuredisplaymath}
\htuse{Gamma104.gn} = \htuse{Gamma104.td}
\end{ensuredisplaymath} & \htuse{DELPHI.Gamma104.pub.ABDALLAH.06A}}%
\htdef{ABREU 92N.cite}{\cite{Abreu:1992gn}}%
\htdef{ABREU 92N.collab}{DELPHI}%
\htdef{ABREU 92N.ref}{ABREU 92N (DELPHI) \cite{Abreu:1992gn}}%
\htdef{ABREU 92N.meas}{%
\begin{ensuredisplaymath}
\htuse{Gamma7.gn} = \htuse{Gamma7.td}
\end{ensuredisplaymath} & \htuse{DELPHI.Gamma7.pub.ABREU.92N}}%
\htdef{ABREU 94K.cite}{\cite{Abreu:1994fi}}%
\htdef{ABREU 94K.collab}{DELPHI}%
\htdef{ABREU 94K.ref}{ABREU 94K (DELPHI) \cite{Abreu:1994fi}}%
\htdef{ABREU 94K.meas}{%
\begin{ensuredisplaymath}
\htuse{Gamma10.gn} = \htuse{Gamma10.td}
\end{ensuredisplaymath} & \htuse{DELPHI.Gamma10.pub.ABREU.94K}
\\
\begin{ensuredisplaymath}
\htuse{Gamma31.gn} = \htuse{Gamma31.td}
\end{ensuredisplaymath} & \htuse{DELPHI.Gamma31.pub.ABREU.94K}}%
\htdef{ABREU 99X.cite}{\cite{Abreu:1999rb}}%
\htdef{ABREU 99X.collab}{DELPHI}%
\htdef{ABREU 99X.ref}{ABREU 99X (DELPHI) \cite{Abreu:1999rb}}%
\htdef{ABREU 99X.meas}{%
\begin{ensuredisplaymath}
\htuse{Gamma3.gn} = \htuse{Gamma3.td}
\end{ensuredisplaymath} & \htuse{DELPHI.Gamma3.pub.ABREU.99X}
\\
\begin{ensuredisplaymath}
\htuse{Gamma5.gn} = \htuse{Gamma5.td}
\end{ensuredisplaymath} & \htuse{DELPHI.Gamma5.pub.ABREU.99X}}%
\htdef{BYLSMA 87.cite}{\cite{Bylsma:1986zy}}%
\htdef{BYLSMA 87.collab}{HRS}%
\htdef{BYLSMA 87.ref}{BYLSMA 87 (HRS) \cite{Bylsma:1986zy}}%
\htdef{BYLSMA 87.meas}{%
\begin{ensuredisplaymath}
\htuse{Gamma102.gn} = \htuse{Gamma102.td}
\end{ensuredisplaymath} & \htuse{HRS.Gamma102.pub.BYLSMA.87}
\\
\begin{ensuredisplaymath}
\htuse{Gamma103.gn} = \htuse{Gamma103.td}
\end{ensuredisplaymath} & \htuse{HRS.Gamma103.pub.BYLSMA.87}}%
\htdef{ACCIARRI 01F.cite}{\cite{Acciarri:2001sg}}%
\htdef{ACCIARRI 01F.collab}{L3}%
\htdef{ACCIARRI 01F.ref}{ACCIARRI 01F (L3) \cite{Acciarri:2001sg}}%
\htdef{ACCIARRI 01F.meas}{%
\begin{ensuredisplaymath}
\htuse{Gamma3.gn} = \htuse{Gamma3.td}
\end{ensuredisplaymath} & \htuse{L3.Gamma3.pub.ACCIARRI.01F}
\\
\begin{ensuredisplaymath}
\htuse{Gamma5.gn} = \htuse{Gamma5.td}
\end{ensuredisplaymath} & \htuse{L3.Gamma5.pub.ACCIARRI.01F}}%
\htdef{ACCIARRI 95.cite}{\cite{Acciarri:1994vr}}%
\htdef{ACCIARRI 95.collab}{L3}%
\htdef{ACCIARRI 95.ref}{ACCIARRI 95 (L3) \cite{Acciarri:1994vr}}%
\htdef{ACCIARRI 95.meas}{%
\begin{ensuredisplaymath}
\htuse{Gamma7.gn} = \htuse{Gamma7.td}
\end{ensuredisplaymath} & \htuse{L3.Gamma7.pub.ACCIARRI.95}
\\
\begin{ensuredisplaymath}
\htuse{Gamma13.gn} = \htuse{Gamma13.td}
\end{ensuredisplaymath} & \htuse{L3.Gamma13.pub.ACCIARRI.95}
\\
\begin{ensuredisplaymath}
\htuse{Gamma19.gn} = \htuse{Gamma19.td}
\end{ensuredisplaymath} & \htuse{L3.Gamma19.pub.ACCIARRI.95}
\\
\begin{ensuredisplaymath}
\htuse{Gamma26.gn} = \htuse{Gamma26.td}
\end{ensuredisplaymath} & \htuse{L3.Gamma26.pub.ACCIARRI.95}}%
\htdef{ACCIARRI 95F.cite}{\cite{Acciarri:1995kx}}%
\htdef{ACCIARRI 95F.collab}{L3}%
\htdef{ACCIARRI 95F.ref}{ACCIARRI 95F (L3) \cite{Acciarri:1995kx}}%
\htdef{ACCIARRI 95F.meas}{%
\begin{ensuredisplaymath}
\htuse{Gamma35.gn} = \htuse{Gamma35.td}
\end{ensuredisplaymath} & \htuse{L3.Gamma35.pub.ACCIARRI.95F}
\\
\begin{ensuredisplaymath}
\htuse{Gamma40.gn} = \htuse{Gamma40.td}
\end{ensuredisplaymath} & \htuse{L3.Gamma40.pub.ACCIARRI.95F}}%
\htdef{ACHARD 01D.cite}{\cite{Achard:2001pk}}%
\htdef{ACHARD 01D.collab}{L3}%
\htdef{ACHARD 01D.ref}{ACHARD 01D (L3) \cite{Achard:2001pk}}%
\htdef{ACHARD 01D.meas}{%
\begin{ensuredisplaymath}
\htuse{Gamma55.gn} = \htuse{Gamma55.td}
\end{ensuredisplaymath} & \htuse{L3.Gamma55.pub.ACHARD.01D}
\\
\begin{ensuredisplaymath}
\htuse{Gamma102.gn} = \htuse{Gamma102.td}
\end{ensuredisplaymath} & \htuse{L3.Gamma102.pub.ACHARD.01D}}%
\htdef{ADEVA 91F.cite}{\cite{Adeva:1991qq}}%
\htdef{ADEVA 91F.collab}{L3}%
\htdef{ADEVA 91F.ref}{ADEVA 91F (L3) \cite{Adeva:1991qq}}%
\htdef{ADEVA 91F.meas}{%
\begin{ensuredisplaymath}
\htuse{Gamma54.gn} = \htuse{Gamma54.td}
\end{ensuredisplaymath} & \htuse{L3.Gamma54.pub.ADEVA.91F}}%
\htdef{ABBIENDI 00C.cite}{\cite{Abbiendi:1999pm}}%
\htdef{ABBIENDI 00C.collab}{OPAL}%
\htdef{ABBIENDI 00C.ref}{ABBIENDI 00C (OPAL) \cite{Abbiendi:1999pm}}%
\htdef{ABBIENDI 00C.meas}{%
\begin{ensuredisplaymath}
\htuse{Gamma35.gn} = \htuse{Gamma35.td}
\end{ensuredisplaymath} & \htuse{OPAL.Gamma35.pub.ABBIENDI.00C}
\\
\begin{ensuredisplaymath}
\htuse{Gamma38.gn} = \htuse{Gamma38.td}
\end{ensuredisplaymath} & \htuse{OPAL.Gamma38.pub.ABBIENDI.00C}
\\
\begin{ensuredisplaymath}
\htuse{Gamma43.gn} = \htuse{Gamma43.td}
\end{ensuredisplaymath} & \htuse{OPAL.Gamma43.pub.ABBIENDI.00C}}%
\htdef{ABBIENDI 00D.cite}{\cite{Abbiendi:1999cq}}%
\htdef{ABBIENDI 00D.collab}{OPAL}%
\htdef{ABBIENDI 00D.ref}{ABBIENDI 00D (OPAL) \cite{Abbiendi:1999cq}}%
\htdef{ABBIENDI 00D.meas}{%
\begin{ensuredisplaymath}
\htuse{Gamma92.gn} = \htuse{Gamma92.td}
\end{ensuredisplaymath} & \htuse{OPAL.Gamma92.pub.ABBIENDI.00D}}%
\htdef{ABBIENDI 01J.cite}{\cite{Abbiendi:2000ee}}%
\htdef{ABBIENDI 01J.collab}{OPAL}%
\htdef{ABBIENDI 01J.ref}{ABBIENDI 01J (OPAL) \cite{Abbiendi:2000ee}}%
\htdef{ABBIENDI 01J.meas}{%
\begin{ensuredisplaymath}
\htuse{Gamma10.gn} = \htuse{Gamma10.td}
\end{ensuredisplaymath} & \htuse{OPAL.Gamma10.pub.ABBIENDI.01J}
\\
\begin{ensuredisplaymath}
\htuse{Gamma31.gn} = \htuse{Gamma31.td}
\end{ensuredisplaymath} & \htuse{OPAL.Gamma31.pub.ABBIENDI.01J}}%
\htdef{ABBIENDI 03.cite}{\cite{Abbiendi:2002jw}}%
\htdef{ABBIENDI 03.collab}{OPAL}%
\htdef{ABBIENDI 03.ref}{ABBIENDI 03 (OPAL) \cite{Abbiendi:2002jw}}%
\htdef{ABBIENDI 03.meas}{%
\begin{ensuredisplaymath}
\htuse{Gamma3.gn} = \htuse{Gamma3.td}
\end{ensuredisplaymath} & \htuse{OPAL.Gamma3.pub.ABBIENDI.03}}%
\htdef{ABBIENDI 04J.cite}{\cite{Abbiendi:2004xa}}%
\htdef{ABBIENDI 04J.collab}{OPAL}%
\htdef{ABBIENDI 04J.ref}{ABBIENDI 04J (OPAL) \cite{Abbiendi:2004xa}}%
\htdef{ABBIENDI 04J.meas}{%
\begin{ensuredisplaymath}
\htuse{Gamma16.gn} = \htuse{Gamma16.td}
\end{ensuredisplaymath} & \htuse{OPAL.Gamma16.pub.ABBIENDI.04J}
\\
\begin{ensuredisplaymath}
\htuse{Gamma85.gn} = \htuse{Gamma85.td}
\end{ensuredisplaymath} & \htuse{OPAL.Gamma85.pub.ABBIENDI.04J}}%
\htdef{ABBIENDI 99H.cite}{\cite{Abbiendi:1998cx}}%
\htdef{ABBIENDI 99H.collab}{OPAL}%
\htdef{ABBIENDI 99H.ref}{ABBIENDI 99H (OPAL) \cite{Abbiendi:1998cx}}%
\htdef{ABBIENDI 99H.meas}{%
\begin{ensuredisplaymath}
\htuse{Gamma5.gn} = \htuse{Gamma5.td}
\end{ensuredisplaymath} & \htuse{OPAL.Gamma5.pub.ABBIENDI.99H}}%
\htdef{ACKERSTAFF 98M.cite}{\cite{Ackerstaff:1997tx}}%
\htdef{ACKERSTAFF 98M.collab}{OPAL}%
\htdef{ACKERSTAFF 98M.ref}{ACKERSTAFF 98M (OPAL) \cite{Ackerstaff:1997tx}}%
\htdef{ACKERSTAFF 98M.meas}{%
\begin{ensuredisplaymath}
\htuse{Gamma8.gn} = \htuse{Gamma8.td}
\end{ensuredisplaymath} & \htuse{OPAL.Gamma8.pub.ACKERSTAFF.98M}
\\
\begin{ensuredisplaymath}
\htuse{Gamma13.gn} = \htuse{Gamma13.td}
\end{ensuredisplaymath} & \htuse{OPAL.Gamma13.pub.ACKERSTAFF.98M}
\\
\begin{ensuredisplaymath}
\htuse{Gamma17.gn} = \htuse{Gamma17.td}
\end{ensuredisplaymath} & \htuse{OPAL.Gamma17.pub.ACKERSTAFF.98M}}%
\htdef{ACKERSTAFF 99E.cite}{\cite{Ackerstaff:1998ia}}%
\htdef{ACKERSTAFF 99E.collab}{OPAL}%
\htdef{ACKERSTAFF 99E.ref}{ACKERSTAFF 99E (OPAL) \cite{Ackerstaff:1998ia}}%
\htdef{ACKERSTAFF 99E.meas}{%
\begin{ensuredisplaymath}
\htuse{Gamma103.gn} = \htuse{Gamma103.td}
\end{ensuredisplaymath} & \htuse{OPAL.Gamma103.pub.ACKERSTAFF.99E}
\\
\begin{ensuredisplaymath}
\htuse{Gamma104.gn} = \htuse{Gamma104.td}
\end{ensuredisplaymath} & \htuse{OPAL.Gamma104.pub.ACKERSTAFF.99E}}%
\htdef{AKERS 94G.cite}{\cite{Akers:1994td}}%
\htdef{AKERS 94G.collab}{OPAL}%
\htdef{AKERS 94G.ref}{AKERS 94G (OPAL) \cite{Akers:1994td}}%
\htdef{AKERS 94G.meas}{%
\begin{ensuredisplaymath}
\htuse{Gamma33.gn} = \htuse{Gamma33.td}
\end{ensuredisplaymath} & \htuse{OPAL.Gamma33.pub.AKERS.94G}}%
\htdef{AKERS 95Y.cite}{\cite{Akers:1995ry}}%
\htdef{AKERS 95Y.collab}{OPAL}%
\htdef{AKERS 95Y.ref}{AKERS 95Y (OPAL) \cite{Akers:1995ry}}%
\htdef{AKERS 95Y.meas}{%
\begin{ensuredisplaymath}
\htuse{Gamma55.gn} = \htuse{Gamma55.td}
\end{ensuredisplaymath} & \htuse{OPAL.Gamma55.pub.AKERS.95Y}
\\
\begin{ensuredisplaymath}
\htuse{Gamma57by55.gn} = \htuse{Gamma57by55.td}
\end{ensuredisplaymath} & \htuse{OPAL.Gamma57by55.pub.AKERS.95Y}}%
\htdef{ALEXANDER 91D.cite}{\cite{Alexander:1991am}}%
\htdef{ALEXANDER 91D.collab}{OPAL}%
\htdef{ALEXANDER 91D.ref}{ALEXANDER 91D (OPAL) \cite{Alexander:1991am}}%
\htdef{ALEXANDER 91D.meas}{%
\begin{ensuredisplaymath}
\htuse{Gamma7.gn} = \htuse{Gamma7.td}
\end{ensuredisplaymath} & \htuse{OPAL.Gamma7.pub.ALEXANDER.91D}}%
\htdef{AIHARA 87B.cite}{\cite{Aihara:1986mw}}%
\htdef{AIHARA 87B.collab}{TPC}%
\htdef{AIHARA 87B.ref}{AIHARA 87B (TPC) \cite{Aihara:1986mw}}%
\htdef{AIHARA 87B.meas}{%
\begin{ensuredisplaymath}
\htuse{Gamma54.gn} = \htuse{Gamma54.td}
\end{ensuredisplaymath} & \htuse{TPC.Gamma54.pub.AIHARA.87B}}%
\htdef{BAUER 94.cite}{\cite{Bauer:1993wn}}%
\htdef{BAUER 94.collab}{TPC}%
\htdef{BAUER 94.ref}{BAUER 94 (TPC) \cite{Bauer:1993wn}}%
\htdef{BAUER 94.meas}{%
\begin{ensuredisplaymath}
\htuse{Gamma82.gn} = \htuse{Gamma82.td}
\end{ensuredisplaymath} & \htuse{TPC.Gamma82.pub.BAUER.94}
\\
\begin{ensuredisplaymath}
\htuse{Gamma92.gn} = \htuse{Gamma92.td}
\end{ensuredisplaymath} & \htuse{TPC.Gamma92.pub.BAUER.94}}%
\htdef{MeasPaper}{%
\multicolumn{2}{l}{\htuse{BARATE 98.ref}} \\
\htuse{BARATE 98.meas} \\\hline
\multicolumn{2}{l}{\htuse{BARATE 98E.ref}} \\
\htuse{BARATE 98E.meas} \\\hline
\multicolumn{2}{l}{\htuse{BARATE 99K.ref}} \\
\htuse{BARATE 99K.meas} \\\hline
\multicolumn{2}{l}{\htuse{BARATE 99R.ref}} \\
\htuse{BARATE 99R.meas} \\\hline
\multicolumn{2}{l}{\htuse{BUSKULIC 96.ref}} \\
\htuse{BUSKULIC 96.meas} \\\hline
\multicolumn{2}{l}{\htuse{BUSKULIC 97C.ref}} \\
\htuse{BUSKULIC 97C.meas} \\\hline
\multicolumn{2}{l}{\htuse{SCHAEL 05C.ref}} \\
\htuse{SCHAEL 05C.meas} \\\hline
\multicolumn{2}{l}{\htuse{ALBRECHT 88B.ref}} \\
\htuse{ALBRECHT 88B.meas} \\\hline
\multicolumn{2}{l}{\htuse{ALBRECHT 92D.ref}} \\
\htuse{ALBRECHT 92D.meas} \\\hline
\multicolumn{2}{l}{\htuse{AUBERT 08.ref}} \\
\htuse{AUBERT 08.meas} \\\hline
\multicolumn{2}{l}{\htuse{AUBERT 10F.ref}} \\
\htuse{AUBERT 10F.meas} \\\hline
\multicolumn{2}{l}{\htuse{DEL-AMO-SANCHEZ 11E.ref}} \\
\htuse{DEL-AMO-SANCHEZ 11E.meas} \\\hline
\multicolumn{2}{l}{\htuse{LEES 12X.ref}} \\
\htuse{LEES 12X.meas} \\\hline
\multicolumn{2}{l}{\htuse{LEES 12Y.ref}} \\
\htuse{LEES 12Y.meas} \\\hline
\multicolumn{2}{l}{\htuse{LEES 18B.ref}} \\
\htuse{LEES 18B.meas} \\\hline
\multicolumn{2}{l}{\htuse{BaBar prelim. ICHEP2018.ref}} \\
\htuse{BaBar prelim. ICHEP2018.meas} \\\hline
\multicolumn{2}{l}{\htuse{FUJIKAWA 08.ref}} \\
\htuse{FUJIKAWA 08.meas} \\\hline
\multicolumn{2}{l}{\htuse{INAMI 09.ref}} \\
\htuse{INAMI 09.meas} \\\hline
\multicolumn{2}{l}{\htuse{LEE 10.ref}} \\
\htuse{LEE 10.meas} \\\hline
\multicolumn{2}{l}{\htuse{RYU 14vpc.ref}} \\
\htuse{RYU 14vpc.meas} \\\hline
\multicolumn{2}{l}{\htuse{BEHREND 89B.ref}} \\
\htuse{BEHREND 89B.meas} \\\hline
\multicolumn{2}{l}{\htuse{ANASTASSOV 01.ref}} \\
\htuse{ANASTASSOV 01.meas} \\\hline
\multicolumn{2}{l}{\htuse{ANASTASSOV 97.ref}} \\
\htuse{ANASTASSOV 97.meas} \\\hline
\multicolumn{2}{l}{\htuse{ARTUSO 92.ref}} \\
\htuse{ARTUSO 92.meas} \\\hline
\multicolumn{2}{l}{\htuse{ARTUSO 94.ref}} \\
\htuse{ARTUSO 94.meas} \\\hline
\multicolumn{2}{l}{\htuse{BALEST 95C.ref}} \\
\htuse{BALEST 95C.meas} \\\hline
\multicolumn{2}{l}{\htuse{BARINGER 87.ref}} \\
\htuse{BARINGER 87.meas} \\\hline
\multicolumn{2}{l}{\htuse{BARTELT 96.ref}} \\
\htuse{BARTELT 96.meas} \\\hline
\multicolumn{2}{l}{\htuse{BATTLE 94.ref}} \\
\htuse{BATTLE 94.meas} \\\hline
\multicolumn{2}{l}{\htuse{BISHAI 99.ref}} \\
\htuse{BISHAI 99.meas} \\\hline
\multicolumn{2}{l}{\htuse{BORTOLETTO 93.ref}} \\
\htuse{BORTOLETTO 93.meas} \\\hline
\multicolumn{2}{l}{\htuse{COAN 96.ref}} \\
\htuse{COAN 96.meas} \\\hline
\multicolumn{2}{l}{\htuse{EDWARDS 00A.ref}} \\
\htuse{EDWARDS 00A.meas} \\\hline
\multicolumn{2}{l}{\htuse{GIBAUT 94B.ref}} \\
\htuse{GIBAUT 94B.meas} \\\hline
\multicolumn{2}{l}{\htuse{PROCARIO 93.ref}} \\
\htuse{PROCARIO 93.meas} \\\hline
\multicolumn{2}{l}{\htuse{RICHICHI 99.ref}} \\
\htuse{RICHICHI 99.meas} \\\hline
\multicolumn{2}{l}{\htuse{ARMS 05.ref}} \\
\htuse{ARMS 05.meas} \\\hline
\multicolumn{2}{l}{\htuse{BRIERE 03.ref}} \\
\htuse{BRIERE 03.meas} \\\hline
\multicolumn{2}{l}{\htuse{ABDALLAH 06A.ref}} \\
\htuse{ABDALLAH 06A.meas} \\\hline
\multicolumn{2}{l}{\htuse{ABREU 92N.ref}} \\
\htuse{ABREU 92N.meas} \\\hline
\multicolumn{2}{l}{\htuse{ABREU 94K.ref}} \\
\htuse{ABREU 94K.meas} \\\hline
\multicolumn{2}{l}{\htuse{ABREU 99X.ref}} \\
\htuse{ABREU 99X.meas} \\\hline
\multicolumn{2}{l}{\htuse{BYLSMA 87.ref}} \\
\htuse{BYLSMA 87.meas} \\\hline
\multicolumn{2}{l}{\htuse{ACCIARRI 01F.ref}} \\
\htuse{ACCIARRI 01F.meas} \\\hline
\multicolumn{2}{l}{\htuse{ACCIARRI 95.ref}} \\
\htuse{ACCIARRI 95.meas} \\\hline
\multicolumn{2}{l}{\htuse{ACCIARRI 95F.ref}} \\
\htuse{ACCIARRI 95F.meas} \\\hline
\multicolumn{2}{l}{\htuse{ACHARD 01D.ref}} \\
\htuse{ACHARD 01D.meas} \\\hline
\multicolumn{2}{l}{\htuse{ADEVA 91F.ref}} \\
\htuse{ADEVA 91F.meas} \\\hline
\multicolumn{2}{l}{\htuse{ABBIENDI 00C.ref}} \\
\htuse{ABBIENDI 00C.meas} \\\hline
\multicolumn{2}{l}{\htuse{ABBIENDI 00D.ref}} \\
\htuse{ABBIENDI 00D.meas} \\\hline
\multicolumn{2}{l}{\htuse{ABBIENDI 01J.ref}} \\
\htuse{ABBIENDI 01J.meas} \\\hline
\multicolumn{2}{l}{\htuse{ABBIENDI 03.ref}} \\
\htuse{ABBIENDI 03.meas} \\\hline
\multicolumn{2}{l}{\htuse{ABBIENDI 04J.ref}} \\
\htuse{ABBIENDI 04J.meas} \\\hline
\multicolumn{2}{l}{\htuse{ABBIENDI 99H.ref}} \\
\htuse{ABBIENDI 99H.meas} \\\hline
\multicolumn{2}{l}{\htuse{ACKERSTAFF 98M.ref}} \\
\htuse{ACKERSTAFF 98M.meas} \\\hline
\multicolumn{2}{l}{\htuse{ACKERSTAFF 99E.ref}} \\
\htuse{ACKERSTAFF 99E.meas} \\\hline
\multicolumn{2}{l}{\htuse{AKERS 94G.ref}} \\
\htuse{AKERS 94G.meas} \\\hline
\multicolumn{2}{l}{\htuse{AKERS 95Y.ref}} \\
\htuse{AKERS 95Y.meas} \\\hline
\multicolumn{2}{l}{\htuse{ALEXANDER 91D.ref}} \\
\htuse{ALEXANDER 91D.meas} \\\hline
\multicolumn{2}{l}{\htuse{AIHARA 87B.ref}} \\
\htuse{AIHARA 87B.meas} \\\hline
\multicolumn{2}{l}{\htuse{BAUER 94.ref}} \\
\htuse{BAUER 94.meas}}%
\htdef{BrStrangeVal}{%
\htQuantLine{Gamma10}{0.6986 \pm 0.0085}{-2} 
\htQuantLine{Gamma16}{0.4904 \pm 0.0092}{-2} 
\htQuantLine{Gamma23}{0.0585 \pm 0.0027}{-2} 
\htQuantLine{Gamma28}{0.0113 \pm 0.0026}{-2} 
\htQuantLine{Gamma35}{0.8378 \pm 0.0139}{-2} 
\htQuantLine{Gamma40}{0.3807 \pm 0.0129}{-2} 
\htQuantLine{Gamma44}{0.0235 \pm 0.0231}{-2} 
\htQuantLine{Gamma53}{0.0222 \pm 0.0202}{-2} 
\htQuantLine{Gamma128}{0.0154 \pm 0.0008}{-2} 
\htQuantLine{Gamma130}{0.0048 \pm 0.0012}{-2} 
\htQuantLine{Gamma132}{0.0094 \pm 0.0015}{-2} 
\htQuantLine{Gamma151}{0.0410 \pm 0.0092}{-2} 
\htQuantLine{Gamma168}{0.0022 \pm 0.0008}{-2} 
\htQuantLine{Gamma169}{0.0015 \pm 0.0006}{-2} 
\htQuantLine{Gamma802}{0.2923 \pm 0.0067}{-2} 
\htQuantLine{Gamma803}{0.0410 \pm 0.0143}{-2} 
\htQuantLine{Gamma822}{0.0001 \pm 0.0001}{-2} 
\htQuantLine{Gamma833}{0.0001 \pm 0.0001}{-2}}%
\htdef{BrStrangeTotVal}{%
\htQuantLine{Gamma110}{2.9307 \pm 0.0412}{-2}}%
\htdef{UnitarityQuants}{%
\htConstrLine{Gamma3}{17.3915 \pm 0.0395}{1.0000}{-2}{0} 
\htConstrLine{Gamma5}{17.8171 \pm 0.0409}{1.0000}{-2}{0} 
\htConstrLine{Gamma9}{10.8036 \pm 0.0519}{1.0000}{-2}{0} 
\htConstrLine{Gamma10}{0.6986 \pm 0.0085}{1.0000}{-2}{0} 
\htConstrLine{Gamma14}{25.4474 \pm 0.0908}{1.0000}{-2}{0} 
\htConstrLine{Gamma16}{0.4904 \pm 0.0092}{1.0000}{-2}{0} 
\htConstrLine{Gamma20}{9.2114 \pm 0.0921}{1.0000}{-2}{0} 
\htConstrLine{Gamma23}{0.0585 \pm 0.0027}{1.0000}{-2}{0} 
\htConstrLine{Gamma27}{1.1381 \pm 0.0292}{1.0000}{-2}{0} 
\htConstrLine{Gamma28}{0.0113 \pm 0.0026}{1.0000}{-2}{0} 
\htConstrLine{Gamma30}{0.0864 \pm 0.0067}{1.0000}{-2}{0} 
\htConstrLine{Gamma35}{0.8378 \pm 0.0139}{1.0000}{-2}{0} 
\htConstrLine{Gamma37}{0.1483 \pm 0.0034}{1.0000}{-2}{0} 
\htConstrLine{Gamma40}{0.3807 \pm 0.0129}{1.0000}{-2}{0} 
\htConstrLine{Gamma42}{0.1494 \pm 0.0070}{1.0000}{-2}{0} 
\htConstrLine{Gamma44}{0.0235 \pm 0.0231}{1.0000}{-2}{0} 
\htConstrLine{Gamma47}{0.0234 \pm 0.0006}{2.0000}{-2}{0} 
\htConstrLine{Gamma48}{0.1048 \pm 0.0247}{1.0000}{-2}{0} 
\htConstrLine{Gamma50}{0.0018 \pm 0.0002}{2.0000}{-2}{0} 
\htConstrLine{Gamma51}{0.0318 \pm 0.0119}{1.0000}{-2}{0} 
\htConstrLine{Gamma53}{0.0222 \pm 0.0202}{1.0000}{-2}{0} 
\htConstrLine{Gamma62}{8.9597 \pm 0.0511}{1.0000}{-2}{0} 
\htConstrLine{Gamma70}{2.7683 \pm 0.0711}{1.0000}{-2}{0} 
\htConstrLine{Gamma77}{0.0977 \pm 0.0356}{1.0000}{-2}{0} 
\htConstrLine{Gamma93}{0.1432 \pm 0.0027}{1.0000}{-2}{0} 
\htConstrLine{Gamma94}{0.0061 \pm 0.0018}{1.0000}{-2}{0} 
\htConstrLine{Gamma126}{0.1386 \pm 0.0072}{1.0000}{-2}{0} 
\htConstrLine{Gamma128}{0.0154 \pm 0.0008}{1.0000}{-2}{0} 
\htConstrLine{Gamma130}{0.0048 \pm 0.0012}{1.0000}{-2}{0} 
\htConstrLine{Gamma132}{0.0094 \pm 0.0015}{1.0000}{-2}{0} 
\htConstrLine{Gamma136}{0.0220 \pm 0.0013}{1.0000}{-2}{0} 
\htConstrLine{Gamma151}{0.0410 \pm 0.0092}{1.0000}{-2}{0} 
\htConstrLine{Gamma152}{0.4066 \pm 0.0419}{1.0000}{-2}{0} 
\htConstrLine{Gamma167}{0.0044 \pm 0.0016}{0.8320}{-2}{0} 
\htConstrLine{Gamma800}{1.9546 \pm 0.0647}{1.0000}{-2}{0} 
\htConstrLine{Gamma802}{0.2923 \pm 0.0067}{1.0000}{-2}{0} 
\htConstrLine{Gamma803}{0.0410 \pm 0.0143}{1.0000}{-2}{0} 
\htConstrLine{Gamma805}{0.0400 \pm 0.0200}{1.0000}{-2}{0} 
\htConstrLine{Gamma811}{0.0071 \pm 0.0016}{1.0000}{-2}{0} 
\htConstrLine{Gamma812}{0.0013 \pm 0.0027}{1.0000}{-2}{0} 
\htConstrLine{Gamma821}{0.0772 \pm 0.0030}{1.0000}{-2}{0} 
\htConstrLine{Gamma822}{0.0001 \pm 0.0001}{1.0000}{-2}{0} 
\htConstrLine{Gamma831}{0.0084 \pm 0.0006}{1.0000}{-2}{0} 
\htConstrLine{Gamma832}{0.0038 \pm 0.0009}{1.0000}{-2}{0} 
\htConstrLine{Gamma833}{0.0001 \pm 0.0001}{1.0000}{-2}{0} 
\htConstrLine{Gamma920}{0.0052 \pm 0.0004}{1.0000}{-2}{0} 
\htConstrLine{Gamma945}{0.0194 \pm 0.0038}{1.0000}{-2}{0} 
\htConstrLine{Gamma998}{0.0298 \pm 0.1026}{1.0000}{-2}{0}}%
\htdef{BaseQuants}{%
\htQuantLine{Gamma3}{17.3915 \pm 0.0395}{-2} 
\htQuantLine{Gamma5}{17.8171 \pm 0.0409}{-2} 
\htQuantLine{Gamma9}{10.8036 \pm 0.0519}{-2} 
\htQuantLine{Gamma10}{0.6986 \pm 0.0085}{-2} 
\htQuantLine{Gamma14}{25.4474 \pm 0.0908}{-2} 
\htQuantLine{Gamma16}{0.4904 \pm 0.0092}{-2} 
\htQuantLine{Gamma20}{9.2114 \pm 0.0921}{-2} 
\htQuantLine{Gamma23}{0.0585 \pm 0.0027}{-2} 
\htQuantLine{Gamma27}{1.1381 \pm 0.0292}{-2} 
\htQuantLine{Gamma28}{0.0113 \pm 0.0026}{-2} 
\htQuantLine{Gamma30}{0.0864 \pm 0.0067}{-2} 
\htQuantLine{Gamma35}{0.8378 \pm 0.0139}{-2} 
\htQuantLine{Gamma37}{0.1483 \pm 0.0034}{-2} 
\htQuantLine{Gamma40}{0.3807 \pm 0.0129}{-2} 
\htQuantLine{Gamma42}{0.1494 \pm 0.0070}{-2} 
\htQuantLine{Gamma44}{0.0235 \pm 0.0231}{-2} 
\htQuantLine{Gamma47}{0.0234 \pm 0.0006}{-2} 
\htQuantLine{Gamma48}{0.1048 \pm 0.0247}{-2} 
\htQuantLine{Gamma50}{0.0018 \pm 0.0002}{-2} 
\htQuantLine{Gamma51}{0.0318 \pm 0.0119}{-2} 
\htQuantLine{Gamma53}{0.0222 \pm 0.0202}{-2} 
\htQuantLine{Gamma62}{8.9597 \pm 0.0511}{-2} 
\htQuantLine{Gamma70}{2.7683 \pm 0.0711}{-2} 
\htQuantLine{Gamma77}{0.0977 \pm 0.0356}{-2} 
\htQuantLine{Gamma93}{0.1432 \pm 0.0027}{-2} 
\htQuantLine{Gamma94}{0.0061 \pm 0.0018}{-2} 
\htQuantLine{Gamma126}{0.1386 \pm 0.0072}{-2} 
\htQuantLine{Gamma128}{0.0154 \pm 0.0008}{-2} 
\htQuantLine{Gamma130}{0.0048 \pm 0.0012}{-2} 
\htQuantLine{Gamma132}{0.0094 \pm 0.0015}{-2} 
\htQuantLine{Gamma136}{0.0220 \pm 0.0013}{-2} 
\htQuantLine{Gamma151}{0.0410 \pm 0.0092}{-2} 
\htQuantLine{Gamma152}{0.4066 \pm 0.0419}{-2} 
\htQuantLine{Gamma167}{0.0044 \pm 0.0016}{-2} 
\htQuantLine{Gamma800}{1.9546 \pm 0.0647}{-2} 
\htQuantLine{Gamma802}{0.2923 \pm 0.0067}{-2} 
\htQuantLine{Gamma803}{0.0410 \pm 0.0143}{-2} 
\htQuantLine{Gamma805}{0.0400 \pm 0.0200}{-2} 
\htQuantLine{Gamma811}{0.0071 \pm 0.0016}{-2} 
\htQuantLine{Gamma812}{0.0013 \pm 0.0027}{-2} 
\htQuantLine{Gamma821}{0.0772 \pm 0.0030}{-2} 
\htQuantLine{Gamma822}{0.0001 \pm 0.0001}{-2} 
\htQuantLine{Gamma831}{0.0084 \pm 0.0006}{-2} 
\htQuantLine{Gamma832}{0.0038 \pm 0.0009}{-2} 
\htQuantLine{Gamma833}{0.0001 \pm 0.0001}{-2} 
\htQuantLine{Gamma920}{0.0052 \pm 0.0004}{-2} 
\htQuantLine{Gamma945}{0.0194 \pm 0.0038}{-2}}%
\htdef{BrCorr}{%
\ifhevea\begin{table}\fi
\begin{center}
\ifhevea
\caption{Basis quantities correlation coefficients in percent, subtable 1.\label{tab:tau:br-fit-corr1}}%
\else
\begin{minipage}{\linewidth}
\begin{center}
\captionof{table}{Basis quantities correlation coefficients in percent, subtable 1.}\label{tab:tau:br-fit-corr1}%
\fi
\begin{envsmall}
\begin{center}
\begin{tabular}{rrrrrrrrrrrrrrr}
\hline
\( \BR_{5} \) &   22 &  &  &  &  &  &  &  &  &  &  &  &  &  \\
\( \BR_{9} \) &    6 &    4 &  &  &  &  &  &  &  &  &  &  &  &  \\
\( \BR_{10} \) &    2 &    4 &    2 &  &  &  &  &  &  &  &  &  &  &  \\
\( \BR_{14} \) &  -13 &  -14 &  -13 &   -7 &  &  &  &  &  &  &  &  &  &  \\
\( \BR_{16} \) &   -2 &   -1 &   -3 &   35 &  -13 &  &  &  &  &  &  &  &  &  \\
\( \BR_{20} \) &   -7 &   -7 &  -12 &   -4 &  -42 &  -16 &  &  &  &  &  &  &  &  \\
\( \BR_{23} \) &   -3 &   -2 &   -5 &   14 &   -9 &   66 &  -18 &  &  &  &  &  &  &  \\
\( \BR_{27} \) &   -4 &   -4 &   -7 &    3 &   -9 &   61 &  -23 &   72 &  &  &  &  &  &  \\
\( \BR_{28} \) &   -2 &   -1 &   -3 &    2 &   -4 &   32 &  -10 &   28 &   37 &  &  &  &  &  \\
\( \BR_{30} \) &   -3 &   -3 &   -6 &   -1 &   -6 &   34 &  -14 &   41 &   52 &   23 &  &  &  &  \\
\( \BR_{35} \) &    0 &    0 &    0 &    0 &    0 &    0 &    0 &    0 &    0 &    0 &    0 &  &  &  \\
\( \BR_{37} \) &    0 &   -1 &    1 &    0 &    0 &    0 &    0 &    0 &    0 &   -1 &    0 &  -15 &  &  \\
\( \BR_{40} \) &    0 &    0 &    0 &    0 &    0 &    0 &    0 &    0 &   -1 &    0 &    0 &  -12 &    2 &  \\
 & \( \BR_{3} \) & \( \BR_{5} \) & \( \BR_{9} \) & \( \BR_{10} \) & \( \BR_{14} \) & \( \BR_{16} \) & \( \BR_{20} \) & \( \BR_{23} \) & \( \BR_{27} \) & \( \BR_{28} \) & \( \BR_{30} \) & \( \BR_{35} \) & \( \BR_{37} \) & \( \BR_{40} \)
\\\hline
\end{tabular}
\end{center}
\end{envsmall}
\ifhevea\else
\end{center}
\end{minipage}
\fi
\end{center}
\ifhevea\end{table}\fi
\ifhevea\begin{table}\fi
\begin{center}
\ifhevea
\caption{Basis quantities correlation coefficients in percent, subtable 2.\label{tab:tau:br-fit-corr2}}%
\else
\begin{minipage}{\linewidth}
\begin{center}
\captionof{table}{Basis quantities correlation coefficients in percent, subtable 2.}\label{tab:tau:br-fit-corr2}%
\fi
\begin{envsmall}
\begin{center}
\begin{tabular}{rrrrrrrrrrrrrrr}
\hline
\( \BR_{42} \) &    0 &    0 &    0 &   -2 &    1 &   -5 &    1 &   -4 &   -4 &   -2 &   -2 &   -1 &  -15 &  -20 \\
\( \BR_{44} \) &    0 &    0 &    0 &    0 &    0 &    0 &    0 &    0 &    0 &    0 &    0 &   -1 &    0 &   -4 \\
\( \BR_{47} \) &    0 &   -1 &    2 &    1 &   -1 &    2 &   -1 &    1 &    1 &    0 &    0 &   -1 &    2 &   -4 \\
\( \BR_{48} \) &    0 &    0 &    0 &    0 &    0 &    0 &    0 &    0 &    0 &    0 &    0 &   -3 &    0 &   -2 \\
\( \BR_{50} \) &    0 &    0 &    0 &    0 &    0 &    0 &    0 &    0 &    0 &    0 &    0 &    1 &    5 &    0 \\
\( \BR_{51} \) &    0 &    0 &    0 &    0 &    0 &    0 &    0 &    0 &    0 &    0 &    0 &   -1 &    0 &   -1 \\
\( \BR_{53} \) &    0 &    0 &    0 &    0 &    0 &    0 &    0 &    0 &    0 &    0 &    0 &    0 &    0 &    0 \\
\( \BR_{62} \) &   -4 &   -5 &    6 &    2 &   -4 &    1 &  -11 &   -1 &   -2 &   -2 &   -3 &   -1 &    3 &    0 \\
\( \BR_{70} \) &   -5 &   -6 &   -7 &   -2 &   -8 &   -1 &   -1 &   -1 &   -1 &    0 &    0 &    0 &   -1 &    0 \\
\( \BR_{77} \) &    0 &    0 &   -2 &    0 &   -2 &    1 &    0 &    1 &    2 &    1 &    1 &    0 &    0 &    0 \\
\( \BR_{93} \) &   -1 &   -1 &    2 &    1 &   -1 &    1 &   -2 &    0 &    0 &    0 &    0 &    0 &    1 &    0 \\
\( \BR_{94} \) &    0 &    0 &    0 &    0 &    0 &    0 &    0 &    0 &    0 &    0 &    0 &    0 &    0 &    0 \\
\( \BR_{126} \) &    0 &    0 &    0 &    0 &    0 &    0 &   -1 &    0 &    0 &    0 &    0 &    0 &    0 &    0 \\
\( \BR_{128} \) &    0 &    0 &    1 &    0 &    0 &    0 &    0 &    0 &    0 &    0 &    0 &    0 &    1 &    0 \\
 & \( \BR_{3} \) & \( \BR_{5} \) & \( \BR_{9} \) & \( \BR_{10} \) & \( \BR_{14} \) & \( \BR_{16} \) & \( \BR_{20} \) & \( \BR_{23} \) & \( \BR_{27} \) & \( \BR_{28} \) & \( \BR_{30} \) & \( \BR_{35} \) & \( \BR_{37} \) & \( \BR_{40} \)
\\\hline
\end{tabular}
\end{center}
\end{envsmall}
\ifhevea\else
\end{center}
\end{minipage}
\fi
\end{center}
\ifhevea\end{table}\fi
\ifhevea\begin{table}\fi
\begin{center}
\ifhevea
\caption{Basis quantities correlation coefficients in percent, subtable 3.\label{tab:tau:br-fit-corr3}}%
\else
\begin{minipage}{\linewidth}
\begin{center}
\captionof{table}{Basis quantities correlation coefficients in percent, subtable 3.}\label{tab:tau:br-fit-corr3}%
\fi
\begin{envsmall}
\begin{center}
\begin{tabular}{rrrrrrrrrrrrrrr}
\hline
\( \BR_{130} \) &    0 &    0 &    0 &    0 &    0 &    0 &    0 &    0 &    0 &    0 &    0 &    0 &    0 &    0 \\
\( \BR_{132} \) &    0 &    0 &    0 &    0 &    0 &    0 &    0 &    0 &    0 &    0 &    0 &    0 &    0 &    0 \\
\( \BR_{136} \) &    0 &    0 &    1 &    1 &    0 &    1 &   -1 &    0 &    0 &    0 &    0 &    0 &    1 &    0 \\
\( \BR_{151} \) &    0 &    0 &    0 &    0 &    0 &    0 &    0 &    0 &    0 &    0 &    0 &    0 &    0 &    0 \\
\( \BR_{152} \) &    0 &    0 &   -3 &    0 &   -2 &    1 &    0 &    1 &    2 &    1 &    2 &    0 &    0 &    0 \\
\( \BR_{167} \) &    0 &    0 &    0 &    0 &    0 &    0 &    0 &    0 &    0 &    0 &    0 &    0 &    0 &    0 \\
\( \BR_{800} \) &   -1 &   -1 &   -2 &    0 &   -3 &    0 &    0 &    0 &    0 &    0 &    0 &    0 &    0 &    0 \\
\( \BR_{802} \) &   -1 &   -1 &    0 &    0 &   -1 &   -1 &   -3 &   -1 &   -2 &   -1 &   -1 &    0 &    0 &    0 \\
\( \BR_{803} \) &    0 &    0 &    0 &    0 &    0 &    0 &    0 &    0 &    0 &    0 &    0 &    0 &    0 &    0 \\
\( \BR_{805} \) &    0 &    0 &    0 &    0 &    0 &    0 &    0 &    0 &    0 &    0 &    0 &    0 &    0 &    0 \\
\( \BR_{811} \) &    0 &    0 &    0 &    0 &    0 &    0 &    0 &    0 &    0 &    0 &    0 &    0 &    0 &    0 \\
\( \BR_{812} \) &    1 &    1 &    0 &    0 &    0 &    0 &    0 &    0 &    0 &    0 &    0 &    0 &    0 &    0 \\
\( \BR_{821} \) &    0 &    0 &    2 &    1 &    0 &    1 &   -2 &    0 &    0 &    0 &    0 &    0 &    1 &    0 \\
\( \BR_{822} \) &    0 &    0 &    0 &    0 &    0 &    0 &    0 &    0 &    0 &    0 &    0 &    0 &    0 &    0 \\
 & \( \BR_{3} \) & \( \BR_{5} \) & \( \BR_{9} \) & \( \BR_{10} \) & \( \BR_{14} \) & \( \BR_{16} \) & \( \BR_{20} \) & \( \BR_{23} \) & \( \BR_{27} \) & \( \BR_{28} \) & \( \BR_{30} \) & \( \BR_{35} \) & \( \BR_{37} \) & \( \BR_{40} \)
\\\hline
\end{tabular}
\end{center}
\end{envsmall}
\ifhevea\else
\end{center}
\end{minipage}
\fi
\end{center}
\ifhevea\end{table}\fi
\ifhevea\begin{table}\fi
\begin{center}
\ifhevea
\caption{Basis quantities correlation coefficients in percent, subtable 4.\label{tab:tau:br-fit-corr4}}%
\else
\begin{minipage}{\linewidth}
\begin{center}
\captionof{table}{Basis quantities correlation coefficients in percent, subtable 4.}\label{tab:tau:br-fit-corr4}%
\fi
\begin{envsmall}
\begin{center}
\begin{tabular}{rrrrrrrrrrrrrrr}
\hline
\( \BR_{831} \) &    0 &    0 &    1 &    0 &    0 &    0 &   -1 &    0 &    0 &    0 &    0 &    0 &    1 &    0 \\
\( \BR_{832} \) &    0 &    0 &    0 &    0 &    0 &    0 &    0 &    0 &    0 &    0 &    0 &    0 &    0 &    0 \\
\( \BR_{833} \) &    0 &    0 &    0 &    0 &    0 &    0 &    0 &    0 &    0 &    0 &    0 &    0 &    0 &    0 \\
\( \BR_{920} \) &    0 &    0 &    1 &    0 &    0 &    0 &   -1 &    0 &    0 &    0 &    0 &    0 &    0 &    0 \\
\( \BR_{945} \) &    0 &    0 &    0 &    0 &    0 &    0 &    0 &    0 &    0 &    0 &    0 &    0 &    0 &    0 \\
 & \( \BR_{3} \) & \( \BR_{5} \) & \( \BR_{9} \) & \( \BR_{10} \) & \( \BR_{14} \) & \( \BR_{16} \) & \( \BR_{20} \) & \( \BR_{23} \) & \( \BR_{27} \) & \( \BR_{28} \) & \( \BR_{30} \) & \( \BR_{35} \) & \( \BR_{37} \) & \( \BR_{40} \)
\\\hline
\end{tabular}
\end{center}
\end{envsmall}
\ifhevea\else
\end{center}
\end{minipage}
\fi
\end{center}
\ifhevea\end{table}\fi
\ifhevea\begin{table}\fi
\begin{center}
\ifhevea
\caption{Basis quantities correlation coefficients in percent, subtable 5.\label{tab:tau:br-fit-corr5}}%
\else
\begin{minipage}{\linewidth}
\begin{center}
\captionof{table}{Basis quantities correlation coefficients in percent, subtable 5.}\label{tab:tau:br-fit-corr5}%
\fi
\begin{envsmall}
\begin{center}
\begin{tabular}{rrrrrrrrrrrrrrr}
\hline
\( \BR_{44} \) &    0 &  &  &  &  &  &  &  &  &  &  &  &  &  \\
\( \BR_{47} \) &    1 &    0 &  &  &  &  &  &  &  &  &  &  &  &  \\
\( \BR_{48} \) &   -1 &   -6 &    0 &  &  &  &  &  &  &  &  &  &  &  \\
\( \BR_{50} \) &    6 &    0 &   -7 &    0 &  &  &  &  &  &  &  &  &  &  \\
\( \BR_{51} \) &    0 &   -3 &    0 &   -6 &    0 &  &  &  &  &  &  &  &  &  \\
\( \BR_{53} \) &    0 &    0 &    0 &    0 &    0 &    0 &  &  &  &  &  &  &  &  \\
\( \BR_{62} \) &   -1 &    0 &    5 &    0 &    1 &    0 &    0 &  &  &  &  &  &  &  \\
\( \BR_{70} \) &    0 &    0 &   -1 &    0 &    0 &    0 &    0 &  -19 &  &  &  &  &  &  \\
\( \BR_{77} \) &    0 &    0 &    0 &    0 &    0 &    0 &    0 &   -1 &   -7 &  &  &  &  &  \\
\( \BR_{93} \) &    0 &    0 &    2 &    0 &    0 &    0 &    0 &   14 &   -4 &    0 &  &  &  &  \\
\( \BR_{94} \) &    0 &    0 &    0 &    0 &    0 &    0 &    0 &    0 &   -2 &    0 &    0 &  &  &  \\
\( \BR_{126} \) &    0 &    0 &    0 &    0 &    0 &    0 &    0 &    0 &    0 &   -5 &    0 &    0 &  &  \\
\( \BR_{128} \) &    0 &    0 &    1 &    0 &    0 &    0 &    0 &    2 &    0 &    0 &    1 &    0 &    4 &  \\
 & \( \BR_{42} \) & \( \BR_{44} \) & \( \BR_{47} \) & \( \BR_{48} \) & \( \BR_{50} \) & \( \BR_{51} \) & \( \BR_{53} \) & \( \BR_{62} \) & \( \BR_{70} \) & \( \BR_{77} \) & \( \BR_{93} \) & \( \BR_{94} \) & \( \BR_{126} \) & \( \BR_{128} \)
\\\hline
\end{tabular}
\end{center}
\end{envsmall}
\ifhevea\else
\end{center}
\end{minipage}
\fi
\end{center}
\ifhevea\end{table}\fi
\ifhevea\begin{table}\fi
\begin{center}
\ifhevea
\caption{Basis quantities correlation coefficients in percent, subtable 6.\label{tab:tau:br-fit-corr6}}%
\else
\begin{minipage}{\linewidth}
\begin{center}
\captionof{table}{Basis quantities correlation coefficients in percent, subtable 6.}\label{tab:tau:br-fit-corr6}%
\fi
\begin{envsmall}
\begin{center}
\begin{tabular}{rrrrrrrrrrrrrrr}
\hline
\( \BR_{130} \) &    0 &    0 &    0 &    0 &    0 &    0 &    0 &    0 &    0 &   -1 &    0 &    0 &    1 &    1 \\
\( \BR_{132} \) &    0 &    0 &    0 &    0 &    0 &    0 &    0 &    0 &    0 &    0 &    0 &    0 &    2 &    1 \\
\( \BR_{136} \) &    0 &    0 &    1 &    0 &    0 &    0 &    0 &    2 &   -1 &    0 &    1 &    0 &    0 &    0 \\
\( \BR_{151} \) &    0 &    0 &    0 &    0 &    0 &    0 &    0 &    0 &   12 &    0 &    0 &    0 &    0 &    0 \\
\( \BR_{152} \) &    0 &    0 &    0 &    0 &    0 &    0 &    0 &   -1 &  -11 &  -64 &    0 &    0 &    0 &    0 \\
\( \BR_{167} \) &    0 &    0 &    0 &    0 &    0 &    0 &    0 &   -1 &    0 &    0 &    1 &    0 &    0 &    0 \\
\( \BR_{800} \) &    0 &    0 &    0 &    0 &    0 &    0 &    0 &   -8 &  -69 &   -2 &   -1 &    0 &    0 &    0 \\
\( \BR_{802} \) &    0 &    0 &    0 &    0 &    0 &    0 &    0 &   16 &   -6 &    0 &    0 &    0 &    0 &    0 \\
\( \BR_{803} \) &    0 &    0 &    0 &    0 &    0 &    0 &    0 &   -1 &  -19 &    0 &    0 &   -2 &    0 &   -1 \\
\( \BR_{805} \) &    0 &    0 &    0 &    0 &    0 &    0 &    0 &    0 &    0 &    0 &    0 &    0 &    0 &    0 \\
\( \BR_{811} \) &    0 &    0 &    0 &    0 &    0 &    0 &    0 &    0 &   -1 &    0 &    0 &    0 &    0 &    0 \\
\( \BR_{812} \) &    0 &    0 &    0 &    0 &   -1 &    0 &    0 &   -1 &   -1 &    0 &    0 &    0 &    0 &    0 \\
\( \BR_{821} \) &    0 &    0 &    2 &    0 &    0 &    0 &    0 &    3 &   -1 &    0 &    1 &    0 &    0 &    1 \\
\( \BR_{822} \) &    0 &    0 &    0 &    0 &    0 &    0 &    0 &    0 &    0 &    0 &    0 &    0 &    0 &    0 \\
 & \( \BR_{42} \) & \( \BR_{44} \) & \( \BR_{47} \) & \( \BR_{48} \) & \( \BR_{50} \) & \( \BR_{51} \) & \( \BR_{53} \) & \( \BR_{62} \) & \( \BR_{70} \) & \( \BR_{77} \) & \( \BR_{93} \) & \( \BR_{94} \) & \( \BR_{126} \) & \( \BR_{128} \)
\\\hline
\end{tabular}
\end{center}
\end{envsmall}
\ifhevea\else
\end{center}
\end{minipage}
\fi
\end{center}
\ifhevea\end{table}\fi
\ifhevea\begin{table}\fi
\begin{center}
\ifhevea
\caption{Basis quantities correlation coefficients in percent, subtable 7.\label{tab:tau:br-fit-corr7}}%
\else
\begin{minipage}{\linewidth}
\begin{center}
\captionof{table}{Basis quantities correlation coefficients in percent, subtable 7.}\label{tab:tau:br-fit-corr7}%
\fi
\begin{envsmall}
\begin{center}
\begin{tabular}{rrrrrrrrrrrrrrr}
\hline
\( \BR_{831} \) &    0 &    0 &    1 &    0 &    0 &    0 &    0 &    1 &   -1 &    0 &    1 &    0 &    0 &    0 \\
\( \BR_{832} \) &    0 &    0 &    0 &    0 &    0 &    0 &    0 &    0 &    0 &    0 &    0 &    0 &    0 &    0 \\
\( \BR_{833} \) &    0 &    0 &    0 &    0 &    0 &    0 &    0 &    0 &    0 &    0 &    0 &    0 &    0 &    0 \\
\( \BR_{920} \) &    0 &    0 &    1 &    0 &    0 &    0 &    0 &    1 &   -1 &    0 &    1 &    0 &    0 &    0 \\
\( \BR_{945} \) &    0 &    0 &    0 &    0 &    0 &    0 &    0 &    0 &    0 &    0 &    0 &    0 &    0 &    0 \\
 & \( \BR_{42} \) & \( \BR_{44} \) & \( \BR_{47} \) & \( \BR_{48} \) & \( \BR_{50} \) & \( \BR_{51} \) & \( \BR_{53} \) & \( \BR_{62} \) & \( \BR_{70} \) & \( \BR_{77} \) & \( \BR_{93} \) & \( \BR_{94} \) & \( \BR_{126} \) & \( \BR_{128} \)
\\\hline
\end{tabular}
\end{center}
\end{envsmall}
\ifhevea\else
\end{center}
\end{minipage}
\fi
\end{center}
\ifhevea\end{table}\fi
\ifhevea\begin{table}\fi
\begin{center}
\ifhevea
\caption{Basis quantities correlation coefficients in percent, subtable 8.\label{tab:tau:br-fit-corr8}}%
\else
\begin{minipage}{\linewidth}
\begin{center}
\captionof{table}{Basis quantities correlation coefficients in percent, subtable 8.}\label{tab:tau:br-fit-corr8}%
\fi
\begin{envsmall}
\begin{center}
\begin{tabular}{rrrrrrrrrrrrrrr}
\hline
\( \BR_{132} \) &    0 &  &  &  &  &  &  &  &  &  &  &  &  &  \\
\( \BR_{136} \) &    0 &    0 &  &  &  &  &  &  &  &  &  &  &  &  \\
\( \BR_{151} \) &    0 &    0 &    0 &  &  &  &  &  &  &  &  &  &  &  \\
\( \BR_{152} \) &    0 &    0 &    0 &    0 &  &  &  &  &  &  &  &  &  &  \\
\( \BR_{167} \) &    0 &    0 &    0 &    0 &    0 &  &  &  &  &  &  &  &  &  \\
\( \BR_{800} \) &    0 &    0 &    0 &  -14 &   -3 &    0 &  &  &  &  &  &  &  &  \\
\( \BR_{802} \) &    0 &    0 &    0 &   -2 &    0 &    1 &   -1 &  &  &  &  &  &  &  \\
\( \BR_{803} \) &    0 &    0 &    0 &  -58 &    0 &    0 &    9 &    1 &  &  &  &  &  &  \\
\( \BR_{805} \) &    0 &    0 &    0 &    0 &    0 &    0 &    0 &    0 &    0 &  &  &  &  &  \\
\( \BR_{811} \) &    0 &   -1 &   20 &    0 &    0 &    0 &    0 &    0 &    0 &    0 &  &  &  &  \\
\( \BR_{812} \) &    0 &   -2 &   -8 &    0 &    0 &    0 &    0 &    0 &    0 &    0 &  -16 &  &  &  \\
\( \BR_{821} \) &    0 &    0 &   46 &    0 &    0 &    0 &    0 &    0 &    0 &    0 &    8 &   -4 &  &  \\
\( \BR_{822} \) &    0 &    0 &   -1 &    0 &    0 &    0 &    0 &    0 &    0 &    0 &    0 &    0 &   -1 &  \\
 & \( \BR_{130} \) & \( \BR_{132} \) & \( \BR_{136} \) & \( \BR_{151} \) & \( \BR_{152} \) & \( \BR_{167} \) & \( \BR_{800} \) & \( \BR_{802} \) & \( \BR_{803} \) & \( \BR_{805} \) & \( \BR_{811} \) & \( \BR_{812} \) & \( \BR_{821} \) & \( \BR_{822} \)
\\\hline
\end{tabular}
\end{center}
\end{envsmall}
\ifhevea\else
\end{center}
\end{minipage}
\fi
\end{center}
\ifhevea\end{table}\fi
\ifhevea\begin{table}\fi
\begin{center}
\ifhevea
\caption{Basis quantities correlation coefficients in percent, subtable 9.\label{tab:tau:br-fit-corr9}}%
\else
\begin{minipage}{\linewidth}
\begin{center}
\captionof{table}{Basis quantities correlation coefficients in percent, subtable 9.}\label{tab:tau:br-fit-corr9}%
\fi
\begin{envsmall}
\begin{center}
\begin{tabular}{rrrrrrrrrrrrrrr}
\hline
\( \BR_{831} \) &    0 &    0 &   38 &    0 &    0 &    0 &    0 &    0 &    0 &    0 &   14 &   -4 &   39 &   -1 \\
\( \BR_{832} \) &    0 &    0 &    3 &    0 &    0 &    0 &    0 &    0 &    0 &    0 &    2 &    0 &    3 &    0 \\
\( \BR_{833} \) &    0 &    0 &   -1 &    0 &    0 &    0 &    0 &    0 &    0 &    0 &    0 &    0 &   -1 &    0 \\
\( \BR_{920} \) &    0 &    0 &   20 &    0 &    0 &    0 &    0 &    0 &    0 &    0 &    3 &   -2 &   34 &   -1 \\
\( \BR_{945} \) &    0 &   -1 &   25 &    0 &    0 &    0 &    0 &    0 &    0 &    0 &   10 &  -11 &   10 &    0 \\
 & \( \BR_{130} \) & \( \BR_{132} \) & \( \BR_{136} \) & \( \BR_{151} \) & \( \BR_{152} \) & \( \BR_{167} \) & \( \BR_{800} \) & \( \BR_{802} \) & \( \BR_{803} \) & \( \BR_{805} \) & \( \BR_{811} \) & \( \BR_{812} \) & \( \BR_{821} \) & \( \BR_{822} \)
\\\hline
\end{tabular}
\end{center}
\end{envsmall}
\ifhevea\else
\end{center}
\end{minipage}
\fi
\end{center}
\ifhevea\end{table}\fi
\ifhevea\begin{table}\fi
\begin{center}
\ifhevea
\caption{Basis quantities correlation coefficients in percent, subtable 10.\label{tab:tau:br-fit-corr10}}%
\else
\begin{minipage}{\linewidth}
\begin{center}
\captionof{table}{Basis quantities correlation coefficients in percent, subtable 10.}\label{tab:tau:br-fit-corr10}%
\fi
\begin{envsmall}
\begin{center}
\begin{tabular}{rrrrrr}
\hline
\( \BR_{832} \) &   -2 &  &  &  &  \\
\( \BR_{833} \) &   -1 &   -1 &  &  &  \\
\( \BR_{920} \) &   17 &    1 &    0 &  &  \\
\( \BR_{945} \) &   17 &    2 &    0 &    4 &  \\
 & \( \BR_{831} \) & \( \BR_{832} \) & \( \BR_{833} \) & \( \BR_{920} \) & \( \BR_{945} \)
\\\hline
\end{tabular}
\end{center}
\end{envsmall}
\ifhevea\else
\end{center}
\end{minipage}
\fi
\end{center}
\ifhevea\end{table}\fi}%
\htconstrdef{Gamma1.c}{\BR_{1}}{\BR_{3} + \BR_{5} + \BR_{9} + \BR_{10} + \BR_{14} + \BR_{16} + \BR_{20} + \BR_{23} + \BR_{27} + \BR_{28} + \BR_{30} + \BR_{35} + \BR_{40} + \BR_{44} + \BR_{37} + \BR_{42} + \BR_{47} + \BR_{48} + \BR_{804} + \BR_{50} + \BR_{51} + \BR_{806} + \BR_{126}\cdot{}\BR_{\eta\to\text{neutral}} + \BR_{128}\cdot{}\BR_{\eta\to\text{neutral}} + \BR_{130}\cdot{}\BR_{\eta\to\text{neutral}} + \BR_{132}\cdot{}\BR_{\eta\to\text{neutral}} + \BR_{800}\cdot{}\BR_{\omega\to\pi^0\gamma} + \BR_{151}\cdot{}\BR_{\omega\to\pi^0\gamma} + \BR_{152}\cdot{}\BR_{\omega\to\pi^0\gamma} + \BR_{167}\cdot{}\BR_{\phi\to K_S K_L}}{\BR_{3} + \BR_{5} + \BR_{9} + \BR_{10} + \BR_{14} + \BR_{16}  \\ 
  {}& + \BR_{20} + \BR_{23} + \BR_{27} + \BR_{28} + \BR_{30} + \BR_{35}  \\ 
  {}& + \BR_{40} + \BR_{44} + \BR_{37} + \BR_{42} + \BR_{47} + \BR_{48}  \\ 
  {}& + \BR_{804} + \BR_{50} + \BR_{51} + \BR_{806} + \BR_{126}\cdot{}\BR_{\eta\to\text{neutral}}  \\ 
  {}& + \BR_{128}\cdot{}\BR_{\eta\to\text{neutral}} + \BR_{130}\cdot{}\BR_{\eta\to\text{neutral}} + \BR_{132}\cdot{}\BR_{\eta\to\text{neutral}}  \\ 
  {}& + \BR_{800}\cdot{}\BR_{\omega\to\pi^0\gamma} + \BR_{151}\cdot{}\BR_{\omega\to\pi^0\gamma} + \BR_{152}\cdot{}\BR_{\omega\to\pi^0\gamma}  \\ 
  {}& + \BR_{167}\cdot{}\BR_{\phi\to K_S K_L}}%
\htconstrdef{Gamma2.c}{\BR_{2}}{\BR_{3} + \BR_{5} + \BR_{9} + \BR_{10} + \BR_{14} + \BR_{16} + \BR_{20} + \BR_{23} + \BR_{27} + \BR_{28} + \BR_{30} + \BR_{35}\cdot{}(\BR_{<\bar{K}^0|K_S>}\cdot{}\BR_{K_S\to\pi^0\pi^0}+\BR_{<\bar{K}^0|K_L>}) + \BR_{40}\cdot{}(\BR_{<\bar{K}^0|K_S>}\cdot{}\BR_{K_S\to\pi^0\pi^0}+\BR_{<\bar{K}^0|K_L>}) + \BR_{44}\cdot{}(\BR_{<\bar{K}^0|K_S>}\cdot{}\BR_{K_S\to\pi^0\pi^0}+\BR_{<\bar{K}^0|K_L>}) + \BR_{37}\cdot{}(\BR_{<\bar{K}^0|K_S>}\cdot{}\BR_{K_S\to\pi^0\pi^0}+\BR_{<\bar{K}^0|K_L>}) + \BR_{42}\cdot{}(\BR_{<\bar{K}^0|K_S>}\cdot{}\BR_{K_S\to\pi^0\pi^0}+\BR_{<\bar{K}^0|K_L>}) + \BR_{47}\cdot{}(\BR_{K_S\to\pi^0\pi^0}\cdot{}\BR_{K_S\to\pi^0\pi^0}) + \BR_{48}\cdot{}\BR_{K_S\to\pi^0\pi^0} + \BR_{804} + \BR_{50}\cdot{}(\BR_{K_S\to\pi^0\pi^0}\cdot{}\BR_{K_S\to\pi^0\pi^0}) + \BR_{51}\cdot{}\BR_{K_S\to\pi^0\pi^0} + \BR_{806} + \BR_{126}\cdot{}\BR_{\eta\to\text{neutral}} + \BR_{128}\cdot{}\BR_{\eta\to\text{neutral}} + \BR_{130}\cdot{}\BR_{\eta\to\text{neutral}} + \BR_{132}\cdot{}(\BR_{\eta\to\text{neutral}}\cdot{}(\BR_{<\bar{K}^0|K_S>}\cdot{}\BR_{K_S\to\pi^0\pi^0}+\BR_{<\bar{K}^0|K_L>})) + \BR_{800}\cdot{}\BR_{\omega\to\pi^0\gamma} + \BR_{151}\cdot{}\BR_{\omega\to\pi^0\gamma} + \BR_{152}\cdot{}\BR_{\omega\to\pi^0\gamma} + \BR_{167}\cdot{}(\BR_{\phi\to K_S K_L}\cdot{}\BR_{K_S\to\pi^0\pi^0})}{\BR_{3} + \BR_{5} + \BR_{9} + \BR_{10} + \BR_{14} + \BR_{16}  \\ 
  {}& + \BR_{20} + \BR_{23} + \BR_{27} + \BR_{28} + \BR_{30} + \BR_{35}\cdot{}(\BR_{<\bar{K}^0|K_S>}\cdot{}\BR_{K_S\to\pi^0\pi^0} \\ 
  {}& +\BR_{<\bar{K}^0|K_L>}) + \BR_{40}\cdot{}(\BR_{<\bar{K}^0|K_S>}\cdot{}\BR_{K_S\to\pi^0\pi^0}+\BR_{<\bar{K}^0|K_L>}) + \BR_{44}\cdot{}(\BR_{<\bar{K}^0|K_S>}\cdot{}\BR_{K_S\to\pi^0\pi^0} \\ 
  {}& +\BR_{<\bar{K}^0|K_L>}) + \BR_{37}\cdot{}(\BR_{<\bar{K}^0|K_S>}\cdot{}\BR_{K_S\to\pi^0\pi^0}+\BR_{<\bar{K}^0|K_L>}) + \BR_{42}\cdot{}(\BR_{<\bar{K}^0|K_S>}\cdot{}\BR_{K_S\to\pi^0\pi^0} \\ 
  {}& +\BR_{<\bar{K}^0|K_L>}) + \BR_{47}\cdot{}(\BR_{K_S\to\pi^0\pi^0}\cdot{}\BR_{K_S\to\pi^0\pi^0}) + \BR_{48}\cdot{}\BR_{K_S\to\pi^0\pi^0}  \\ 
  {}& + \BR_{804} + \BR_{50}\cdot{}(\BR_{K_S\to\pi^0\pi^0}\cdot{}\BR_{K_S\to\pi^0\pi^0}) + \BR_{51}\cdot{}\BR_{K_S\to\pi^0\pi^0}  \\ 
  {}& + \BR_{806} + \BR_{126}\cdot{}\BR_{\eta\to\text{neutral}} + \BR_{128}\cdot{}\BR_{\eta\to\text{neutral}} + \BR_{130}\cdot{}\BR_{\eta\to\text{neutral}}  \\ 
  {}& + \BR_{132}\cdot{}(\BR_{\eta\to\text{neutral}}\cdot{}(\BR_{<\bar{K}^0|K_S>}\cdot{}\BR_{K_S\to\pi^0\pi^0}+\BR_{<\bar{K}^0|K_L>})) + \BR_{800}\cdot{}\BR_{\omega\to\pi^0\gamma}  \\ 
  {}& + \BR_{151}\cdot{}\BR_{\omega\to\pi^0\gamma} + \BR_{152}\cdot{}\BR_{\omega\to\pi^0\gamma} + \BR_{167}\cdot{}(\BR_{\phi\to K_S K_L}\cdot{}\BR_{K_S\to\pi^0\pi^0})}%
\htconstrdef{Gamma3by5.c}{\frac{\BR_{3}}{\BR_{5}}}{\frac{\BR_{3}}{\BR_{5}}}{\frac{\BR_{3}}{\BR_{5}}}%
\htconstrdef{Gamma7.c}{\BR_{7}}{\BR_{35}\cdot{}\BR_{<\bar{K}^0|K_L>} + \BR_{9} + \BR_{804} + \BR_{37}\cdot{}\BR_{<K^0|K_L>} + \BR_{10}}{\BR_{35}\cdot{}\BR_{<\bar{K}^0|K_L>} + \BR_{9} + \BR_{804} + \BR_{37}\cdot{}\BR_{<K^0|K_L>}  \\ 
  {}& + \BR_{10}}%
\htconstrdef{Gamma8.c}{\BR_{8}}{\BR_{9} + \BR_{10}}{\BR_{9} + \BR_{10}}%
\htconstrdef{Gamma8by5.c}{\frac{\BR_{8}}{\BR_{5}}}{\frac{\BR_{8}}{\BR_{5}}}{\frac{\BR_{8}}{\BR_{5}}}%
\htconstrdef{Gamma9by5.c}{\frac{\BR_{9}}{\BR_{5}}}{\frac{\BR_{9}}{\BR_{5}}}{\frac{\BR_{9}}{\BR_{5}}}%
\htconstrdef{Gamma10by5.c}{\frac{\BR_{10}}{\BR_{5}}}{\frac{\BR_{10}}{\BR_{5}}}{\frac{\BR_{10}}{\BR_{5}}}%
\htconstrdef{Gamma10by9.c}{\frac{\BR_{10}}{\BR_{9}}}{\frac{\BR_{10}}{\BR_{9}}}{\frac{\BR_{10}}{\BR_{9}}}%
\htconstrdef{Gamma11.c}{\BR_{11}}{\BR_{14} + \BR_{16} + \BR_{20} + \BR_{23} + \BR_{27} + \BR_{28} + \BR_{30} + \BR_{35}\cdot{}(\BR_{<K^0|K_S>}\cdot{}\BR_{K_S\to\pi^0\pi^0}) + \BR_{37}\cdot{}(\BR_{<K^0|K_S>}\cdot{}\BR_{K_S\to\pi^0\pi^0}) + \BR_{40}\cdot{}(\BR_{<K^0|K_S>}\cdot{}\BR_{K_S\to\pi^0\pi^0}) + \BR_{42}\cdot{}(\BR_{<K^0|K_S>}\cdot{}\BR_{K_S\to\pi^0\pi^0}) + \BR_{47}\cdot{}(\BR_{K_S\to\pi^0\pi^0}\cdot{}\BR_{K_S\to\pi^0\pi^0}) + \BR_{50}\cdot{}(\BR_{K_S\to\pi^0\pi^0}\cdot{}\BR_{K_S\to\pi^0\pi^0}) + \BR_{126}\cdot{}\BR_{\eta\to\text{neutral}} + \BR_{128}\cdot{}\BR_{\eta\to\text{neutral}} + \BR_{130}\cdot{}\BR_{\eta\to\text{neutral}} + \BR_{132}\cdot{}(\BR_{<K^0|K_S>}\cdot{}\BR_{K_S\to\pi^0\pi^0}\cdot{}\BR_{\eta\to\text{neutral}}) + \BR_{151}\cdot{}\BR_{\omega\to\pi^0\gamma} + \BR_{152}\cdot{}\BR_{\omega\to\pi^0\gamma} + \BR_{800}\cdot{}\BR_{\omega\to\pi^0\gamma}}{\BR_{14} + \BR_{16} + \BR_{20} + \BR_{23} + \BR_{27} + \BR_{28}  \\ 
  {}& + \BR_{30} + \BR_{35}\cdot{}(\BR_{<K^0|K_S>}\cdot{}\BR_{K_S\to\pi^0\pi^0}) + \BR_{37}\cdot{}(\BR_{<K^0|K_S>}\cdot{}\BR_{K_S\to\pi^0\pi^0})  \\ 
  {}& + \BR_{40}\cdot{}(\BR_{<K^0|K_S>}\cdot{}\BR_{K_S\to\pi^0\pi^0}) + \BR_{42}\cdot{}(\BR_{<K^0|K_S>}\cdot{}\BR_{K_S\to\pi^0\pi^0})  \\ 
  {}& + \BR_{47}\cdot{}(\BR_{K_S\to\pi^0\pi^0}\cdot{}\BR_{K_S\to\pi^0\pi^0}) + \BR_{50}\cdot{}(\BR_{K_S\to\pi^0\pi^0}\cdot{}\BR_{K_S\to\pi^0\pi^0})  \\ 
  {}& + \BR_{126}\cdot{}\BR_{\eta\to\text{neutral}} + \BR_{128}\cdot{}\BR_{\eta\to\text{neutral}} + \BR_{130}\cdot{}\BR_{\eta\to\text{neutral}}  \\ 
  {}& + \BR_{132}\cdot{}(\BR_{<K^0|K_S>}\cdot{}\BR_{K_S\to\pi^0\pi^0}\cdot{}\BR_{\eta\to\text{neutral}}) + \BR_{151}\cdot{}\BR_{\omega\to\pi^0\gamma}  \\ 
  {}& + \BR_{152}\cdot{}\BR_{\omega\to\pi^0\gamma} + \BR_{800}\cdot{}\BR_{\omega\to\pi^0\gamma}}%
\htconstrdef{Gamma12.c}{\BR_{12}}{\BR_{128}\cdot{}\BR_{\eta\to3\pi^0} + \BR_{30} + \BR_{23} + \BR_{28} + \BR_{14} + \BR_{16} + \BR_{20} + \BR_{27} + \BR_{126}\cdot{}\BR_{\eta\to3\pi^0} + \BR_{130}\cdot{}\BR_{\eta\to3\pi^0}}{\BR_{128}\cdot{}\BR_{\eta\to3\pi^0} + \BR_{30} + \BR_{23} + \BR_{28} + \BR_{14}  \\ 
  {}& + \BR_{16} + \BR_{20} + \BR_{27} + \BR_{126}\cdot{}\BR_{\eta\to3\pi^0} + \BR_{130}\cdot{}\BR_{\eta\to3\pi^0}}%
\htconstrdef{Gamma13.c}{\BR_{13}}{\BR_{14} + \BR_{16}}{\BR_{14} + \BR_{16}}%
\htconstrdef{Gamma17.c}{\BR_{17}}{\BR_{128}\cdot{}\BR_{\eta\to3\pi^0} + \BR_{30} + \BR_{23} + \BR_{28} + \BR_{35}\cdot{}(\BR_{<K^0|K_S>}\cdot{}\BR_{K_S\to\pi^0\pi^0}) + \BR_{40}\cdot{}(\BR_{<K^0|K_S>}\cdot{}\BR_{K_S\to\pi^0\pi^0}) + \BR_{42}\cdot{}(\BR_{<K^0|K_S>}\cdot{}\BR_{K_S\to\pi^0\pi^0}) + \BR_{20} + \BR_{27} + \BR_{47}\cdot{}(\BR_{K_S\to\pi^0\pi^0}\cdot{}\BR_{K_S\to\pi^0\pi^0}) + \BR_{50}\cdot{}(\BR_{K_S\to\pi^0\pi^0}\cdot{}\BR_{K_S\to\pi^0\pi^0}) + \BR_{126}\cdot{}\BR_{\eta\to3\pi^0} + \BR_{37}\cdot{}(\BR_{<K^0|K_S>}\cdot{}\BR_{K_S\to\pi^0\pi^0}) + \BR_{130}\cdot{}\BR_{\eta\to3\pi^0}}{\BR_{128}\cdot{}\BR_{\eta\to3\pi^0} + \BR_{30} + \BR_{23} + \BR_{28} + \BR_{35}\cdot{}(\BR_{<K^0|K_S>}\cdot{}\BR_{K_S\to\pi^0\pi^0})  \\ 
  {}& + \BR_{40}\cdot{}(\BR_{<K^0|K_S>}\cdot{}\BR_{K_S\to\pi^0\pi^0}) + \BR_{42}\cdot{}(\BR_{<K^0|K_S>}\cdot{}\BR_{K_S\to\pi^0\pi^0})  \\ 
  {}& + \BR_{20} + \BR_{27} + \BR_{47}\cdot{}(\BR_{K_S\to\pi^0\pi^0}\cdot{}\BR_{K_S\to\pi^0\pi^0}) + \BR_{50}\cdot{}(\BR_{K_S\to\pi^0\pi^0}\cdot{}\BR_{K_S\to\pi^0\pi^0})  \\ 
  {}& + \BR_{126}\cdot{}\BR_{\eta\to3\pi^0} + \BR_{37}\cdot{}(\BR_{<K^0|K_S>}\cdot{}\BR_{K_S\to\pi^0\pi^0}) + \BR_{130}\cdot{}\BR_{\eta\to3\pi^0}}%
\htconstrdef{Gamma18.c}{\BR_{18}}{\BR_{23} + \BR_{35}\cdot{}(\BR_{<K^0|K_S>}\cdot{}\BR_{K_S\to\pi^0\pi^0}) + \BR_{20} + \BR_{37}\cdot{}(\BR_{<K^0|K_S>}\cdot{}\BR_{K_S\to\pi^0\pi^0})}{\BR_{23} + \BR_{35}\cdot{}(\BR_{<K^0|K_S>}\cdot{}\BR_{K_S\to\pi^0\pi^0}) + \BR_{20} + \BR_{37}\cdot{}(\BR_{<K^0|K_S>}\cdot{}\BR_{K_S\to\pi^0\pi^0})}%
\htconstrdef{Gamma19.c}{\BR_{19}}{\BR_{23} + \BR_{20}}{\BR_{23} + \BR_{20}}%
\htconstrdef{Gamma19by13.c}{\frac{\BR_{19}}{\BR_{13}}}{\frac{\BR_{19}}{\BR_{13}}}{\frac{\BR_{19}}{\BR_{13}}}%
\htconstrdef{Gamma24.c}{\BR_{24}}{\BR_{27} + \BR_{28} + \BR_{30} + \BR_{40}\cdot{}(\BR_{<K^0|K_S>}\cdot{}\BR_{K_S\to\pi^0\pi^0}) + \BR_{42}\cdot{}(\BR_{<K^0|K_S>}\cdot{}\BR_{K_S\to\pi^0\pi^0}) + \BR_{47}\cdot{}(\BR_{K_S\to\pi^0\pi^0}\cdot{}\BR_{K_S\to\pi^0\pi^0}) + \BR_{50}\cdot{}(\BR_{K_S\to\pi^0\pi^0}\cdot{}\BR_{K_S\to\pi^0\pi^0}) + \BR_{126}\cdot{}\BR_{\eta\to3\pi^0} + \BR_{128}\cdot{}\BR_{\eta\to3\pi^0} + \BR_{130}\cdot{}\BR_{\eta\to3\pi^0} + \BR_{132}\cdot{}(\BR_{<K^0|K_S>}\cdot{}\BR_{K_S\to\pi^0\pi^0}\cdot{}\BR_{\eta\to3\pi^0})}{\BR_{27} + \BR_{28} + \BR_{30} + \BR_{40}\cdot{}(\BR_{<K^0|K_S>}\cdot{}\BR_{K_S\to\pi^0\pi^0})  \\ 
  {}& + \BR_{42}\cdot{}(\BR_{<K^0|K_S>}\cdot{}\BR_{K_S\to\pi^0\pi^0}) + \BR_{47}\cdot{}(\BR_{K_S\to\pi^0\pi^0}\cdot{}\BR_{K_S\to\pi^0\pi^0})  \\ 
  {}& + \BR_{50}\cdot{}(\BR_{K_S\to\pi^0\pi^0}\cdot{}\BR_{K_S\to\pi^0\pi^0}) + \BR_{126}\cdot{}\BR_{\eta\to3\pi^0} + \BR_{128}\cdot{}\BR_{\eta\to3\pi^0}  \\ 
  {}& + \BR_{130}\cdot{}\BR_{\eta\to3\pi^0} + \BR_{132}\cdot{}(\BR_{<K^0|K_S>}\cdot{}\BR_{K_S\to\pi^0\pi^0}\cdot{}\BR_{\eta\to3\pi^0})}%
\htconstrdef{Gamma25.c}{\BR_{25}}{\BR_{128}\cdot{}\BR_{\eta\to3\pi^0} + \BR_{30} + \BR_{28} + \BR_{27} + \BR_{126}\cdot{}\BR_{\eta\to3\pi^0} + \BR_{130}\cdot{}\BR_{\eta\to3\pi^0}}{\BR_{128}\cdot{}\BR_{\eta\to3\pi^0} + \BR_{30} + \BR_{28} + \BR_{27} + \BR_{126}\cdot{}\BR_{\eta\to3\pi^0}  \\ 
  {}& + \BR_{130}\cdot{}\BR_{\eta\to3\pi^0}}%
\htconstrdef{Gamma26.c}{\BR_{26}}{\BR_{128}\cdot{}\BR_{\eta\to3\pi^0} + \BR_{28} + \BR_{40}\cdot{}(\BR_{<K^0|K_S>}\cdot{}\BR_{K_S\to\pi^0\pi^0}) + \BR_{42}\cdot{}(\BR_{<K^0|K_S>}\cdot{}\BR_{K_S\to\pi^0\pi^0}) + \BR_{27}}{\BR_{128}\cdot{}\BR_{\eta\to3\pi^0} + \BR_{28} + \BR_{40}\cdot{}(\BR_{<K^0|K_S>}\cdot{}\BR_{K_S\to\pi^0\pi^0})  \\ 
  {}& + \BR_{42}\cdot{}(\BR_{<K^0|K_S>}\cdot{}\BR_{K_S\to\pi^0\pi^0}) + \BR_{27}}%
\htconstrdef{Gamma26by13.c}{\frac{\BR_{26}}{\BR_{13}}}{\frac{\BR_{26}}{\BR_{13}}}{\frac{\BR_{26}}{\BR_{13}}}%
\htconstrdef{Gamma29.c}{\BR_{29}}{\BR_{30} + \BR_{126}\cdot{}\BR_{\eta\to3\pi^0} + \BR_{130}\cdot{}\BR_{\eta\to3\pi^0}}{\BR_{30} + \BR_{126}\cdot{}\BR_{\eta\to3\pi^0} + \BR_{130}\cdot{}\BR_{\eta\to3\pi^0}}%
\htconstrdef{Gamma31.c}{\BR_{31}}{\BR_{128}\cdot{}\BR_{\eta\to\text{neutral}} + \BR_{23} + \BR_{28} + \BR_{42} + \BR_{16} + \BR_{37} + \BR_{10} + \BR_{167}\cdot{}(\BR_{\phi\to K_S K_L}\cdot{}\BR_{K_S\to\pi^0\pi^0})}{\BR_{128}\cdot{}\BR_{\eta\to\text{neutral}} + \BR_{23} + \BR_{28} + \BR_{42} + \BR_{16}  \\ 
  {}& + \BR_{37} + \BR_{10} + \BR_{167}\cdot{}(\BR_{\phi\to K_S K_L}\cdot{}\BR_{K_S\to\pi^0\pi^0})}%
\htconstrdef{Gamma32.c}{\BR_{32}}{\BR_{16} + \BR_{23} + \BR_{28} + \BR_{37} + \BR_{42} + \BR_{128}\cdot{}\BR_{\eta\to\text{neutral}} + \BR_{130}\cdot{}\BR_{\eta\to\text{neutral}} + \BR_{167}\cdot{}(\BR_{\phi\to K_S K_L}\cdot{}\BR_{K_S\to\pi^0\pi^0})}{\BR_{16} + \BR_{23} + \BR_{28} + \BR_{37} + \BR_{42} + \BR_{128}\cdot{}\BR_{\eta\to\text{neutral}}  \\ 
  {}& + \BR_{130}\cdot{}\BR_{\eta\to\text{neutral}} + \BR_{167}\cdot{}(\BR_{\phi\to K_S K_L}\cdot{}\BR_{K_S\to\pi^0\pi^0})}%
\htconstrdef{Gamma33.c}{\BR_{33}}{\BR_{35}\cdot{}\BR_{<\bar{K}^0|K_S>} + \BR_{40}\cdot{}\BR_{<\bar{K}^0|K_S>} + \BR_{42}\cdot{}\BR_{<K^0|K_S>} + \BR_{47} + \BR_{48} + \BR_{50} + \BR_{51} + \BR_{37}\cdot{}\BR_{<K^0|K_S>} + \BR_{132}\cdot{}(\BR_{<\bar{K}^0|K_S>}\cdot{}\BR_{\eta\to\text{neutral}}) + \BR_{44}\cdot{}\BR_{<\bar{K}^0|K_S>} + \BR_{167}\cdot{}\BR_{\phi\to K_S K_L}}{\BR_{35}\cdot{}\BR_{<\bar{K}^0|K_S>} + \BR_{40}\cdot{}\BR_{<\bar{K}^0|K_S>} + \BR_{42}\cdot{}\BR_{<K^0|K_S>}  \\ 
  {}& + \BR_{47} + \BR_{48} + \BR_{50} + \BR_{51} + \BR_{37}\cdot{}\BR_{<K^0|K_S>}  \\ 
  {}& + \BR_{132}\cdot{}(\BR_{<\bar{K}^0|K_S>}\cdot{}\BR_{\eta\to\text{neutral}}) + \BR_{44}\cdot{}\BR_{<\bar{K}^0|K_S>} + \BR_{167}\cdot{}\BR_{\phi\to K_S K_L}}%
\htconstrdef{Gamma34.c}{\BR_{34}}{\BR_{35} + \BR_{37}}{\BR_{35} + \BR_{37}}%
\htconstrdef{Gamma38.c}{\BR_{38}}{\BR_{42} + \BR_{37}}{\BR_{42} + \BR_{37}}%
\htconstrdef{Gamma39.c}{\BR_{39}}{\BR_{40} + \BR_{42}}{\BR_{40} + \BR_{42}}%
\htconstrdef{Gamma43.c}{\BR_{43}}{\BR_{40} + \BR_{44}}{\BR_{40} + \BR_{44}}%
\htconstrdef{Gamma46.c}{\BR_{46}}{\BR_{48} + \BR_{47} + \BR_{804}}{\BR_{48} + \BR_{47} + \BR_{804}}%
\htconstrdef{Gamma49.c}{\BR_{49}}{\BR_{50} + \BR_{51} + \BR_{806}}{\BR_{50} + \BR_{51} + \BR_{806}}%
\htconstrdef{Gamma54.c}{\BR_{54}}{\BR_{35}\cdot{}(\BR_{<K^0|K_S>}\cdot{}\BR_{K_S\to\pi^+\pi^-}) + \BR_{37}\cdot{}(\BR_{<K^0|K_S>}\cdot{}\BR_{K_S\to\pi^+\pi^-}) + \BR_{40}\cdot{}(\BR_{<K^0|K_S>}\cdot{}\BR_{K_S\to\pi^+\pi^-}) + \BR_{42}\cdot{}(\BR_{<K^0|K_S>}\cdot{}\BR_{K_S\to\pi^+\pi^-}) + \BR_{47}\cdot{}(2\cdot{}\BR_{K_S\to\pi^+\pi^-}\cdot{}\BR_{K_S\to\pi^0\pi^0}) + \BR_{48}\cdot{}\BR_{K_S\to\pi^+\pi^-} + \BR_{50}\cdot{}(2\cdot{}\BR_{K_S\to\pi^+\pi^-}\cdot{}\BR_{K_S\to\pi^0\pi^0}) + \BR_{51}\cdot{}\BR_{K_S\to\pi^+\pi^-} + \BR_{53}\cdot{}(\BR_{<\bar{K}^0|K_S>}\cdot{}\BR_{K_S\to\pi^0\pi^0}+\BR_{<\bar{K}^0|K_L>}) + \BR_{62} + \BR_{70} + \BR_{77} + \BR_{78} + \BR_{93} + \BR_{94} + \BR_{126}\cdot{}\BR_{\eta\to\text{charged}} + \BR_{128}\cdot{}\BR_{\eta\to\text{charged}} + \BR_{130}\cdot{}\BR_{\eta\to\text{charged}} + \BR_{132}\cdot{}(\BR_{<\bar{K}^0|K_L>}\cdot{}\BR_{\eta\to\pi^+\pi^-\pi^0} + \BR_{<\bar{K}^0|K_S>}\cdot{}\BR_{K_S\to\pi^0\pi^0}\cdot{}\BR_{\eta\to\pi^+\pi^-\pi^0} + \BR_{<\bar{K}^0|K_S>}\cdot{}\BR_{K_S\to\pi^+\pi^-}\cdot{}\BR_{\eta\to3\pi^0}) + \BR_{151}\cdot{}(\BR_{\omega\to\pi^+\pi^-\pi^0}+\BR_{\omega\to\pi^+\pi^-}) + \BR_{152}\cdot{}(\BR_{\omega\to\pi^+\pi^-\pi^0}+\BR_{\omega\to\pi^+\pi^-}) + \BR_{167}\cdot{}(\BR_{\phi\to K^+K^-} + \BR_{\phi\to K_S K_L}\cdot{}\BR_{K_S\to\pi^+\pi^-}) + \BR_{802} + \BR_{803} + \BR_{800}\cdot{}(\BR_{\omega\to\pi^+\pi^-\pi^0}+\BR_{\omega\to\pi^+\pi^-})}{\BR_{35}\cdot{}(\BR_{<K^0|K_S>}\cdot{}\BR_{K_S\to\pi^+\pi^-}) + \BR_{37}\cdot{}(\BR_{<K^0|K_S>}\cdot{}\BR_{K_S\to\pi^+\pi^-})  \\ 
  {}& + \BR_{40}\cdot{}(\BR_{<K^0|K_S>}\cdot{}\BR_{K_S\to\pi^+\pi^-}) + \BR_{42}\cdot{}(\BR_{<K^0|K_S>}\cdot{}\BR_{K_S\to\pi^+\pi^-})  \\ 
  {}& + \BR_{47}\cdot{}(2\cdot{}\BR_{K_S\to\pi^+\pi^-}\cdot{}\BR_{K_S\to\pi^0\pi^0}) + \BR_{48}\cdot{}\BR_{K_S\to\pi^+\pi^-}  \\ 
  {}& + \BR_{50}\cdot{}(2\cdot{}\BR_{K_S\to\pi^+\pi^-}\cdot{}\BR_{K_S\to\pi^0\pi^0}) + \BR_{51}\cdot{}\BR_{K_S\to\pi^+\pi^-}  \\ 
  {}& + \BR_{53}\cdot{}(\BR_{<\bar{K}^0|K_S>}\cdot{}\BR_{K_S\to\pi^0\pi^0}+\BR_{<\bar{K}^0|K_L>}) + \BR_{62} + \BR_{70}  \\ 
  {}& + \BR_{77} + \BR_{78} + \BR_{93} + \BR_{94} + \BR_{126}\cdot{}\BR_{\eta\to\text{charged}}  \\ 
  {}& + \BR_{128}\cdot{}\BR_{\eta\to\text{charged}} + \BR_{130}\cdot{}\BR_{\eta\to\text{charged}} + \BR_{132}\cdot{}(\BR_{<\bar{K}^0|K_L>}\cdot{}\BR_{\eta\to\pi^+\pi^-\pi^0}  \\ 
  {}& + \BR_{<\bar{K}^0|K_S>}\cdot{}\BR_{K_S\to\pi^0\pi^0}\cdot{}\BR_{\eta\to\pi^+\pi^-\pi^0} + \BR_{<\bar{K}^0|K_S>}\cdot{}\BR_{K_S\to\pi^+\pi^-}\cdot{}\BR_{\eta\to3\pi^0})  \\ 
  {}& + \BR_{151}\cdot{}(\BR_{\omega\to\pi^+\pi^-\pi^0}+\BR_{\omega\to\pi^+\pi^-}) + \BR_{152}\cdot{}(\BR_{\omega\to\pi^+\pi^-\pi^0}+\BR_{\omega\to\pi^+\pi^-})  \\ 
  {}& + \BR_{167}\cdot{}(\BR_{\phi\to K^+K^-} + \BR_{\phi\to K_S K_L}\cdot{}\BR_{K_S\to\pi^+\pi^-}) + \BR_{802} + \BR_{803}  \\ 
  {}& + \BR_{800}\cdot{}(\BR_{\omega\to\pi^+\pi^-\pi^0}+\BR_{\omega\to\pi^+\pi^-})}%
\htconstrdef{Gamma55.c}{\BR_{55}}{\BR_{128}\cdot{}\BR_{\eta\to\text{charged}} + \BR_{152}\cdot{}(\BR_{\omega\to\pi^+\pi^-\pi^0}+\BR_{\omega\to\pi^+\pi^-}) + \BR_{78} + \BR_{77} + \BR_{94} + \BR_{62} + \BR_{70} + \BR_{93} + \BR_{126}\cdot{}\BR_{\eta\to\text{charged}} + \BR_{802} + \BR_{803} + \BR_{800}\cdot{}(\BR_{\omega\to\pi^+\pi^-\pi^0}+\BR_{\omega\to\pi^+\pi^-}) + \BR_{151}\cdot{}(\BR_{\omega\to\pi^+\pi^-\pi^0}+\BR_{\omega\to\pi^+\pi^-}) + \BR_{130}\cdot{}\BR_{\eta\to\text{charged}} + \BR_{168}}{\BR_{128}\cdot{}\BR_{\eta\to\text{charged}} + \BR_{152}\cdot{}(\BR_{\omega\to\pi^+\pi^-\pi^0}+\BR_{\omega\to\pi^+\pi^-}) + \BR_{78}  \\ 
  {}& + \BR_{77} + \BR_{94} + \BR_{62} + \BR_{70} + \BR_{93} + \BR_{126}\cdot{}\BR_{\eta\to\text{charged}}  \\ 
  {}& + \BR_{802} + \BR_{803} + \BR_{800}\cdot{}(\BR_{\omega\to\pi^+\pi^-\pi^0}+\BR_{\omega\to\pi^+\pi^-}) + \BR_{151}\cdot{}(\BR_{\omega\to\pi^+\pi^-\pi^0} \\ 
  {}& +\BR_{\omega\to\pi^+\pi^-}) + \BR_{130}\cdot{}\BR_{\eta\to\text{charged}} + \BR_{168}}%
\htconstrdef{Gamma56.c}{\BR_{56}}{\BR_{35}\cdot{}(\BR_{<K^0|K_S>}\cdot{}\BR_{K_S\to\pi^+\pi^-}) + \BR_{62} + \BR_{93} + \BR_{37}\cdot{}(\BR_{<K^0|K_S>}\cdot{}\BR_{K_S\to\pi^+\pi^-}) + \BR_{802} + \BR_{800}\cdot{}\BR_{\omega\to\pi^+\pi^-} + \BR_{151}\cdot{}\BR_{\omega\to\pi^+\pi^-} + \BR_{168}}{\BR_{35}\cdot{}(\BR_{<K^0|K_S>}\cdot{}\BR_{K_S\to\pi^+\pi^-}) + \BR_{62} + \BR_{93} + \BR_{37}\cdot{}(\BR_{<K^0|K_S>}\cdot{}\BR_{K_S\to\pi^+\pi^-})  \\ 
  {}& + \BR_{802} + \BR_{800}\cdot{}\BR_{\omega\to\pi^+\pi^-} + \BR_{151}\cdot{}\BR_{\omega\to\pi^+\pi^-} + \BR_{168}}%
\htconstrdef{Gamma57.c}{\BR_{57}}{\BR_{62} + \BR_{93} + \BR_{802} + \BR_{800}\cdot{}\BR_{\omega\to\pi^+\pi^-} + \BR_{151}\cdot{}\BR_{\omega\to\pi^+\pi^-} + \BR_{167}\cdot{}\BR_{\phi\to K^+K^-}}{\BR_{62} + \BR_{93} + \BR_{802} + \BR_{800}\cdot{}\BR_{\omega\to\pi^+\pi^-} + \BR_{151}\cdot{}\BR_{\omega\to\pi^+\pi^-}  \\ 
  {}& + \BR_{167}\cdot{}\BR_{\phi\to K^+K^-}}%
\htconstrdef{Gamma57by55.c}{\frac{\BR_{57}}{\BR_{55}}}{\frac{\BR_{57}}{\BR_{55}}}{\frac{\BR_{57}}{\BR_{55}}}%
\htconstrdef{Gamma58.c}{\BR_{58}}{\BR_{62} + \BR_{93} + \BR_{802} + \BR_{167}\cdot{}\BR_{\phi\to K^+K^-}}{\BR_{62} + \BR_{93} + \BR_{802} + \BR_{167}\cdot{}\BR_{\phi\to K^+K^-}}%
\htconstrdef{Gamma59.c}{\BR_{59}}{\BR_{35}\cdot{}(\BR_{<K^0|K_S>}\cdot{}\BR_{K_S\to\pi^+\pi^-}) + \BR_{62} + \BR_{800}\cdot{}\BR_{\omega\to\pi^+\pi^-}}{\BR_{35}\cdot{}(\BR_{<K^0|K_S>}\cdot{}\BR_{K_S\to\pi^+\pi^-}) + \BR_{62} + \BR_{800}\cdot{}\BR_{\omega\to\pi^+\pi^-}}%
\htconstrdef{Gamma60.c}{\BR_{60}}{\BR_{62} + \BR_{800}\cdot{}\BR_{\omega\to\pi^+\pi^-}}{\BR_{62} + \BR_{800}\cdot{}\BR_{\omega\to\pi^+\pi^-}}%
\htconstrdef{Gamma63.c}{\BR_{63}}{\BR_{40}\cdot{}(\BR_{<K^0|K_S>}\cdot{}\BR_{K_S\to\pi^+\pi^-}) + \BR_{42}\cdot{}(\BR_{<K^0|K_S>}\cdot{}\BR_{K_S\to\pi^+\pi^-}) + \BR_{47}\cdot{}(2\cdot{}\BR_{K_S\to\pi^+\pi^-}\cdot{}\BR_{K_S\to\pi^0\pi^0}) + \BR_{50}\cdot{}(2\cdot{}\BR_{K_S\to\pi^+\pi^-}\cdot{}\BR_{K_S\to\pi^0\pi^0}) + \BR_{70} + \BR_{77} + \BR_{78} + \BR_{94} + \BR_{126}\cdot{}\BR_{\eta\to\text{charged}} + \BR_{128}\cdot{}\BR_{\eta\to\text{charged}} + \BR_{130}\cdot{}\BR_{\eta\to\text{charged}} + \BR_{132}\cdot{}(\BR_{<\bar{K}^0|K_S>}\cdot{}\BR_{K_S\to\pi^+\pi^-}\cdot{}\BR_{\eta\to\text{neutral}} + \BR_{<\bar{K}^0|K_S>}\cdot{}\BR_{K_S\to\pi^0\pi^0}\cdot{}\BR_{\eta\to\text{charged}}) + \BR_{151}\cdot{}\BR_{\omega\to\pi^+\pi^-\pi^0} + \BR_{152}\cdot{}(\BR_{\omega\to\pi^+\pi^-\pi^0}+\BR_{\omega\to\pi^+\pi^-}) + \BR_{800}\cdot{}\BR_{\omega\to\pi^+\pi^-\pi^0} + \BR_{803}}{\BR_{40}\cdot{}(\BR_{<K^0|K_S>}\cdot{}\BR_{K_S\to\pi^+\pi^-}) + \BR_{42}\cdot{}(\BR_{<K^0|K_S>}\cdot{}\BR_{K_S\to\pi^+\pi^-})  \\ 
  {}& + \BR_{47}\cdot{}(2\cdot{}\BR_{K_S\to\pi^+\pi^-}\cdot{}\BR_{K_S\to\pi^0\pi^0}) + \BR_{50}\cdot{}(2\cdot{}\BR_{K_S\to\pi^+\pi^-}\cdot{}\BR_{K_S\to\pi^0\pi^0})  \\ 
  {}& + \BR_{70} + \BR_{77} + \BR_{78} + \BR_{94} + \BR_{126}\cdot{}\BR_{\eta\to\text{charged}}  \\ 
  {}& + \BR_{128}\cdot{}\BR_{\eta\to\text{charged}} + \BR_{130}\cdot{}\BR_{\eta\to\text{charged}} + \BR_{132}\cdot{}(\BR_{<\bar{K}^0|K_S>}\cdot{}\BR_{K_S\to\pi^+\pi^-}\cdot{}\BR_{\eta\to\text{neutral}}  \\ 
  {}& + \BR_{<\bar{K}^0|K_S>}\cdot{}\BR_{K_S\to\pi^0\pi^0}\cdot{}\BR_{\eta\to\text{charged}}) + \BR_{151}\cdot{}\BR_{\omega\to\pi^+\pi^-\pi^0} + \BR_{152}\cdot{}(\BR_{\omega\to\pi^+\pi^-\pi^0} \\ 
  {}& +\BR_{\omega\to\pi^+\pi^-}) + \BR_{800}\cdot{}\BR_{\omega\to\pi^+\pi^-\pi^0} + \BR_{803}}%
\htconstrdef{Gamma64.c}{\BR_{64}}{\BR_{78} + \BR_{77} + \BR_{94} + \BR_{70} + \BR_{126}\cdot{}\BR_{\eta\to\pi^+\pi^-\pi^0} + \BR_{128}\cdot{}\BR_{\eta\to\pi^+\pi^-\pi^0} + \BR_{130}\cdot{}\BR_{\eta\to\pi^+\pi^-\pi^0} + \BR_{800}\cdot{}\BR_{\omega\to\pi^+\pi^-\pi^0} + \BR_{151}\cdot{}\BR_{\omega\to\pi^+\pi^-\pi^0} + \BR_{152}\cdot{}(\BR_{\omega\to\pi^+\pi^-\pi^0}+\BR_{\omega\to\pi^+\pi^-}) + \BR_{803}}{\BR_{78} + \BR_{77} + \BR_{94} + \BR_{70} + \BR_{126}\cdot{}\BR_{\eta\to\pi^+\pi^-\pi^0}  \\ 
  {}& + \BR_{128}\cdot{}\BR_{\eta\to\pi^+\pi^-\pi^0} + \BR_{130}\cdot{}\BR_{\eta\to\pi^+\pi^-\pi^0} + \BR_{800}\cdot{}\BR_{\omega\to\pi^+\pi^-\pi^0}  \\ 
  {}& + \BR_{151}\cdot{}\BR_{\omega\to\pi^+\pi^-\pi^0} + \BR_{152}\cdot{}(\BR_{\omega\to\pi^+\pi^-\pi^0}+\BR_{\omega\to\pi^+\pi^-}) + \BR_{803}}%
\htconstrdef{Gamma65.c}{\BR_{65}}{\BR_{40}\cdot{}(\BR_{<K^0|K_S>}\cdot{}\BR_{K_S\to\pi^+\pi^-}) + \BR_{42}\cdot{}(\BR_{<K^0|K_S>}\cdot{}\BR_{K_S\to\pi^+\pi^-}) + \BR_{70} + \BR_{94} + \BR_{128}\cdot{}\BR_{\eta\to\pi^+\pi^-\pi^0} + \BR_{151}\cdot{}\BR_{\omega\to\pi^+\pi^-\pi^0} + \BR_{152}\cdot{}\BR_{\omega\to\pi^+\pi^-} + \BR_{800}\cdot{}\BR_{\omega\to\pi^+\pi^-\pi^0} + \BR_{803}}{\BR_{40}\cdot{}(\BR_{<K^0|K_S>}\cdot{}\BR_{K_S\to\pi^+\pi^-}) + \BR_{42}\cdot{}(\BR_{<K^0|K_S>}\cdot{}\BR_{K_S\to\pi^+\pi^-})  \\ 
  {}& + \BR_{70} + \BR_{94} + \BR_{128}\cdot{}\BR_{\eta\to\pi^+\pi^-\pi^0} + \BR_{151}\cdot{}\BR_{\omega\to\pi^+\pi^-\pi^0}  \\ 
  {}& + \BR_{152}\cdot{}\BR_{\omega\to\pi^+\pi^-} + \BR_{800}\cdot{}\BR_{\omega\to\pi^+\pi^-\pi^0} + \BR_{803}}%
\htconstrdef{Gamma66.c}{\BR_{66}}{\BR_{70} + \BR_{94} + \BR_{128}\cdot{}\BR_{\eta\to\pi^+\pi^-\pi^0} + \BR_{151}\cdot{}\BR_{\omega\to\pi^+\pi^-\pi^0} + \BR_{152}\cdot{}\BR_{\omega\to\pi^+\pi^-} + \BR_{800}\cdot{}\BR_{\omega\to\pi^+\pi^-\pi^0} + \BR_{803}}{\BR_{70} + \BR_{94} + \BR_{128}\cdot{}\BR_{\eta\to\pi^+\pi^-\pi^0} + \BR_{151}\cdot{}\BR_{\omega\to\pi^+\pi^-\pi^0}  \\ 
  {}& + \BR_{152}\cdot{}\BR_{\omega\to\pi^+\pi^-} + \BR_{800}\cdot{}\BR_{\omega\to\pi^+\pi^-\pi^0} + \BR_{803}}%
\htconstrdef{Gamma67.c}{\BR_{67}}{\BR_{70} + \BR_{94} + \BR_{128}\cdot{}\BR_{\eta\to\pi^+\pi^-\pi^0} + \BR_{803}}{\BR_{70} + \BR_{94} + \BR_{128}\cdot{}\BR_{\eta\to\pi^+\pi^-\pi^0} + \BR_{803}}%
\htconstrdef{Gamma68.c}{\BR_{68}}{\BR_{40}\cdot{}(\BR_{<K^0|K_S>}\cdot{}\BR_{K_S\to\pi^+\pi^-}) + \BR_{70} + \BR_{152}\cdot{}\BR_{\omega\to\pi^+\pi^-} + \BR_{800}\cdot{}\BR_{\omega\to\pi^+\pi^-\pi^0}}{\BR_{40}\cdot{}(\BR_{<K^0|K_S>}\cdot{}\BR_{K_S\to\pi^+\pi^-}) + \BR_{70} + \BR_{152}\cdot{}\BR_{\omega\to\pi^+\pi^-}  \\ 
  {}& + \BR_{800}\cdot{}\BR_{\omega\to\pi^+\pi^-\pi^0}}%
\htconstrdef{Gamma69.c}{\BR_{69}}{\BR_{152}\cdot{}\BR_{\omega\to\pi^+\pi^-} + \BR_{70} + \BR_{800}\cdot{}\BR_{\omega\to\pi^+\pi^-\pi^0}}{\BR_{152}\cdot{}\BR_{\omega\to\pi^+\pi^-} + \BR_{70} + \BR_{800}\cdot{}\BR_{\omega\to\pi^+\pi^-\pi^0}}%
\htconstrdef{Gamma74.c}{\BR_{74}}{\BR_{152}\cdot{}\BR_{\omega\to\pi^+\pi^-\pi^0} + \BR_{78} + \BR_{77} + \BR_{126}\cdot{}\BR_{\eta\to\pi^+\pi^-\pi^0} + \BR_{130}\cdot{}\BR_{\eta\to\pi^+\pi^-\pi^0}}{\BR_{152}\cdot{}\BR_{\omega\to\pi^+\pi^-\pi^0} + \BR_{78} + \BR_{77} + \BR_{126}\cdot{}\BR_{\eta\to\pi^+\pi^-\pi^0}  \\ 
  {}& + \BR_{130}\cdot{}\BR_{\eta\to\pi^+\pi^-\pi^0}}%
\htconstrdef{Gamma75.c}{\BR_{75}}{\BR_{152}\cdot{}\BR_{\omega\to\pi^+\pi^-\pi^0} + \BR_{47}\cdot{}(2\cdot{}\BR_{K_S\to\pi^+\pi^-}\cdot{}\BR_{K_S\to\pi^0\pi^0}) + \BR_{77} + \BR_{126}\cdot{}\BR_{\eta\to\pi^+\pi^-\pi^0} + \BR_{130}\cdot{}\BR_{\eta\to\pi^+\pi^-\pi^0}}{\BR_{152}\cdot{}\BR_{\omega\to\pi^+\pi^-\pi^0} + \BR_{47}\cdot{}(2\cdot{}\BR_{K_S\to\pi^+\pi^-}\cdot{}\BR_{K_S\to\pi^0\pi^0})  \\ 
  {}& + \BR_{77} + \BR_{126}\cdot{}\BR_{\eta\to\pi^+\pi^-\pi^0} + \BR_{130}\cdot{}\BR_{\eta\to\pi^+\pi^-\pi^0}}%
\htconstrdef{Gamma76.c}{\BR_{76}}{\BR_{152}\cdot{}\BR_{\omega\to\pi^+\pi^-\pi^0} + \BR_{77} + \BR_{126}\cdot{}\BR_{\eta\to\pi^+\pi^-\pi^0} + \BR_{130}\cdot{}\BR_{\eta\to\pi^+\pi^-\pi^0}}{\BR_{152}\cdot{}\BR_{\omega\to\pi^+\pi^-\pi^0} + \BR_{77} + \BR_{126}\cdot{}\BR_{\eta\to\pi^+\pi^-\pi^0} + \BR_{130}\cdot{}\BR_{\eta\to\pi^+\pi^-\pi^0}}%
\htconstrdef{Gamma76by54.c}{\frac{\BR_{76}}{\BR_{54}}}{\frac{\BR_{76}}{\BR_{54}}}{\frac{\BR_{76}}{\BR_{54}}}%
\htconstrdef{Gamma78.c}{\BR_{78}}{\BR_{810} + \BR_{50}\cdot{}(2\cdot{}\BR_{K_S\to\pi^+\pi^-}\cdot{}\BR_{K_S\to\pi^0\pi^0}) + \BR_{132}\cdot{}(\BR_{<\bar{K}^0|K_S>}\cdot{}\BR_{K_S\to\pi^+\pi^-}\cdot{}\BR_{\eta\to3\pi^0})}{\BR_{810} + \BR_{50}\cdot{}(2\cdot{}\BR_{K_S\to\pi^+\pi^-}\cdot{}\BR_{K_S\to\pi^0\pi^0}) + \BR_{132}\cdot{}(\BR_{<\bar{K}^0|K_S>}\cdot{}\BR_{K_S\to\pi^+\pi^-}\cdot{}\BR_{\eta\to3\pi^0})}%
\htconstrdef{Gamma79.c}{\BR_{79}}{\BR_{37}\cdot{}(\BR_{<K^0|K_S>}\cdot{}\BR_{K_S\to\pi^+\pi^-}) + \BR_{42}\cdot{}(\BR_{<K^0|K_S>}\cdot{}\BR_{K_S\to\pi^+\pi^-}) + \BR_{93} + \BR_{94} + \BR_{128}\cdot{}\BR_{\eta\to\text{charged}} + \BR_{151}\cdot{}(\BR_{\omega\to\pi^+\pi^-\pi^0}+\BR_{\omega\to\pi^+\pi^-}) + \BR_{168} + \BR_{802} + \BR_{803}}{\BR_{37}\cdot{}(\BR_{<K^0|K_S>}\cdot{}\BR_{K_S\to\pi^+\pi^-}) + \BR_{42}\cdot{}(\BR_{<K^0|K_S>}\cdot{}\BR_{K_S\to\pi^+\pi^-})  \\ 
  {}& + \BR_{93} + \BR_{94} + \BR_{128}\cdot{}\BR_{\eta\to\text{charged}} + \BR_{151}\cdot{}(\BR_{\omega\to\pi^+\pi^-\pi^0} \\ 
  {}& +\BR_{\omega\to\pi^+\pi^-}) + \BR_{168} + \BR_{802} + \BR_{803}}%
\htconstrdef{Gamma80.c}{\BR_{80}}{\BR_{93} + \BR_{802} + \BR_{151}\cdot{}\BR_{\omega\to\pi^+\pi^-}}{\BR_{93} + \BR_{802} + \BR_{151}\cdot{}\BR_{\omega\to\pi^+\pi^-}}%
\htconstrdef{Gamma80by60.c}{\frac{\BR_{80}}{\BR_{60}}}{\frac{\BR_{80}}{\BR_{60}}}{\frac{\BR_{80}}{\BR_{60}}}%
\htconstrdef{Gamma81.c}{\BR_{81}}{\BR_{128}\cdot{}\BR_{\eta\to\pi^+\pi^-\pi^0} + \BR_{94} + \BR_{803} + \BR_{151}\cdot{}\BR_{\omega\to\pi^+\pi^-\pi^0}}{\BR_{128}\cdot{}\BR_{\eta\to\pi^+\pi^-\pi^0} + \BR_{94} + \BR_{803} + \BR_{151}\cdot{}\BR_{\omega\to\pi^+\pi^-\pi^0}}%
\htconstrdef{Gamma81by69.c}{\frac{\BR_{81}}{\BR_{69}}}{\frac{\BR_{81}}{\BR_{69}}}{\frac{\BR_{81}}{\BR_{69}}}%
\htconstrdef{Gamma82.c}{\BR_{82}}{\BR_{128}\cdot{}\BR_{\eta\to\text{charged}} + \BR_{42}\cdot{}(\BR_{<K^0|K_S>}\cdot{}\BR_{K_S\to\pi^+\pi^-}) + \BR_{802} + \BR_{803} + \BR_{151}\cdot{}(\BR_{\omega\to\pi^+\pi^-\pi^0}+\BR_{\omega\to\pi^+\pi^-}) + \BR_{37}\cdot{}(\BR_{<K^0|K_S>}\cdot{}\BR_{K_S\to\pi^+\pi^-})}{\BR_{128}\cdot{}\BR_{\eta\to\text{charged}} + \BR_{42}\cdot{}(\BR_{<K^0|K_S>}\cdot{}\BR_{K_S\to\pi^+\pi^-}) + \BR_{802}  \\ 
  {}& + \BR_{803} + \BR_{151}\cdot{}(\BR_{\omega\to\pi^+\pi^-\pi^0}+\BR_{\omega\to\pi^+\pi^-}) + \BR_{37}\cdot{}(\BR_{<K^0|K_S>}\cdot{}\BR_{K_S\to\pi^+\pi^-})}%
\htconstrdef{Gamma83.c}{\BR_{83}}{\BR_{128}\cdot{}\BR_{\eta\to\pi^+\pi^-\pi^0} + \BR_{802} + \BR_{803} + \BR_{151}\cdot{}(\BR_{\omega\to\pi^+\pi^-\pi^0}+\BR_{\omega\to\pi^+\pi^-})}{\BR_{128}\cdot{}\BR_{\eta\to\pi^+\pi^-\pi^0} + \BR_{802} + \BR_{803} + \BR_{151}\cdot{}(\BR_{\omega\to\pi^+\pi^-\pi^0} \\ 
  {}& +\BR_{\omega\to\pi^+\pi^-})}%
\htconstrdef{Gamma84.c}{\BR_{84}}{\BR_{802} + \BR_{151}\cdot{}\BR_{\omega\to\pi^+\pi^-} + \BR_{37}\cdot{}(\BR_{<K^0|K_S>}\cdot{}\BR_{K_S\to\pi^+\pi^-})}{\BR_{802} + \BR_{151}\cdot{}\BR_{\omega\to\pi^+\pi^-} + \BR_{37}\cdot{}(\BR_{<K^0|K_S>}\cdot{}\BR_{K_S\to\pi^+\pi^-})}%
\htconstrdef{Gamma85.c}{\BR_{85}}{\BR_{802} + \BR_{151}\cdot{}\BR_{\omega\to\pi^+\pi^-}}{\BR_{802} + \BR_{151}\cdot{}\BR_{\omega\to\pi^+\pi^-}}%
\htconstrdef{Gamma85by60.c}{\frac{\BR_{85}}{\BR_{60}}}{\frac{\BR_{85}}{\BR_{60}}}{\frac{\BR_{85}}{\BR_{60}}}%
\htconstrdef{Gamma87.c}{\BR_{87}}{\BR_{42}\cdot{}(\BR_{<K^0|K_S>}\cdot{}\BR_{K_S\to\pi^+\pi^-}) + \BR_{128}\cdot{}\BR_{\eta\to\pi^+\pi^-\pi^0} + \BR_{151}\cdot{}\BR_{\omega\to\pi^+\pi^-\pi^0} + \BR_{803}}{\BR_{42}\cdot{}(\BR_{<K^0|K_S>}\cdot{}\BR_{K_S\to\pi^+\pi^-}) + \BR_{128}\cdot{}\BR_{\eta\to\pi^+\pi^-\pi^0} + \BR_{151}\cdot{}\BR_{\omega\to\pi^+\pi^-\pi^0}  \\ 
  {}& + \BR_{803}}%
\htconstrdef{Gamma88.c}{\BR_{88}}{\BR_{128}\cdot{}\BR_{\eta\to\pi^+\pi^-\pi^0} + \BR_{803} + \BR_{151}\cdot{}\BR_{\omega\to\pi^+\pi^-\pi^0}}{\BR_{128}\cdot{}\BR_{\eta\to\pi^+\pi^-\pi^0} + \BR_{803} + \BR_{151}\cdot{}\BR_{\omega\to\pi^+\pi^-\pi^0}}%
\htconstrdef{Gamma89.c}{\BR_{89}}{\BR_{803} + \BR_{151}\cdot{}\BR_{\omega\to\pi^+\pi^-\pi^0}}{\BR_{803} + \BR_{151}\cdot{}\BR_{\omega\to\pi^+\pi^-\pi^0}}%
\htconstrdef{Gamma92.c}{\BR_{92}}{\BR_{94} + \BR_{93}}{\BR_{94} + \BR_{93}}%
\htconstrdef{Gamma93by60.c}{\frac{\BR_{93}}{\BR_{60}}}{\frac{\BR_{93}}{\BR_{60}}}{\frac{\BR_{93}}{\BR_{60}}}%
\htconstrdef{Gamma94by69.c}{\frac{\BR_{94}}{\BR_{69}}}{\frac{\BR_{94}}{\BR_{69}}}{\frac{\BR_{94}}{\BR_{69}}}%
\htconstrdef{Gamma96.c}{\BR_{96}}{\BR_{167}\cdot{}\BR_{\phi\to K^+K^-}}{\BR_{167}\cdot{}\BR_{\phi\to K^+K^-}}%
\htconstrdef{Gamma102.c}{\BR_{102}}{\BR_{103} + \BR_{104}}{\BR_{103} + \BR_{104}}%
\htconstrdef{Gamma103.c}{\BR_{103}}{\BR_{820} + \BR_{822} + \BR_{831}\cdot{}\BR_{\omega\to\pi^+\pi^-}}{\BR_{820} + \BR_{822} + \BR_{831}\cdot{}\BR_{\omega\to\pi^+\pi^-}}%
\htconstrdef{Gamma104.c}{\BR_{104}}{\BR_{830} + \BR_{833}}{\BR_{830} + \BR_{833}}%
\htconstrdef{Gamma106.c}{\BR_{106}}{\BR_{30} + \BR_{44}\cdot{}\BR_{<\bar{K}^0|K_S>} + \BR_{47} + \BR_{53}\cdot{}\BR_{<K^0|K_S>} + \BR_{77} + \BR_{103} + \BR_{126}\cdot{}(\BR_{\eta\to3\pi^0}+\BR_{\eta\to\pi^+\pi^-\pi^0}) + \BR_{152}\cdot{}\BR_{\omega\to\pi^+\pi^-\pi^0}}{\BR_{30} + \BR_{44}\cdot{}\BR_{<\bar{K}^0|K_S>} + \BR_{47} + \BR_{53}\cdot{}\BR_{<K^0|K_S>}  \\ 
  {}& + \BR_{77} + \BR_{103} + \BR_{126}\cdot{}(\BR_{\eta\to3\pi^0}+\BR_{\eta\to\pi^+\pi^-\pi^0}) + \BR_{152}\cdot{}\BR_{\omega\to\pi^+\pi^-\pi^0}}%
\htconstrdef{Gamma110.c}{\BR_{110}}{\BR_{10} + \BR_{16} + \BR_{23} + \BR_{28} + \BR_{35} + \BR_{40} + \BR_{128} + \BR_{802} + \BR_{803} + \BR_{151} + \BR_{130} + \BR_{132} + \BR_{44} + \BR_{53} + \BR_{168} + \BR_{169} + \BR_{822} + \BR_{833}}{\BR_{10} + \BR_{16} + \BR_{23} + \BR_{28} + \BR_{35} + \BR_{40}  \\ 
  {}& + \BR_{128} + \BR_{802} + \BR_{803} + \BR_{151} + \BR_{130} + \BR_{132}  \\ 
  {}& + \BR_{44} + \BR_{53} + \BR_{168} + \BR_{169} + \BR_{822} + \BR_{833}}%
\htconstrdef{Gamma149.c}{\BR_{149}}{\BR_{152} + \BR_{800} + \BR_{151}}{\BR_{152} + \BR_{800} + \BR_{151}}%
\htconstrdef{Gamma150.c}{\BR_{150}}{\BR_{800} + \BR_{151}}{\BR_{800} + \BR_{151}}%
\htconstrdef{Gamma150by66.c}{\frac{\BR_{150}}{\BR_{66}}}{\frac{\BR_{150}}{\BR_{66}}}{\frac{\BR_{150}}{\BR_{66}}}%
\htconstrdef{Gamma152by54.c}{\frac{\BR_{152}}{\BR_{54}}}{\frac{\BR_{152}}{\BR_{54}}}{\frac{\BR_{152}}{\BR_{54}}}%
\htconstrdef{Gamma152by76.c}{\frac{\BR_{152}}{\BR_{76}}}{\frac{\BR_{152}}{\BR_{76}}}{\frac{\BR_{152}}{\BR_{76}}}%
\htconstrdef{Gamma168.c}{\BR_{168}}{\BR_{167}\cdot{}\BR_{\phi\to K^+K^-}}{\BR_{167}\cdot{}\BR_{\phi\to K^+K^-}}%
\htconstrdef{Gamma169.c}{\BR_{169}}{\BR_{167}\cdot{}\BR_{\phi\to K_S K_L}}{\BR_{167}\cdot{}\BR_{\phi\to K_S K_L}}%
\htconstrdef{Gamma804.c}{\BR_{804}}{\BR_{47} \cdot{} ((\BR_{<K^0|K_L>}\cdot{}\BR_{<\bar{K}^0|K_L>}) / (\BR_{<K^0|K_S>}\cdot{}\BR_{<\bar{K}^0|K_S>}))}{\BR_{47} \cdot{} ((\BR_{<K^0|K_L>}\cdot{}\BR_{<\bar{K}^0|K_L>}) / (\BR_{<K^0|K_S>}\cdot{}\BR_{<\bar{K}^0|K_S>}))}%
\htconstrdef{Gamma806.c}{\BR_{806}}{\BR_{50} \cdot{} ((\BR_{<K^0|K_L>}\cdot{}\BR_{<\bar{K}^0|K_L>}) / (\BR_{<K^0|K_S>}\cdot{}\BR_{<\bar{K}^0|K_S>}))}{\BR_{50} \cdot{} ((\BR_{<K^0|K_L>}\cdot{}\BR_{<\bar{K}^0|K_L>}) / (\BR_{<K^0|K_S>}\cdot{}\BR_{<\bar{K}^0|K_S>}))}%
\htconstrdef{Gamma809.c}{\BR_{809}}{\BR_{30}}{\BR_{30}}%
\htconstrdef{Gamma810.c}{\BR_{810}}{\BR_{910} + \BR_{911} + \BR_{811}\cdot{}\BR_{\omega\to\pi^+\pi^-\pi^0} + \BR_{812}}{\BR_{910} + \BR_{911} + \BR_{811}\cdot{}\BR_{\omega\to\pi^+\pi^-\pi^0} + \BR_{812}}%
\htconstrdef{Gamma820.c}{\BR_{820}}{\BR_{920} + \BR_{821}}{\BR_{920} + \BR_{821}}%
\htconstrdef{Gamma830.c}{\BR_{830}}{\BR_{930} + \BR_{831}\cdot{}\BR_{\omega\to\pi^+\pi^-\pi^0} + \BR_{832}}{\BR_{930} + \BR_{831}\cdot{}\BR_{\omega\to\pi^+\pi^-\pi^0} + \BR_{832}}%
\htconstrdef{Gamma910.c}{\BR_{910}}{\BR_{136}\cdot{}\BR_{\eta\to3\pi^0}}{\BR_{136}\cdot{}\BR_{\eta\to3\pi^0}}%
\htconstrdef{Gamma911.c}{\BR_{911}}{\BR_{945}\cdot{}\BR_{\eta\to\pi^+\pi^-\pi^0}}{\BR_{945}\cdot{}\BR_{\eta\to\pi^+\pi^-\pi^0}}%
\htconstrdef{Gamma930.c}{\BR_{930}}{\BR_{136}\cdot{}\BR_{\eta\to\pi^+\pi^-\pi^0}}{\BR_{136}\cdot{}\BR_{\eta\to\pi^+\pi^-\pi^0}}%
\htconstrdef{Gamma944.c}{\BR_{944}}{\BR_{136}\cdot{}\BR_{\eta\to\gamma\gamma}}{\BR_{136}\cdot{}\BR_{\eta\to\gamma\gamma}}%
\htconstrdef{GammaAll.c}{\BR_{\text{All}}}{\BR_{3} + \BR_{5} + \BR_{9} + \BR_{10} + \BR_{14} + \BR_{16} + \BR_{20} + \BR_{23} + \BR_{27} + \BR_{28} + \BR_{30} + \BR_{35} + \BR_{37} + \BR_{40} + \BR_{42} + \BR_{47}\cdot{}(1 + ((\BR_{<K^0|K_L>}\cdot{}\BR_{<\bar{K}^0|K_L>}) / (\BR_{<K^0|K_S>}\cdot{}\BR_{<\bar{K}^0|K_S>}))) + \BR_{48} + \BR_{62} + \BR_{70} + \BR_{77} + \BR_{811} + \BR_{812} + \BR_{93} + \BR_{94} + \BR_{832} + \BR_{833} + \BR_{126} + \BR_{128} + \BR_{802} + \BR_{803} + \BR_{800} + \BR_{151} + \BR_{130} + \BR_{132} + \BR_{44} + \BR_{53} + \BR_{50}\cdot{}(1 + ((\BR_{<K^0|K_L>}\cdot{}\BR_{<\bar{K}^0|K_L>}) / (\BR_{<K^0|K_S>}\cdot{}\BR_{<\bar{K}^0|K_S>}))) + \BR_{51} + \BR_{167}\cdot{}(\BR_{\phi\to K^+K^-}+\BR_{\phi\to K_S K_L}) + \BR_{152} + \BR_{920} + \BR_{821} + \BR_{822} + \BR_{831} + \BR_{136} + \BR_{945} + \BR_{805}}{\BR_{3} + \BR_{5} + \BR_{9} + \BR_{10} + \BR_{14} + \BR_{16}  \\ 
  {}& + \BR_{20} + \BR_{23} + \BR_{27} + \BR_{28} + \BR_{30} + \BR_{35}  \\ 
  {}& + \BR_{37} + \BR_{40} + \BR_{42} + \BR_{47}\cdot{}(1 + ((\BR_{<K^0|K_L>}\cdot{}\BR_{<\bar{K}^0|K_L>}) / (\BR_{<K^0|K_S>}\cdot{}\BR_{<\bar{K}^0|K_S>})))  \\ 
  {}& + \BR_{48} + \BR_{62} + \BR_{70} + \BR_{77} + \BR_{811} + \BR_{812}  \\ 
  {}& + \BR_{93} + \BR_{94} + \BR_{832} + \BR_{833} + \BR_{126} + \BR_{128}  \\ 
  {}& + \BR_{802} + \BR_{803} + \BR_{800} + \BR_{151} + \BR_{130} + \BR_{132}  \\ 
  {}& + \BR_{44} + \BR_{53} + \BR_{50}\cdot{}(1 + ((\BR_{<K^0|K_L>}\cdot{}\BR_{<\bar{K}^0|K_L>}) / (\BR_{<K^0|K_S>}\cdot{}\BR_{<\bar{K}^0|K_S>})))  \\ 
  {}& + \BR_{51} + \BR_{167}\cdot{}(\BR_{\phi\to K^+K^-}+\BR_{\phi\to K_S K_L}) + \BR_{152} + \BR_{920}  \\ 
  {}& + \BR_{821} + \BR_{822} + \BR_{831} + \BR_{136} + \BR_{945} + \BR_{805}}%
\htconstrdef{Unitarity}{1}{\BR_{\text{All}} + \BR_{998}}{\BR_{\text{All}} + \BR_{998}}%
\htdef{ConstrEqs}{%
\begin{align*}
\htuse{Gamma1.c.left} ={}& \htuse{Gamma1.c.right.split}
\end{align*}
\begin{align*}
\htuse{Gamma2.c.left} ={}& \htuse{Gamma2.c.right.split}
\end{align*}
\begin{align*}
\htuse{Gamma3by5.c.left} ={}& \htuse{Gamma3by5.c.right.split}
\end{align*}
\begin{align*}
\htuse{Gamma7.c.left} ={}& \htuse{Gamma7.c.right.split}
\end{align*}
\begin{align*}
\htuse{Gamma8.c.left} ={}& \htuse{Gamma8.c.right.split}
\end{align*}
\begin{align*}
\htuse{Gamma8by5.c.left} ={}& \htuse{Gamma8by5.c.right.split}
\end{align*}
\begin{align*}
\htuse{Gamma9by5.c.left} ={}& \htuse{Gamma9by5.c.right.split}
\end{align*}
\begin{align*}
\htuse{Gamma10by5.c.left} ={}& \htuse{Gamma10by5.c.right.split}
\end{align*}
\begin{align*}
\htuse{Gamma10by9.c.left} ={}& \htuse{Gamma10by9.c.right.split}
\end{align*}
\begin{align*}
\htuse{Gamma11.c.left} ={}& \htuse{Gamma11.c.right.split}
\end{align*}
\begin{align*}
\htuse{Gamma12.c.left} ={}& \htuse{Gamma12.c.right.split}
\end{align*}
\begin{align*}
\htuse{Gamma13.c.left} ={}& \htuse{Gamma13.c.right.split}
\end{align*}
\begin{align*}
\htuse{Gamma17.c.left} ={}& \htuse{Gamma17.c.right.split}
\end{align*}
\begin{align*}
\htuse{Gamma18.c.left} ={}& \htuse{Gamma18.c.right.split}
\end{align*}
\begin{align*}
\htuse{Gamma19.c.left} ={}& \htuse{Gamma19.c.right.split}
\end{align*}
\begin{align*}
\htuse{Gamma19by13.c.left} ={}& \htuse{Gamma19by13.c.right.split}
\end{align*}
\begin{align*}
\htuse{Gamma24.c.left} ={}& \htuse{Gamma24.c.right.split}
\end{align*}
\begin{align*}
\htuse{Gamma25.c.left} ={}& \htuse{Gamma25.c.right.split}
\end{align*}
\begin{align*}
\htuse{Gamma26.c.left} ={}& \htuse{Gamma26.c.right.split}
\end{align*}
\begin{align*}
\htuse{Gamma26by13.c.left} ={}& \htuse{Gamma26by13.c.right.split}
\end{align*}
\begin{align*}
\htuse{Gamma29.c.left} ={}& \htuse{Gamma29.c.right.split}
\end{align*}
\begin{align*}
\htuse{Gamma31.c.left} ={}& \htuse{Gamma31.c.right.split}
\end{align*}
\begin{align*}
\htuse{Gamma32.c.left} ={}& \htuse{Gamma32.c.right.split}
\end{align*}
\begin{align*}
\htuse{Gamma33.c.left} ={}& \htuse{Gamma33.c.right.split}
\end{align*}
\begin{align*}
\htuse{Gamma34.c.left} ={}& \htuse{Gamma34.c.right.split}
\end{align*}
\begin{align*}
\htuse{Gamma38.c.left} ={}& \htuse{Gamma38.c.right.split}
\end{align*}
\begin{align*}
\htuse{Gamma39.c.left} ={}& \htuse{Gamma39.c.right.split}
\end{align*}
\begin{align*}
\htuse{Gamma43.c.left} ={}& \htuse{Gamma43.c.right.split}
\end{align*}
\begin{align*}
\htuse{Gamma46.c.left} ={}& \htuse{Gamma46.c.right.split}
\end{align*}
\begin{align*}
\htuse{Gamma49.c.left} ={}& \htuse{Gamma49.c.right.split}
\end{align*}
\begin{align*}
\htuse{Gamma54.c.left} ={}& \htuse{Gamma54.c.right.split}
\end{align*}
\begin{align*}
\htuse{Gamma55.c.left} ={}& \htuse{Gamma55.c.right.split}
\end{align*}
\begin{align*}
\htuse{Gamma56.c.left} ={}& \htuse{Gamma56.c.right.split}
\end{align*}
\begin{align*}
\htuse{Gamma57.c.left} ={}& \htuse{Gamma57.c.right.split}
\end{align*}
\begin{align*}
\htuse{Gamma57by55.c.left} ={}& \htuse{Gamma57by55.c.right.split}
\end{align*}
\begin{align*}
\htuse{Gamma58.c.left} ={}& \htuse{Gamma58.c.right.split}
\end{align*}
\begin{align*}
\htuse{Gamma59.c.left} ={}& \htuse{Gamma59.c.right.split}
\end{align*}
\begin{align*}
\htuse{Gamma60.c.left} ={}& \htuse{Gamma60.c.right.split}
\end{align*}
\begin{align*}
\htuse{Gamma63.c.left} ={}& \htuse{Gamma63.c.right.split}
\end{align*}
\begin{align*}
\htuse{Gamma64.c.left} ={}& \htuse{Gamma64.c.right.split}
\end{align*}
\begin{align*}
\htuse{Gamma65.c.left} ={}& \htuse{Gamma65.c.right.split}
\end{align*}
\begin{align*}
\htuse{Gamma66.c.left} ={}& \htuse{Gamma66.c.right.split}
\end{align*}
\begin{align*}
\htuse{Gamma67.c.left} ={}& \htuse{Gamma67.c.right.split}
\end{align*}
\begin{align*}
\htuse{Gamma68.c.left} ={}& \htuse{Gamma68.c.right.split}
\end{align*}
\begin{align*}
\htuse{Gamma69.c.left} ={}& \htuse{Gamma69.c.right.split}
\end{align*}
\begin{align*}
\htuse{Gamma74.c.left} ={}& \htuse{Gamma74.c.right.split}
\end{align*}
\begin{align*}
\htuse{Gamma75.c.left} ={}& \htuse{Gamma75.c.right.split}
\end{align*}
\begin{align*}
\htuse{Gamma76.c.left} ={}& \htuse{Gamma76.c.right.split}
\end{align*}
\begin{align*}
\htuse{Gamma76by54.c.left} ={}& \htuse{Gamma76by54.c.right.split}
\end{align*}
\begin{align*}
\htuse{Gamma78.c.left} ={}& \htuse{Gamma78.c.right.split}
\end{align*}
\begin{align*}
\htuse{Gamma79.c.left} ={}& \htuse{Gamma79.c.right.split}
\end{align*}
\begin{align*}
\htuse{Gamma80.c.left} ={}& \htuse{Gamma80.c.right.split}
\end{align*}
\begin{align*}
\htuse{Gamma80by60.c.left} ={}& \htuse{Gamma80by60.c.right.split}
\end{align*}
\begin{align*}
\htuse{Gamma81.c.left} ={}& \htuse{Gamma81.c.right.split}
\end{align*}
\begin{align*}
\htuse{Gamma81by69.c.left} ={}& \htuse{Gamma81by69.c.right.split}
\end{align*}
\begin{align*}
\htuse{Gamma82.c.left} ={}& \htuse{Gamma82.c.right.split}
\end{align*}
\begin{align*}
\htuse{Gamma83.c.left} ={}& \htuse{Gamma83.c.right.split}
\end{align*}
\begin{align*}
\htuse{Gamma84.c.left} ={}& \htuse{Gamma84.c.right.split}
\end{align*}
\begin{align*}
\htuse{Gamma85.c.left} ={}& \htuse{Gamma85.c.right.split}
\end{align*}
\begin{align*}
\htuse{Gamma85by60.c.left} ={}& \htuse{Gamma85by60.c.right.split}
\end{align*}
\begin{align*}
\htuse{Gamma87.c.left} ={}& \htuse{Gamma87.c.right.split}
\end{align*}
\begin{align*}
\htuse{Gamma88.c.left} ={}& \htuse{Gamma88.c.right.split}
\end{align*}
\begin{align*}
\htuse{Gamma89.c.left} ={}& \htuse{Gamma89.c.right.split}
\end{align*}
\begin{align*}
\htuse{Gamma92.c.left} ={}& \htuse{Gamma92.c.right.split}
\end{align*}
\begin{align*}
\htuse{Gamma93by60.c.left} ={}& \htuse{Gamma93by60.c.right.split}
\end{align*}
\begin{align*}
\htuse{Gamma94by69.c.left} ={}& \htuse{Gamma94by69.c.right.split}
\end{align*}
\begin{align*}
\htuse{Gamma96.c.left} ={}& \htuse{Gamma96.c.right.split}
\end{align*}
\begin{align*}
\htuse{Gamma102.c.left} ={}& \htuse{Gamma102.c.right.split}
\end{align*}
\begin{align*}
\htuse{Gamma103.c.left} ={}& \htuse{Gamma103.c.right.split}
\end{align*}
\begin{align*}
\htuse{Gamma104.c.left} ={}& \htuse{Gamma104.c.right.split}
\end{align*}
\begin{align*}
\htuse{Gamma106.c.left} ={}& \htuse{Gamma106.c.right.split}
\end{align*}
\begin{align*}
\htuse{Gamma110.c.left} ={}& \htuse{Gamma110.c.right.split}
\end{align*}
\begin{align*}
\htuse{Gamma149.c.left} ={}& \htuse{Gamma149.c.right.split}
\end{align*}
\begin{align*}
\htuse{Gamma150.c.left} ={}& \htuse{Gamma150.c.right.split}
\end{align*}
\begin{align*}
\htuse{Gamma150by66.c.left} ={}& \htuse{Gamma150by66.c.right.split}
\end{align*}
\begin{align*}
\htuse{Gamma152by54.c.left} ={}& \htuse{Gamma152by54.c.right.split}
\end{align*}
\begin{align*}
\htuse{Gamma152by76.c.left} ={}& \htuse{Gamma152by76.c.right.split}
\end{align*}
\begin{align*}
\htuse{Gamma168.c.left} ={}& \htuse{Gamma168.c.right.split}
\end{align*}
\begin{align*}
\htuse{Gamma169.c.left} ={}& \htuse{Gamma169.c.right.split}
\end{align*}
\begin{align*}
\htuse{Gamma804.c.left} ={}& \htuse{Gamma804.c.right.split}
\end{align*}
\begin{align*}
\htuse{Gamma806.c.left} ={}& \htuse{Gamma806.c.right.split}
\end{align*}
\begin{align*}
\htuse{Gamma809.c.left} ={}& \htuse{Gamma809.c.right.split}
\end{align*}
\begin{align*}
\htuse{Gamma810.c.left} ={}& \htuse{Gamma810.c.right.split}
\end{align*}
\begin{align*}
\htuse{Gamma820.c.left} ={}& \htuse{Gamma820.c.right.split}
\end{align*}
\begin{align*}
\htuse{Gamma830.c.left} ={}& \htuse{Gamma830.c.right.split}
\end{align*}
\begin{align*}
\htuse{Gamma910.c.left} ={}& \htuse{Gamma910.c.right.split}
\end{align*}
\begin{align*}
\htuse{Gamma911.c.left} ={}& \htuse{Gamma911.c.right.split}
\end{align*}
\begin{align*}
\htuse{Gamma930.c.left} ={}& \htuse{Gamma930.c.right.split}
\end{align*}
\begin{align*}
\htuse{Gamma944.c.left} ={}& \htuse{Gamma944.c.right.split}
\end{align*}
\begin{align*}
\htuse{GammaAll.c.left} ={}& \htuse{GammaAll.c.right.split}
\end{align*}}%
\htdef{NumMeasALEPH}{39}%
\htdef{NumMeasARGUS}{2}%
\htdef{NumMeasBaBar}{29}%
\htdef{NumMeasBelle}{15}%
\htdef{NumMeasCELLO}{1}%
\htdef{NumMeasCLEO}{35}%
\htdef{NumMeasCLEO3}{6}%
\htdef{NumMeasDELPHI}{14}%
\htdef{NumMeasHRS}{2}%
\htdef{NumMeasL3}{11}%
\htdef{NumMeasOPAL}{19}%
\htdef{NumMeasTPC}{3}%

%% file: hflav-2018-upd2021-elab-vadirect.tex
\htquantdef{B_tau_had_fit}{B_tau_had_fit}{}{64.76 \pm 0.10}{64.76}{0.10}%
\htquantdef{B_tau_had_uni}{B_tau_had_uni}{}{64.79 \pm 0.06}{64.79}{0.06}%
\htquantdef{B_tau_had_uni_SM}{B_tau_had_uni_SM}{}{64.86 \pm 0.04}{64.86}{0.04}%
\htquantdef{B_tau_s_fit}{B_tau_s_fit}{}{2.931 \pm 0.041}{2.931}{0.041}%
\htquantdef{B_tau_s_unitarity}{B_tau_s_unitarity}{}{(2.961 \pm 0.098) \cdot 10^{-2}}{2.961\cdot 10^{-2}}{0.098\cdot 10^{-2}}%
\htquantdef{B_tau_VA}{B_tau_VA}{}{0.61831 \pm 0.00099}{0.61831}{0.00099}%
\htquantdef{B_tau_VA_fit}{B_tau_VA_fit}{}{61.83 \pm 0.10}{61.83}{0.10}%
\htquantdef{B_tau_VA_unitarity}{B_tau_VA_unitarity}{}{0.61861 \pm 0.00075}{0.61861}{0.00075}%
\htquantdef{Be_by_tau_tau}{Be_by_tau_tau}{}{(6.1248 \pm 0.0021) \cdot 10^{11}}{6.1248\cdot 10^{11}}{0.0021\cdot 10^{11}}%
\htquantdef{Be_fit}{Be_fit}{}{0.17817 \pm 0.00041}{0.17817}{0.00041}%
\htquantdef{Be_from_Bmu}{Be_from_Bmu}{}{0.17882 \pm 0.00041}{0.17882}{0.00041}%
\htquantdef{Be_from_taulife}{Be_from_taulife}{}{0.17780 \pm 0.00031}{0.17780}{0.00031}%
\htquantdef{Be_lept}{Be_lept}{}{17.850 \pm 0.032}{17.850}{0.032}%
\htquantdef{Be_lept_unc_perc}{Be_lept_unc_perc}{}{0.178}{0.178}{0}%
\htquantdef{Be_unitarity}{Be_unitarity}{}{0.1785 \pm 0.0010}{0.1785}{0.0010}%
\htquantdef{Be_univ}{Be_univ}{}{17.814 \pm 0.022}{17.814}{0.022}%
\htquantdef{Bmu_by_Be_th}{Bmu_by_Be_th}{}{0.9725606 \pm 0.0000036}{0.9725606}{0.0000036}%
\htquantdef{Bmu_fit}{Bmu_fit}{}{0.17392 \pm 0.00039}{0.17392}{0.00039}%
\htquantdef{Bmu_from_taulife}{Bmu_from_taulife}{}{0.17293 \pm 0.00030}{0.17293}{0.00030}%
\htquantdef{Bmu_unitarity}{Bmu_unitarity}{}{0.1742 \pm 0.0010}{0.1742}{0.0010}%
\htquantdef{BR_a1_pigamma}{BR_a1_pigamma}{}{0.2100\cdot 10^{-2}}{0.2100\cdot 10^{-2}}{0}%
\htquantdef{BR_eta_2gam}{BR_eta_2gam}{}{0.3941}{0.3941}{0}%
\htquantdef{BR_eta_3piz}{BR_eta_3piz}{}{0.3268}{0.3268}{0}%
\htquantdef{BR_eta_charged}{BR_eta_charged}{}{0.2789}{0.2789}{0}%
\htquantdef{BR_eta_neutral}{BR_eta_neutral}{}{0.7212}{0.7212}{0}%
\htquantdef{BR_eta_pimpipgamma}{BR_eta_pimpipgamma}{}{4.220\cdot 10^{-2}}{4.220\cdot 10^{-2}}{0}%
\htquantdef{BR_eta_pimpippiz}{BR_eta_pimpippiz}{}{0.2292}{0.2292}{0}%
\htquantdef{BR_f1_2pip2pim}{BR_f1_2pip2pim}{}{0.1090}{0.1090}{0}%
\htquantdef{BR_f1_2pizpippim}{BR_f1_2pizpippim}{}{0.2180}{0.2180}{0}%
\htquantdef{BR_KS_2piz}{BR_KS_2piz}{}{0.3069}{0.3069}{0}%
\htquantdef{BR_KS_pimpip}{BR_KS_pimpip}{}{0.6920}{0.6920}{0}%
\htquantdef{BR_om_pimpip}{BR_om_pimpip}{}{1.530\cdot 10^{-2}}{1.530\cdot 10^{-2}}{0}%
\htquantdef{BR_om_pimpippiz}{BR_om_pimpippiz}{}{0.8930}{0.8930}{0}%
\htquantdef{BR_om_pizgamma}{BR_om_pizgamma}{}{8.400\cdot 10^{-2}}{8.400\cdot 10^{-2}}{0}%
\htquantdef{BR_phi_KmKp}{BR_phi_KmKp}{}{0.4920}{0.4920}{0}%
\htquantdef{BR_phi_KSKL}{BR_phi_KSKL}{}{0.3400}{0.3400}{0}%
\htquantdef{BRA_Kz_KL_KET}{BRA_Kz_KL_KET}{}{0.5000}{0.5000}{0}%
\htquantdef{BRA_Kz_KS_KET}{BRA_Kz_KS_KET}{}{0.5000}{0.5000}{0}%
\htquantdef{BRA_Kzbar_KL_KET}{BRA_Kzbar_KL_KET}{}{0.5000}{0.5000}{0}%
\htquantdef{BRA_Kzbar_KS_KET}{BRA_Kzbar_KS_KET}{}{0.5000}{0.5000}{0}%
\htquantdef{c0}{c0}{CODATA 2018 speed of light in vacuum}{2.998\cdot 10^{8}}{2.998\cdot 10^{8}}{0}%
\htquantdef{corr_dRrad_pi_munu_K_munu}{corr_dRrad_pi_munu_K_munu}{}{0.67}{0.67}{0}%
\htquantdef{corr_dRrad_pi_munu_K_munu_dicarlo2019}{corr_dRrad_pi_munu_K_munu_dicarlo2019}{}{0.63}{0.63}{0}%
\htquantdef{deltaR_su3break}{deltaR_su3break}{}{0.238 \pm 0.033}{0.238}{0.033}%
\htquantdef{deltaR_su3break_d2pert}{deltaR_su3break_d2pert}{}{9.300 \pm 3.400}{9.300}{3.400}%
\htquantdef{deltaR_su3break_pheno}{deltaR_su3break_pheno}{}{0.1544 \pm 0.0037}{0.1544}{0.0037}%
\htquantdef{deltaR_su3break_remain}{deltaR_su3break_remain}{}{(0.3400 \pm 0.2800) \cdot 10^{-2}}{0.3400\cdot 10^{-2}}{0.2800\cdot 10^{-2}}%
\htquantdef{dRrad_K_munu}{dRrad_K_munu}{}{1.07 \pm 0.21}{1.07}{0.21}%
\htquantdef{dRrad_K_munu_dicarlo2019}{dRrad_K_munu_dicarlo2019}{}{0.24 \pm 0.10}{0.24}{0.10}%
\htquantdef{dRrad_kmunu_by_pimunu}{dRrad_kmunu_by_pimunu}{}{-0.69 \pm 0.17}{-0.69}{0.17}%
\htquantdef{dRrad_kmunu_by_pimunu_dicarlo2019}{dRrad_kmunu_by_pimunu_dicarlo2019}{}{-1.26 \pm 0.14}{-1.26}{0.14}%
\htquantdef{dRrad_pi_munu}{dRrad_pi_munu}{}{1.76 \pm 0.21}{1.76}{0.21}%
\htquantdef{dRrad_pi_munu_dicarlo2019}{dRrad_pi_munu_dicarlo2019}{}{1.53 \pm 0.19}{1.53}{0.19}%
\htquantdef{dRrad_tauK_by_Kmu}{dRrad_tauK_by_Kmu}{}{0.90 \pm 0.22}{0.90}{0.22}%
\htquantdef{dRrad_taupi_by_pimu}{dRrad_taupi_by_pimu}{}{0.16 \pm 0.14}{0.16}{0.14}%
\htquantdef{EmNuNumb}{EmNuNumb}{}{0.1782}{0.1782}{0}%
\htquantdef{f_K_by_f_pi_dicarlo2019}{f_K_by_f_pi_dicarlo2019}{}{1.1966 \pm 0.0018}{1.1966}{0.0018}%
\htquantdef{f_K_dicarlo2019}{f_K_dicarlo2019}{}{156.1 \pm 0.2}{156.1}{0.2}%
\htquantdef{f_Kpm}{f_Kpm}{}{155.7 \pm 0.3}{155.7}{0.3}%
\htquantdef{f_Kpm_by_f_pipm}{f_Kpm_by_f_pipm}{}{1.1932 \pm 0.0021}{1.1932}{0.0021}%
\htquantdef{f_pipm}{f_pipm}{}{130.2 \pm 0.8}{130.2}{0.8}%
\htquantdef{Gamma1}{\BR_{1}}{\BRF{\tau^-}{particle^- \ge{} 0\, \text{neutrals} \ge{} 0\,  K^0\, \nu_\tau}}{0.8521 \pm 0.0011}{0.8521}{0.0011}%
\htquantdef{Gamma10}{\BR_{10}}{\BRF{\tau^-}{K^- \nu_\tau}}{(0.6986 \pm 0.0085) \cdot 10^{-2}}{0.6986\cdot 10^{-2}}{0.0085\cdot 10^{-2}}%
\htquantdef{Gamma102}{\BR_{102}}{\BRF{\tau^-}{3h^- 2h^+ \ge{} 0\,  \text{neutrals}\, \nu_\tau\;(\text{ex.~} K^0)}}{(9.902 \pm 0.368) \cdot 10^{-4}}{9.902\cdot 10^{-4}}{0.368\cdot 10^{-4}}%
\htquantdef{Gamma103}{\BR_{103}}{\BRF{\tau^-}{3h^- 2h^+ \nu_\tau\;(\text{ex.~} K^0)}}{(8.260 \pm 0.314) \cdot 10^{-4}}{8.260\cdot 10^{-4}}{0.314\cdot 10^{-4}}%
\htquantdef{Gamma104}{\BR_{104}}{\BRF{\tau^-}{3h^- 2h^+ \pi^0 \nu_\tau\;(\text{ex.~} K^0)}}{(1.642 \pm 0.114) \cdot 10^{-4}}{1.642\cdot 10^{-4}}{0.114\cdot 10^{-4}}%
\htquantdef{Gamma106}{\BR_{106}}{\BRF{\tau^-}{(5\pi)^- \nu_\tau}}{(0.7532 \pm 0.0356) \cdot 10^{-2}}{0.7532\cdot 10^{-2}}{0.0356\cdot 10^{-2}}%
\htquantdef{Gamma10by5}{\frac{\BR_{10}}{\BR_{5}}}{\frac{\BRF{\tau^-}{K^- \nu_\tau}}{\BRF{\tau^-}{e^- \bar{\nu}_e \nu_\tau}}}{(3.921 \pm 0.048) \cdot 10^{-2}}{3.921\cdot 10^{-2}}{0.048\cdot 10^{-2}}%
\htquantdef{Gamma10by9}{\frac{\BR_{10}}{\BR_{9}}}{\frac{\BRF{\tau^-}{K^- \nu_\tau}}{\BRF{\tau^-}{\pi^- \nu_\tau}}}{(6.467 \pm 0.084) \cdot 10^{-2}}{6.467\cdot 10^{-2}}{0.084\cdot 10^{-2}}%
\htquantdef{Gamma11}{\BR_{11}}{\BRF{\tau^-}{h^- \ge{} 1\,  \text{neutrals}\, \nu_\tau}}{0.36996 \pm 0.00094}{0.36996}{0.00094}%
\htquantdef{Gamma110}{\BR_{110}}{\BRF{\tau^-}{X_s^- \nu_\tau}}{(2.931 \pm 0.041) \cdot 10^{-2}}{2.931\cdot 10^{-2}}{0.041\cdot 10^{-2}}%
\htquantdef{Gamma110_pdg09}{\BR_{110}_pdg09}{}{(2.863 \pm 0.028) \cdot 10^{-2}}{2.863\cdot 10^{-2}}{0.028\cdot 10^{-2}}%
\htquantdef{Gamma12}{\BR_{12}}{\BRF{\tau^-}{h^- \ge{} 1\, \pi^0\, \nu_\tau\;(\text{ex.~} K^0)}}{0.36495 \pm 0.00094}{0.36495}{0.00094}%
\htquantdef{Gamma126}{\BR_{126}}{\BRF{\tau^-}{\pi^- \pi^0 \eta \nu_\tau}}{(0.1386 \pm 0.0072) \cdot 10^{-2}}{0.1386\cdot 10^{-2}}{0.0072\cdot 10^{-2}}%
\htquantdef{Gamma128}{\BR_{128}}{\BRF{\tau^-}{K^- \eta \nu_\tau}}{(1.543 \pm 0.080) \cdot 10^{-4}}{1.543\cdot 10^{-4}}{0.080\cdot 10^{-4}}%
\htquantdef{Gamma13}{\BR_{13}}{\BRF{\tau^-}{h^- \pi^0 \nu_\tau}}{0.25938 \pm 0.00090}{0.25938}{0.00090}%
\htquantdef{Gamma130}{\BR_{130}}{\BRF{\tau^-}{K^- \pi^0 \eta \nu_\tau}}{(4.825 \pm 1.161) \cdot 10^{-5}}{4.825\cdot 10^{-5}}{1.161\cdot 10^{-5}}%
\htquantdef{Gamma132}{\BR_{132}}{\BRF{\tau^-}{\pi^- \bar{K}^0 \eta \nu_\tau}}{(9.364 \pm 1.491) \cdot 10^{-5}}{9.364\cdot 10^{-5}}{1.491\cdot 10^{-5}}%
\htquantdef{Gamma136}{\BR_{136}}{\BRF{\tau^-}{\pi^- \pi^+ \pi^- \eta \nu_\tau\;(\text{ex.~} K^0)}}{(2.196 \pm 0.129) \cdot 10^{-4}}{2.196\cdot 10^{-4}}{0.129\cdot 10^{-4}}%
\htquantdef{Gamma14}{\BR_{14}}{\BRF{\tau^-}{\pi^- \pi^0 \nu_\tau}}{0.25447 \pm 0.00091}{0.25447}{0.00091}%
\htquantdef{Gamma149}{\BR_{149}}{\BRF{\tau^-}{h^- \omega \ge{} 0\,  \text{neutrals}\, \nu_\tau}}{(2.402 \pm 0.075) \cdot 10^{-2}}{2.402\cdot 10^{-2}}{0.075\cdot 10^{-2}}%
\htquantdef{Gamma150}{\BR_{150}}{\BRF{\tau^-}{h^- \omega \nu_\tau}}{(1.996 \pm 0.064) \cdot 10^{-2}}{1.996\cdot 10^{-2}}{0.064\cdot 10^{-2}}%
\htquantdef{Gamma150by66}{\frac{\BR_{150}}{\BR_{66}}}{\frac{\BRF{\tau^-}{h^- \omega \nu_\tau}}{\BRF{\tau^-}{h^- h^- h^+ \pi^0 \nu_\tau\;(\text{ex.~} K^0)}}}{0.4331 \pm 0.0139}{0.4331}{0.0139}%
\htquantdef{Gamma151}{\BR_{151}}{\BRF{\tau^-}{K^- \omega \nu_\tau}}{(4.100 \pm 0.922) \cdot 10^{-4}}{4.100\cdot 10^{-4}}{0.922\cdot 10^{-4}}%
\htquantdef{Gamma152}{\BR_{152}}{\BRF{\tau^-}{h^- \pi^0 \omega \nu_\tau}}{(0.4066 \pm 0.0419) \cdot 10^{-2}}{0.4066\cdot 10^{-2}}{0.0419\cdot 10^{-2}}%
\htquantdef{Gamma152by54}{\frac{\BR_{152}}{\BR_{54}}}{\frac{\BRF{\tau^-}{h^- \omega \pi^0 \nu_\tau}}{\BRF{\tau^-}{h^- h^- h^+ \ge{} 0\, \text{neutrals} \ge{} 0\,  K_L^0\, \nu_\tau}}}{(2.674 \pm 0.275) \cdot 10^{-2}}{2.674\cdot 10^{-2}}{0.275\cdot 10^{-2}}%
\htquantdef{Gamma152by76}{\frac{\BR_{152}}{\BR_{76}}}{\frac{\BRF{\tau^-}{h^- \omega \pi^0 \nu_\tau}}{\BRF{\tau^-}{h^- h^- h^+ 2\pi^0 \nu_\tau\;(\text{ex.~} K^0)}}}{0.8236 \pm 0.0757}{0.8236}{0.0757}%
\htquantdef{Gamma16}{\BR_{16}}{\BRF{\tau^-}{K^- \pi^0 \nu_\tau}}{(0.4904 \pm 0.0092) \cdot 10^{-2}}{0.4904\cdot 10^{-2}}{0.0092\cdot 10^{-2}}%
\htquantdef{Gamma167}{\BR_{167}}{\BRF{\tau^-}{K^- \phi \nu_\tau}}{(4.409 \pm 1.626) \cdot 10^{-5}}{4.409\cdot 10^{-5}}{1.626\cdot 10^{-5}}%
\htquantdef{Gamma168}{\BR_{168}}{\BRF{\tau^-}{K^- \phi(K^+ K^-) \nu_\tau}}{(2.169 \pm 0.800) \cdot 10^{-5}}{2.169\cdot 10^{-5}}{0.800\cdot 10^{-5}}%
\htquantdef{Gamma169}{\BR_{169}}{\BRF{\tau^-}{K^- \phi(K_S^0 K_L^0) \nu_\tau}}{(1.499 \pm 0.553) \cdot 10^{-5}}{1.499\cdot 10^{-5}}{0.553\cdot 10^{-5}}%
\htquantdef{Gamma17}{\BR_{17}}{\BRF{\tau^-}{h^- \ge{} 2\,  \pi^0\, \nu_\tau}}{0.10793 \pm 0.00091}{0.10793}{0.00091}%
\htquantdef{Gamma18}{\BR_{18}}{\BRF{\tau^-}{h^- 2\pi^0 \nu_\tau}}{(9.421 \pm 0.092) \cdot 10^{-2}}{9.421\cdot 10^{-2}}{0.092\cdot 10^{-2}}%
\htquantdef{Gamma19}{\BR_{19}}{\BRF{\tau^-}{h^- 2\pi^0 \nu_\tau\;(\text{ex.~} K^0)}}{(9.270 \pm 0.092) \cdot 10^{-2}}{9.270\cdot 10^{-2}}{0.092\cdot 10^{-2}}%
\htquantdef{Gamma19by13}{\frac{\BR_{19}}{\BR_{13}}}{\frac{\BRF{\tau^-}{h^- 2\pi^0 \nu_\tau\;(\text{ex.~} K^0)}}{\BRF{\tau^-}{h^- \pi^0 \nu_\tau}}}{0.3574 \pm 0.0042}{0.3574}{0.0042}%
\htquantdef{Gamma2}{\BR_{2}}{\BRF{\tau^-}{particle^- \ge{} 0\, \text{neutrals} \ge{} 0\,  K_L^0\, \nu_\tau}}{0.8455 \pm 0.0010}{0.8455}{0.0010}%
\htquantdef{Gamma20}{\BR_{20}}{\BRF{\tau^-}{\pi^- 2\pi^0 \nu_\tau\;(\text{ex.~} K^0)}}{(9.211 \pm 0.092) \cdot 10^{-2}}{9.211\cdot 10^{-2}}{0.092\cdot 10^{-2}}%
\htquantdef{Gamma23}{\BR_{23}}{\BRF{\tau^-}{K^- 2\pi^0 \nu_\tau\;(\text{ex.~} K^0)}}{(5.850 \pm 0.272) \cdot 10^{-4}}{5.850\cdot 10^{-4}}{0.272\cdot 10^{-4}}%
\htquantdef{Gamma24}{\BR_{24}}{\BRF{\tau^-}{h^- \ge{} 3\, \pi^0\, \nu_\tau}}{(1.372 \pm 0.034) \cdot 10^{-2}}{1.372\cdot 10^{-2}}{0.034\cdot 10^{-2}}%
\htquantdef{Gamma25}{\BR_{25}}{\BRF{\tau^-}{h^- \ge{} 3\, \pi^0\, \nu_\tau\;(\text{ex.~} K^0)}}{(1.288 \pm 0.034) \cdot 10^{-2}}{1.288\cdot 10^{-2}}{0.034\cdot 10^{-2}}%
\htquantdef{Gamma26}{\BR_{26}}{\BRF{\tau^-}{h^- 3\pi^0 \nu_\tau}}{(1.236 \pm 0.030) \cdot 10^{-2}}{1.236\cdot 10^{-2}}{0.030\cdot 10^{-2}}%
\htquantdef{Gamma26by13}{\frac{\BR_{26}}{\BR_{13}}}{\frac{\BRF{\tau^-}{h^- 3\pi^0 \nu_\tau}}{\BRF{\tau^-}{h^- \pi^0 \nu_\tau}}}{(4.764 \pm 0.118) \cdot 10^{-2}}{4.764\cdot 10^{-2}}{0.118\cdot 10^{-2}}%
\htquantdef{Gamma27}{\BR_{27}}{\BRF{\tau^-}{\pi^- 3\pi^0 \nu_\tau\;(\text{ex.~} K^0)}}{(1.138 \pm 0.029) \cdot 10^{-2}}{1.138\cdot 10^{-2}}{0.029\cdot 10^{-2}}%
\htquantdef{Gamma28}{\BR_{28}}{\BRF{\tau^-}{K^- 3\pi^0 \nu_\tau\;(\text{ex.~} K^0, \eta)}}{(1.127 \pm 0.263) \cdot 10^{-4}}{1.127\cdot 10^{-4}}{0.263\cdot 10^{-4}}%
\htquantdef{Gamma29}{\BR_{29}}{\BRF{\tau^-}{h^- 4\pi^0 \nu_\tau\;(\text{ex.~} K^0)}}{(0.1333 \pm 0.0071) \cdot 10^{-2}}{0.1333\cdot 10^{-2}}{0.0071\cdot 10^{-2}}%
\htquantdef{Gamma3}{\BR_{3}}{\BRF{\tau^-}{\mu^- \bar{\nu}_\mu \nu_\tau}}{0.17392 \pm 0.00039}{0.17392}{0.00039}%
\htquantdef{Gamma30}{\BR_{30}}{\BRF{\tau^-}{h^- 4\pi^0 \nu_\tau\;(\text{ex.~} K^0, \eta)}}{(8.640 \pm 0.670) \cdot 10^{-4}}{8.640\cdot 10^{-4}}{0.670\cdot 10^{-4}}%
\htquantdef{Gamma31}{\BR_{31}}{\BRF{\tau^-}{K^- \ge{} 0\, \pi^0 \ge{} 0\, K^0 \ge{} 0\, \gamma \nu_\tau}}{(1.568 \pm 0.018) \cdot 10^{-2}}{1.568\cdot 10^{-2}}{0.018\cdot 10^{-2}}%
\htquantdef{Gamma32}{\BR_{32}}{\BRF{\tau^-}{K^- \ge{} 1\, (\pi^0\,\text{or}\,K^0\,\text{or}\,\gamma) \nu_\tau}}{(0.8729 \pm 0.0141) \cdot 10^{-2}}{0.8729\cdot 10^{-2}}{0.0141\cdot 10^{-2}}%
\htquantdef{Gamma33}{\BR_{33}}{\BRF{\tau^-}{K_S^0 (\text{particles})^- \nu_\tau}}{(0.9366 \pm 0.0292) \cdot 10^{-2}}{0.9366\cdot 10^{-2}}{0.0292\cdot 10^{-2}}%
\htquantdef{Gamma34}{\BR_{34}}{\BRF{\tau^-}{h^- \bar{K}^0 \nu_\tau}}{(0.9860 \pm 0.0138) \cdot 10^{-2}}{0.9860\cdot 10^{-2}}{0.0138\cdot 10^{-2}}%
\htquantdef{Gamma35}{\BR_{35}}{\BRF{\tau^-}{\pi^- \bar{K}^0 \nu_\tau}}{(0.8378 \pm 0.0139) \cdot 10^{-2}}{0.8378\cdot 10^{-2}}{0.0139\cdot 10^{-2}}%
\htquantdef{Gamma37}{\BR_{37}}{\BRF{\tau^-}{K^- K^0 \nu_\tau}}{(0.1483 \pm 0.0034) \cdot 10^{-2}}{0.1483\cdot 10^{-2}}{0.0034\cdot 10^{-2}}%
\htquantdef{Gamma38}{\BR_{38}}{\BRF{\tau^-}{K^- K^0 \ge{} 0\,  \pi^0\, \nu_\tau}}{(0.2977 \pm 0.0073) \cdot 10^{-2}}{0.2977\cdot 10^{-2}}{0.0073\cdot 10^{-2}}%
\htquantdef{Gamma39}{\BR_{39}}{\BRF{\tau^-}{h^- \bar{K}^0 \pi^0 \nu_\tau}}{(0.5302 \pm 0.0134) \cdot 10^{-2}}{0.5302\cdot 10^{-2}}{0.0134\cdot 10^{-2}}%
\htquantdef{Gamma3by5}{\frac{\BR_{3}}{\BR_{5}}}{\frac{\BRF{\tau^-}{\mu^- \bar{\nu}_\mu \nu_\tau}}{\BRF{\tau^-}{e^- \bar{\nu}_e \nu_\tau}}}{0.9761 \pm 0.0028}{0.9761}{0.0028}%
\htquantdef{Gamma40}{\BR_{40}}{\BRF{\tau^-}{\pi^- \bar{K}^0 \pi^0 \nu_\tau}}{(0.3807 \pm 0.0129) \cdot 10^{-2}}{0.3807\cdot 10^{-2}}{0.0129\cdot 10^{-2}}%
\htquantdef{Gamma42}{\BR_{42}}{\BRF{\tau^-}{K^- K^0 \pi^0 \nu_\tau}}{(0.1494 \pm 0.0070) \cdot 10^{-2}}{0.1494\cdot 10^{-2}}{0.0070\cdot 10^{-2}}%
\htquantdef{Gamma43}{\BR_{43}}{\BRF{\tau^-}{\pi^- \bar{K}^0 \ge{} 1\,  \pi^0\, \nu_\tau}}{(0.4042 \pm 0.0260) \cdot 10^{-2}}{0.4042\cdot 10^{-2}}{0.0260\cdot 10^{-2}}%
\htquantdef{Gamma44}{\BR_{44}}{\BRF{\tau^-}{\pi^- \bar{K}^0 2\pi^0 \nu_\tau\;(\text{ex.~} K^0)}}{(2.346 \pm 2.306) \cdot 10^{-4}}{2.346\cdot 10^{-4}}{2.306\cdot 10^{-4}}%
\htquantdef{Gamma46}{\BR_{46}}{\BRF{\tau^-}{\pi^- K^0 \bar{K}^0 \nu_\tau}}{(0.1516 \pm 0.0247) \cdot 10^{-2}}{0.1516\cdot 10^{-2}}{0.0247\cdot 10^{-2}}%
\htquantdef{Gamma47}{\BR_{47}}{\BRF{\tau^-}{\pi^- K_S^0 K_S^0 \nu_\tau}}{(2.342 \pm 0.065) \cdot 10^{-4}}{2.342\cdot 10^{-4}}{0.065\cdot 10^{-4}}%
\htquantdef{Gamma48}{\BR_{48}}{\BRF{\tau^-}{\pi^- K_S^0 K_L^0 \nu_\tau}}{(0.1048 \pm 0.0247) \cdot 10^{-2}}{0.1048\cdot 10^{-2}}{0.0247\cdot 10^{-2}}%
\htquantdef{Gamma49}{\BR_{49}}{\BRF{\tau^-}{\pi^- \pi^0 K^0 \bar{K}^0 \nu_\tau}}{(3.543 \pm 1.193) \cdot 10^{-4}}{3.543\cdot 10^{-4}}{1.193\cdot 10^{-4}}%
\htquantdef{Gamma5}{\BR_{5}}{\BRF{\tau^-}{e^- \bar{\nu}_e \nu_\tau}}{0.17817 \pm 0.00041}{0.17817}{0.00041}%
\htquantdef{Gamma50}{\BR_{50}}{\BRF{\tau^-}{\pi^- K_S^0 K_S^0 \pi^0 \nu_\tau}}{(1.816 \pm 0.207) \cdot 10^{-5}}{1.816\cdot 10^{-5}}{0.207\cdot 10^{-5}}%
\htquantdef{Gamma51}{\BR_{51}}{\BRF{\tau^-}{\pi^- K_S^0 K_L^0 \pi^0 \nu_\tau}}{(3.179 \pm 1.192) \cdot 10^{-4}}{3.179\cdot 10^{-4}}{1.192\cdot 10^{-4}}%
\htquantdef{Gamma53}{\BR_{53}}{\BRF{\tau^-}{\bar{K}^0 h^- h^- h^+ \nu_\tau}}{(2.220 \pm 2.024) \cdot 10^{-4}}{2.220\cdot 10^{-4}}{2.024\cdot 10^{-4}}%
\htquantdef{Gamma54}{\BR_{54}}{\BRF{\tau^-}{h^- h^- h^+ \ge{} 0\, \text{neutrals} \ge{} 0\,  K_L^0\, \nu_\tau}}{0.15206 \pm 0.00061}{0.15206}{0.00061}%
\htquantdef{Gamma55}{\BR_{55}}{\BRF{\tau^-}{h^- h^- h^+ \ge{} 0\,  \text{neutrals}\, \nu_\tau\;(\text{ex.~} K^0)}}{0.14558 \pm 0.00056}{0.14558}{0.00056}%
\htquantdef{Gamma56}{\BR_{56}}{\BRF{\tau^-}{h^- h^- h^+ \nu_\tau}}{(9.769 \pm 0.053) \cdot 10^{-2}}{9.769\cdot 10^{-2}}{0.053\cdot 10^{-2}}%
\htquantdef{Gamma57}{\BR_{57}}{\BRF{\tau^-}{h^- h^- h^+ \nu_\tau\;(\text{ex.~} K^0)}}{(9.428 \pm 0.053) \cdot 10^{-2}}{9.428\cdot 10^{-2}}{0.053\cdot 10^{-2}}%
\htquantdef{Gamma57by55}{\frac{\BR_{57}}{\BR_{55}}}{\frac{\BRF{\tau^-}{h^- h^- h^+ \nu_\tau\;(\text{ex.~} K^0)}}{\BRF{\tau^-}{h^- h^- h^+ \ge{} 0\,  \text{neutrals}\, \nu_\tau\;(\text{ex.~} K^0)}}}{0.6476 \pm 0.0029}{0.6476}{0.0029}%
\htquantdef{Gamma58}{\BR_{58}}{\BRF{\tau^-}{h^- h^- h^+ \nu_\tau\;(\text{ex.~} K^0, \omega)}}{(9.397 \pm 0.053) \cdot 10^{-2}}{9.397\cdot 10^{-2}}{0.053\cdot 10^{-2}}%
\htquantdef{Gamma59}{\BR_{59}}{\BRF{\tau^-}{\pi^- \pi^+ \pi^- \nu_\tau}}{(9.279 \pm 0.051) \cdot 10^{-2}}{9.279\cdot 10^{-2}}{0.051\cdot 10^{-2}}%
\htquantdef{Gamma60}{\BR_{60}}{\BRF{\tau^-}{\pi^- \pi^+ \pi^- \nu_\tau\;(\text{ex.~} K^0)}}{(8.990 \pm 0.051) \cdot 10^{-2}}{8.990\cdot 10^{-2}}{0.051\cdot 10^{-2}}%
\htquantdef{Gamma62}{\BR_{62}}{\BRF{\tau^-}{\pi^- \pi^+ \pi^- \nu_\tau\;(\text{ex.~} K^0, \omega)}}{(8.960 \pm 0.051) \cdot 10^{-2}}{8.960\cdot 10^{-2}}{0.051\cdot 10^{-2}}%
\htquantdef{Gamma63}{\BR_{63}}{\BRF{\tau^-}{h^- h^- h^+ \ge{} 1\,  \text{neutrals}\, \nu_\tau}}{(5.327 \pm 0.049) \cdot 10^{-2}}{5.327\cdot 10^{-2}}{0.049\cdot 10^{-2}}%
\htquantdef{Gamma64}{\BR_{64}}{\BRF{\tau^-}{h^- h^- h^+ \ge{} 1\,  \pi^0\, \nu_\tau\;(\text{ex.~} K^0)}}{(5.122 \pm 0.049) \cdot 10^{-2}}{5.122\cdot 10^{-2}}{0.049\cdot 10^{-2}}%
\htquantdef{Gamma65}{\BR_{65}}{\BRF{\tau^-}{h^- h^- h^+ \pi^0 \nu_\tau}}{(4.791 \pm 0.052) \cdot 10^{-2}}{4.791\cdot 10^{-2}}{0.052\cdot 10^{-2}}%
\htquantdef{Gamma66}{\BR_{66}}{\BRF{\tau^-}{h^- h^- h^+ \pi^0 \nu_\tau\;(\text{ex.~} K^0)}}{(4.607 \pm 0.051) \cdot 10^{-2}}{4.607\cdot 10^{-2}}{0.051\cdot 10^{-2}}%
\htquantdef{Gamma67}{\BR_{67}}{\BRF{\tau^-}{h^- h^- h^+ \pi^0 \nu_\tau\;(\text{ex.~} K^0, \omega)}}{(2.819 \pm 0.070) \cdot 10^{-2}}{2.819\cdot 10^{-2}}{0.070\cdot 10^{-2}}%
\htquantdef{Gamma68}{\BR_{68}}{\BRF{\tau^-}{\pi^- \pi^+ \pi^- \pi^0 \nu_\tau}}{(4.652 \pm 0.053) \cdot 10^{-2}}{4.652\cdot 10^{-2}}{0.053\cdot 10^{-2}}%
\htquantdef{Gamma69}{\BR_{69}}{\BRF{\tau^-}{\pi^- \pi^+ \pi^- \pi^0 \nu_\tau\;(\text{ex.~} K^0)}}{(4.520 \pm 0.052) \cdot 10^{-2}}{4.520\cdot 10^{-2}}{0.052\cdot 10^{-2}}%
\htquantdef{Gamma7}{\BR_{7}}{\BRF{\tau^-}{h^- \ge{} 0\,  K_L^0\, \nu_\tau}}{0.12019 \pm 0.00053}{0.12019}{0.00053}%
\htquantdef{Gamma70}{\BR_{70}}{\BRF{\tau^-}{\pi^- \pi^+ \pi^- \pi^0 \nu_\tau\;(\text{ex.~} K^0, \omega)}}{(2.768 \pm 0.071) \cdot 10^{-2}}{2.768\cdot 10^{-2}}{0.071\cdot 10^{-2}}%
\htquantdef{Gamma74}{\BR_{74}}{\BRF{\tau^-}{h^- h^- h^+ \ge{} 2\, \pi^0\, \nu_\tau\;(\text{ex.~} K^0)}}{(0.5149 \pm 0.0311) \cdot 10^{-2}}{0.5149\cdot 10^{-2}}{0.0311\cdot 10^{-2}}%
\htquantdef{Gamma75}{\BR_{75}}{\BRF{\tau^-}{h^- h^- h^+ 2\pi^0 \nu_\tau}}{(0.5037 \pm 0.0309) \cdot 10^{-2}}{0.5037\cdot 10^{-2}}{0.0309\cdot 10^{-2}}%
\htquantdef{Gamma76}{\BR_{76}}{\BRF{\tau^-}{h^- h^- h^+ 2\pi^0 \nu_\tau\;(\text{ex.~} K^0)}}{(0.4937 \pm 0.0309) \cdot 10^{-2}}{0.4937\cdot 10^{-2}}{0.0309\cdot 10^{-2}}%
\htquantdef{Gamma76by54}{\frac{\BR_{76}}{\BR_{54}}}{\frac{\BRF{\tau^-}{h^- h^- h^+ 2\pi^0 \nu_\tau\;(\text{ex.~} K^0)}}{\BRF{\tau^-}{h^- h^- h^+ \ge{} 0\, \text{neutrals} \ge{} 0\,  K_L^0\, \nu_\tau}}}{(3.247 \pm 0.202) \cdot 10^{-2}}{3.247\cdot 10^{-2}}{0.202\cdot 10^{-2}}%
\htquantdef{Gamma77}{\BR_{77}}{\BRF{\tau^-}{h^- h^- h^+ 2\pi^0 \nu_\tau\;(\text{ex.~} K^0, \omega, \eta)}}{(9.772 \pm 3.558) \cdot 10^{-4}}{9.772\cdot 10^{-4}}{3.558\cdot 10^{-4}}%
\htquantdef{Gamma78}{\BR_{78}}{\BRF{\tau^-}{h^- h^- h^+ 3\pi^0 \nu_\tau}}{(2.115 \pm 0.299) \cdot 10^{-4}}{2.115\cdot 10^{-4}}{0.299\cdot 10^{-4}}%
\htquantdef{Gamma79}{\BR_{79}}{\BRF{\tau^-}{K^- h^- h^+ \ge{} 0\,  \text{neutrals}\, \nu_\tau}}{(0.6293 \pm 0.0140) \cdot 10^{-2}}{0.6293\cdot 10^{-2}}{0.0140\cdot 10^{-2}}%
\htquantdef{Gamma8}{\BR_{8}}{\BRF{\tau^-}{h^- \nu_\tau}}{0.11502 \pm 0.00053}{0.11502}{0.00053}%
\htquantdef{Gamma80}{\BR_{80}}{\BRF{\tau^-}{K^- \pi^- h^+ \nu_\tau\;(\text{ex.~} K^0)}}{(0.4361 \pm 0.0072) \cdot 10^{-2}}{0.4361\cdot 10^{-2}}{0.0072\cdot 10^{-2}}%
\htquantdef{Gamma800}{\BR_{800}}{\BRF{\tau^-}{\pi^- \omega \nu_\tau}}{(1.955 \pm 0.065) \cdot 10^{-2}}{1.955\cdot 10^{-2}}{0.065\cdot 10^{-2}}%
\htquantdef{Gamma802}{\BR_{802}}{\BRF{\tau^-}{K^- \pi^- \pi^+ \nu_\tau\;(\text{ex.~} K^0, \omega)}}{(0.2923 \pm 0.0067) \cdot 10^{-2}}{0.2923\cdot 10^{-2}}{0.0067\cdot 10^{-2}}%
\htquantdef{Gamma803}{\BR_{803}}{\BRF{\tau^-}{K^- \pi^- \pi^+ \pi^0 \nu_\tau\;(\text{ex.~} K^0, \omega, \eta)}}{(4.101 \pm 1.429) \cdot 10^{-4}}{4.101\cdot 10^{-4}}{1.429\cdot 10^{-4}}%
\htquantdef{Gamma804}{\BR_{804}}{\BRF{\tau^-}{\pi^- K_L^0 K_L^0 \nu_\tau}}{(2.342 \pm 0.065) \cdot 10^{-4}}{2.342\cdot 10^{-4}}{0.065\cdot 10^{-4}}%
\htquantdef{Gamma805}{\BR_{805}}{\BRF{\tau^-}{a1^-(\pi^- \gamma) \nu_\tau}}{(4.000 \pm 2.000) \cdot 10^{-4}}{4.000\cdot 10^{-4}}{2.000\cdot 10^{-4}}%
\htquantdef{Gamma806}{\BR_{806}}{\BRF{\tau^-}{\pi^- K_L^0 K_L^0 \pi^0 \nu_\tau}}{(1.816 \pm 0.207) \cdot 10^{-5}}{1.816\cdot 10^{-5}}{0.207\cdot 10^{-5}}%
\htquantdef{Gamma809}{\BR_{809}}{\BRF{\tau^-}{\pi^- 4\pi^0 \nu_\tau\;(\text{ex.~} K^0, \eta)}}{(8.640 \pm 0.670) \cdot 10^{-4}}{8.640\cdot 10^{-4}}{0.670\cdot 10^{-4}}%
\htquantdef{Gamma80by60}{\frac{\BR_{80}}{\BR_{60}}}{\frac{\BRF{\tau^-}{K^- \pi^- h^+ \nu_\tau\;(\text{ex.~} K^0)}}{\BRF{\tau^-}{\pi^- \pi^+ \pi^- \nu_\tau\;(\text{ex.~} K^0)}}}{(4.851 \pm 0.080) \cdot 10^{-2}}{4.851\cdot 10^{-2}}{0.080\cdot 10^{-2}}%
\htquantdef{Gamma81}{\BR_{81}}{\BRF{\tau^-}{K^- \pi^- h^+ \pi^0 \nu_\tau\;(\text{ex.~} K^0)}}{(8.727 \pm 1.177) \cdot 10^{-4}}{8.727\cdot 10^{-4}}{1.177\cdot 10^{-4}}%
\htquantdef{Gamma810}{\BR_{810}}{\BRF{\tau^-}{2\pi^- \pi^+ 3\pi^0 \nu_\tau\;(\text{ex.~} K^0)}}{(1.932 \pm 0.298) \cdot 10^{-4}}{1.932\cdot 10^{-4}}{0.298\cdot 10^{-4}}%
\htquantdef{Gamma811}{\BR_{811}}{\BRF{\tau^-}{\pi^- 2\pi^0 \omega \nu_\tau}}{(7.139 \pm 1.586) \cdot 10^{-5}}{7.139\cdot 10^{-5}}{1.586\cdot 10^{-5}}%
\htquantdef{Gamma812}{\BR_{812}}{\BRF{\tau^-}{2\pi^- \pi^+ 3\pi^0 \nu_\tau\;(\text{ex.~} K^0, \eta, \omega, f1)}}{(1.323 \pm 2.683) \cdot 10^{-5}}{1.323\cdot 10^{-5}}{2.683\cdot 10^{-5}}%
\htquantdef{Gamma81by69}{\frac{\BR_{81}}{\BR_{69}}}{\frac{\BRF{\tau^-}{K^- \pi^- h^+ \pi^0 \nu_\tau\;(\text{ex.~} K^0)}}{\BRF{\tau^-}{\pi^- \pi^+ \pi^- \pi^0 \nu_\tau\;(\text{ex.~} K^0)}}}{(1.931 \pm 0.266) \cdot 10^{-2}}{1.931\cdot 10^{-2}}{0.266\cdot 10^{-2}}%
\htquantdef{Gamma82}{\BR_{82}}{\BRF{\tau^-}{K^- \pi^- \pi^+ \ge{} 0\,  \text{neutrals}\, \nu_\tau}}{(0.4779 \pm 0.0137) \cdot 10^{-2}}{0.4779\cdot 10^{-2}}{0.0137\cdot 10^{-2}}%
\htquantdef{Gamma820}{\BR_{820}}{\BRF{\tau^-}{3\pi^- 2\pi^+ \nu_\tau\;(\text{ex.~} K^0, \omega)}}{(8.241 \pm 0.313) \cdot 10^{-4}}{8.241\cdot 10^{-4}}{0.313\cdot 10^{-4}}%
\htquantdef{Gamma821}{\BR_{821}}{\BRF{\tau^-}{3\pi^- 2\pi^+ \nu_\tau\;(\text{ex.~} K^0, \omega, f1)}}{(7.719 \pm 0.295) \cdot 10^{-4}}{7.719\cdot 10^{-4}}{0.295\cdot 10^{-4}}%
\htquantdef{Gamma822}{\BR_{822}}{\BRF{\tau^-}{K^- 2\pi^- 2\pi^+ \nu_\tau\;(\text{ex.~} K^0)}}{(0.594 \pm 1.208) \cdot 10^{-6}}{0.594\cdot 10^{-6}}{1.208\cdot 10^{-6}}%
\htquantdef{Gamma83}{\BR_{83}}{\BRF{\tau^-}{K^- \pi^- \pi^+ \ge{} 0\,  \pi^0\, \nu_\tau\;(\text{ex.~} K^0)}}{(0.3741 \pm 0.0135) \cdot 10^{-2}}{0.3741\cdot 10^{-2}}{0.0135\cdot 10^{-2}}%
\htquantdef{Gamma830}{\BR_{830}}{\BRF{\tau^-}{3\pi^- 2\pi^+ \pi^0 \nu_\tau\;(\text{ex.~} K^0)}}{(1.631 \pm 0.113) \cdot 10^{-4}}{1.631\cdot 10^{-4}}{0.113\cdot 10^{-4}}%
\htquantdef{Gamma831}{\BR_{831}}{\BRF{\tau^-}{2\pi^- \pi^+ \omega \nu_\tau\;(\text{ex.~} K^0)}}{(8.399 \pm 0.624) \cdot 10^{-5}}{8.399\cdot 10^{-5}}{0.624\cdot 10^{-5}}%
\htquantdef{Gamma832}{\BR_{832}}{\BRF{\tau^-}{3\pi^- 2\pi^+ \pi^0 \nu_\tau\;(\text{ex.~} K^0, \eta, \omega, f1)}}{(3.774 \pm 0.874) \cdot 10^{-5}}{3.774\cdot 10^{-5}}{0.874\cdot 10^{-5}}%
\htquantdef{Gamma833}{\BR_{833}}{\BRF{\tau^-}{K^- 2\pi^- 2\pi^+ \pi^0 \nu_\tau\;(\text{ex.~} K^0)}}{(1.107 \pm 0.566) \cdot 10^{-6}}{1.107\cdot 10^{-6}}{0.566\cdot 10^{-6}}%
\htquantdef{Gamma84}{\BR_{84}}{\BRF{\tau^-}{K^- \pi^- \pi^+ \nu_\tau}}{(0.3442 \pm 0.0068) \cdot 10^{-2}}{0.3442\cdot 10^{-2}}{0.0068\cdot 10^{-2}}%
\htquantdef{Gamma85}{\BR_{85}}{\BRF{\tau^-}{K^- \pi^+ \pi^- \nu_\tau\;(\text{ex.~} K^0)}}{(0.2929 \pm 0.0067) \cdot 10^{-2}}{0.2929\cdot 10^{-2}}{0.0067\cdot 10^{-2}}%
\htquantdef{Gamma85by60}{\frac{\BR_{85}}{\BR_{60}}}{\frac{\BRF{\tau^-}{K^- \pi^+ \pi^- \nu_\tau\;(\text{ex.~}K^0)}}{\BRF{\tau^-}{\pi^- \pi^+ \pi^- \nu_\tau\;(\text{ex.~}K^0)}}}{(3.258 \pm 0.074) \cdot 10^{-2}}{3.258\cdot 10^{-2}}{0.074\cdot 10^{-2}}%
\htquantdef{Gamma87}{\BR_{87}}{\BRF{\tau^-}{K^- \pi^- \pi^+ \pi^0 \nu_\tau}}{(0.1329 \pm 0.0119) \cdot 10^{-2}}{0.1329\cdot 10^{-2}}{0.0119\cdot 10^{-2}}%
\htquantdef{Gamma88}{\BR_{88}}{\BRF{\tau^-}{K^- \pi^- \pi^+ \pi^0 \nu_\tau\;(\text{ex.~} K^0)}}{(8.116 \pm 1.168) \cdot 10^{-4}}{8.116\cdot 10^{-4}}{1.168\cdot 10^{-4}}%
\htquantdef{Gamma89}{\BR_{89}}{\BRF{\tau^-}{K^- \pi^- \pi^+ \pi^0 \nu_\tau\;(\text{ex.~} K^0, \eta)}}{(7.762 \pm 1.168) \cdot 10^{-4}}{7.762\cdot 10^{-4}}{1.168\cdot 10^{-4}}%
\htquantdef{Gamma8by5}{\frac{\BR_{8}}{\BR_{5}}}{\frac{\BRF{\tau^-}{h^- \nu_\tau}}{\BRF{\tau^-}{e^- \bar{\nu}_e \nu_\tau}}}{0.6456 \pm 0.0033}{0.6456}{0.0033}%
\htquantdef{Gamma9}{\BR_{9}}{\BRF{\tau^-}{\pi^- \nu_\tau}}{0.10804 \pm 0.00052}{0.10804}{0.00052}%
\htquantdef{Gamma910}{\BR_{910}}{\BRF{\tau^-}{2\pi^- \pi^+ \eta(3\pi^0) \nu_\tau\;(\text{ex.~} K^0)}}{(7.176 \pm 0.422) \cdot 10^{-5}}{7.176\cdot 10^{-5}}{0.422\cdot 10^{-5}}%
\htquantdef{Gamma911}{\BR_{911}}{\BRF{\tau^-}{\pi^- 2\pi^0 \eta(\pi^+ \pi^- \pi^0) \nu_\tau\;(\text{ex.~} K^0)}}{(4.443 \pm 0.867) \cdot 10^{-5}}{4.443\cdot 10^{-5}}{0.867\cdot 10^{-5}}%
\htquantdef{Gamma92}{\BR_{92}}{\BRF{\tau^-}{\pi^- K^- K^+ \ge{} 0\,  \text{neutrals}\, \nu_\tau}}{(0.1493 \pm 0.0033) \cdot 10^{-2}}{0.1493\cdot 10^{-2}}{0.0033\cdot 10^{-2}}%
\htquantdef{Gamma920}{\BR_{920}}{\BRF{\tau^-}{\pi^- f1(2\pi^- 2\pi^+) \nu_\tau}}{(5.225 \pm 0.444) \cdot 10^{-5}}{5.225\cdot 10^{-5}}{0.444\cdot 10^{-5}}%
\htquantdef{Gamma93}{\BR_{93}}{\BRF{\tau^-}{\pi^- K^- K^+ \nu_\tau}}{(0.1432 \pm 0.0027) \cdot 10^{-2}}{0.1432\cdot 10^{-2}}{0.0027\cdot 10^{-2}}%
\htquantdef{Gamma930}{\BR_{930}}{\BRF{\tau^-}{2\pi^- \pi^+ \eta(\pi^+ \pi^- \pi^0) \nu_\tau\;(\text{ex.~} K^0)}}{(5.033 \pm 0.296) \cdot 10^{-5}}{5.033\cdot 10^{-5}}{0.296\cdot 10^{-5}}%
\htquantdef{Gamma93by60}{\frac{\BR_{93}}{\BR_{60}}}{\frac{\BRF{\tau^-}{\pi^- K^- K^+ \nu_\tau}}{\BRF{\tau^-}{\pi^- \pi^+ \pi^- \nu_\tau\;(\text{ex.~} K^0)}}}{(1.592 \pm 0.030) \cdot 10^{-2}}{1.592\cdot 10^{-2}}{0.030\cdot 10^{-2}}%
\htquantdef{Gamma94}{\BR_{94}}{\BRF{\tau^-}{\pi^- K^- K^+ \pi^0 \nu_\tau}}{(6.114 \pm 1.829) \cdot 10^{-5}}{6.114\cdot 10^{-5}}{1.829\cdot 10^{-5}}%
\htquantdef{Gamma944}{\BR_{944}}{\BRF{\tau^-}{2\pi^- \pi^+ \eta(\gamma \gamma) \nu_\tau\;(\text{ex.~} K^0)}}{(8.653 \pm 0.509) \cdot 10^{-5}}{8.653\cdot 10^{-5}}{0.509\cdot 10^{-5}}%
\htquantdef{Gamma945}{\BR_{945}}{\BRF{\tau^-}{\pi^- 2\pi^0 \eta \nu_\tau\;(\text{ex.~} K^0)}}{(1.939 \pm 0.378) \cdot 10^{-4}}{1.939\cdot 10^{-4}}{0.378\cdot 10^{-4}}%
\htquantdef{Gamma94by69}{\frac{\BR_{94}}{\BR_{69}}}{\frac{\BRF{\tau^-}{\pi^- K^- K^+ \pi^0 \nu_\tau}}{\BRF{\tau^-}{\pi^- \pi^+ \pi^- \pi^0 \nu_\tau\;(\text{ex.~} K^0)}}}{(0.1353 \pm 0.0405) \cdot 10^{-2}}{0.1353\cdot 10^{-2}}{0.0405\cdot 10^{-2}}%
\htquantdef{Gamma96}{\BR_{96}}{\BRF{\tau^-}{K^- K^- K^+ \nu_\tau}}{(2.169 \pm 0.800) \cdot 10^{-5}}{2.169\cdot 10^{-5}}{0.800\cdot 10^{-5}}%
\htquantdef{Gamma998}{\BR_{998}}{1 - \BR_{\text{All}}}{(0.0298 \pm 0.1026) \cdot 10^{-2}}{0.0298\cdot 10^{-2}}{0.1026\cdot 10^{-2}}%
\htquantdef{Gamma9by5}{\frac{\BR_{9}}{\BR_{5}}}{\frac{\BRF{\tau^-}{\pi^- \nu_\tau}}{\BRF{\tau^-}{e^- \bar{\nu}_e \nu_\tau}}}{0.6064 \pm 0.0032}{0.6064}{0.0032}%
\htquantdef{GammaAll}{\BR_{\text{all}}}{\BRF{\tau^-}{\text{all}}}{0.9997 \pm 0.0010}{0.9997}{0.0010}%
\htquantdef{gf}{gf}{CODATA 2018 Fermi coupling constant}{(1.16637870 \pm 0.00000060) \cdot 10^{-5}}{1.16637870\cdot 10^{-5}}{0.00000060\cdot 10^{-5}}%
\htquantdef{gmubyge_tau}{gmubyge_tau}{}{1.0018 \pm 0.0014}{1.0018}{0.0014}%
\htquantdef{gtaubyge_tau}{gtaubyge_tau}{}{1.0029 \pm 0.0014}{1.0029}{0.0014}%
\htquantdef{gtaubygemu_tau}{gtaubygemu_tau}{}{1.0020 \pm 0.0013}{1.0020}{0.0013}%
\htquantdef{gtaubygemu_tau_syst_Be_lept_perc}{gtaubygemu_tau_syst_Be_lept_perc}{}{0.089}{0.089}{0}%
\htquantdef{gtaubygemu_tau_syst_m_tau_perc}{gtaubygemu_tau_syst_m_tau_perc}{}{0.017}{0.017}{0}%
\htquantdef{gtaubygemu_tau_syst_tau_tau_perc}{gtaubygemu_tau_syst_tau_tau_perc}{}{0.086}{0.086}{0}%
\htquantdef{gtaubygemu_tau_unc_perc}{gtaubygemu_tau_unc_perc}{}{0.125}{0.125}{0}%
\htquantdef{gtaubygmu_fit}{gtaubygmu_fit}{}{0.9999 \pm 0.0014}{0.9999}{0.0014}%
\htquantdef{gtaubygmu_K}{gtaubygmu_K}{}{0.9879 \pm 0.0063}{0.9879}{0.0063}%
\htquantdef{gtaubygmu_pi}{gtaubygmu_pi}{}{0.9958 \pm 0.0026}{0.9958}{0.0026}%
\htquantdef{gtaubygmu_piK_fit}{gtaubygmu_piK_fit}{}{0.9947 \pm 0.0025}{0.9947}{0.0025}%
\htquantdef{gtaubygmu_tau}{gtaubygmu_tau}{}{1.0010 \pm 0.0014}{1.0010}{0.0014}%
\htquantdef{hbarev}{hbarev}{CODATA 2018 reduced Planck constant in eV s}{6.582\cdot 10^{-16}}{6.582\cdot 10^{-16}}{0}%
\htquantdef{KmKzsNu}{KmKzsNu}{}{7.430\cdot 10^{-4}}{7.430\cdot 10^{-4}}{0}%
\htquantdef{KmPizKzsNu}{KmPizKzsNu}{}{7.500\cdot 10^{-4}}{7.500\cdot 10^{-4}}{0}%
\htquantdef{KtoENu}{KtoENu}{}{(1.5820 \pm 0.0070) \cdot 10^{-5}}{1.5820\cdot 10^{-5}}{0.0070\cdot 10^{-5}}%
\htquantdef{KtoMuNu}{KtoMuNu}{}{0.6356 \pm 0.0011}{0.6356}{0.0011}%
\htquantdef{m_e}{m_e}{}{0.5109989461 \pm 0.0000000031}{0.5109989461}{0.0000000031}%
\htquantdef{m_K}{m_K}{}{493.677 \pm 0.016}{493.677}{0.016}%
\htquantdef{m_mu}{m_mu}{}{105.6583745 \pm 0.0000024}{105.6583745}{0.0000024}%
\htquantdef{m_pi}{m_pi}{}{139.57039 \pm 0.00018}{139.57039}{0.00018}%
\htquantdef{m_s}{m_s}{}{93.00 \pm 8.54}{93.00}{8.54}%
\htquantdef{m_tau}{m_tau}{}{(1.77686 \pm 0.00012) \cdot 10^{3}}{1.77686\cdot 10^{3}}{0.00012\cdot 10^{3}}%
\htquantdef{m_tau_unc_perc}{m_tau_unc_perc}{}{0.007}{0.007}{0}%
\htquantdef{m_W}{m_W}{}{(8.0379 \pm 0.0012) \cdot 10^{4}}{8.0379\cdot 10^{4}}{0.0012\cdot 10^{4}}%
\htquantdef{MumNuNumb}{MumNuNumb}{}{0.1739}{0.1739}{0}%
\htquantdef{phspf_mebymmu}{phspf_mebymmu}{}{0.9998129491711 \pm 0.0000000000088}{0.9998129491711}{0.0000000000088}%
\htquantdef{phspf_mebymtau}{phspf_mebymtau}{}{0.999999338359 \pm 0.000000000089}{0.999999338359}{0.000000000089}%
\htquantdef{phspf_mmubymtau}{phspf_mmubymtau}{}{0.9725600 \pm 0.0000036}{0.9725600}{0.0000036}%
\htquantdef{PimKmKpNu}{PimKmKpNu}{}{0.1435\cdot 10^{-2}}{0.1435\cdot 10^{-2}}{0}%
\htquantdef{PimKmPipNu}{PimKmPipNu}{}{0.2930\cdot 10^{-2}}{0.2930\cdot 10^{-2}}{0}%
\htquantdef{PimKzsKzlNu}{PimKzsKzlNu}{}{0.1080\cdot 10^{-2}}{0.1080\cdot 10^{-2}}{0}%
\htquantdef{PimKzsKzsNu}{PimKzsKzsNu}{}{2.330\cdot 10^{-4}}{2.330\cdot 10^{-4}}{0}%
\htquantdef{PimPimPipNu}{PimPimPipNu}{}{8.990\cdot 10^{-2}}{8.990\cdot 10^{-2}}{0}%
\htquantdef{PimPimPipPizNu}{PimPimPipPizNu}{}{4.620\cdot 10^{-2}}{4.620\cdot 10^{-2}}{0}%
\htquantdef{PimPizKzsNu}{PimPizKzsNu}{}{0.1910\cdot 10^{-2}}{0.1910\cdot 10^{-2}}{0}%
\htquantdef{PimPizNu}{PimPizNu}{}{0.2549}{0.2549}{0}%
\htquantdef{pitoENu}{pitoENu}{}{(1.2300 \pm 0.0040) \cdot 10^{-4}}{1.2300\cdot 10^{-4}}{0.0040\cdot 10^{-4}}%
\htquantdef{pitoMuNu}{pitoMuNu}{}{0.99987700 \pm 0.00000040}{0.99987700}{0.00000040}%
\htquantdef{R_tau}{R_tau}{}{3.6353 \pm 0.0081}{3.6353}{0.0081}%
\htquantdef{R_tau_leptonly}{R_tau_leptonly}{}{3.6370 \pm 0.0075}{3.6370}{0.0075}%
\htquantdef{R_tau_leptuniv}{R_tau_leptuniv}{}{3.6409 \pm 0.0070}{3.6409}{0.0070}%
\htquantdef{R_tau_s}{R_tau_s}{}{0.1645 \pm 0.0023}{0.1645}{0.0023}%
\htquantdef{R_tau_uni}{R_tau_uni}{}{3.6370 \pm 0.0075}{3.6370}{0.0075}%
\htquantdef{R_tau_uni_SM}{R_tau_uni_SM}{}{3.6409 \pm 0.0070}{3.6409}{0.0070}%
\htquantdef{R_tau_VA}{R_tau_VA}{}{3.4708 \pm 0.0078}{3.4708}{0.0078}%
\htquantdef{Rrad_mu_gamma}{Rrad_mu_gamma}{}{0.9958}{0.9958}{0}%
\htquantdef{Rrad_mu_W}{Rrad_mu_W}{}{1.00000103682 \pm 0.00000000031}{1.00000103682}{0.00000000031}%
\htquantdef{Rrad_SEW_tau_Knu}{Rrad_SEW_tau_Knu}{}{1.02010 \pm 0.00030}{1.02010}{0.00030}%
\htquantdef{Rrad_tau_gamma}{Rrad_tau_gamma}{}{0.9957}{0.9957}{0}%
\htquantdef{Rrad_tau_W}{Rrad_tau_W}{}{1.000296316 \pm 0.000000097}{1.000296316}{0.000000097}%
\htquantdef{sigmataupmy4s}{sigmataupmy4s}{}{0.9190}{0.9190}{0}%
\htquantdef{tau_K}{tau_K}{}{(1.2380 \pm 0.0020) \cdot 10^{-8}}{1.2380\cdot 10^{-8}}{0.0020\cdot 10^{-8}}%
\htquantdef{tau_mu}{tau_mu}{}{(2.196981 \pm 0.000022) \cdot 10^{-6}}{2.196981\cdot 10^{-6}}{0.000022\cdot 10^{-6}}%
\htquantdef{tau_pi}{tau_pi}{}{(2.60330 \pm 0.00050) \cdot 10^{-8}}{2.60330\cdot 10^{-8}}{0.00050\cdot 10^{-8}}%
\htquantdef{tau_tau}{tau_tau}{}{290.3 \pm 0.5}{290.3}{0.5}%
\htquantdef{tau_tau_unc_perc}{tau_tau_unc_perc}{}{0.172}{0.172}{0}%
\htquantdef{Vub}{Vub}{}{(0.3820 \pm 0.0240) \cdot 10^{-2}}{0.3820\cdot 10^{-2}}{0.0240\cdot 10^{-2}}%
\htquantdef{Vud}{Vud}{}{0.97373 \pm 0.00031}{0.97373}{0.00031}%
\htquantdef{Vus}{Vus}{}{0.2192 \pm 0.0019}{0.2192}{0.0019}%
\htquantdef{Vus_by_Vud_tauKpi}{Vus_by_Vud_tauKpi}{}{0.2295 \pm 0.0016}{0.2295}{0.0016}%
\htquantdef{Vus_by_Vud_tauKpi_dicarlo2019}{Vus_by_Vud_tauKpi_dicarlo2019}{}{0.2295 \pm 0.0016}{0.2295}{0.0016}%
\htquantdef{Vus_err_exp}{Vus_err_exp}{}{0.1575\cdot 10^{-2}}{0.1575\cdot 10^{-2}}{0}%
\htquantdef{Vus_err_exp_perc}{Vus_err_exp_perc}{}{0.7185}{0.7185}{0}%
\htquantdef{Vus_err_perc}{Vus_err_perc}{}{0.8672}{0.8672}{0}%
\htquantdef{Vus_err_th}{Vus_err_th}{}{0.0011}{0.0011}{0}%
\htquantdef{Vus_err_th_perc}{Vus_err_th_perc}{}{0.49}{0.49}{0}%
\htquantdef{Vus_mism}{Vus_mism}{}{(-0.8425 \pm 0.2360) \cdot 10^{-2}}{-0.8425\cdot 10^{-2}}{0.2360\cdot 10^{-2}}%
\htquantdef{Vus_mism_sigma}{Vus_mism_sigma}{}{-3.6}{-3.6}{0}%
\htquantdef{Vus_mism_sigma_abs}{Vus_mism_sigma_abs}{}{3.6}{3.6}{0}%
\htquantdef{Vus_tau}{Vus_tau}{}{0.2217 \pm 0.0013}{0.2217}{0.0013}%
\htquantdef{Vus_tau_dicarlo2019}{Vus_tau_dicarlo2019}{}{0.2219 \pm 0.0013}{0.2219}{0.0013}%
\htquantdef{Vus_tau_mism}{Vus_tau_mism}{}{(-0.5976 \pm 0.1866) \cdot 10^{-2}}{-0.5976\cdot 10^{-2}}{0.1866\cdot 10^{-2}}%
\htquantdef{Vus_tau_mism_dicarlo2019}{Vus_tau_mism_dicarlo2019}{}{(-0.5816 \pm 0.1852) \cdot 10^{-2}}{-0.5816\cdot 10^{-2}}{0.1852\cdot 10^{-2}}%
\htquantdef{Vus_tau_mism_sigma}{Vus_tau_mism_sigma}{}{-3.2}{-3.2}{0}%
\htquantdef{Vus_tau_mism_sigma_abs}{Vus_tau_mism_sigma_abs}{}{3.2}{3.2}{0}%
\htquantdef{Vus_tau_mism_sigma_abs_dicarlo2019}{Vus_tau_mism_sigma_abs_dicarlo2019}{}{3.1}{3.1}{0}%
\htquantdef{Vus_tau_mism_sigma_dicarlo2019}{Vus_tau_mism_sigma_dicarlo2019}{}{-3.1}{-3.1}{0}%
\htquantdef{Vus_tau2}{Vus_tau2}{}{0.2219 \pm 0.0013}{0.2219}{0.0013}%
\htquantdef{Vus_tau2_mism}{Vus_tau2_mism}{}{(-0.5976 \pm 0.1866) \cdot 10^{-2}}{-0.5976\cdot 10^{-2}}{0.1866\cdot 10^{-2}}%
\htquantdef{Vus_tau2_mism_sigma}{Vus_tau2_mism_sigma}{}{-3.2}{-3.2}{0}%
\htquantdef{Vus_tau2_mism_sigma_abs}{Vus_tau2_mism_sigma_abs}{}{3.2}{3.2}{0}%
\htquantdef{Vus_tauKnu}{Vus_tauKnu}{}{0.2226 \pm 0.0015}{0.2226}{0.0015}%
\htquantdef{Vus_tauKnu_dicarlo2019}{Vus_tauKnu_dicarlo2019}{}{0.2229 \pm 0.0014}{0.2229}{0.0014}%
\htquantdef{Vus_tauKnu_err_th_perc}{Vus_tauKnu_err_th_perc}{}{0.2445}{0.2445}{0}%
\htquantdef{Vus_tauKnu_err_th_perc_dicarlo2019}{Vus_tauKnu_err_th_perc_dicarlo2019}{}{0.1090}{0.1090}{0}%
\htquantdef{Vus_tauKnu_mism}{Vus_tauKnu_mism}{}{(-0.5064 \pm 0.1983) \cdot 10^{-2}}{-0.5064\cdot 10^{-2}}{0.1983\cdot 10^{-2}}%
\htquantdef{Vus_tauKnu_mism_dicarlo2019}{Vus_tauKnu_mism_dicarlo2019}{}{(-0.5064 \pm 0.1983) \cdot 10^{-2}}{-0.5064\cdot 10^{-2}}{0.1983\cdot 10^{-2}}%
\htquantdef{Vus_tauKnu_mism_sigma}{Vus_tauKnu_mism_sigma}{}{-2.6}{-2.6}{0}%
\htquantdef{Vus_tauKnu_mism_sigma_abs}{Vus_tauKnu_mism_sigma_abs}{}{2.6}{2.6}{0}%
\htquantdef{Vus_tauKnu_mism_sigma_abs_dicarlo2019}{Vus_tauKnu_mism_sigma_abs_dicarlo2019}{}{2.6}{2.6}{0}%
\htquantdef{Vus_tauKnu_mism_sigma_dicarlo2019}{Vus_tauKnu_mism_sigma_dicarlo2019}{}{-2.6}{-2.6}{0}%
\htquantdef{Vus_tauKpi}{Vus_tauKpi}{}{0.2234 \pm 0.0015}{0.2234}{0.0015}%
\htquantdef{Vus_tauKpi_dicarlo2019}{Vus_tauKpi_dicarlo2019}{}{0.2235 \pm 0.0015}{0.2235}{0.0015}%
\htquantdef{Vus_tauKpi_err_th_perc}{Vus_tauKpi_err_th_perc}{}{0.2347}{0.2347}{0}%
\htquantdef{Vus_tauKpi_err_th_perc_dicarlo2019}{Vus_tauKpi_err_th_perc_dicarlo2019}{}{0.2108}{0.2108}{0}%
\htquantdef{Vus_tauKpi_err_th_perc_dRrad_kmunu_by_pimunu}{Vus_tauKpi_err_th_perc_dRrad_kmunu_by_pimunu}{}{8.559\cdot 10^{-2}}{8.559\cdot 10^{-2}}{0}%
\htquantdef{Vus_tauKpi_err_th_perc_dRrad_kmunu_by_pimunu_dicarlo2019_dicarlo2019}{Vus_tauKpi_err_th_perc_dRrad_kmunu_by_pimunu_dicarlo2019_dicarlo2019}{}{7.089\cdot 10^{-2}}{7.089\cdot 10^{-2}}{0}%
\htquantdef{Vus_tauKpi_err_th_perc_dRrad_tauK_by_Kmu}{Vus_tauKpi_err_th_perc_dRrad_tauK_by_Kmu}{}{0.1090}{0.1090}{0}%
\htquantdef{Vus_tauKpi_err_th_perc_dRrad_tauK_by_Kmu_dicarlo2019}{Vus_tauKpi_err_th_perc_dRrad_tauK_by_Kmu_dicarlo2019}{}{0.1090}{0.1090}{0}%
\htquantdef{Vus_tauKpi_err_th_perc_dRrad_taupi_by_pimu}{Vus_tauKpi_err_th_perc_dRrad_taupi_by_pimu}{}{6.989\cdot 10^{-2}}{6.989\cdot 10^{-2}}{0}%
\htquantdef{Vus_tauKpi_err_th_perc_dRrad_taupi_by_pimu_dicarlo2019}{Vus_tauKpi_err_th_perc_dRrad_taupi_by_pimu_dicarlo2019}{}{6.989\cdot 10^{-2}}{6.989\cdot 10^{-2}}{0}%
\htquantdef{Vus_tauKpi_err_th_perc_f_K_by_f_pi_dicarlo2019_dicarlo2019}{Vus_tauKpi_err_th_perc_f_K_by_f_pi_dicarlo2019_dicarlo2019}{}{0.1504}{0.1504}{0}%
\htquantdef{Vus_tauKpi_err_th_perc_f_Kpm_by_f_pipm}{Vus_tauKpi_err_th_perc_f_Kpm_by_f_pipm}{}{0.1760}{0.1760}{0}%
\htquantdef{Vus_tauKpi_mism}{Vus_tauKpi_mism}{}{(-0.4227 \pm 0.2085) \cdot 10^{-2}}{-0.4227\cdot 10^{-2}}{0.2085\cdot 10^{-2}}%
\htquantdef{Vus_tauKpi_mism_dicarlo2019}{Vus_tauKpi_mism_dicarlo2019}{}{(-0.4220 \pm 0.2072) \cdot 10^{-2}}{-0.4220\cdot 10^{-2}}{0.2072\cdot 10^{-2}}%
\htquantdef{Vus_tauKpi_mism_sigma}{Vus_tauKpi_mism_sigma}{}{-2.0}{-2.0}{0}%
\htquantdef{Vus_tauKpi_mism_sigma_abs}{Vus_tauKpi_mism_sigma_abs}{}{2.0}{2.0}{0}%
\htquantdef{Vus_tauKpi_mism_sigma_abs_dicarlo2019}{Vus_tauKpi_mism_sigma_abs_dicarlo2019}{}{2.0}{2.0}{0}%
\htquantdef{Vus_tauKpi_mism_sigma_dicarlo2019}{Vus_tauKpi_mism_sigma_dicarlo2019}{}{-2.0}{-2.0}{0}%
\htquantdef{Vus_uni}{Vus_uni}{}{0.2277 \pm 0.0013}{0.2277}{0.0013}%
\htdef{couplingsCorr}{%
$\left( \frac{g_\tau}{g_e} \right)$ &   51\\
$\left( \frac{g_\mu}{g_e} \right)$ &  -50 &   49\\
$\left( \frac{g_\tau}{g_\mu} \right)_\pi$ &   23 &   25 &    2\\
$\left( \frac{g_\tau}{g_\mu} \right)_K$ &   11 &   10 &   -1 &    6\\
 & $\left( \frac{g_\tau}{g_\mu} \right)$ & $\left( \frac{g_\tau}{g_e} \right)$ & $\left( \frac{g_\mu}{g_e} \right)$ & $\left( \frac{g_\tau}{g_\mu} \right)_\pi$}%